\documentclass[a4paper,12pt,twoside]{report}
\pdfoutput=1
\usepackage[english]{babel}

\usepackage{PhD} 

\usepackage[centertags]{amsmath}
\usepackage{graphicx}
\usepackage{epigraph,amsfonts,amssymb,amsthm,newlfont,textcomp,array,cite,dsfont,feynmf,microtype}
\usepackage{color,tabularx,booktabs,lscape,afterpage,longtable,dcolumn,cite,truncate}
\usepackage[font={footnotesize}]{caption}
\usepackage[perpage,marginal]{footmisc}
\usepackage[charter]{mathdesign}
\usepackage{wrapfig,floatflt,hvfloat,underscore,tikz}

\definecolor{DarkBlue}{rgb}{0,0,0.5}
\DefineFNsymbols{numbers}{123456789{10}{11}{12}{13}{14}{15}{16}{17}{18}{19}{20}}
\setfnsymbol{numbers}
\DeclareSymbolFont{operators}{OT1}{cmss}{m}{n}
\SetMathAlphabet{\mathcal}{normal}{OT1}{pzc}{m}{it}

\newcommand{\fint}{\text{\raisebox{4.5pt}{\rotatebox{-60}{\footnotesize|}}}\kern-12pt\int}
\renewcommand{\iint}{\int\kern-7pt\int}
\renewcommand{\iiint}{\int\kern-7pt\int\kern-7pt\int}

\newcommand{\Ast}{\text{\raisebox{-0.5pt}{\scalebox{1.5}{$\ast$}}}}

\newcommand{\champagne}{\kern-1.1ex
\raisebox{-0.5pt}{
\resizebox{1.8ex}{1.68ex}{
\begin{tikzpicture}
\draw[thick,rounded corners=3pt] (0,0) -- (0,0.2) -- (-0.12,0.3); %
\draw[thick,rounded corners=3pt] (0,0) -- (0,0.2) -- (0.12,0.3); %
\end{tikzpicture}}}\kern-1.5pt}

\newcommand{\waterfalls}{\kern-1.1ex
\raisebox{-0.5pt}{
\resizebox{1.8ex}{1.68ex}{
\begin{tikzpicture}
\draw[thick,rounded corners=2pt] (-0.04,0) -- (-0.04,0.22) -- (-0.12,0.3); %
\draw[thick,rounded corners=2pt] (0.04,0) -- (0.04,0.22) -- (0.12,0.3); %
\end{tikzpicture}}}\kern-1pt}

\usepackage[colorlinks,plainpages=false,linkcolor=black,urlcolor=blue,citecolor=black,pdfpagemode=UseNone,pdfstartview=FitBH,
            pdftitle={Angle-Resolved Photoelectron Spectroscopy Studies of the Many-Body Effects in the Electronic
            Structure of High-Tc Cuprates},
            pdfsubject={ARPES}, pdfauthor={Dmytro S. Inosov}, pdfcreator={Dmytro S. Inosov},
            pdfkeywords={ARPES, high-Tc, HTSC, superconductivity, superconductors, cuprates}]{hyperref}

\begin{document} \phd
\nolistoffigures \nolistoftables 
\graphicspath{{Images/}} \sloppy
\renewcommand{\thefootnote}{\fnsymbol{footnote}} 
\newcolumntype{.}{D{.}{.}{-1}} 
\title{Angle-Resolved Photoelectron Spectroscopy Studies of the Many-Body Effects in the Electronic Structure of
High-$T_\textup{c}$ Cuprates}

\author{Dmytro S. Inosov}

\email{\href{mailto:d.inosov@ifw-dresden.de}{d.inosov@ifw-dresden.de}}

\copyrightyear{2005--2008} \submitdate{\today}

\examiner{Prof. Dr. H.-H. Klau\ss}

\supervisor{Prof. Dr. Bernd Büchner}

\firstreader{Prof. Dr. Ilya Eremin}

\secondreader{Dr. Luc Patthey}

\university{Leibnitz~Institut~für~Festkörper-~und~Werkstoffforschung\bigskip\\Fakultät Mathematik und
Naturwissenschaften\\der Technischen Universität Dresden}

\address{Dresden, Germany}

\dept{Mathematics and Natural Sciences, TU Dresden}
\beforepreface

\setlength{\epigraphwidth}{0.618\textwidth}
\renewcommand{\textflush}{flushepinormal}

\prefacesection{Introduction} \fancyhead[LO]{Introduction} \thispagestyle{fancy}

\epigraph{Physical reality remains so mysterious even to physicists because of the extreme improbability that it was
constructed to be understood by the human mind.}{\vspace{0.3em}Steven Weinberg}

For more than twenty years now the phenomenon of the high-temperature superconductivity \cite{BednorzMueller86} has
been
standing fast against all efforts to find a suitable theory that would describe the behavior of the doped copper
oxides
in all its diversity. On the one hand, the layered cuprates exhibit a number of anomalous electronic properties in
both
superconducting and normal states, which seem to slip out of the scope of any conventional theory, therefore making
the
understanding of these systems unprecedentedly difficult. On the other hand, high-temperature superconductors can be
called simple in many respects. First of all, these are layered materials with quasi-two-dimensional electronic
structure, which simplifies the treatment of experimental results. Second, some of these materials, like the Bi-based
family of cuprates, are easily cleavable and have already become model materials for surface-sensitive experiments,
such
as photoelectron spectroscopy and tunneling electron microscopy. Third, they become metallic at sufficient doping,
which
is also essential for the above-mentioned experimental techniques. Such a unique combination of interesting physical
properties with experimental accessibility has been stimulating an unceasing interest to the high-temperature
superconductors since their discovery in 1986. Since then, the whole branches of condensed matter physics and
experimental research have been pushed forward by the endeavor to understand the physics of unconventional
superconductors and optimize their critical temperature. Among them is the angle-resolved photoelectron spectroscopy
(ARPES), which is the topic of the present work.

In spite of the failures to find an ultimate theory of unconventional superconductivity, after many years of research
the scientific community possesses a considerable store of theoretical knowledge about the problem. Over time, the
focus
is gradually shifted from finding a theoretical description of an experimentally observed phenomenon to distinguishing
between multiple models that offer comparably reasonable descriptions. From the point of view of an experimentalist,
this means that any qualitative understanding of an experimental observation would no longer suffice. Instead, the
emphasis in the experimental research should be shifted to accurate quantification of observations, which becomes
possible only if the results available from all the available experimental methods are connected together by the
theoretical glue. Among the methods that are to be unified, ARPES plays a central role. The reason for this is that it
gives access to the single-particle excitation spectrum of the material as a function of both momentum and energy with
very high resolution. Other experimental techniques, such as inelastic neutron scattering (INS), Raman spectroscopy,
or the newly established Fourier-transform scanning tunneling spectroscopy (FT-STS) probe more complicated two-particle
spectra of the electrons and up to now can not achieve the momentum resolution comparable with that of ARPES. Such reasoning
serves as the motivation for the present work, in which some steps are done towards understanding the anomalous
effects observed in the single-particle excitation spectra of cuprates and relating the ARPES technique to other experimental
methods.

\textbf{Structure of this thesis}. In the \textit{first chapter}, some aspects of the condensed matter physics and
many-body quantum theory are summarized. In particular, the notions of the self-energy, one- and two-particle Green's
functions, and the Bardeen-Cooper-Schrieffer theory of conventional superconductivity are briefly reviewed. The scope of
this chapter is kept to the minimum necessary for understanding the main results presented in the subsequent chapters.
Where possible, references to the literature containing more extensive information are provided. In spite of
conciseness, all attempts were made to keep this chapter in line with the strict theoretical constructions found in
modern literature.

The \textit{second chapter} is an introduction to the angle-resolved photoemission. It consists of two sections covering
the theory of photoemission and experimental aspects of ARPES respectively. In the first part, the so-called one- and
three-step models of photoemission are discussed in relation to the many-particle systems of increased complexity:
non-interacting, interacting, and superconducting. In the experimental part, most attention is paid to the specifics of
the experimental setup, experimental geometry, and synchrotron light sources used in the present work. This chapter also
contains a description of the sample preparation techniques used.

The \textit{third chapter} is focused on the electronic properties of Bi$_2$Sr$_2$CaCu$_2$O$_{8+\delta}$\,---\,the
superconducting cuprate most studied by surface sensitive methods. The chapter starts from the description of its
crystal structure and its place among other families of cuprates. Then a report of the recent progress in understanding
the electronic structure of this material is presented, focusing mainly on the understanding of the many-body effects
(renormalization) and their manifestation in the ARPES spectra. The main result of this chapter is a model of the
Green's function that is later used in chapter 5 for calculating the two-particle excitation spectrum.

The \textit{fourth chapter} is devoted to the matrix element effects in the photoemission spectra of cuprates. After a
general introduction to the problem, is focuses on the recently discovered anomalous behavior of the ARPES spectra that
partially originates from the momentum-dependent photoemission matrix element. The momentum- and excitation energy
dependence of the anomalous high-energy dispersion, termed ``waterfalls'', is covered in full detail. Understanding the
role of the matrix element effects in this phenomenon proves crucial, as they obstruct the view of the underlying
excitation spectrum that is of indisputable interest.

Finally, the \textit{fifth chapter} is devoted to the relation of ARPES with other experimental methods, with the
special focus on the INS spectroscopy. For the optimally doped bilayer Bi-based cuprate, the renormalized two-particle
correlation function in the superconducting state is calculated from ARPES data within an itinerant model based on the
random phase approximation (RPA). The results are compared with the experimental INS data on
Bi$_2$Sr$_2$CaCu$_2$O$_{8+\delta}$ and YBa$_2$Cu$_3$O$_{6.85}$. The calculation is based on numerical models for the
normal and anomalous Green's functions fitted to the experimental single-particle spectra. The renormalization is
taken
into account both in the single-particle Green's function by means of the self-energy, and in the two-particle
correlation function by RPA. In the last sections of the same chapter, two other applications of the same approach are
briefly sketched:
the relation of ARPES to FT-STS, and the nesting properties of Fermi surfaces in two-dimensional charge density wave
systems.

\cleardoublepage

\smallskip

\afterpreface

\fancyhead[LO]{\truncate[...]{0.95\textwidth}{\nouppercase\rightmark}}
\fancyhead[RE]{\truncate[...]{0.95\textwidth}{\hfill\nouppercase\leftmark}}


\chapter{Excerpts From the Many-Body Quantum Theory}\label{Chap:Theory}

\section{The many-body system}

\subsection{The $N$-particle quantum states}

The many-body quantum theory \cite{BruusFlensberg81, Mahan00, Marder00, FetterWalecka71, AtlandSimons06, Matuck92,
DoniachSondheimer99, AbrikosovGorkov63} rests upon the simple mathematical fact that if
$\{\psi_\nu(\mathbf{r})\}\!=\!\{|\kern1pt\nu\rangle\}$ is a basis set in the single-particle state space $\Psi^1$, then
the set of all possible products
$\bigl\{\psi_{\nu_1}(\mathbf{r}_1)\,\psi_{\nu_2}(\mathbf{r}_2)\ldots\psi_{\nu_N}(\mathbf{r}_N)\bigr\}$ will be a basis
in the $N$-particle state space $\Psi^N$, i.e. $\Psi^N = \Psi^1 \otimes \ldots \otimes \Psi^1$. The single-particle
basis set $\{\psi_\nu(\mathbf{r})\}$ is usually chosen to be the proper basis of the single-particle Hamiltonian, and
any many-particle state is represented as a linear combination of the form
$\,\sum_{\,\nu_1,\,\ldots\,,\nu_N}A_{\,\nu_1,\,\ldots\,,\nu_N}\psi_{\nu_1}(\mathbf{r})\,\psi_{\nu_2}(\mathbf{r})\ldots\psi_{\nu_N}(\mathbf{r})$.
This formal expansion is actually the only relationship between a many-particle system and single particles. For
convenience, the many-particle basis set is symmetrized (for bosons) or anti-symmetrized (for fermions), being replaced
by a set of Slater determinants
$|\kern1pt\nu\rangle=\psi_\nu(\mathbf{r}_1,\,\ldots\,,\mathbf{r}_N)=\displaystyle\sum\nolimits_{\,p\in
S_N}\textstyle\Bigl(\prod_{j=1}^N\psi_{\nu_j}(\mathbf{r}_{p_j})\Bigr)\!\cdot\!\Bigl\{{}^{1,\phantom{\mathrm{sign}(p)}
\text{bosons,}}_{\mathrm{sign}(p),\phantom{1} \text{fermions.}}$ Here $S_N$ is the group of $N$-element permutations and
$\nu=\{\nu_1,\,\ldots\,,\nu_N\}\in S_N$. In this new basis the coefficients $B_\nu$ in the expansion of any
many-particle state $\psi(\mathbf{r}_1,\,\ldots\,,\mathbf{r}_N)=\sum_\nu B_\nu |\kern1pt\nu\rangle$ obtain their natural
(fermionic or bosonic) symmetry with respect to the permutation of indices. Unfortunately, the possibility to consider
the interacting many-particle system as a combination of single particles (as dictated by the reality of the experiment)
is achieved here at the expense of the new basis being no longer a proper basis of the many-particle Hamiltonian.

Each element in the $N$-particle basis set is uniquely defined by specifying the indices of the single-particle states
that enter the Slater determinant. This allows one to simplify the notation by writing the basis elements in the
so-called
occupation number representation: $|n_1,\,n_2,\,\ldots\,\rangle$, where $\sum_j n_j=N.$ The complete linear hull of
all
these basis vectors for all $N$ is called the Fock space.

\subsection{Second quantization}

Having the Fock space defined, one can introduce the creation and annihilation operators acting on it. These are
denoted
as $\hat{b}^\dag_i$ and $\hat{b}_i$ respectively for bosons and as $\hat{c}^\dag_i$ and $\hat{c}_i$ for fermions. By
definition,
\begin{subequations}
\begin{align}
\hat{b}^\dag_i |\ldots\,,\,n_i\,,\,\ldots\,\rangle &= |\ldots\,,\,n_i+1\,,\,\ldots\,\rangle\,\sqrt{n_i+1}\\
\hat{b}_i |\ldots\,,\,n_i\,,\,\ldots\,\rangle &= |\ldots\,,\,n_i-1\,,\,\ldots\,\rangle\,\sqrt{\vphantom{1}n_i}\\
\hat{c}^\dag_i |\ldots\,,\,n_i\,,\,\ldots\,\rangle &=
|\ldots\,,\,n_i+1\,,\,\ldots\,\rangle\,(1-n_i)\,(-1)^{\sum_{j<i}\,n_j}\\
\hat{c}_i |\ldots\,,\,n_i\,,\,\ldots\,\rangle &= |\ldots\,,\,n_i-1\,,\,\ldots\,\rangle\,n_i\,(-1)^{\sum_{j<i}\,n_j}
\end{align}
\end{subequations}
Both pairs of creation and annihilation operators are mutually Hermitian conjugate with the following
properties:\footnote{The $\hat{c}^\dag$ and $\hat{c}$ operators form a Banach algebra with a norm-preserving
involution
operator $\dag$, making it a $\textrm{C}^\ast$-algebra \cite{HagenRoch01}.}
\begin{subequations}
\begin{align}
[\hat{b}^\dag_i, \hat{b}^\dag_j] &= 0\text{,} &[\hat{b}_i, \hat{b}_j] &= 0\text{,} &[\hat{b}_i, \hat{b}^\dag_j] &=
\delta_{ij}\text{\,;}&&\\
\{\hat{c}^\dag_i, \hat{c}^\dag_j\} &= 0\text{,} &\{\hat{c}_i, \hat{c}_j\} &= 0\text{,} &\{\hat{c}_i,
\hat{c}^\dag_j\} &= \delta_{ij}{,} & (\hat{c}^\dag_i)^2 = (\hat{c}_i)^2 = 0\text{\,.}
\end{align}
\end{subequations} From now on, for simplicity of notation we will denote by $\hat{a}^\dag$ either a boson operator
$\hat{b}^\dag$ or a fermion operator $\hat{c}^\dag$. In this notation a basis state $|\kern1pt\nu\rangle$ will be
written as $\hat{a}_{\nu_1}^\dag\ldots\,\hat{a}_{\nu_N}^\dag|0\rangle$.

A special case of the creation operator is the so-called \textit{quantum field operator}
$\kern.5pt\hat{\kern-2.2pt\mathit{\Psi}}^\dag(\mathbf{r})$, which creates a delta-function $|\mathbf{r}\rangle$ at
point
$\mathbf{r}$.

As a rule, every term entering the Hamiltonian of a physical system is either a local single-particle operator
$\hat{T}_i= \hat{T}(\mathbf{r}_i,\nabla_{\mathbf{r}_i})$, such as the kinetic energy operator
$-\frac{\hslash^2}{2m}\nabla^2_{\mathbf{r}_i}$, or a two-particle operator $\hat{V}_{ij}=
\hat{V}(\mathbf{r}_i,\,\mathbf{r}_j,\nabla_{\mathbf{r}_i},\nabla_{\mathbf{r}_j})$, such as the Coulomb interaction
$V(\mathbf{r}_i-\mathbf{r}_j)=\frac{e^2}{4\piup\varepsilon_0}\frac{1}{|\mathbf{r}_i-\mathbf{r}_j|}$. In second
quantization notation, such operators can be rewritten as:\footnote{The proof is straightforward and can be found in
Ref. \citenum{BruusFlensberg81}, p.\,14.}
\begin{align}
\hat{T}_i &= \sum_{\,j} T_{ij}\,\hat{a}^\dag_i\hat{a}_j=\sum_{j}
\raisebox{-4.2ex}{\hbox{\scalebox{0.5}{
\begin{fmffile}{fynm1}
\begin{fmfgraph*}(100,70)
\fmfleft{l}
\fmfright{r}
\fmftop{t}
\fmf{fermion,label=\scalebox{1.5}{$\kern-5ex \smash{j}$},label.side=left}{l,m}
\fmf{fermion,label=\scalebox{1.5}{$\kern 5ex \smash{i}$},label.side=left}{m,r}
\fmf{dashes,label=\scalebox{1.5}{$\kern-.3ex T\!\smash{_{ij}}$}}{m,t}
\fmffreeze\fmfshift{0pt,-6pt}{m}
\fmfv{decor.shape=pentagram,decor.filled=30}{t}
\fmfdot{m}
\end{fmfgraph*}
\end{fmffile}}}}\\
\hat{V}_{ij} &= \text{\scalebox{0.8}{$\frac{1}{2}$}}\sum_{k,\,l}
V_{ijkl}\,\hat{a}^\dag_i\hat{a}^\dag_j\hat{a}_l\kern1pt\hat{a}_k
=\text{\scalebox{0.8}{$\frac{1}{2}$}}\sum_{\,k,\,l}\kern-1.5ex
\vcenter{\hbox{\scalebox{0.5}{
\begin{fmffile}{fynm2}
\begin{fmfgraph*}(100,70)
\fmfleft{l1,l2}
\fmfright{r1,r2}
\fmf{fermion}{l1,m1,r1}
\fmf{fermion}{l2,m2,r2}
\fmf{photon,label=\scalebox{1.5}{$V\!\smash{_{ijkl}}$}}{m1,m2}
\fmflabel{\scalebox{1.5}{$l\kern-1ex$}}{l1}
\fmflabel{\scalebox{1.5}{$\kern-1ex j$}}{r1}
\fmflabel{\scalebox{1.5}{$k\kern-1ex$}}{l2}
\fmflabel{\scalebox{1.5}{$\kern-1ex i$}}{r2}
\fmfdot{m1,m2}
\end{fmfgraph*}
\end{fmffile}}}}
\end{align}
\vspace{-1.7ex}\\* The order of indices in the coefficients is chosen in such a way that $V_{ijkl}$ corresponds to a
transition from the initial state $|kl\rangle$ to the final state $|ij\rangle$. Explicit expansions of some commonly
used operators in second quantization are collected in table \ref{Table:Operators}.

\begin{table}[t]
\begin{tabular}{ll}
\toprule\addlinespace[0.8pt]
Electromagnetic
field:&$\hat{\mathbf{A}}(\mathbf{r},t)=\kern-1pt\sqrt{\frac{\hslash}{2\varepsilon_0\omega_\mathbf{k}}}\kern1pt\text{\scalebox{0.8}{$\displaystyle\int$}}\!\frac{\mathrm{d}\mathbf{k}}{(2\piup)^d}\kern-3pt\sum\limits_{^{\lambda=1,\kern1pt2}}\Bigl(\hat{a}^{\phantom{\dag}}_{\mathbf{k},\lambda}\mathrm{e}^{\mathrm{i}\kern1pt(\mathbf{k}\cdot\mathbf{r}-\omega_\mathbf{k}t)}+\hat{a}^\dag_{\mathbf{k},\lambda}\mathrm{e}^{-\mathrm{i}\kern1pt(\mathbf{k}\cdot\mathbf{r}-\omega_\mathbf{k}t)}\Bigr)\mathbf{p}_\lambda$,\vspace{-.5em}\hfill\\
&where $\mathbf{p}_\lambda$ is the polarization vector.\\
Kinetic energy
operator:&$\hat{T}=\text{\scalebox{0.7}{$\displaystyle\int$}}\!\frac{\mathrm{d}\mathbf{k}}{(2\piup)^d}\frac{\hslash^2k^2}{2m}\sum\limits_\sigma\hat{a}^\dag_{\textbf{k}\sigma}\hat{a}^{\phantom{\dag}}_{\textbf{k}\sigma}$\\
Spin
operator:&$\hat{\mathbf{S}}=\frac{\hslash}{2}\sum\limits_\mu\bigl([\hat{c}^\dag_{\mu\downarrow}\hat{c}^{\phantom{\dag}}_{\mu\uparrow}+\hat{c}^\dag_{\mu\uparrow}\hat{c}^{\phantom{\dag}}_{\mu\downarrow}],\,\mathrm{i}[\hat{c}^\dag_{\mu\downarrow}\hat{c}^{\phantom{\dag}}_{\mu\uparrow}-\hat{c}^\dag_{\mu\uparrow}\hat{c}^{\phantom{\dag}}_{\mu\downarrow}],\,[\hat{c}^\dag_{\mu\uparrow}\hat{c}^{\phantom{\dag}}_{\mu\uparrow}-\hat{c}^\dag_{\mu\downarrow}\hat{c}^{\phantom{\dag}}_{\mu\downarrow}]\bigr)$\\
Coulomb
interaction:&$\hat{V}=\frac{1}{2}\text{\scalebox{0.7}{$\displaystyle\iiint$}}\frac{\mathrm{d}\mathbf{k}_1\mathrm{d}\mathbf{k}_2\mathrm{d}\mathbf{q}}{(2\piup)^{3d}}\,\frac{4\piup
e^2}{q^2}\!\sum\limits_{\sigma_1\sigma_2}\hat{a}^\dag_{\mathbf{k}_1+\mathbf{q}\,\sigma_1}\hat{a}^\dag_{\mathbf{k}_2-\mathbf{q}\,\sigma_2}\hat{a}^{\phantom{\dag}}_{\mathbf{k}_2\,\sigma_2}\hat{a}^{\phantom{\dag}}_{\mathbf{k}_1\,\sigma_1}$\vspace{-0.3ex}\\
Charge density
operator:&$\hat{\rho}_\sigma(\mathbf{q})=\text{\scalebox{0.7}{$\displaystyle\int$}}\frac{\mathrm{d}\mathbf{k}}{(2\piup)^d}\,\hat{a}^\dag_{\mathbf{k}\,\sigma}\hat{a}^{\phantom{\dag}}_{\mathbf{k}+\mathbf{q}\,\sigma}$\\
Spin density
operator:&$\hat{S}_\sigma(\mathbf{q})=\text{\scalebox{0.7}{$\displaystyle\int$}}\frac{\mathrm{d}\mathbf{k}}{(2\piup)^d}\,\hat{a}^\dag_{\mathbf{k}\,\sigma}\hat{a}^{\phantom{\dag}}_{\mathbf{k}+\mathbf{q};\,-\sigma}$\vspace{0.5ex}\\
\bottomrule
\end{tabular}
\caption{Several commonly used operators in second quantization representation.} \label{Table:Operators}
\end{table}

\subsection{Non-interacting electron gas}

The simplest possible generalization of a single-electron problem is the non-interacting electron gas within the
so-called \textsl{jellium model}, which means that the external periodic potential is replaced by its average (a
constant) and is therefore neglected. The word `non-interacting' here means that no particle-particle interaction
terms
enter the Hamiltonian:
\begin{equation}
{\hat{\mathcal{H}\,}\kern-2pt}_\text{jel}=\int\!\frac{\mathrm{d}\mathbf{k}}{(2\piup)^d}\frac{\hslash^2k^2}{2m}\sum\limits_\sigma\hat{c}^\dag_{\textbf{k}\sigma}\hat{c}^{\phantom{\dag}}_{\textbf{k}\sigma}\text{,}
\label{Eq:JelliumHamiltonian}
\end{equation}
but the electrons still feel each other through Pauli exclusion principle and are therefore not independent. The
ground
state of such a system is a \textsl{Fermi sphere}:
$\hat{c}^\dag_{\mathbf{k}_{N/2}\uparrow}\hat{c}^\dag_{\mathbf{k}_{N/2}\downarrow}\ldots\,\hat{c}^\dag_{\mathbf{k}_1\uparrow}\hat{c}^\dag_{\mathbf{k}_1\downarrow}|0\rangle$.
At zero temperature it consists of plane waves filling all states with $k\leq
k_\text{F}=\!\sqrt{2m\epsilon_\text{F}}/\hslash$, $\epsilon_\text{F}$ being the energy of the topmost occupied state
(\textsl{Fermi energy}). At finite temperatures, the occupation number is given by the Fermi-Dirac distribution
$n_\text{F}(\epsilon)=1/[\mathrm{e}^{\beta(\epsilon-\mu)}+1]$, where $\beta=1/k_\text{B}T$.

The same problem can be also considered in the periodic potential, with the only difference that the natural basis in
this case is a set of Bloch waves. But in case of localized electrons (tightly bound to their parent atom), it is
reasonable however to change the basis to localized wave functions that would approach atomic orbitals of individual
atoms in the \textit{atomic limit} \cite[p.\,194]{Marder00}.

Wannier functions \cite{Wannier37} centered at lattice sites $\mathbf{R}$ are defined to be
\begin{equation}
w_\mathbf{R}^\nu(\mathbf{r}) = N^{-1/2} \sum_\mathbf{k}
\psi_\mathbf{k}^\nu(\mathbf{r})\,\mathrm{e}^{-\mathrm{i}\mathbf{k}\mathbf{R}}\text{,}
\label{Eq:WannierBasis}
\end{equation}
where index $\mathbf{k}$ enumerates the states within some energy band, and $\nu$ incorporates the rest of quantum
numbers, thus enumerating different energy bands. The Wannier functions for all $\mathbf{k}$ and $\nu$ form an
orthonormal basis, and the Bloch functions can be recovered from them by computing
\begin{equation}
\psi_\mathbf{k}^\nu(\mathbf{r}) = N^{-1/2} \sum_\mathbf{R}
w_\mathbf{R}^\nu(\mathbf{r})\,\mathrm{e}^{\mathrm{i}\mathbf{k}\mathbf{R}}\text{.}
\label{Eq:BlochFromWannier}
\end{equation}
As each Bloch function is defined only up to an overall phase factor, the Wannier functions can be rewritten as
\begin{equation}
w_\mathbf{R}^\nu(\mathbf{r}) = N^{-1/2} \sum_\mathbf{k}
\psi_\mathbf{k}^\nu(\mathbf{r})\,\mathrm{e}^{-\mathrm{i}\mathbf{k}\mathbf{R}+\mathrm{i}\varphi(\mathbf{k})}\text{,}
\end{equation}
where $\varphi(\mathbf{k})$ is an arbitrary real function.

By manipulating $\varphi(\mathbf{k})$, one can try to optimize the Wannier functions so that they drop off as fast as
possible when $\mathbf{r}$ moves away from $\mathbf{R}$. Such localization is possible whenever there is an energy gap
separating the band from any other one. When two bands have a point at which they are degenerate, it is not possible
to
construct localized wave functions out of the Bloch states \cite[p.\,197]{Marder00}. For an insulator, the localized
Wannier functions are approaching the atomic orbitals, which makes it possible in many applications to approximate the
Wannier functions $w_\mathbf{R}^\nu(\mathbf{r})$ by the corresponding atomic orbitals
$\phi^\nu(\mathbf{r}-\mathbf{R})$.

Eq. (\ref{Eq:BlochFromWannier}), restricted to the $\nu$-th band, can be rewritten in the second quantization as
\begin{equation}
\hat{c}^\dag_\mathbf{k} = N^{-1/2} \sum_\mathbf{R}
\hat{c}^\dag_\mathbf{R}\,\mathrm{e}^{\mathrm{i}\mathbf{k}\mathbf{R}}\text{,}
\label{Eq:TightBindingBasisSecQuant}
\end{equation}
where the operators $\hat{c}^\dag_{\mathstrut\mathbf{k}}$ and $\hat{c}^\dag_{\mathstrut\mathbf{R}}$ create an electron
in the states $\psi_{\mathstrut\mathbf{k}}^\nu(\mathbf{r})$ and $w_\mathbf{R}^\nu(\mathbf{r})$ respectively.

By analogy with (\ref{Eq:JelliumHamiltonian}) the system's Hamiltonian will be
\begin{equation}\label{Eq:TightBindingHamiltonian}
{\hat{\mathcal{H}\,}\kern-2pt}_0=\int\!\frac{\mathrm{d}\mathbf{k}}{(2\piup)^d}\,\epsilon_\mathbf{k}\sum\limits_\sigma\hat{c}^\dag_{\textbf{k}\sigma}\hat{c}^{\phantom{\dag}}_{\textbf{k}\sigma}\text{,}
\end{equation}
where the matrix element $\epsilon_\mathbf{k}=\langle\mathbf{k}|{\hat{\mathcal{H}\,}\kern-2pt}|\mathbf{k}\rangle$ is
the
energy of the state $|\mathbf{k}\rangle$. By substitution of (\ref{Eq:TightBindingBasisSecQuant}) it can be rewritten
as
\begin{equation}
\epsilon_\mathbf{k}=N^{-1}\sum_\mathbf{R_1R_2}\langle\mathbf{R}_1|{\hat{\mathcal{H}\,}\kern-2pt}|\mathbf{R}_2\rangle\,\mathrm{e}^{\mathrm{i}\mathbf{k}(\mathbf{R}_1-\mathbf{R}_2)}\text{.}
\label{Eq:TightBindingDispersionFull}
\end{equation}
If the Wannier functions are localized around their parent atoms, the sum (\ref{Eq:TightBindingDispersionFull}) can be
restricted to several nearest neighbors only, which is called the \textit{tight binding approximation}:
\begin{equation}
\epsilon_\mathbf{k}\approx \sum_{\langle ij
\rangle}t_{ij}\,\mathrm{e}^{\mathrm{i}\mathbf{k}(\mathbf{R}_i-\mathbf{R}_j)}\text{,}
\label{Eq:TightBindingDispersion}
\end{equation}
where $\langle ij \rangle$ indicates a sum over distinct nearest neighbor pairs, and $t_{ij}=
\langle\mathbf{R}_i|{\hat{\mathcal{H}\,}\kern-2pt}|\mathbf{R}_j\rangle$ are the so-called tight binding parameters
that
characterize the overlap of the Wannier functions on different sites and are often called \textit{hopping
integrals}.\footnote{According to the physical folklore, the electron ``hops'' from site $i$ to site $j$. This notion
is
somewhat misleading, because ``hopping'' suggests that there is some time-dependence in the electron's behavior which
contradicts the fact that we are dealing with a time-independent Schrödinger equation.}

\subsection{Interacting electron gas}

The general form of the Hamiltonian for the interacting electron gas in the tight binding model is
\begin{equation}
{\hat{\mathcal{H}\,}\kern-2pt}=\sum_{ij\sigma}T_{ij}\hat{c}^\dag_{i\sigma}\hat{c}^{\phantom{\dag}}_{j\sigma}+\text{\scalebox{0.8}{$\,\frac{1}{2}\,$}}{\sum_{\substack{ijkl\\\sigma_1\sigma_2}}}^{\kern1pt\textstyle\prime}
V_{ijkl}\,\hat{c}^\dag_{i\sigma_1}\hat{c}^\dag_{j\sigma_2}\hat{c}^{\phantom{\dag}}_{l\sigma_2}\hat{c}^{\phantom{\dag}}_{k\sigma_1}
\end{equation}
The first (bilinear) term includes the kinetic energy operator and interaction with the external potential, while the
second term represents the electron-electron interaction. The prime over the summation sign indicates that no equal
electronic states should appear in the creation and annihilation parts of each summation term (in order to exclude the
bilinear terms already included in the first sum).

A simplification to this Hamiltonian was proposed by Hubbard in 1963 \cite{Hubbard63} to study the electronic
correlations leading to magnetism and was solved exactly in one dimension by Lieb and Wu in 1968 \cite{LiebWu68}. In
the
so-called Hubbard model, only several nearest-neighbor bilinear terms are usually considered and the Coulomb
interaction
between electrons on different sites is neglected:
\begin{equation}
{\hat{\mathcal{H}\,}\kern-2pt}_\text{Hub}=\sum_{\langle
ij\rangle\sigma}t_{ij}\hat{c}^\dag_{i\sigma}\hat{c}^{\phantom{\dag}}_{j\sigma}+U\sum_i
n_{i\uparrow}n_{i\downarrow}\text{.}
\end{equation}
Here $U$ is the so-called \textsl{on-site repulsion} (energy lost if two electrons with opposite spins occupy the same
orbital), also known as \textsl{Hubbard $U$}, and
$n_{i\sigma}=\hat{c}^\dag_{i\sigma}\hat{c}^{\phantom{\dag}}_{i\sigma}$
is the occupation number of the state $|i\sigma\rangle$. Thus, the Hubbard model is the simplest possible extension of
the tight-binding model that introduces an energy penalty $U$ for any atomic site occupied by more than one electron.
The two limiting cases of the Hubbard model are $U=0$ (\textsl{nearest neighbor tight-binding model}) and $t_{ij}=0$
(\textsl{atomic limit}).

If one considers the Hubbard model with a single (most significant) hopping term and treats this term as a
perturbation
(assuming $t\ll U$) \cite{Anderson59}, in the second order of the perturbation theory the Hamiltonian will take the
form, which is referred to as the $t$-$J$ \textit{model}:
\begin{equation}\label{Eq:tJHamiltonian}
{\hat{\mathcal{H}\,}\kern-2pt}_\text{$t$-$J$}=-t\sum_{\langle
ij\rangle\sigma}\hat{c}^\dag_{i\sigma}\hat{c}^{\phantom{\dag}}_{j\sigma}+J_\text{eff}\sum_{\langle
ij\rangle}\hat{\mathbf{S}}_i\hat{\mathbf{S}}_j\text{,}
\end{equation}
where $J_\text{eff}=4t^2/U$ is the effective interatomic exchange interaction constant.

In the particular case of half-filling, the $t$-$J$ Hamiltonian does no longer explicitly contain $t$, and only the
exchange interaction term is preserved:
\begin{equation}\label{Eq:HeisenbergHamiltonian}
{\hat{\mathcal{H}\,}\kern-2pt}_\text{ex}=J_\text{eff}\sum_{\langle
ij\rangle}\bigl(\hat{\mathbf{S}}_i\hat{\mathbf{S}}_j-\textstyle\frac{1}{4}\displaystyle\bigr)\text{.}
\end{equation}
This Hamiltonian was first considered by Heisenberg \cite{Heisenberg28} and describes the so-called \textit{localized
electron model}, appropriate for the description of magnetic insulators. We will return to the discussion of this
model
in \S\ref{SubSec:LocalizedItinerant}, where it will be compared with the alternative \textit{itinerant electron
model}.

\section{The concept of the Green's function}

\subsection{Single-particle Green's function}\label{SubSec:SingleParticleGF}

Consider the following Schrödinger equation, where $\hat{V}$ is a perturbation:
\begin{equation}
[E-{\hat{\mathcal{H}\,}\kern-2pt}_0-\hat{V}]\psi=0\text{.}
\label{Eq:GreensSchroedinger}
\end{equation}
By definition, the unperturbed Green's operator (or Green's function) of this equation is
\begin{equation}
\hat{G}_0=[E-{\hat{\mathcal{H}\,}\kern-2pt}_0]^{-1}\text{.}
\end{equation}
The solution to the Schrödinger equation then satisfies the following relationship:
\begin{equation}
\psi=\psi_0+\hat{G}_0\hat{V}\psi\text{, given $[E-{\hat{\mathcal{H}\,}\kern-2pt}_0]\psi_0=0$.}
\label{Eq:DysonFunctions}
\end{equation}
This allows solving the Eq. (\ref{Eq:GreensSchroedinger}) iteratively, given that the following limit exists:
\begin{equation}
\psi=\lim_{n\rightarrow\infty}\psi_n\text{, where~}\psi_n=\psi_{n-1}+\hat{G}_0\hat{V}\psi_{n-1}\text{.}
\end{equation}
One can also introduce the Green's operator $\hat{G}$ for the perturbed Hamiltonian:
\begin{equation}
\hat{G}=[E-{\hat{\mathcal{H}\,}\kern-2pt}_0-\hat{V}]^{-1}\text{.}
\end{equation}
By analogy with Eq. (\ref{Eq:DysonFunctions}), the two Green's operators can be related via the so-called Dyson
equation:
\begin{equation}
\hat{G}=\hat{G}_0+\hat{G}_0\hat{V}\hat{G}_0+\hat{G}_0\hat{V}\hat{G}_0\hat{V}\hat{G}_0+\,\cdots\,=\hat{G}_0+\hat{G}_0\hat{V}\hat{G}\text{.}
\end{equation}
It is instructive to rewrite the same equation in Feynman diagrams:
\begin{equation}\label{Eq:DysonFeynman}
\vcenter{\hbox{\scalebox{0.5}{
\begin{fmffile}{feyG}
\begin{fmfgraph*}(60,50)
\fmfleft{l}
\fmfright{r}
\fmf{heavy}{l,r}
\end{fmfgraph*}
\end{fmffile}}}}=\!\!
\vcenter{\hbox{\scalebox{0.5}{
\begin{fmffile}{feyG0}
\begin{fmfgraph*}(60,50)
\fmfleft{l}
\fmfright{r}
\fmf{fermion}{l,r}
\end{fmfgraph*}
\end{fmffile}}}}+\!\!
\vcenter{\hbox{\scalebox{0.5}{
\begin{fmffile}{feyGSG}
\begin{fmfgraph*}(90,50)
\fmfleft{l}
\fmfright{r}
\fmf{fermion}{l,m,r}
\fmfv{decor.shape=circle,decor.filled=shaded}{m}
\end{fmfgraph*}
\end{fmffile}}}}+\!\!
\vcenter{\hbox{\scalebox{0.5}{
\begin{fmffile}{feyGSGSG}
\begin{fmfgraph*}(120,50)
\fmfleft{l}
\fmfright{r}
\fmf{fermion}{l,m,n,r}
\fmfv{decor.shape=circle,decor.filled=shaded}{m}
\fmfv{decor.shape=circle,decor.filled=shaded}{n}
\end{fmfgraph*}
\end{fmffile}}}}+\cdots=\!\!
\vcenter{\hbox{\scalebox{0.5}{
\begin{fmffile}{feyG0SG}
\begin{fmfgraph*}(120,50)
\fmfleft{l}
\fmfright{r}
\fmf{fermion}{l,m}
\fmf{heavy}{m,r}
\fmfv{decor.shape=circle,decor.filled=shaded}{m}
\end{fmfgraph*}
\end{fmffile}}}}\text{.}\vspace{-1em}
\end{equation}
Here $\vcenter{\hbox{\scalebox{0.5}{\begin{fmffile}{feyG}\begin{fmfgraph*}(60,50) \fmfleft{l} \fmfright{r}
\fmf{heavy}{l,r}\end{fmfgraph*}\end{fmffile}}}}$, $\vcenter{\hbox{\scalebox{0.5}{
\begin{fmffile}{feyG0}\begin{fmfgraph*}(60,50)\fmfleft{l} \fmfright{r} \fmf{fermion}{l,r}
\end{fmfgraph*}\end{fmffile}}}}$, and $~\vcenter{\hbox{\scalebox{0.5}{\begin{fmffile}{feyS}\begin{fmfgraph*}(10,50)
\fmfkeep{feyS}\fmfleft{l}\fmfv{decor.shape=circle,decor.filled=shaded}{l}\end{fmfgraph*}\end{fmffile}}}}$ denote
$\hat{G}$, $\hat{G}_0$, and the perturbation $\hat{V}$ respectively. Such notation is meant to visualize the analogy
between this simple single-particle equation and the self-energy expansion of the single-particle Green's function of
a
many-body system that will be discussed below.

\subsection{Single-particle Green's function of a many-body electron system}\label{SubSec:SingleParticleGFint}

In a many-body system, the single-particle Green's functions no longer carry the full information about the system's
wave function. Nevertheless, they are still extremely useful for treating the results of experiments, which are to a
large degree defined by one- and two-particle processes.

There are several types of single-particle Green's functions \cite{BruusFlensberg81,Mahan00}:
\begin{subequations}\label{Eq:GreensFuncDef}
\begin{align}
&\kern-1.2ex\text{greater:}&\kern-1ex G^\text{>}(\mathbf{r},\kern1pt\sigma,\kern1pt
t;\,\mathbf{r\,}'\!,\kern1pt\sigma'\!,\kern1pt
t')&=-\mathrm{i}\,\bigl\langle\kern.5pt\hat{\kern-2.2pt\mathit{\Psi}}(\mathbf{r},\kern1pt\sigma,\kern1pt
t)\,\kern1pt\hat{\kern-2.2pt\mathit{\Psi}}^\dag(\mathbf{r\,}'\!,\kern1pt\sigma'\!,\kern1pt t')\bigr\rangle\\
&\kern-1.2ex\text{lesser:}&\kern-1ex G^\text{<}(\mathbf{r},\kern1pt\sigma,\kern1pt
t;\,\mathbf{r\,}'\!,\kern1pt\sigma'\!,\kern1pt
t')&=\mathrm{i}\,\bigl\langle\kern.5pt\hat{\kern-2.2pt\mathit{\Psi}}^\dag(\mathbf{r\,}'\!,\kern1pt\sigma'\!,\kern1pt
t')\,\kern1pt\hat{\kern-2.2pt\mathit{\Psi}}(\mathbf{r},\kern1pt\sigma,\kern1pt t)\bigr\rangle\\
&\kern-1.2ex\text{retarded:}&\kern-1ex G^\text{R}(\mathbf{r},\kern1pt\sigma,\kern1pt
t;\,\mathbf{r\,}'\!,\kern1pt\sigma'\!,\kern1pt
t')&=-\mathrm{i}\theta(t-t')\bigl\langle\bigl\{\kern.5pt\hat{\kern-2.2pt\mathit{\Psi}}(\mathbf{r},\kern1pt\sigma,\kern1pt
t),\kern1pt\hat{\kern-2.2pt\mathit{\Psi}}^\dag(\mathbf{r\,}'\!,\kern1pt\sigma'\!,\kern1pt t')\bigr\}\bigr\rangle\\
&\kern-1.2ex\text{advanced:}&\kern-1ex G^\text{A}(\mathbf{r},\kern1pt\sigma,\kern1pt
t;\,\mathbf{r\,}'\!,\kern1pt\sigma'\!,\kern1pt
t')&=-\mathrm{i}\theta(t'-t)\bigl\langle\bigl\{\kern.5pt\hat{\kern-2.2pt\mathit{\Psi}}(\mathbf{r},\kern1pt\sigma,\kern1pt
t),\kern1pt\hat{\kern-2.2pt\mathit{\Psi}}^\dag(\mathbf{r\,}'\!,\kern1pt\sigma'\!,\kern1pt t')\bigr\}\bigr\rangle\\
&\kern-1.2ex\text{time-ordered:}&\kern-1ex G^\text{t}(\mathbf{r},\kern1pt\sigma,\kern1pt
t;\,\mathbf{r\,}'\!,\kern1pt\sigma'\!,\kern1pt t')&=\theta(t-t')\,G^\text{>}+\theta(t'-t)\,G^\text{<}\\
&\kern-1.2ex\text{anti-time-ordered:}&\kern-1ex G^\text{\=t}(\mathbf{r},\kern1pt\sigma,\kern1pt
t;\,\mathbf{r\,}'\!,\kern1pt\sigma'\!,\kern1pt t')&=\theta(t'-t)\,G^\text{>}+\theta(t-t')\,G^\text{<}
\end{align}
\end{subequations}
As directly follows from this definition, the retarded and advanced Green's functions can be also expressed in terms
of
the greater and lesser Green's functions:
\begin{equation}
G^\text{R}(\mathbf{r},\,\sigma,\,t;\,\mathbf{r\,}'\!,\,\sigma'\!,\,t')=\theta(t-t')\bigl[G^\text{>}(\mathbf{r},\,\sigma,\,t;\,\mathbf{r\,}'\!,\,\sigma'\!,\,t')-G^\text{<}(\mathbf{r},\,\sigma,\,t;\,\mathbf{r\,}'\!,\,\sigma'\!,\,t')\bigr]
\end{equation}
Here the definitions were given in real space representation, but they can also be rewritten in any
$|\kern1pt\nu\rangle$-basis:
\begin{equation}
G^\text{R}(\nu,\,\sigma,\,t;\,\nu'\!,\,\sigma'\!,\,t')=-\mathrm{i}\theta(t-t')\bigl\langle\bigl\{\hat{c}^{\phantom{\dag}}_{\nu\sigma}(t),\,\hat{c}^\dag_{\nu'\sigma'}(t')\bigr\}\bigr\rangle
\end{equation}

For translation-invariant systems, such as an infinite or semi-infinite crystal, which are of interest in the current
work, the $|\mathbf{k}\rangle$-representation is the most natural one. In this case $G(\mathbf{r},\mathbf{r}')$
becomes
$G(\mathbf{r}-\mathbf{r}')$ due to translation invariance, and in the $|\mathbf{k}\rangle$-representation
$G(\mathbf{k},\mathbf{k}')=\delta_{\mathbf{k}\mathbf{k}'}G(\mathbf{k})$, i.e. the Green's function depends on a single
$\mathbf{k}$ argument:
\begin{equation}\label{Eq:DressedGDef}
G^\text{R}(\mathbf{k};\,\sigma,\,t;\,\,\sigma'\!,\,t')=-\mathrm{i}\theta(t-t')\bigl\langle\bigl\{\hat{c}^{\phantom{\dag}}_{\mathbf{k}\sigma}(t),\,\hat{c}^\dag_{\mathbf{k}\sigma'}(t')\bigr\}\bigr\rangle
\end{equation}

As an example, let us consider the greater Green's function $G^\text{>}_0$ of the non-interacting electron system
described by the quadratic Hamiltonian
\begin{equation}
{\hat{\mathcal{H}\,}\kern-2pt}_0=\int\!\frac{\mathrm{d}\mathbf{k}}{(2\piup)^d}\,\epsilon_{\mathbf{k}}\sum\limits_\sigma\,\hat{c}^\dag_{\textbf{k}\sigma}\hat{c}^{\phantom{\dag}}_{\textbf{k}\sigma}\text{,}
\label{Eq:NoninteractingHamiltonian}
\end{equation}
for which the $\hat{c}$ operators have a simple time dependence:
$\hat{c}^{\phantom{\dag}}_{\mathbf{k}\sigma}(t)=\mathrm{e}^{\mathrm{i}{\hat{\mathcal{H}\,}\kern-2pt}_0
t}\hat{c}^{\phantom{\dag}}_{\mathbf{k}\sigma}\mathrm{e}^{-\mathrm{i}{\hat{\mathcal{H}\,}\kern-2pt}_0
t}=\hat{c}^{\phantom{\dag}}_{\mathbf{k}\sigma}\mathrm{e}^{-\mathrm{i}\kern.3pt\epsilon_\mathbf{k}t}$ and
$\hat{c}^\dag_{\mathbf{k}\sigma}(t)=\hat{c}^\dag_{\mathbf{k}\sigma}\mathrm{e}^{\mathrm{i}\kern.3pt\epsilon_\mathbf{k}t}$
(the so-called \textit{bare Green's function}). Because the Hamiltonian is diagonal in $\sigma$, so is the Green's
function:
\begin{subequations}
\begin{multline}
G^\text{>}_0(\mathbf{k},\,\sigma,\,t-t')=-\mathrm{i}\bigl\langle
\hat{c}^{\phantom{\dag}}_{\mathbf{k}\sigma}(t)\,\hat{c}^\dag_{\mathbf{k}\sigma}(t')\bigr\rangle\\ = -\mathrm{i}\langle
\hat{c}^{\phantom{\dag}}_{\mathbf{k}\sigma}\,\hat{c}^\dag_{\mathbf{k}\sigma}\rangle\mathrm{e}^{-\mathrm{i}\kern.3pt\epsilon_\mathbf{k}(t-t')}
= -\mathrm{i}[1-n_\text{F}(\epsilon_\mathbf{k})]\mathrm{e}^{-\mathrm{i}\kern.3pt\epsilon_\mathbf{k}(t-t')}\text{.}
\label{Eq:Ggreater0}
\end{multline}
Similarly,\vspace{-1em}
\begin{align}
G^\text{<}_0(\mathbf{k},\,\sigma,\,t-t')&=\mathrm{i}\,n_\text{F}(\epsilon_\mathbf{k})\,\mathrm{e}^{-\mathrm{i}\kern.3pt\epsilon_\mathbf{k}(t-t')}\label{Eq:Glesser0}\\
G^\text{R}_0(\mathbf{k},\,\sigma,\,t-t')&=-\mathrm{i}\kern.5pt\theta(t-t')\,\mathrm{e}^{-\mathrm{i}\kern.3pt\epsilon_\mathbf{k}(t-t')}
\end{align}
\end{subequations}

By applying Fourier transform to (\ref{Eq:Ggreater0}) or (\ref{Eq:Glesser0}), one can immediately see that in the
frequency domain the greater and lesser Green's functions of a non-interacting system are localized along the
dispersion
$\epsilon_\mathbf{k}$:
\begin{subequations}
\begin{align}
G^\text{>}_0(\mathbf{k},\,\sigma,\,\omega)&=-2\piup\kern.5pt\mathrm{i}\,[1-n_\text{F}(\epsilon_\mathbf{k})]\,\delta(\epsilon_\mathbf{k}-\omega)\text{.}\\
G^\text{<}_0(\mathbf{k},\,\sigma,\,\omega)&=2\piup\kern.5pt\mathrm{i}\,\,n_\text{F}(\epsilon_\mathbf{k})\,\delta(\epsilon_\mathbf{k}-\omega)\text{.}
\end{align}
For the retarded Green's function the situation is a bit different. Here because of the $\theta(t-t')$ function the
Fourier transform will result in
\begin{equation}\label{Eq:G0}
G^\text{R}_0(\mathbf{k},\,\sigma,\,\omega)=\frac{1}{\omega-\epsilon_\mathbf{k}+\mathrm{i}\,0^{+}}\text{.}
\end{equation}
\end{subequations}
Hence, $G^\text{R}_0(\mathbf{k},\,\sigma,\,\omega)$ is an analytical function of $\omega$ except for one pole at
$\epsilon_\mathbf{k}$. The corresponding \textit{bare spectral function}
$A_0(\mathbf{k},\,\sigma,\,\omega)\stackrel{\text{def}}{=}-\piup^{-1}\,\mathrm{Im}\,
G_0^\text{R}(\mathbf{k},\,\sigma,\,\omega)=\delta(\omega-\epsilon_\mathbf{k})$, which will be important for treating
the
results of photoemission experiments (see section \ref{Sec:TheoryOfPhotoemission}), is still localized at
$\epsilon_\mathbf{k}$.

\subsection{Renormalization and the concept of self-energy.}\label{SubSec:SelfEnergy}

In \S\ref{SubSec:SingleParticleGF} it was shown how the Green's operator of a perturbed Hamiltonian can be expressed
in
terms of the ``bare'' Green's operator via the Dyson equation. This was demonstrated in the simplest case of a
single-particle Hamiltonian. A similar result holds for the retarded Green's function (\ref{Eq:DressedGDef}) of a
many-body system with small interactions that can be considered as a perturbation. In Feynman notation
\cite{Feynman49},
the perturbation series for the Green's function\footnote{The Feynman notation is usually used only for the
time-ordered
Matsubara Green's functions, which will be introduced in \S\ref{SubSec:Matsubara}, but as long as we do not explicitly
write the Feynman rules here, we can formally denote the perturbation expansion for any real-time Green's function by
the same diagrams.} is [\citenum{AtlandSimons06}, p.\,227; \citenum{DoniachSondheimer99}, p.\,138]\vspace{-1ex}
\begin{multline}\label{Eq:FeynSeriesG}
\vcenter{\hbox{\scalebox{0.5}{
\begin{fmffile}{feyG}
\begin{fmfgraph*}(60,50)
\fmfleft{l} \fmfright{r} \fmf{heavy}{l,r}
\end{fmfgraph*}
\end{fmffile}}}}=\!\!
\vcenter{\hbox{\scalebox{0.5}{
\begin{fmffile}{feyG0}
\begin{fmfgraph*}(60,50)
\fmfleft{l} \fmfright{r} \fmf{fermion}{l,r}
\end{fmfgraph*}
\end{fmffile}}}}+\!\!
\vcenter{\hbox{\scalebox{0.5}{
\begin{fmffile}{fynm3}
\begin{fmfgraph*}(60,50)
\fmfleft{l}\fmfright{r}\fmftop{t} \fmf{fermion}{l,m,r}\fmffreeze \fmf{photon}{m,t}\fmf{fermion}{t,t}
\end{fmfgraph*}
\end{fmffile}}}}+\!\!
\vcenter{\hbox{\scalebox{0.5}{
\begin{fmffile}{fynm4}
\begin{fmfgraph*}(90,50)
\fmfleft{l}\fmfright{r} \fmf{fermion}{l,m,n,r}\fmffreeze \fmf{photon,left}{m,n}
\end{fmfgraph*}
\end{fmffile}}}}+\!\!
\vcenter{\hbox{\scalebox{0.5}{
\begin{fmffile}{fynm5}
\begin{fmfgraph*}(120,50)
\fmfleft{l}\fmfright{r}\fmftop{t} \fmf{fermion}{l,m,n,k,r}\fmffreeze \fmf{photon}{m,t}\fmf{fermion}{t,t}
\fmf{photon,left}{n,k} \fmfshift{-0.25w,0}{t}
\end{fmfgraph*}
\end{fmffile}}}}+\!\!
\vcenter{\hbox{\scalebox{0.5}{
\begin{fmffile}{fynm6}
\begin{fmfgraph*}(120,50)
\fmfleft{l}\fmfright{r}\fmftop{t} \fmf{fermion}{l,m,n,k,r}\fmffreeze \fmf{photon}{n,t}\fmf{fermion}{t,t}
\fmf{photon,right=0.5}{m,k}
\end{fmfgraph*}
\end{fmffile}}}}+\!\!
\vcenter{\hbox{\scalebox{0.5}{
\begin{fmffile}{fynm7}
\begin{fmfgraph*}(90,50)
\fmfleft{l}\fmfright{r}\fmftop{t1}\fmftop{t2} \fmf{fermion}{l,m,n,r}\fmffreeze \fmf{photon}{m,t1}\fmf{photon}{n,t2}
\fmf{fermion}{t1,t1}\fmf{fermion}{t2,t2} \fmfshift{-0.1667w,0}{t1}\fmfshift{0.1667w,0}{t2}
\end{fmfgraph*}
\end{fmffile}}}}+\!\!
\vcenter{\hbox{\scalebox{0.5}{
\begin{fmffile}{fynm8}
\begin{fmfgraph*}(60,100)
\fmfleft{l}\fmfright{r}\fmftop{t}\fmf{fermion}{l,m,r}\fmffreeze
\fmf{photon}{m,t1}\fmf{fermion,left=0.7,tension=0.25}{t1,t2}\fmf{fermion,left=0.7,tension=0.25}{t2,t1}\fmf{photon}{t2,t}
\fmf{fermion,tension=0.5}{t,t}
\end{fmfgraph*}
\end{fmffile}}}}\\+\!\!
\vcenter{\hbox{\scalebox{0.5}{
\begin{fmffile}{fynm9}
\begin{fmfgraph*}(120,50)
\fmfleft{l}\fmfright{r} \fmf{fermion}{l,m,n1}\fmf{plain,tension=3}{n1,n2}\fmf{fermion}{n2,k,r}\fmffreeze
\fmf{photon,left}{m,n2}\fmf{photon,right}{n1,k}
\end{fmfgraph*}
\end{fmffile}}}}+\!\!
\vcenter{\hbox{\scalebox{0.5}{
\begin{fmffile}{fynm10}
\begin{fmfgraph*}(150,50)
\fmfleft{l}\fmfright{r}\fmf{fermion}{l,m1,m2,m3,m4,r}\fmffreeze \fmf{photon,left}{m1,m2}\fmf{photon,left}{m3,m4}
\end{fmfgraph*}
\end{fmffile}}}}+\!\!
\vcenter{\hbox{\scalebox{0.5}{
\begin{fmffile}{fynm11}
\begin{fmfgraph*}(150,50)
\fmfleft{l}\fmfright{r}\fmf{fermion}{l,m1,m2,m3,m4,r}\fmffreeze \fmf{photon,left=0.5}{m1,m4}\fmf{photon,right}{m2,m3}
\end{fmfgraph*}
\end{fmffile}}}}+\!\!
\vcenter{\hbox{\scalebox{0.5}{
\begin{fmffile}{fynm12}
\begin{fmfgraph*}(90,50)
\fmfleft{l}\fmfright{r}\fmftop{t1}\fmftop{t2}\fmf{fermion}{l,m,n,r}\fmffreeze
\fmf{photon}{m,t1}\fmf{photon}{n,t2}\fmf{fermion,left=0.55}{t1,t2}\fmf{fermion,left=0.55}{t2,t1}
\fmfshift{-0.1667w,0}{t1}\fmfshift{0.1667w,0}{t2}
\end{fmfgraph*}
\end{fmffile}}}}+\,\cdots\text{.}\vspace{-1em}
\end{multline}
The series (\ref{Eq:FeynSeriesG}) contains repeating patterns of diagrams that allow simplification by denoting
\begin{equation}\label{Eq:FeynSeriesS}
\vcenter{\hbox{\scalebox{0.5}{\fmfreuse{feyS}}}}~=
\vcenter{\hbox{\scalebox{0.5}{
\begin{fmffile}{fynm13}
\begin{fmfgraph*}(10,25)\fmfkeep{feyHartreeS}
\fmfbottom{m}\fmftop{t}\fmffreeze\fmf{photon}{m,t}\fmf{fermion}{t,t}\fmfshift{0,0.5h}{m,t}
\end{fmfgraph*}
\end{fmffile}}}}~+
\vcenter{\hbox{\scalebox{0.5}{
\begin{fmffile}{fynm14}
\begin{fmfgraph*}(30,50)\fmfkeep{feyFockS}
\fmfleft{l}\fmfright{r}\fmf{fermion}{l,r}\fmffreeze\fmf{photon,left}{l,r}
\end{fmfgraph*}
\end{fmffile}}}}~+
\vcenter{\hbox{\scalebox{0.5}{
\begin{fmffile}{fynm15}
\begin{fmfgraph*}(60,50)
\fmfleft{l}\fmfright{r}\fmftop{t}\fmf{fermion}{l,m,r}\fmffreeze\fmf{photon,right=0.5}{l,r}\fmf{photon}{m,t}\fmf{fermion}{t,t}
\end{fmfgraph*}
\end{fmffile}}}}~+
\vcenter{\hbox{\scalebox{0.5}{
\begin{fmffile}{fynm16}
\begin{fmfgraph*}(60,50)
\fmfleft{l}\fmfright{r}\fmf{fermion}{l,n1}\fmf{plain,tension=3}{n1,n2}\fmf{fermion}{n2,r}\fmffreeze
\fmf{photon,left}{l,n2}\fmf{photon,right}{n1,r}
\end{fmfgraph*}
\end{fmffile}}}}~+
\vcenter{\hbox{\scalebox{0.5}{
\begin{fmffile}{fynm17}
\begin{fmfgraph*}(10,50)
\fmfbottom{m}\fmftop{t}\fmf{photon}{m,t1}\fmf{fermion,left=0.7,tension=0.25}{t1,t2}\fmf{fermion,left=0.7,tension=0.25}{t2,t1}
\fmf{photon}{t2,t}\fmf{fermion,tension=0.5}{t,t}\fmffreeze\fmfshift{0,0.5h}{m,t1,t2,t}
\end{fmfgraph*}
\end{fmffile}}}}~+
\vcenter{\hbox{\scalebox{0.5}{
\begin{fmffile}{fynm18}
\begin{fmfgraph*}(30,50)
\fmfleft{l}\fmfright{r}\fmftop{t1}\fmftop{t2}\fmf{fermion}{l,r}\fmffreeze
\fmf{photon}{l,t1}\fmf{photon}{r,t2}\fmf{fermion,left=0.55}{t1,t2}\fmf{fermion,left=0.55}{t2,t1}
\fmfshift{-0.5w,0}{t1}\fmfshift{0.5w,0}{t2}
\end{fmfgraph*}
\end{fmffile}}}}~+
\vcenter{\hbox{\scalebox{0.5}{
\begin{fmffile}{fynm19}
\begin{fmfgraph*}(90,50)
\fmfleft{l}\fmfright{r} \fmf{fermion}{l,m,n,r}\fmffreeze
\fmf{photon,left=0.5}{l,r}\fmf{photon,right}{m,n}
\end{fmfgraph*}
\end{fmffile}}}}~+\,\cdots\text{.}
\end{equation}
The sum (\ref{Eq:FeynSeriesS}) is called the \textit{self-energy} and includes all \textit{irreducible diagrams} (i.e.
those that cannot be disconnected by cutting an internal fermion line). Substitution of (\ref{Eq:FeynSeriesS}) into
(\ref{Eq:FeynSeriesG}) yields the Dyson equation of the form (\ref{Eq:DysonFeynman}). Thus the self-energy function is
the measure of the perturbation introduced to the bare Hamiltonian by the interaction terms. In Fourier space the
Dyson
equation turns into a simple algebraic form with the self-energy $\mathit{\Sigma}^\text{R}(\mathbf{k},\omega)$ being a
complex-valued function of $\omega$:
\begin{equation}\label{Eq:DysonEquation}
   G^\text{R}(\mathbf{k},\omega)=G_0^\text{R}(\mathbf{k},\omega)+G_0^\text{R}(\mathbf{k},\omega)\mathit{\Sigma}^\text{R}(\mathbf{k},\omega)G^\text{R}(\mathbf{k},\omega)\text{.}
\end{equation}
The same equation holds also for the advanced Green's functions, the retarded and advanced self-energies being complex
conjugate to each other:
$\mathit{\Sigma}^\text{R}(\mathbf{k},\omega)=\mathit{\Sigma}^\text{A}(\mathbf{k},\omega)^\Ast$.
From now on we will work mainly with the retarded Green's function and self-energy, omitting the upper index
$\scriptstyle\text{R}$.

Dyson's equation (\ref{Eq:DysonEquation}) can be rewritten in a slightly different but equivalent form by using the
algebraic form for $G_0$ (\ref{Eq:G0}):
\begin{equation}\label{Eq:RenormG}
   G(\mathbf{k},\,\omega)=\frac{1}{\omega-\epsilon_\mathbf{k}-\mathit{\Sigma}(\mathbf{k},\omega)}\text{.}
\end{equation}

Here we see that the perturbation of the quadratic Hamiltonian causes a shift of the electron dispersion by
$\mathit{\Sigma}'=\mathrm{Re}\,\mathit{\Sigma}(\mathbf{k},\omega)$ (so-called \textit{renormalization} of the bare
band)
and Lorentzian broadening of the \textit{spectral function} $A\stackrel{\text{def}}{=}-\piup^{-1}\,\mathrm{Im}\,
G^\text{R}$ caused by $\mathit{\Sigma}''=\mathrm{Im}\,\mathit{\Sigma}(\mathbf{k},\omega)$. This broadening can be
interpreted as finite ``lifetime'' of the single-particle states that are destroyed by the interactions, and the
imaginary part of the self-energy $\mathit{\Sigma}''$ is therefore termed \textit{scattering rate}. It is remarkable
that all kinds of interactions (such as electron-electron, electron-phonon, electron-magnon, impurity scattering,
etc.)
enter as additive terms into the self-energy, making it a universal tool for describing weakly interacting
many-particle
systems.\footnote{In spite of these seemingly simple basic ideas, the renormalization group theory is a complicated
branch of quantum field theory and mathematics, still being actively explored. The self-energy is represented as a sum
of infinitely many Feynman diagrams, which is impossible to calculate exactly. Thus, limiting the sums like
(\ref{Eq:FeynSeriesS}) only to the most significant terms and formalization of the summation procedure utilizing the
self-similarities (scaling properties) of the series are the main challenges. For a detailed introduction into the
field, see Ref. \citenum{BruusFlensberg81, Mahan00, Moriya85, Matuck92, AtlandSimons06, AbrikosovGorkov63}.}

Due to the analytical properties of the self-energy, its real and imaginary parts are bound by the following
\textit{Kramers-Kronig relations}:
\begin{subequations}\label{Eq:KK}
\begin{equation}\label{Eq:KKRe}
   \mathrm{Re}\kern.5pt\mathit{\Sigma}(\mathbf{k},\omega)=\phantom{-}\frac{1}{\piup}\,\fint_{-\infty}^{\infty}\frac{\mathrm{Im}\kern.5pt\mathit{\Sigma}(\mathbf{k},\omega')\,\mathrm{d}\omega'}{\omega'-\omega}\text{;}
\end{equation}
\begin{equation}\label{Eq:KKIm}
   \mathrm{Im}\kern.5pt\mathit{\Sigma}(\mathbf{k},\omega)=-\frac{1}{\piup}\,\fint_{-\infty}^{\infty}\frac{\mathrm{Re}\kern.5pt\mathit{\Sigma}(\mathbf{k},\omega')\,\mathrm{d}\omega'}{\omega'-\omega}\text{.}
\end{equation}
\end{subequations}
These important relations are discussed in detail in appendix \ref{Appendix:KK}.

\subsection{The spectral function}\label{SubSec:SpectralFunction}

In the previous paragraph the spectral function of an interacting electron gas has been introduced. Here we will
derive
the expression for the spectral function at finite temperature in Lehmann representation that will be useful later in
section \ref{Sec:TheoryOfPhotoemission}.

We start with the definition of the retarded Green's function (\ref{Eq:GreensFuncDef}c) rewritten in the one-particle
basis $|\nu\rangle$ (note that we retain the off-diagonal elements $\nu_1\neq\nu_2$):
\begin{equation}
G^\text{R}_{\nu_1,\,\nu_2}(t)=\frac{-\mathrm{i}\,\theta(t)}{Z}\sum_n\mathrm{e}^{-\beta E_n}\bigl\langle
n\big|\hat{c}^{\phantom{\dag}}_{\nu_2}(t)\hat{c}^\dag_{\nu_1}(0)+\hat{c}^\dag_{\nu_1}(0)\hat{c}^{\phantom{\dag}}_{\nu_2}(t)\big|n\bigr\rangle\text{,~where~}Z=\sum_n\mathrm{e}^{-\beta
E_n}\text{.}
\end{equation}
Inserting the identity matrix $\sum_m|m\rangle\langle m|$ and writing down explicitly the time evolution of the
$\hat{c}$-operators, we obtain the so-called Lehmann representation of the retarded Green's function
\cite[p.\,128]{BruusFlensberg81}:
\begin{equation}
G^\text{R}_{\nu_1,\,\nu_2}(t)=\frac{-\mathrm{i}\,\theta(t)}{Z}\sum_{n,\,m}\mathrm{e}^{-\beta E_n}\bigl(\langle
n|\hat{c}^{\phantom{\dag}}_{\nu_2}|m\rangle\langle
m|\hat{c}^\dag_{\nu_1}|n\rangle\mathrm{e}^{\mathrm{i}(E_n-E_m)t}+\langle n|\hat{c}^\dag_{\nu_1}|m\rangle\langle
m|\hat{c}^{\phantom{\dag}}_{\nu_2}|n\rangle\mathrm{e}^{-\mathrm{i}(E_n-E_m)t}\bigr)\text{,}
\end{equation}
which after taking the Fourier transform becomes
\begin{equation}
G^\text{R}_{\nu_1,\,\nu_2}\text{(}\omega\text{)}=\frac{1}{Z}\sum_{n,\,m}\,(\mathrm{e}^{-\beta E_n}+\mathrm{e}^{-\beta
E_m})\,\frac{\langle n|\hat{c}^\dag_{\nu_1}|m\rangle\langle
m|\hat{c}^{\phantom{\dag}}_{\nu_2}|n\rangle}{\omega+E_m-E_n+\mathrm{i}0^+}\text{.}
\end{equation}
\mbox{If we define the \textit{spectral function} as
$A_{\nu_1,\,\nu_2}\!\text{(}\omega\text{)}\kern-1pt=\kern-1pt-\piup^{-1}\,\mathrm{Im}\,G^\text{R}_{\nu_1,\,\nu_2}\!\text{(}\omega\text{)}\kern-1pt=\kern-1pt-\frac{\mathrm{i}}{2\piup}\kern.3pt
G^{<}_{\nu_1,\,\nu_2}\!\text{(}\omega\text{)}/n_\text{F}\text{(}\omega\text{)}\kern-1pt=$}
$\frac{\mathrm{i}}{2\piup}\kern.3pt
G^{>}_{\nu_1,\,\nu_2}\!\text{(}\omega\text{)}/(1-n_\text{F}\text{(}\omega\text{)})$\footnote{In some sources
\cite{BruusFlensberg81,Mahan00}, the spectral function is normalized to $2\piup$, rather than to unity, i.e.
$A_0(\mathbf{k},\,\sigma,\,\omega)$ is defined as $-2\,\mathrm{Im}\,
G_0^\text{R}(\mathbf{k},\,\sigma,\,\omega)=2\piup\,\delta(\omega-\epsilon_\mathbf{k})$.} and use the Dirac identity
$\mathrm{Im}\,\text{(}\omega+\mathrm{i}0^+\text{)}^{-1}=-\piup\delta\text{(}\omega\text{)}$, we have
\begin{equation}\label{Eq:SpectralFunction}
A_{\kern.3pt\nu_1,\,\nu_2}\text{(}\omega\text{)}=\frac{1}{Z}\sum_{n,\,m}\,\langle
n|\hat{c}^\dag_{\nu_1}|m\rangle\langle m|\hat{c}^{\phantom{\dag}}_{\nu_2}|n\rangle\,\mathrm{e}^{-\beta
E_n}\underset{1/n_\text{F}\text{(}\omega\text{)}}{\underbrace{(1+\mathrm{e}^{\beta\omega})}}\,\delta(\omega+E_m-E_n)\text{.}
\end{equation}
In many applications, the spectral function is considered diagonal, and we will also neglect the off-diagonal
components
($\nu_1\neq\nu_2$) when treating the photoelectron spectroscopy experiments (\S\ref{SubSec:SuddenApprox}).

\subsection{Imaginary time Green's functions}\label{SubSec:Matsubara}

The definition of the Green's functions (\ref{Eq:GreensFuncDef}) includes thermodynamic averaging, which at a finite
temperature $T$ can be explicitly written as
\begin{equation}\label{Eq:GreensFuncExplicit}
\bigl\langle\kern.5pt\hat{\kern-2.2pt\mathit{\Psi}}(\mathbf{r},\,t)\,\kern.5pt\hat{\kern-2.2pt\mathit{\Psi}}^\dag(\mathbf{r\,}',\,t')\bigr\rangle=\frac{\mathrm{Tr}\bigl(\mathrm{e}^{-\beta
H}\,\hat{\kern-2.2pt\mathit{\Psi}}(\mathbf{r},\,t)\,\kern.5pt\hat{\kern-2.2pt\mathit{\Psi}}^\dag(\mathbf{r}',\,t')\bigr)}{\mathrm{Tr}\bigl(\mathrm{e}^{-\beta
H}\bigr)}\text{, where~}
~\hat{\kern-2.2pt\mathit{\Psi}}(\mathbf{r},\,t)=\mathrm{e}^{\mathrm{i}tH}\,\hat{\kern-2.2pt\mathit{\Psi}}(\mathbf{r})\,\mathrm{e}^{-\mathrm{i}tH}\text{.}
\end{equation}
Because of the similarity between the two exponents $\mathrm{e}^{\pm\mathrm{i}tH}$ and $\mathrm{e}^{-\beta H}$, it
appears convenient to extend the time $t$ to complex values and consider a special case, when $t$ is purely imaginary,
so that $\tau=\mathrm{i}t$ is real.

The single-particle \textit{Matsubara Green's function}, or imaginary-time Green's function, is defined as
\begin{equation}\label{Eq:GreensFuncMatsubaraDef}
\mathcal{G}(\mathbf{r},\,\sigma,\,\tau;\,\mathbf{r\,}'\!,\,\sigma'\!,\,\tau')=-\bigl\langle
T_\tau\,\hat{\kern-2.2pt\mathit{\Psi}}(\mathbf{r},\,\sigma,\,\tau)\,\hat{\kern-2.2pt\mathit{\Psi}}^\dag(\mathbf{r\,}'\!,\,\sigma'\!,\,\tau')\bigr\rangle\text{,}
\end{equation}
where $T_\tau$ is the imaginary time ordering operator:
\begin{equation}
T_\tau\,\hat{A}(\tau)\kern.5pt\hat{B}(\tau')=\theta(\tau-\tau')\,\hat{A}(\tau)\kern.5pt\hat{B}(\tau')\pm\theta(\tau'-\tau)\,\hat{B}(\tau')\kern.5pt\hat{A}(\tau)\text{.}
\end{equation}

One can show that the convergence of (\ref{Eq:GreensFuncMatsubaraDef}) is guaranteed only for $-\beta<\tau<\beta$.
Moreover, the function $\mathcal{G}(\tau,\kern.5pt\tau')$ depends on time difference only, i.e.
$\mathcal{G}(\tau,\kern.5pt\tau')=\mathcal{G}(\tau-\tau')$, and has the symmetry property
$\mathcal{G}(\tau)=\pm\mathcal{G}(\tau+\beta)$ for $\tau<0$. With these properties taken into account, it can be
expanded into a Fourier series in the interval $-\beta<\tau<\beta$:
\begin{equation}
\mathcal{G}(\tau)=\frac{1}{\beta}\sum_{n=-\infty}^\infty\mathrm{e}^{-\mathrm{i}\omega_n\tau}\mathcal{G}(\mathrm{i}\omega_n)\text{,\quad}
\mathcal{G}(\mathrm{i}\omega_n)=\int_0^\beta\kern-4pt\,\mathrm{e}^{\mathrm{i}\omega_n\tau}\mathcal{G}(\tau)\,\mathrm{d}\kern-.5pt\tau\text{,}
\end{equation}
where $\omega_n=2n\piup/\beta$ for bosons and $\omega_n=(2n+1)\piup/\beta$ for fermions are the so-called
\textit{Matsubara frequencies}.

By using the Lehmann representation, it is possible to show that the newly introduced function
$\mathcal{G}(\mathrm{i}\omega_n)$ and the retarded Green's function $G^\text{R}\text{(}\omega\text{)}$ have a common
analytical continuation.\footnote{The proof can be found in Ref. \citenum{BruusFlensberg81}, p.\,189.} In other words,
there exists a function $G\text{(}z\text{)}$, $z\in\mathds{C}$, analytic in the upper half-plane, which coincides with
$G^\text{R}\text{(}\omega\text{)}$ on the real axis and with $\mathcal{G}(\mathrm{i}\omega_n)$ on the imaginary axis.
This means that as soon as one of the functions is known analytically, the other one can be immediately obtained by a
formal change of variables: $\mathrm{i}\omega_n \leftrightarrow \omega+\mathrm{i}\kern.5pt 0^+$. Nevertheless, if only
the numerical representation of one of the functions is known, computing the other one by analytical continuation
requires an extra level of integration. In most cases, when dealing with perturbation series, the calculations
involving
Matsubara functions are easier and lead to more compact expressions, which finally need to be continued analytically
to
the real axis to make any comparison with experimental data possible.

\mbox{When\,the\,Matsubara\,functions\,are\,considered\,as\,functions\,of\,momentum\,$\mathbf{k}$\,and\,frequency}
$\mathrm{i}\omega_n$, it is convenient to treat both arguments as components of a single \mbox{4-dimensional} argument,
called \textit{4-momentum}, with one imaginary and three real components. In this sense, the frequency and momentum
components within the same \mbox{4-momentum} can be denoted by the same letter, e.g.
$\{\mathrm{i}k_n,\,\mathbf{k}\}\equiv\tilde{\mathbf{k}}$, which makes it easier to distinguish the corresponding
arguments in formulae containing more than one \mbox{4-momentum}.\footnote{Note that the summation indices like
$\mathrm{i}k_n$ are used as single symbols instead of $n$. This allows distinguishing different Matsubara frequency
arguments, which would otherwise require using multiple subindices $n_1, n_2, ...$} Integration over all 4-momentum
arguments is denoted as
$\int\kern-2pt\mathrm{d}\tilde{\mathbf{k}}\,\stackrel{\text{def}}{=}\,\sum_{\,\mathrm{i}k_n}\int\kern-2pt\mathrm{d}\mathbf{k}$.

\section{Two-particle correlation functions}\label{Sec:TwoParticleCorrelation}

\subsection{Lindhard function}

The single-particle Green's functions describe the properties of individual particles, renormalized by excitations of
the weakly interacting many-body system, within the linear response regime. Still, many experiments, such as
resistivity
measurements, inelastic light scattering (Raman scattering) or inelastic neutron scattering (INS) are described in the
linear approximation by higher order response functions\footnote{This connection follows from the Kubo formula,
describing the change of any observable $X$ under a perturbation $\hat{\mathcal{H}'}$ of the Hamiltonian within linear
approximation as $\delta\langle\hat{X}(t)\rangle= -\mathrm{i}\int_{t_0}^\infty\mathrm{d}t'\theta(t-t')
\bigl\langle[\,\hat{X}(t),\,{\hat{\mathcal{H}'}}(t')]\bigr\rangle$. The details can be found in Ref.
\citenum{BruusFlensberg81, Mahan00}.}---\,the so-called \textit{two-particle correlation functions} of the form
\begin{equation}
C_{A\,A}(t,\,t')=-\mathrm{i}\theta(t-t')\bigl\langle[\,A\kern.3pt(t),\,A\kern.3pt(t')\,]\bigr\rangle\text{,}
\end{equation}
where $A$ is some two-particle operator, e.g. the charge or spin density operator, current operator, etc. Here we will
be mostly dealing with the spin and charge density correlation functions, which describe the dynamic electromagnetic
susceptibility $\chi(\mathbf{k},\omega)$ of the electron system.

Let us consider the retarded charge density correlation function of a non-interacting electron gas \cite{Lindhard54,
BruusFlensberg81, Gorbachenko80}, also known as \textit{bare susceptibility}, or the \textit{Lindhard function}:
\begin{equation}\label{Eq:Hi0Def}
\chi_0^\text{R}(\mathbf{q},t-t')=-\mathrm{i}\theta(t-t')\bigl\langle[\,\hat{\rho}(\mathbf{q},t),\,\hat{\rho}(-\mathbf{q},t')\,]\bigr\rangle\text{.}
\end{equation}
For free electrons the time dependence of the charge density operator is given by
\begin{equation}\label{Eq:ChargeDensityTimeDep}
\hat{\rho}(\mathbf{q},t)=\int\!\!\frac{\mathrm{d}\mathbf{k}}{(2\piup)^d}\sum\limits_\sigma\,\hat{c}^\dag_{\textbf{k}\sigma}\hat{c}^{\phantom{\dag}}_{\textbf{k}+\textbf{q}\sigma}\mathrm{e}^{\mathrm{i}(\epsilon_{\mathbf{k}}-\epsilon_{\mathbf{k}+\mathbf{q}})t}\text{.}
\end{equation}
Substituting it into (\ref{Eq:Hi0Def}) and using the identities
$[\,\hat{c}^\dag_{\nu\vphantom{'}}\hat{c}^{\phantom{\dag}}_{\mu\vphantom{'}},\,\hat{c}^\dag_{\nu'}\hat{c}^{\phantom{\dag}}_{\mu'}]=
\hat{c}^\dag_{\nu\vphantom{'}}\hat{c}^{\phantom{\dag}}_{\mu'}\delta_{\mu\nu'}-\hat{c}^\dag_{\nu'}\hat{c}^{\phantom{\dag}}_{\mu\vphantom{'}}\delta_{\mu'\nu}$
and $n_\text{F}(\epsilon_\mathbf{k})=\langle\hat{c}^\dag_\textbf{k}\hat{c}^{\phantom{\dag}}_\textbf{k}\rangle$, we get
(again omitting the index $\scriptstyle\text{R}$ for brevity)
\begin{equation}
\chi_0(\mathbf{q},t-t')=-2\kern.3pt\mathrm{i}\theta(t-t')\int\!\!\frac{\mathrm{d}\mathbf{k}}{(2\piup)^d}\,[\kern.5pt
n_\text{F}(\epsilon_\mathbf{k})-n_\text{F}(\epsilon_{\mathbf{k}+\mathbf{q}})]\mathrm{e}^{\mathrm{i}(\epsilon_{\mathbf{k}}-\epsilon_{\mathbf{k}+\mathbf{q}})(t-t')}\text{,}
\end{equation}
or, after taking the Fourier transform,
\begin{equation}\label{Eq:Hi0Bare}
\chi_0(\mathbf{q},\omega)=2\int\!\!\frac{\mathrm{d}\mathbf{k}}{(2\piup)^d}\,\frac{n_\text{F}(\epsilon_\mathbf{k})-n_\text{F}(\epsilon_{\mathbf{k}+\mathbf{q}})}{\epsilon_\mathbf{k}-\epsilon_{\mathbf{k}+\mathbf{q}}+\omega+\mathrm{i}\,0^+}\text{.}
\end{equation}

The Lindhard function is a good starting point towards calculating the two-particle correlation functions of
interacting
electron systems, if the interactions are small enough to be considered as perturbations. This becomes clear if we
rewrite the expression (\ref{Eq:Hi0Bare}) in terms of bare Green's functions. To do this, one needs to apply the
Wick's
theorem \cite{BruusFlensberg81,AtlandSimons06} to (\ref{Eq:Hi0Def}), which yields
\begin{equation}\label{Eq:Hi0BareMatsubara}
\chi_0(\tilde{\mathbf{q}})=-\frac{1}{\beta}\int\!\!\frac{\mathrm{d}\tilde{\mathbf{k}}}{(2\piup)^d}\sum\limits_\sigma\,\mathcal{G}^{\phantom{*}}_0(\tilde{\mathbf{k}}+\tilde{\mathbf{q}},\,\sigma)\mathcal{G}^*_0(\tilde{\mathbf{k}},\,\sigma)=
\kern-3pt\vcenter{\hbox{\scalebox{0.5}{
\begin{fmffile}{feyHi0}
\begin{fmfgraph*}(60,30)\fmfkeep{feyHi0}
\fmfleft{l}\fmfright{r}\fmf{fermion,left=0.5}{l,r,l}
\end{fmfgraph*}
\end{fmffile}}}}\text{.}
\end{equation}

Here $\mathcal{G}_0(\tilde{\mathbf{k}},\,\sigma)=1/(\mathrm{i}k_n-\epsilon_\mathbf{k})$ is the bare Green's function
in
Matsubara notation, so the Lindhard's function turns out to be simply an autocorrelation of the Green's function taken
over the 4-momentum, which is denoted by a bubble-like Feynman diagram. This clearly demonstrates the convenience of
the
Matsubara formalism.

For the spin density correlation function the situation is very similar. The time dependence of the spin density
operator, similar to (\ref{Eq:ChargeDensityTimeDep}), is given by
\begin{equation}\label{Eq:SpinDensityTimeDep}
\hat{S}(\mathbf{q},t)=\int\!\!\frac{\mathrm{d}\mathbf{k}}{(2\piup)^d}\sum\limits_\sigma\,\hat{c}^\dag_{\textbf{k}\sigma}\hat{c}^{\phantom{\dag}}_{\textbf{k}+\textbf{q};\,-\sigma}\mathrm{e}^{\mathrm{i}(\epsilon_{\mathbf{k}}-\epsilon_{\mathbf{k}+\mathbf{q}})t}\text{,}
\end{equation}
which finally results in exactly the same expression for the Lindhard function (\ref{Eq:Hi0Bare}).

\subsection{Random phase approximation}

Let us now consider the perturbation series for the charge density correlation function $\chi(\tilde{\mathbf{q}})$ in
the presence of some weak interaction:
\begin{equation}\label{Eq:FeynSeriesHi}
\chi(\tilde{\mathbf{q}})=\kern-3pt
\vcenter{\hbox{\scalebox{0.5}{
\begin{fmffile}{feyHi}
\begin{fmfgraph*}(60,33)\fmfkeep{feyHi}
\fmfpoly{filled=hatched}{m1,m2,m3,m4}\fmfsurround{r,m1,m2,l,m3,m4}
\fmf{fermion,left=0.2,tension=0.5}{l,m2}\fmf{fermion,left=0.2}{m3,l}
\fmf{fermion,left=0.2,tension=0.5}{m1,r}\fmf{fermion,left=0.2}{r,m4}
\fmffreeze\fmfshift{-0.1w,0}{m1}\fmfshift{0.1w,0}{m2}\fmfshift{0.1w,0}{m3}\fmfshift{-0.1w,0}{m4}
\end{fmfgraph*}
\end{fmffile}}}}=\,\vcenter{\hbox{\scalebox{0.5}{\fmfreuse{feyHi0}}}}\,+\kern-3pt
\vcenter{\hbox{\scalebox{0.5}{
\begin{fmffile}{fynm20}
\begin{fmfgraph*}(60,30)
\fmfsurround{r,t,l,b}\fmf{fermion,left=0.23}{l,t,r,b,l}\fmf{photon}{t,b}
\end{fmfgraph*}
\end{fmffile}}}}+\kern-3pt
\vcenter{\hbox{\scalebox{0.5}{
\begin{fmffile}{fynm21}
\begin{fmfgraph*}(60,27)
\fmfsurround{r,t1,t2,l,b1,b2}\fmf{fermion,left=0.2}{l,t2,t1,r}\fmf{fermion,left=0.5}{r,l}\fmf{photon,left=0.5}{t1,t2}
\end{fmfgraph*}
\end{fmffile}}}}+\kern-3pt
\vcenter{\hbox{\scalebox{0.5}{
\begin{fmffile}{fynm22}
\begin{fmfgraph*}(60,29)
\fmfsurround{r,t0,t1,t2,t3,l,b0,b1,b2,b3}\fmf{fermion,left=0.2}{b1,l,t2}\fmf{fermion,left=0.2}{t1,r,b2}
\fmf{fermion,left=0.1}{t2,t1}\fmf{fermion,left=0.1}{b2,b1}\fmf{photon}{t2,b1}\fmf{photon}{t1,b2}
\end{fmfgraph*}
\end{fmffile}}}}+\kern-3pt
\vcenter{\hbox{\scalebox{0.5}{
\begin{fmffile}{feyHiOa}
\begin{fmfgraph*}(140,30)\fmfkeep{feyHiOa}
\fmfleft{l}\fmfright{r}\fmf{fermion,left=0.5}{l,m1,l}\fmf{photon,tension=6}{m1,m2}\fmf{fermion,left=0.5}{m2,r,m2}
\end{fmfgraph*}
\end{fmffile}}}}+\cdots\text{.}
\end{equation}
In analogy with the self-energy diagrams (\ref{Eq:FeynSeriesS}), the expression can be simplified by collecting the
irreducible diagrams into $\chi^{\text{irr}}(\tilde{\mathbf{q}})=\kern-3pt \vcenter{\hbox{\scalebox{0.5}{
\begin{fmffile}{feyHiIrr}
\begin{fmfgraph*}(60,33)\fmfkeep{feyHiIrr}
\fmfpoly{filled=shaded}{m1,m2,m3,m4}\fmfsurround{r,m1,m2,l,m3,m4}
\fmf{fermion,left=0.2,tension=0.5}{l,m2}\fmf{fermion,left=0.2}{m3,l}
\fmf{fermion,left=0.2,tension=0.5}{m1,r}\fmf{fermion,left=0.2}{r,m4}
\fmffreeze\fmfshift{-0.1w,0}{m1}\fmfshift{0.1w,0}{m2}\fmfshift{0.1w,0}{m3}\fmfshift{-0.1w,0}{m4}
\end{fmfgraph*}
\end{fmffile}}}}$ and rewriting the sum (\ref{Eq:FeynSeriesHi}) in the form of a Dyson equation
\begin{equation}
\chi(\tilde{\mathbf{q}})=\,
\vcenter{\hbox{\scalebox{0.5}{\fmfreuse{feyHi}}}}\,=\,
\vcenter{\hbox{\scalebox{0.5}{\fmfreuse{feyHiIrr}}}}\,+\,
\vcenter{\hbox{\scalebox{0.5}{\fmfreuse{feyHiIrr}}}}\kern-5pt
\vcenter{\hbox{\scalebox{0.5}{
\begin{fmffile}{feyPhoton}
\begin{fmfgraph*}(20,33)\fmfkeep{feyPhoton}
\fmfleft{l}\fmfright{r}\fmf{photon}{l,r}
\end{fmfgraph*}
\end{fmffile}}}}\kern-1.5pt
\vcenter{\hbox{\scalebox{0.5}{\fmfreuse{feyHi}}}}\,=\chi^{\text{irr}}(\tilde{\mathbf{q}})+\chi^{\text{irr}}(\tilde{\mathbf{q}})W(\tilde{\mathbf{q}})\chi(\tilde{\mathbf{q}})
\end{equation}
with the solution
\begin{equation}\label{Eq:HiItinerant}
\chi(\tilde{\mathbf{q}})=\,\vcenter{\hbox{\scalebox{0.5}{\fmfreuse{feyHi}}}}\,=
\frac{\vcenter{\hbox{\scalebox{0.5}{
\fmfreuse{feyHiIrr}}}}}{1-\,\vcenter{\hbox{\scalebox{0.5}{
\fmfreuse{feyHiIrr}}}}\kern-1.5pt\vcenter{\hbox{\scalebox{0.5}{
\fmfreuse{feyPhoton}}}}}\,=\frac{\chi^{\text{irr}}(\tilde{\mathbf{q}})}{1-W(\tilde{\mathbf{q}})\chi^{\text{irr}}(\tilde{\mathbf{q}})}\text{,}
\end{equation}
where $W(\tilde{\mathbf{q}})=\frac{4\piup e^2}{q^2}$ is the unscreened Coulomb interaction (in case we include only
Coulomb interaction as a perturbation into the Hamiltonian). Note that according to the Feynman rules, each fermion
loop
in (\ref{Eq:FeynSeriesHi}) is defined with a minus sign, as in Eq. (\ref{Eq:Hi0BareMatsubara}). This holds true as
long
as the total spin of the loop is zero. If we consider a similar expansion for the spin density correlation function,
an
additional minus sign should appear in front of each fermion loop, which is equivalent to the change of sign in the
denominator of (\ref{Eq:HiItinerant}).

The simplest possible approximation for $\chi^{\text{irr}}(\tilde{\mathbf{q}})$, known as \textit{random phase
approximation} (RPA), is the pair-bubble diagram:
$\chi^{\text{irr}}(\tilde{\mathbf{q}})\approx\vcenter{\hbox{\scalebox{0.5}{
\fmfreuse{feyHi0}}}}=\chi_0(\tilde{\mathbf{q}})$. In this approximation the density correlation function becomes the
sum
of the following multi-loop diagrams, which can be shown to provide the highest order contribution in the high-density
limit \cite{BruusFlensberg81,FetterWalecka71}:
\begin{equation}\label{Eq:MultiLoopSeries}
\vcenter{\hbox{\scalebox{0.5}{
\begin{fmffile}{feyHiRPA}
\begin{fmfgraph*}(60,34)\fmfkeep{feyHiRPA}
\fmfpoly{filled=30,label=\scalebox{1.2}{RPA}}{m1,m2,m3,m4}\fmfsurround{r,m1,m2,l,m3,m4}
\fmf{fermion,left=0.2,tension=0.5}{l,m2}\fmf{fermion,left=0.2}{m3,l}
\fmf{fermion,left=0.2,tension=0.5}{m1,r}\fmf{fermion,left=0.2}{r,m4}
\fmffreeze\fmfshift{-0.0w,0}{m1}\fmfshift{0.0w,0}{m2}\fmfshift{0.0w,0}{m3}\fmfshift{-0.0w,0}{m4}
\end{fmfgraph*}
\end{fmffile}}}}=\,\vcenter{\hbox{\scalebox{0.5}{\fmfreuse{feyHi0}}}}\,+\,
\vcenter{\hbox{\scalebox{0.5}{\fmfreuse{feyHiOa}}}}\,+\,
\vcenter{\hbox{\scalebox{0.5}{\begin{fmffile}{feyHiOb}\begin{fmfgraph*}(220,30)
\fmfleft{l}\fmfright{r}\fmf{fermion,left=0.5}{l,m1,l}\fmf{photon,tension=6}{m1,m2}\fmf{fermion,left=0.5}{m2,m3,m2}
\fmf{photon,tension=6}{m3,m4}\fmf{fermion,left=0.5}{m4,r,m4}
\end{fmfgraph*}
\end{fmffile}}}}+\,\cdots.
\end{equation}

A further improvement to this approximation can be made by performing partial summation of (\ref{Eq:FeynSeriesHi}) to
account for the renormalization of the Green's functions entering the Lindhard function, but still ignoring the
diagrams
with interaction lines connecting the two Green's functions (\textit{vertex corrections}), which results in
\begin{equation}\label{Eq:RenormHi0}
\chi^{\text{irr}}(\tilde{\mathbf{q}})\approx\kern-3pt\vcenter{\hbox{\scalebox{0.5}{
\begin{fmffile}{feyHi0dr}\begin{fmfgraph*}(60,30)\fmfkeep{feyHi0dr}\fmfleft{l}\fmfright{r}\fmf{dbl_plain_arrow,left=0.5}{l,r,l}
\end{fmfgraph*}
\end{fmffile}}}}=\frac{1}{\beta}\int\!\!\frac{\mathrm{d}\tilde{\mathbf{k}}}{(2\piup)^d}\sum\limits_\sigma\,
\mathcal{G}(\tilde{\mathbf{k}}+\tilde{\mathbf{q}},\,\sigma)\mathcal{G}^*(\tilde{\mathbf{k}},\,\sigma)\text{,}
\end{equation}
where $\mathcal{G}(\tilde{\mathbf{k}},\,\sigma)$ is the Matsubara representation of the renormalized Green's function
(\ref{Eq:RenormG}). This particular modification of the RPA will be used in section \ref{Sec:BSCCO_RPA}.

Up to now we have neglected non-Coulomb interactions in the electron subsystem. Let us see how the result
(\ref{Eq:HiItinerant}) changes in the presence of some other interaction (e.g. impurity or phonon scattering)
characterized by the Fourier component $W'(\tilde{\mathbf{q}})$. Again, if we consider only multi-loop diagrams of the
form (\ref{Eq:MultiLoopSeries}), either the $W$- or $W'$-interaction lines will interleave between the ``bubbles'',
and
Eq. (\ref{Eq:HiItinerant}) takes the form
\begin{equation}\label{Eq:Dyson2Interactions}
\chi(\tilde{\mathbf{q}})=\frac{\chi^{\text{irr}}(\tilde{\mathbf{q}})}
{1-W(\tilde{\mathbf{q}})\chi^{\text{irr}}(\tilde{\mathbf{q}})-W'(\tilde{\mathbf{q}})\chi^{\text{irr}}(\tilde{\mathbf{q}})}\text{.}
\end{equation}
Now we see that any new interaction simply adds up with the Coulomb term, so after denoting
$W_\text{eff}(\tilde{\mathbf{q}})=W(\tilde{\mathbf{q}})+W'(\tilde{\mathbf{q}})$ (which can be done for as many
interactions present in the system as needed), we again arrive at Eq. (\ref{Eq:HiItinerant}), with the effective
interaction $W_\text{eff}(\tilde{\mathbf{q}})$ in place of the original $W(\tilde{\mathbf{q}})$. Of course, the
irreducible susceptibility itself changes after another interaction is switched on. But if we assume we know it, e.g.
by
approximating it with the Lindhard function in RPA, then the calculation of $\chi(\tilde{\mathbf{q}})$ reduces to
finding the effective interaction $W_\text{eff}(\tilde{\mathbf{q}})$, which is in some cases easier than calculation
of
each interaction term separately.

\section{Superconductivity}

\subsection{Bardeen-Cooper-Schrieffer theory}\label{SubSec:BCS}

The idea behind the Bardeen-Cooper-Schrieffer (BCS) theory of superconductivity \cite{Cooper56, BCS57Feb, BCS57Jul} is
that the many-particle state consisting of independent single-particle Bloch waves (Sommerfeld-Bloch free electron
model) is unstable to the formation of correlations between states with opposite momenta and spins (so-called
\textit{Cooper pairs}), i.e. if a state $\mathbf{k}\!\uparrow$ is occupied, so is $-\mathbf{k}\!\downarrow$, making the
ground-state expectation value $\langle\hat{c}^\dag_{\mathbf{k}\uparrow}\hat{c}^\dag_{-\mathbf{k}\downarrow}\rangle$
essentially nonzero. The Cooper instability is due to the effective phonon-mediated electron-electron interaction, which
turns out to be attractive for energies $|\epsilon_\mathbf{k}|<\omega_\mathrm{D}\ll\epsilon_\text{F}$. Creation and
annihilation operators for such pairs are defined as
$\hat{b}^\dag_\mathbf{k}=\hat{c}^\dag_{\mathbf{k}\uparrow}\hat{c}^\dag_{-\mathbf{k}\downarrow}$ and
$\hat{b}^{\phantom{\dag}}_\mathbf{k}=\hat{c}^{\phantom{\dag}}_{\mathbf{-k}\downarrow}\hat{c}^{\phantom{\dag}}_{\mathbf{k}\uparrow}$.
The $2N$-particle wave function satisfying the pairing condition would thus be
$\sum_{\mathbf{k}_1\ldots\mathbf{k}_N}g_{\mathbf{k}_1\ldots\mathbf{k}_N}\hat{b}^\dag_{\mathbf{k}_1}\ldots\hat{b}^\dag_{\mathbf{k}_N}|\text{FS}\rangle$,
where $|\text{FS}\rangle$ is the vacuum state\footnote{The vacuum state of some operator algebra
$\{\hat{\alpha}^{\phantom{\dag}}_{\mathbf{k}},\,\hat{\alpha}^\dag_{\mathbf{k}}\}$ is the unique state annihilated by any
annihilation operator $\hat{\alpha}_{\mathbf{k}}$.} of the
$\{\hat{c}^{\phantom{\dag}}_{\mathbf{k}},\,\hat{c}^\dag_{\mathbf{k}}\}$ operators (FS stands for `Fermi sphere'), and
the Hamiltonian incorporating the pairing condition can be written as \cite[p.\,329]{BruusFlensberg81}
\begin{equation}
{\hat{\mathcal{H}\,}\kern-2pt}_\text{BCS}=\sum_{\mathbf{k}\sigma}\epsilon_\mathbf{k}\,\hat{c}^\dag_{\mathbf{k}\sigma}\hat{c}^{\phantom{\dag}}_{\mathbf{k}\sigma}+\sum_{\mathbf{k}\mathbf{k'}}
V_{\mathbf{k}\mathbf{k}\smash{'}}\,\hat{b}^\dag_{\mathbf{k}}\hat{b}^{\phantom{\dag}}_{\mathbf{k}\smash{'}}
\end{equation}
The BCS theory treats this Hamiltonian within the mean-field approximation \cite{BruusFlensberg81}:
\begin{equation}\label{Eq:BCSMF}
{\hat{\mathcal{H}\,}\kern-2pt}^\text{MF}_\text{BCS}=\sum_{\mathbf{k}\sigma}\epsilon_\mathbf{k}\,\hat{c}^\dag_{\mathbf{k}\sigma}\hat{c}^{\phantom{\dag}}_{\mathbf{k}\sigma}-\sum_\mathbf{k}\mathit{\Delta^{\phantom{*}}_\mathbf{k}}\hat{b}^\dag_\mathbf{k}-\sum_\mathbf{k}\mathit{\Delta^*_\mathbf{k}}\hat{b}^{\phantom{\dag}}_\mathbf{k}\text{,}
\end{equation}
where $\mathit{\Delta}_\mathbf{k}=-\sum_\mathbf{k'}V_{\mathbf{k}\mathbf{k'}}\langle \hat{b}_\mathbf{k}\rangle$ is the
\textit{BCS order parameter}. The mean-field Hamiltonian is quadratic in the electron operators, but not diagonal, so
it
is natural to introduce a unitary
($u^*_\mathbf{k}u^{\phantom{*}}_\mathbf{k}+v^*_\mathbf{k}v^{\phantom{*}}_\mathbf{k}=1$) transformation which
diagonalizes (\ref{Eq:BCSMF}):
\begin{equation}\label{Eq:BogoliubovTransform}
\text{\raisebox{-0.7ex}{$\Biggl($}}\begin{matrix}\hat{\gamma}^{\phantom{\dag}}_{\mathbf{k}\uparrow}\\\hat{\gamma}^\dag_{-\mathbf{k}\downarrow}\end{matrix}\text{\raisebox{-0.7ex}{$\Biggr)$}}\text{\raisebox{-0.7ex}{$=$}}
\text{\raisebox{-0.7ex}{$\Biggl($}}\begin{matrix}u^{\phantom{*}}_\mathbf{k}&\kern-.8ex-v^{\phantom{*}}_\mathbf{k}\\v^*_\mathbf{k}&u^*_\mathbf{k}\end{matrix}\text{\raisebox{-0.7ex}{$\Biggr)$}}
\text{\raisebox{-0.7ex}{$\Biggl($}}\begin{matrix}\hat{c}^{\phantom{\dag}}_{\mathbf{k}\uparrow}\\\hat{c}^\dag_{-\mathbf{k}\downarrow}\end{matrix}\text{\raisebox{-0.7ex}{$\Biggr)$}}
\end{equation}
The newly introduced operators $\hat{\gamma}^\dag_{-\mathbf{k}\downarrow}$ and
$\hat{\gamma}^{\phantom{\dag}}_{\mathbf{k}\uparrow}$ create and annihilate the so-called \textit{Bogoliubov
quasiparticles} (bogoliubons) \cite{Bogoliubov58}\,---\,special linear combinations of the particle and hole states,
diagonalizing the BCS Hamiltonian. Writing down explicitly the coefficients of the transformation matrix, we arrive at
the so-called \textit{gap equations}:
\begin{equation}\label{Eq:GapEquations}
u^*_\mathbf{k}u^{\phantom{*}}_\mathbf{k}=\text{\scalebox{0.9}{$\frac{1}{2}$}}\biggl(1+\frac{\epsilon_\mathbf{k}}{E_\mathbf{k}}\biggr)\text{;~~}
v^*_\mathbf{k}v^{\phantom{*}}_\mathbf{k}=\text{\scalebox{0.9}{$\frac{1}{2}$}}\biggl(1-\frac{\epsilon_\mathbf{k}}{E_\mathbf{k}}\biggr)\text{;~~}
E_\mathbf{k}^2=\epsilon_\mathbf{k}^2+|\mathit{\Delta}_\mathbf{k}|^2\text{.}
\end{equation}
Assuming that the $\mathbf{k}$-dependence of $\mathit{\Delta}_\mathbf{k}$ is much weaker than that of the bare
dispersion $\epsilon_\mathbf{k}$, the Bogoliubov quasiparticles will have a finite minimal energy
$\mathit{\Delta}=\mathit{\Delta}_{\mathbf{k}_\text{F}}$, known as the \textit{energy gap}. Rewriting the Hamiltonian
(\ref{Eq:BCSMF}) in terms of the newly defined operators, we get
\begin{equation}
{\hat{\mathcal{H}\,}\kern-2pt}^\text{MF}_\text{BCS}=\sum_\mathbf{k}E_\mathbf{k}\bigl(\hat{\gamma}^\dag_{\mathbf{k}\uparrow}\hat{\gamma}^{\phantom{\dag}}_{\mathbf{k}\uparrow}+\hat{\gamma}^\dag_{\mathbf{k}\downarrow}\hat{\gamma}^{\phantom{\dag}}_{\mathbf{k}\downarrow}\bigr)+E_0\text{,}
\end{equation}
where
$E_0=\bigl\langle\psi^0_\text{BCS}\big|{\hat{\mathcal{H}\,}\kern-2pt}^\text{MF}_\text{BCS}\big|\psi^0_\text{BCS}\bigr\rangle$
is the ground-state energy. The BCS ground state itself, i.e. the vacuum state of the Bogoliubov operators, can be
written as
\begin{equation}
|\psi^0_\text{BCS}\rangle=\prod_\mathbf{k}\hat{\gamma}^{\phantom{\dag}}_{\mathbf{k}\uparrow}\hat{\gamma}^{\phantom{\dag}}_{\mathbf{-k}\downarrow}|\text{FS}\rangle
\bigg/\Bigl\|\prod_\mathbf{k}\hat{\gamma}^{\phantom{\dag}}_{\mathbf{k}\uparrow}\hat{\gamma}^{\phantom{\dag}}_{\mathbf{-k}\downarrow}|\text{FS}\rangle\Bigr\|
=\prod_\mathbf{k}\bigl(u^{\phantom{\dag}}_\mathbf{k}+v^{\phantom{\dag}}_\mathbf{k}\hat{b}^\dag_\mathbf{k}\bigr)|\text{FS}\rangle\text{.}
\end{equation}

\subsection{Green's functions of the superconducting state}\label{SubSec:SupercondG}

As already mentioned above, for the Cooper pair condensate the expectation value
$\langle\hat{c}^\dag_{\mathbf{k}\uparrow}\hat{c}^\dag_{-\mathbf{k}\downarrow}\rangle$ is nonzero because of the
correlations between states $\mathbf{k}\!\uparrow$ and $-\mathbf{k}\!\downarrow$. Therefore the following two
Matsubara
Green's functions can be defined \cite[p.\,336]{BruusFlensberg81}:
\begin{subequations}\label{Eq:GFMatsubaraSupercondDef}
\begin{align}
\hfill&\text{normal:}&\mathcal{G}_{\uparrow\uparrow}(\mathbf{k},\,\tau)&=-\bigl\langle
T_\tau\,\hat{c}^{\phantom{\dag}}_{\mathbf{k}\uparrow}(\tau)\,\hat{c}^\dag_{\mathbf{k}\uparrow}(0)\bigr\rangle\text{,}&\hfill\\
\hfill&\text{anomalous:}&\mathcal{F}_{\downarrow\uparrow}(\mathbf{k},\,\tau)&=-\bigl\langle
T_\tau\,\hat{c}^\dag_{-\mathbf{k}\downarrow}(\tau)\,\hat{c}^\dag_{\mathbf{k}\uparrow}(0)\bigr\rangle\text{.}&\hfill
\end{align}
\end{subequations}
By writing the equations of motion for the mean-field BCS Hamiltonian (\ref{Eq:BCSMF}) with subsequent Fourier
transform
to Matsubara frequency space \cite{BruusFlensberg81}, one gets
\begin{equation}\label{Eq:GFMatsubaraSupercond}
\mathcal{G}_{\uparrow\uparrow}(\tilde{\mathbf{k}})=\frac{\mathrm{i}\mathit{k}_n+\epsilon_\mathbf{k}}{(\mathrm{i}\mathit{k}_n)^2-E_\mathbf{k}^2}\text{;\quad}\mathcal{F}_{\downarrow\uparrow}(\tilde{\mathbf{k}})=\frac{-\mathit{\Delta}^*_\mathbf{k}}{(\mathrm{i}\mathit{k}_n)^2-E_\mathbf{k}^2}
\end{equation}
If we combine the electron operators into a single \textit{Nambu spinor} \cite{Nambu59}
$\hat{\alpha}^\dag_\mathbf{k}=\bigl(\text{\raisebox{0.5ex}{$\begin{matrix}\hat{c}^\dag_{\mathbf{k}\uparrow}&\kern-1ex\hat{c}^{\phantom{\dag}}_{-\mathbf{k}\downarrow}\end{matrix}$}}\bigr)$,
similarly to (\ref{Eq:BogoliubovTransform}), the normal and anomalous Green's functions become elements of a
2$\kern.5pt\times$2 matrix called the \textit{Nambu Green's function}:
\begin{equation}
\stackrel{=}{\smash[t]{\mathcal{G}\,}\vphantom{x}}\!\!(\mathbf{k},\,\tau)=-\bigl\langle
T_\tau\,\hat{\alpha}^{\phantom{\dag}}_{\mathbf{k}\uparrow}(\tau)\,\hat{\alpha}^\dag_{\mathbf{k}\uparrow}(0)\bigr\rangle
=\Biggl(\text{\raisebox{0.5ex}{$\begin{matrix}\mathcal{G}^{\phantom{*}}_{\uparrow\uparrow}(\mathbf{k},\,\tau)&\mathcal{F}^*_{\downarrow\uparrow}(\mathbf{k},\,\tau)\phantom{-}\\\mathcal{F}^{\phantom{*}}_{\downarrow\uparrow}(\mathbf{k},\,\tau)&\mathcal{G}^*_{\downarrow\downarrow}(-\mathbf{k},\,\tau)\end{matrix}$}}\Biggr)
\end{equation}
or, in the frequency domain,\vspace{-1ex}
\begin{equation}\label{Eq:GreensFuncTensor}
\stackrel{=}{\smash[t]{\mathcal{G}\,}\vphantom{x}}\!\!(\tilde{\mathbf{k}})=\frac{1}{(\mathrm{i}k_n)^2-E_\mathbf{k}^2}\,
\Biggl(\text{\raisebox{0.5ex}{$\begin{matrix}\mathrm{i}k_n+\epsilon_\mathbf{k}&-\mathit{\Delta}_\mathbf{k}\\-\mathit{\Delta}^*_\mathbf{k}&\mathrm{i}k_n-\epsilon_\mathbf{k}\end{matrix}$}}\Biggr)\text{.}
\end{equation}
Let us now see how the expression for the renormalized Lindhard function (\ref{Eq:RenormHi0}) changes if written for
the
Nambu Green's function. According to the Feynman rules, summation over all internal variables has to be performed, in
particular the Nambu indices. In the assumption that the Green's functions are independent of the spin, we arrive at
an
important expression to be used in section \ref{Sec:BSCCO_RPA} for calculating the dynamic spin susceptibility of a
high-temperature superconductor:
\begin{multline}\label{Eq:RenormHi0Nambu}
\vcenter{\hbox{\scalebox{0.5}{
\begin{fmffile}{feyHi0Nambu}\begin{fmfgraph*}(60,60)\fmfleft{l}\fmfright{r}
\fmf{dbl_plain_arrow,left=0.5,label=\scalebox{1.5}{$\stackrel{=}{\smash[t]{\mathcal{G}\kern1pt}\vphantom{x}}\!\!(\tilde{\mathbf{k}}+\tilde{\mathbf{q}})$}}{l,r}
\fmf{dbl_plain_arrow,left=0.5,label=\scalebox{1.5}{$\stackrel{=}{\smash[t]{\mathcal{G}\kern1pt}\vphantom{x}}\!\!(\tilde{\mathbf{k}})$}}{r,l}
\end{fmfgraph*}\end{fmffile}}}}
=-\frac{1}{(2\piup)^d\beta}\int\!\!\,
\mathrm{Tr}\bigl[\stackrel{=}{\smash[t]{\mathcal{G}\,}\vphantom{x}}\!\!(\tilde{\mathbf{k}}+\tilde{\mathbf{q}})
\stackrel{=}{\smash[t]{\mathcal{G}\,}\vphantom{x}}\!\!\smash{^*}(\tilde{\mathbf{k}})\bigr]\mathrm{d}\tilde{\mathbf{k}}\\
=-\frac{2}{(2\piup)^d\beta}\int\!\!\,\bigl[\mathcal{G}^{\phantom{*}}\!\!(\tilde{\mathbf{k}}+\tilde{\mathbf{q}})\,\mathcal{G}^*(\tilde{\mathbf{k}})
+\mathcal{F}^{\phantom{*}}\!\!(\tilde{\mathbf{k}}+\tilde{\mathbf{q}})\,\mathcal{F}^*(\tilde{\mathbf{k}})\bigr]\mathrm{d}\tilde{\mathbf{k}}\text{.}
\end{multline}

\section{Self-energy manifestation in two physical models}

\subsection{The notion of a Fermi liquid}\label{SubSec:NotionFL}

\textbf{General idea.} In a Fermi gas (non-interacting fermion system) all thermodynamic properties, such as specific
heat or susceptibility, are analytic functions of both energy $\omega$ and temperature $T$
\cite[p.\,7]{MaslovLectureNotes05}. The main idea behind the Fermi liquid theory \cite{Landau57, Landau59} is to
consider interacting systems in an approximation which conserves this analyticity condition, making the interacting
system similar in this respect to a non-interacting Fermi gas. In a weakly interacting regime, when only the leading
terms in the power expansions of the thermodynamic properties in $T$ and $\omega$ are preserved, the Fermi liquid is
often a good approximation.

If the retarded self-energy $\mathit{\Sigma}(\omega,\,\epsilon_\mathbf{k},\,T)$ is an analytic function of $\omega$
and
$T$, it can be expanded into a power series around zero. From the causality of the retarded self-energy it follows
that
its real part $\mathit{\Sigma}'$ is odd, and the imaginary part $\mathit{\Sigma}''$ is even in $\omega$ (see
appendix~\ref{Appendix:KK}). This lets us write down the following expansion for the self-energy\footnote{For the
derivation of the $\piup^2(k_\text{B}T)^2$ term, see Ref.\,\citenum{Hewson97},\,p.\,124.} \cite{AbrikosovGorkov63,
MaslovLectureNotes05, MaslovLectureNotes07, ChubukovMaslov04, ChubukovMaslov03, FetterWalecka71, Hewson97}:
\begin{equation}\label{Eq:SelfEnergyExpansion}
\mathit{\Sigma}(\omega,\,\epsilon_\mathbf{k})=\mathit{\Sigma}_0-\lambda_\omega\omega+\lambda_{\epsilon_\mathbf{k}}\epsilon_\mathbf{k}-\mathrm{i}(1+\lambda_\omega)\gamma\bigl[\omega^2+\piup^2(k_\text{B}T)^2\bigr]+\text{higher
order terms}\text{,}
\end{equation}
where $\lambda_\omega$ is called the coupling constant, and $\gamma$ is the Sommerfeld coefficient. The real part of the
zero-order term $\mathit{\Sigma}_0$ can be absorbed in a shift of the chemical potential, so it will be skipped. Its
imaginary part is zero in the two models that we are to consider. The higher-order terms in Eq.
(\ref{Eq:SelfEnergyExpansion}) can be neglected in the weak interaction limit, therefore we end up with the following
expression for the Fermi liquid self-energy:
\begin{equation}\label{Eq:SelfEnergyFL}
\mathit{\Sigma}(\omega,\,\epsilon_\mathbf{k})=-\lambda_\omega\omega+\lambda_{\epsilon_\mathbf{k}}\epsilon_\mathbf{k}-\mathrm{i}\kern.3pt(1+\lambda_\omega)\gamma\bigl[\omega^2+\piup^2(k_\text{B}T)^2\bigr]\text{.}
\end{equation}
The coefficient $\lambda_{\epsilon_\mathbf{k}}$ determines the $\mathbf{k}$-dependence of the self-energy in the
direction orthogonal to the Fermi surface. At low energies, where the spectral function is well localized in momentum,
it can be simply combined with the first term by considering the self-energy dependence only along the dispersion and
neglecting the change in the orthogonal direction. Such approximation will replace the $\lambda_\omega$ coefficient
with
$\tilde{\lambda}_\omega=\lambda_\omega-\lambda_{\epsilon_\mathbf{k}}(1+\lambda_\omega)/(1+\lambda_{\epsilon_\mathbf{k}})$
without affecting the analytic properties of the self-energy. On the other hand, the coefficients
$\tilde{\lambda}_\omega$ and $\gamma$ may also be $\mathbf{k}$-dependent, resulting in a non-negligible variation of
the
self-energy along the Fermi surface.

Substituting (\ref{Eq:SelfEnergyFL}) into the Green's function (\ref{Eq:RenormG}) yields \cite{MaslovLectureNotes07,
Hewson97}
\begin{equation}\label{Eq:GreensFunctionFL}
   G(\mathbf{k},\,\omega)=\frac{(1+\lambda_\omega)^{-1}}{\omega-\epsilon_\mathbf{k}(1\!+\!\lambda_{\epsilon_\mathbf{k}})/(1\!+\!\lambda_\omega)+\mathrm{i}\gamma\bigl[\omega^2\!+\!\piup^2(k_\text{B}T)^2\bigr]}
   =\frac{Z}{\omega-\tilde{\epsilon}_\mathbf{k}+\mathrm{i}/2\tilde{\tau}_\mathbf{k}}\text{,}
\end{equation}
where $Z=\bigl(1+\partial\!\mathit{\Sigma}/\partial\!\omega\big|_{\omega=0}\bigr)^{-1}$ is called the coherence
(renormalization) factor \cite{Lundqvist69}, $\tilde{\epsilon}_\mathbf{k}$ is the renormalized dispersion and
$\tilde{\tau}_\mathbf{k}$ is
the effective ``lifetime'' of the quasiparticle. In a Fermi liquid, the leading order term in the interaction does not
explicitly depend on $\omega$, which corresponds to $Z=1$, but the higher order non-analytic terms may lead to $0 < Z
<
1$. From (\ref{Eq:GreensFunctionFL}) it is clear that in the Fermi liquid regime renormalization changes the
dispersion
from $\epsilon_\mathbf{k}$ to
$\tilde{\epsilon}_\mathbf{k}=\epsilon_\mathbf{k}(1+\lambda_{\epsilon_\mathbf{k}})/(1+\lambda_\omega)$, therefore both
\textit{Fermi velocity}
$v_\text{F}\stackrel{\text{def}}{=}\|\mathrm{grad}\,\epsilon_\mathbf{k}\|_{\mathbf{k}=\mathbf{k}_\text{F}}$ and the
\textit{effective mass} $m\stackrel{\text{def}}{=}k_\text{F}/v_\text{F}$ are renormalized respectively:
\begin{equation}\label{Eq:RenormalizedVM}
\tilde{v}_\text{F}=v_\text{F}\,\frac{1+\lambda_{\epsilon_\mathbf{k}}}{1+\lambda_\omega}=\frac{v_\text{F}}{1+\tilde{\lambda}_\omega}\text{;}\quad
\tilde{m}=m\,\frac{1+\lambda_\omega}{1+\lambda_{\epsilon_\mathbf{k}}}=m\,(1+\tilde{\lambda}_\omega)\text{.}
\end{equation}
At this point we see that the analyticity condition for the self-energy indeed results in a system similar to a Fermi
gas of non-interacting \textit{quasiparticles}. The spectral function is still peaked along a well-defined dispersion
$\tilde{\epsilon}_\mathbf{k}$, and the quasiparticles are characterized by a finite effective mass, which means, among
other things, that the specific heat remains linear in $T$, as for a Fermi gas
\cite[p.\,23]{MaslovLectureNotes07}. Note that for $Z\neq1$ the Green's function (\ref{Eq:GreensFunctionFL}) no longer
satisfies the sum rule $\int_{-\infty}^\infty A(\mathbf{k},\,\omega)\,\mathrm{d}\omega=1$ (which holds for a Fermi
gas),
as the integral amounts only to $Z$ \cite[p.\,278]{BruusFlensberg81}. To preserve the normalization, the coherence
factor is usually omitted in the Green's function definition \cite[p.\,115]{Hewson97}.

In reality the self-energy needs not be an analytic function. It usually includes a non-analytic part
$\mathit{\Sigma}_\text{n.\,a.}$, which can drastically affect the behavior of the system. As a rule, the analytic
terms
dominate in three-dimensional systems, whereas the non-analytic terms dominate in low-dimensional systems (1D or 2D).
Here the asymptotics of the scattering rate $\mathit{\Sigma}''_\text{n.\,a.}$ at small $\omega$ becomes crucial. (i)
If
$\mathit{\Sigma}''_\text{n.\,a.}$ vanishes as $o(\omega^2)$ at $\omega\rightarrow 0$, it can be neglected altogether
at
low energies. This case is called a \textit{normal Fermi liquid} and holds, for example, in a three-dimensional
Coulomb
gas. The higher-order non-analytic terms are present in every real system \cite{MaslovLectureNotes05,
ChubukovMaslov04,
ChubukovMaslov03}, so a normal Fermi liquid is similar to a Fermi gas only in the leading-order approximation. At
higher
energies, the non-analytic sub-leading terms may need to be accounted for. (ii) If the scattering rate behaves
asymptotically as $O(\omega^\alpha)$ with $1 < \alpha < 2$, the non-analytic terms become dominant and can no longer
be
neglected. Still, the quasiparticles are well defined (effective mass remains finite) and the resemblance to the Fermi
gas is retained. The two limiting scenarios of such behavior typically occur in two- and one-dimensional systems. In
2D,
$\mathit{\Sigma}''=O(\omega^2)+O(\omega^2\,\mathrm{ln}|\omega|)$ \cite{CoffeyBedell93, Gangadharaiah05,
ChubukovMaslov05}, so the analytic and non-analytic leading terms have similar asymptotics at low energies, and the
system behaves almost like a normal Fermi liquid. The other limiting case is the so-called \textit{marginal Fermi
liquid} \cite{VarmaLittlewood89, Castellani94, KrotovChubukov06}, where the asymptotics is given by
$\mathit{\Sigma}'\propto\omega\,\mathrm{ln}|\omega|$ and $\mathit{\Sigma}''\propto|\omega|$. At this point the
effective
mass of the quasiparticles starts to diverge logarithmically. The marginal Fermi liquid is a generic many-particle
state
of one-dimensional systems (\textit{Luttinger liquids}) [\citenum{BruusFlensberg81}, p.\,347; \citenum{Mahan00},
p.\,257; \citenum{Luttinger63, Tomonaga50, Schonhammer98}], such as carbon nanotubes \cite{BockrathCobden99},
nanowires
\cite{Auslaender02, Auslaender05}, and quasi-one-dimensional crystals \cite{DagottoRice96, MullerRice99,
HagiwaraTsujii06}. (iii) An even slower asymptotic decay of the scattering rate with $0 < \alpha < 1$ corresponds to
the
essentially non-Fermi-liquid behavior with a sturdy power-law divergence of the effective mass \cite{MathoMueller99,
VekhterChubukov04, MathoNotes}.

\subsection{Self-energy of a normal Fermi liquid}\label{SubSec:SelfEnergyFL}

Up to now, our consideration of the Fermi-liquid self-energy was rather phenomenological.

\noindent In fact, a rigorous microscopic theory starting from the definition of the self-energy
(\ref{Eq:FeynSeriesS})
can be applied to produce the same results. Let us consider the case of a weakly interacting Coulomb gas in a periodic
potential. There are two first-order terms in the series (\ref{Eq:FeynSeriesS}):
\hbox{\hspace{-1ex}\rotatebox{-90}{$\kern-1.2ex\vcenter{\hbox{\scalebox{0.5}{\fmfreuse{feyHartreeS}}}}$}}\hbox{\hspace{3ex}~(Hartree
term)} and \raisebox{-0.5ex}{$\,\vcenter{\hbox{\scalebox{0.5}{\fmfreuse{feyFockS}}}}$} (Fock term). The Hartree term
can
be set to zero by introducing the positive background (jellium) to compensate the electronic charge (otherwise this
term
is divergent) \cite[p.\,94]{Mahan00}. The Fock term, also called \textit{unscreened exchange energy}, is renormalized
by
replacing the Coulomb interaction line by the screened Coulomb interaction line \cite[p.\,250]{BruusFlensberg81} to
obtain the \textit{screened exchange energy}:
\begin{equation}
\widetilde{W}(\tilde{\mathbf{q}})\equiv\vcenter{\hbox{\scalebox{0.5}{\begin{fmffile}{feyScrPhot}\begin{fmfgraph*}(60,50)
\fmfleft{l} \fmfright{r}
\fmf{dbl_wiggly}{l,r}\end{fmfgraph*}\end{fmffile}}}}\,\stackrel{\text{def}}{=}W(\tilde{\mathbf{q}})\bigl(1+\chi(\tilde{\mathbf{q}})
W(\tilde{\mathbf{q}})\bigr)\stackrel{\text{RPA}}{\approx}W(\tilde{\mathbf{q}})\big/\bigl(1-W(\tilde{\mathbf{q}})\chi_0(\tilde{\mathbf{q}})\bigr)\text{.}
\end{equation}
The leading term of the self-energy (including the most divergent diagrams in the high-density limit) will be then
given
by
\begin{equation}\label{Eq:SelfEnergyEl}
\mathit{\Sigma}_1^\text{RPA}(\tilde{\mathbf{k}},\,\sigma)\equiv \vcenter{\hbox{\scalebox{0.5}{
\begin{fmffile}{feyScrS}\begin{fmfgraph*}(60,50)
\fmfleft{l}\fmfright{r}\fmf{fermion}{l,r}\fmffreeze\fmf{dbl_wiggly,left=0.7}{l,r}
\end{fmfgraph*}\end{fmffile}}}}~=
-\frac{1}{\beta}\int\!\!\frac{\mathrm{d}\tilde{\mathbf{q}}}{(2\piup)^d}\,\widetilde{W}(\tilde{\mathbf{q}})\mathcal{G}_0(\tilde{\mathbf{k}}+\tilde{\mathbf{q}})\text{.}
\end{equation}
After performing the Matsubara summation, substituting the explicit form of the Lindhard function (\ref{Eq:Hi0Bare})
into $\widetilde{W}$, and taking the imaginary part, this expression becomes\footnote{For the details of the
derivation
see Ref.~\citenum{BruusFlensberg81}, p.\,280; Ref.~\citenum{Mahan00}, p.\,92 and 297; Ref.~\citenum{FetterWalecka71},
p.\,268; or Ref.~\citenum{Luttinger61}.}
\begin{multline}\label{Eq:SelfEnergyIm}
\mathrm{Im}\,\mathit{\Sigma}_1^\text{RPA}(\epsilon_\mathbf{k},\,\sigma)=-\!\!\iint\!\!\frac{\mathrm{d}\mathbf{k}'\,\mathrm{d}\mathbf{q}}{(2\piup)^{2d-1}}
\Bigl[n_\text{F}(\epsilon_{\mathbf{k}'})\bigl(1-n_\text{F}(\epsilon_{\mathbf{k}+\mathbf{q}})\bigr)\bigl(1-n_\text{F}(\epsilon_{\mathbf{k}'-\mathbf{q}})\bigr)\\
+\bigl(1-n_\text{F}(\epsilon_{\mathbf{k}'})\bigr)n_\text{F}(\epsilon_{\mathbf{k}+\mathbf{q}})n_\text{F}(\epsilon_{\mathbf{k}'-\mathbf{q}})\Bigr]\\
\times\biggl|\frac{W(q)}{1-W(q)\chi_0(\mathbf{q},\,\epsilon_{\mathbf{k}+\mathbf{q}}-\epsilon_{\mathbf{k}})}\biggr|^2
\delta(\epsilon_{\mathbf{k}'}-\epsilon_{\mathbf{k}'-\mathbf{q}}-\epsilon_{\mathbf{k}+\mathbf{q}}+\epsilon_\mathbf{k})\text{.}
\end{multline}\vspace{-0.8em}

\noindent The following consideration essentially relies upon a linear approximation for the dispersion\footnote{If
the
curvature of the dispersion is taken into account, Eq.~(\ref{Eq:SelfEnergyIm}) no longer produces the correct result,
because the energy and momentum conservation conditions will be impossible to satisfy simultaneously for almost all
pairs of $\mathbf{k}'$ and $\mathbf{q}$. One has to solve the problem self-consistently, taking into account the
broadening of the dispersion due to the self-energy effects, to find out that the asymptotic behavior at
$\omega\rightarrow0$ actually does not depend on the curvature of the bare dispersion, and therefore the treatment in
the linear approximation given here is actually valid. The curvature of the dispersion can however change the behavior
of the self-energy away from zero, making it less steep than in the linear approximation.} $\epsilon_\mathbf{k}\approx
\mathbf{v}_\mathbf{F}(\mathbf{k}_\text{F})\cdot(\mathbf{k}-\mathbf{k}_\text{F})$ in the limit $\omega\rightarrow0$.
The
first term in the square brackets corresponds to the scattering of an electron in state $|\mathbf{k}\rangle$ into the
previously unoccupied state $|\mathbf{k}+\mathbf{q}\rangle$ by an electron in state $|\mathbf{k}'\rangle$, which is in
turn scattered into a previously unoccupied state $|\mathbf{k}'-\mathbf{q}\rangle$, as illustrated in
Fig.\,\ref{Fig:FL-scatter} (a). The second term corresponds to the filling of a hole in state $|\mathbf{k}\rangle$ by
an
electron scattered from an occupied state $|\mathbf{k}+\mathbf{q}\rangle$ by an electron in state
$|\mathbf{k}'-\mathbf{q}\rangle$, which after the scattering event fills the hole in state $|\mathbf{k}'\rangle$, as
shown in panel (b) of the same figure. The $\delta$-function takes care of the energy conservation. As seen from the
figure, the phase space available for $\mathbf{q}$ is given by $|\mathbf{k}-\mathbf{k}_\text{F}|$, whereas
$\mathbf{k}'$
can be arbitrarily chosen between $\mathbf{k}_\text{F}$ and $\mathbf{k}_\text{F}+\mathbf{q}$. Therefore, at zero
temperature the scattering probability scales as $\epsilon_\mathbf{k}^2/2$.

\begin{floatingfigure}[p]{0.5\textwidth}
\noindent\includegraphics[width=0.5\textwidth]{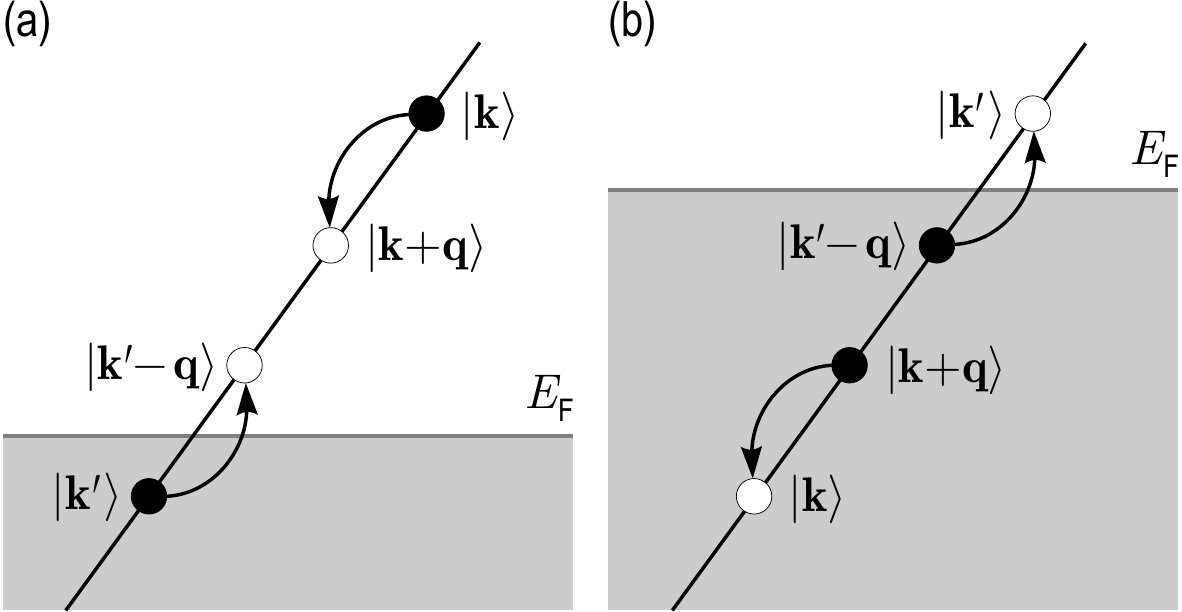}%
\caption{Graphical illustration of the first (a) and second (b) terms of Eq.~(\ref{Eq:SelfEnergyIm}).\vspace{-1.5em}}
\label{Fig:FL-scatter}
\end{floatingfigure}%

In 1958, J.\,J.\,Quinn and R.\,A.\,Ferrell \cite{QuinnFerrell58} have evaluated expression (\ref{Eq:SelfEnergyIm})
explicitly for a homogeneous three-dimensional electron gas in the zero-temperature limit. Their result was
$\tau_\mathbf{k}^{-1}\equiv-2\mathit{\Sigma}''(\epsilon_\mathbf{k},\,\sigma)=\frac{\sqrt{3}\piup^2}{128}\omega_\text{p}(\epsilon_\mathbf{k}/\epsilon_\text{F})^2$,
where $\omega_\text{p}$ is the plasma frequency. Again, as we see, in this simple model the leading order term of the
self-energy behaves quadratically in $\epsilon_\mathbf{k}$ and vanishes at zero energy. In 1960 J.\,M.\,Luttinger
\cite{Luttinger61} has rigorously shown that this behavior holds to all orders of perturbation theory in the
interaction, the higher-order diagrams corresponding to the sub-leading (nonanalytic) corrections.

\subsection{Self-energy due to the coupling to a bosonic mode}\label{SubSec:SelfEnergyMode}

Let us assume that there exists a collective excitation in the system, for example a phonon or a spin resonance mode,
with energy $\mathit{\Omega}$, which we will for simplicity consider to be $\mathbf{k}$-independent. The self-energy
term originating from such an interaction was originally considered by S.\,Engelsberg and J.\,R.\,Schrieffer in 1963
\cite{EngelsbergSchrieffer63} in a coupled electron-phonon system, but coupling to any other bosonic mode is fully
analogous \cite{FongKeimer95, FongKeimer96, IvanovLoktev98, ManskeEremin03}. Similarly to the Eq.
(\ref{Eq:SelfEnergyEl}), the leading order contribution to the self-energy due to the one-phonon interaction can be
written as \cite[p.\,476]{Mahan00}
\begin{equation}\label{Eq:SelfEnergyPh}
\mathit{\Sigma}_1^\text{ph}(\tilde{\mathbf{k}},\,\sigma)\equiv \vcenter{\hbox{\scalebox{0.5}{
\begin{fmffile}{feyScrSPh}\begin{fmfgraph*}(60,50)
\fmfleft{l}\fmfright{r}\fmf{fermion}{l,r}\fmffreeze\fmf{dbl_curly,right=0.7}{r,l}\fmffreeze\fmfv{decor.shape=circle,decor.filled=full,decor.size=60}{l,r}
\end{fmfgraph*}\end{fmffile}}}}~=
-\frac{1}{\beta}\int\!\!\frac{\mathrm{d}\tilde{\mathbf{q}}}{(2\piup)^d}\,
\frac{V_\mathbf{q}^2}{\varepsilon(\tilde{\mathbf{q}})^2}\,
\mathcal{D}(\tilde{\mathbf{q}})\mathcal{G}_0(\tilde{\mathbf{k}}+\tilde{\mathbf{q}})\text{,}
\end{equation}
where $V_\mathbf{q}\equiv\bullet$ is the matrix element determining the coupling of an electron with momentum
$\mathbf{q}$ to the boson, $\mathcal{D}(\tilde{\mathbf{q}})$ is a boson propagator, and the dielectric function
$\varepsilon(\tilde{\mathbf{q}})$ determines the screening of the electron-boson interaction, which is denoted by a
double curly line in the Feynman notation.

According to the Einstein model, the collective mode in a solid can be treated as an undamped oscillator with the
(unscreened) propagator $\mathcal{D}_0(\mathrm{i}\omega_n)\equiv
\vcenter{\hbox{\scalebox{0.5}{\begin{fmffile}{feyScrPhon}
\begin{fmfgraph*}(60,50)\fmfleft{l}\fmfright{r}\fmf{curly}{r,l}\end{fmfgraph*}\end{fmffile}}}}\,
=-2\mathit{\Omega}\big/(\omega_n^2+\mathit{\Omega}^2)$, which can be used as a first-order approximation for
$\mathcal{D}(\tilde{\mathbf{q}})$. The matrix element $V_\mathbf{q}$ can be combined with the dielectric function into
a
screened interaction $\widetilde{V}(\tilde{\mathbf{q}})\equiv V_\mathbf{q}^2/\varepsilon(\tilde{\mathbf{q}})^2$. If
the
boson frequency is much less than the plasmon frequency, the dielectric function can be approximated by its static
value
$\varepsilon(\mathbf{q})$, neglecting the $\omega$-dependence of $\widetilde{V}(\tilde{\mathbf{q}})$. The simplest model
for the dielectric function is the so-called Thomas-Fermi model \cite{Thomas27, Fermi28}
$\varepsilon(q)=1+q_\text{TF}^2/q^2$, where $q_\text{TF}=\sqrt{6\piup e^2 n_0/E_\text{F}}$ is the Thomas-Fermi
screening
wave vector ($n_0$ stands for the equilibrium charge density). In these approximations, after the Matsubara summation
\cite[p.\,136]{Mahan00}, the self-energy (\ref{Eq:SelfEnergyPh}) simplifies to
\begin{equation}\label{Eq:SelfEnergyPh2}
\mathit{\Sigma}_1^\text{ph}(\mathbf{k},\,\omega)=\int\!\!\frac{\mathrm{d}\mathbf{q}}{(2\piup)^d}\,
\widetilde{V}(\mathbf{k}-\mathbf{q})\Bigl[\frac{1+n_{\mathit{\Omega}}-n_\text{F}(\epsilon_\mathbf{q})}{\omega-\epsilon_\mathbf{q}-\mathit{\Omega}+\mathrm{i}0^+}
+\frac{n_{\mathit{\Omega}}+n_\text{F}(\epsilon_\mathbf{q})}{\omega-\epsilon_\mathbf{q}+\mathit{\Omega}+\mathrm{i}0^+}\Bigr]\text{,}
\end{equation}
where $n_{\mathit{\Omega}}=1/(\mathrm{e}^{\beta\mathit{\Omega}}-1)$ is the thermal occupation factor of the boson.

At zero temperature, the integral (\ref{Eq:SelfEnergyPh2}) can be evaluated explicitly, which results in the following
form of the imaginary part:
\begin{equation}\label{Eq:SelfEnergyPh3}
\mathrm{Im}\,\mathit{\Sigma}_1^\text{ph}(\mathbf{k},\,\omega)\propto\theta(\omega-\mathit{\Omega})g_\mathbf{k}(\omega-\mathit{\Omega})
+\theta(-\omega-\mathit{\Omega})g_\mathbf{k}(\omega+\mathit{\Omega})\text{,}
\end{equation}
where $g_\mathbf{k}\text{(}\omega\text{)}$ is a sufficiently smooth function such that in most applications it can be
considered constant (sometimes called \textit{renormalized coupling constant}) by taking the limits
$\omega\rightarrow0$
and $\mathbf{k}\rightarrow0$.\footnote{For more details about this approximation and for the explicit form of
$g_\mathbf{k}\text{(}\omega\text{)}$, see Ref.~\citenum{Mahan00},~p.\,477\,--\,479.}

If from the very beginning of our considerations the screening of the electron-phonon interaction and the momentum
dependence of the coupling constant were neglected, as was originally done in Ref.~\citenum{EngelsbergSchrieffer63},
we
would end up with $g_\mathbf{k}\text{(}\omega\text{)}$ strictly independent of both $\mathbf{k}$ and $\omega$, but
this
constant would be, generally speaking, different from the renormalized one.

By looking at (\ref{Eq:SelfEnergyPh3}) we see that the scattering rate has a sharp onset at $\pm\mathit{\Omega}$ and
is
zero inside this interval. The real part of the self-energy, which can be calculated by the Kramers-Kronig transform
(see appendix \ref{Appendix:KK}), has a logarithmic divergence at $\pm\mathit{\Omega}$:
\begin{equation}\label{Eq:SelfEnergyT0}
\mathit{\Sigma}^\text{ph}(\omega)\approx
-\frac{g}{\piup}\,\mathrm{ln}\biggl|\frac{\omega+\mathit{\Omega}}{\omega-\mathit{\Omega}}\biggr|
-\mathrm{i}g\bigl[\theta(\omega-\mathit{\Omega})-\theta(-\omega-\mathit{\Omega})\bigr]\text{,}
\end{equation}
where $g$ is the coupling constant.

In the superconducting state, the corresponding Green's function (\ref{Eq:GFMatsubaraSupercond}) is to be substituted
into (\ref{Eq:SelfEnergyPh}), which according to Ref.~\citenum{FinkKoitzsch06} yields for $\omega<0$ the following
scattering rate:
\begin{equation}\label{Eq:SelfEnergyT0Im}
\mathrm{Im}\,\mathit{\Sigma}_{\text{SC}}^\text{ph}(\omega)\propto
-\frac{1}{\mathit{\Omega}}\,\mathrm{Re}\,\frac{\omega+\mathit{\Omega}}{\sqrt{(\omega+\mathit{\Omega})^2-\mathit{\Delta}^2}}\text{.}
\end{equation}

\begin{figure}[t]
\!\includegraphics[width=1.01\textwidth]{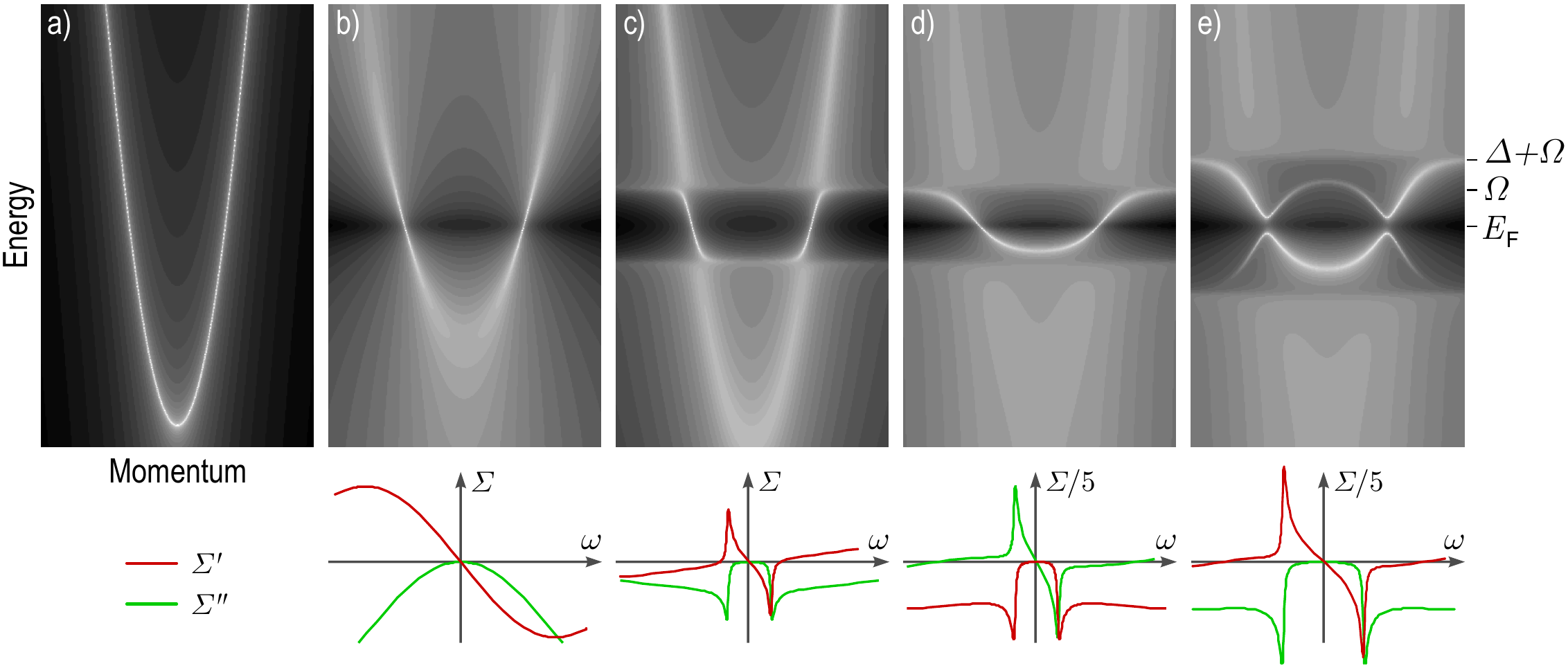}\vspace{-.5em}%
\caption{Spectral function plotted in logarithmic color scale assuming \textbf{(a)}~no $\omega$-dependence of the
self-energy; \textbf{(b)}~Fermi-liquid-like scattering rate; \textbf{(c)} weak coupling to a bosonic mode
$\mathit{\Omega}$; \textbf{(d)}~both Fermi liquid contribution and strong coupling to a bosonic mode; \textbf{(e)}~the
same as (d)
but in the superconducting state with a $\mathbf{k}$-independent gap $\mathit{\Delta}$. The real and imaginary parts
of
the corresponding self-energies are schematically shown below each image.}
\label{Fig:SE_Model}
\end{figure}

Fig.\,\ref{Fig:SE_Model} summarizes the results of this section by showing the spectral function calculated using
different models of the self-energy. In panel (a), no $\omega$-dependence of the self-energy is assumed, which
corresponds to a noninteracting system (the scattering rate is set to a small constant to introduce some broadening of
the delta-function). The maxima of the spectral function, seen as a white line in this panel, follow the
quasi-parabolic
bare dispersion. In panel (b), the $\omega^2$-like asymptotic behavior of the scattering rate is assumed, as expected
for a normal Fermi liquid. In the vicinity of the Fermi level the picture is similar to the non-interacting case, but
with a different slope of the dispersion. At higher energies, however, there is a significant broadening of the
features, and the apparent band width is smaller than in the first panel. Next we consider the weak coupling to a
boson
mode, which is illustrated in panel (c). The electron-electron interaction is now switched off, and in the energy
interval $\pm\mathit{\Omega}$ the spectrum looks similar to the non-interacting picture. At the energy of the
collective
mode there is a visible cusp in the dispersion, after which the spectrum again follows the bare band, but is
broadened.
Notice the horizontal feature connecting two branches of the dispersion at $-\mathit{\Omega}$. In panel (d), the
strength of the coupling is increased by a factor of seven and the electron-electron interaction is switched on again.
At this point the spectral weight is well localized around a new, drastically narrower band, which is located fully
within the $\pm\mathit{\Omega}$ interval. Outside of the interval the spectral weight is so broadened, that almost no
traces of the dispersion are preserved. Finally, the last panel (e) corresponds to the same coupling strength, but in
the superconducting state. The finite value of the superconducting gap results in a shift of the cusp in the
dispersion
from $\pm\mathit{\Omega}$ to $\pm(\mathit{\Omega}+\mathit{\Delta})$. The energy gap opens around $E_\text{F}$, and the
backfolding results in a second weaker branch of the dispersion. These particular features of the superconducting
spectral function will be also discussed later in \S\ref{Sec:PhotoemissionBCS}.


\chapter{Angle Resolved Photoemission}

\section{Theory of photoemission}\label{Sec:TheoryOfPhotoemission}

\subsection{The photoelectric effect}

\hvFloat[floatPos=t, capWidth=1.0, capPos=b, capVPos=t, objectAngle=0]{figure}
        {\includegraphics[width=0.85\textwidth]{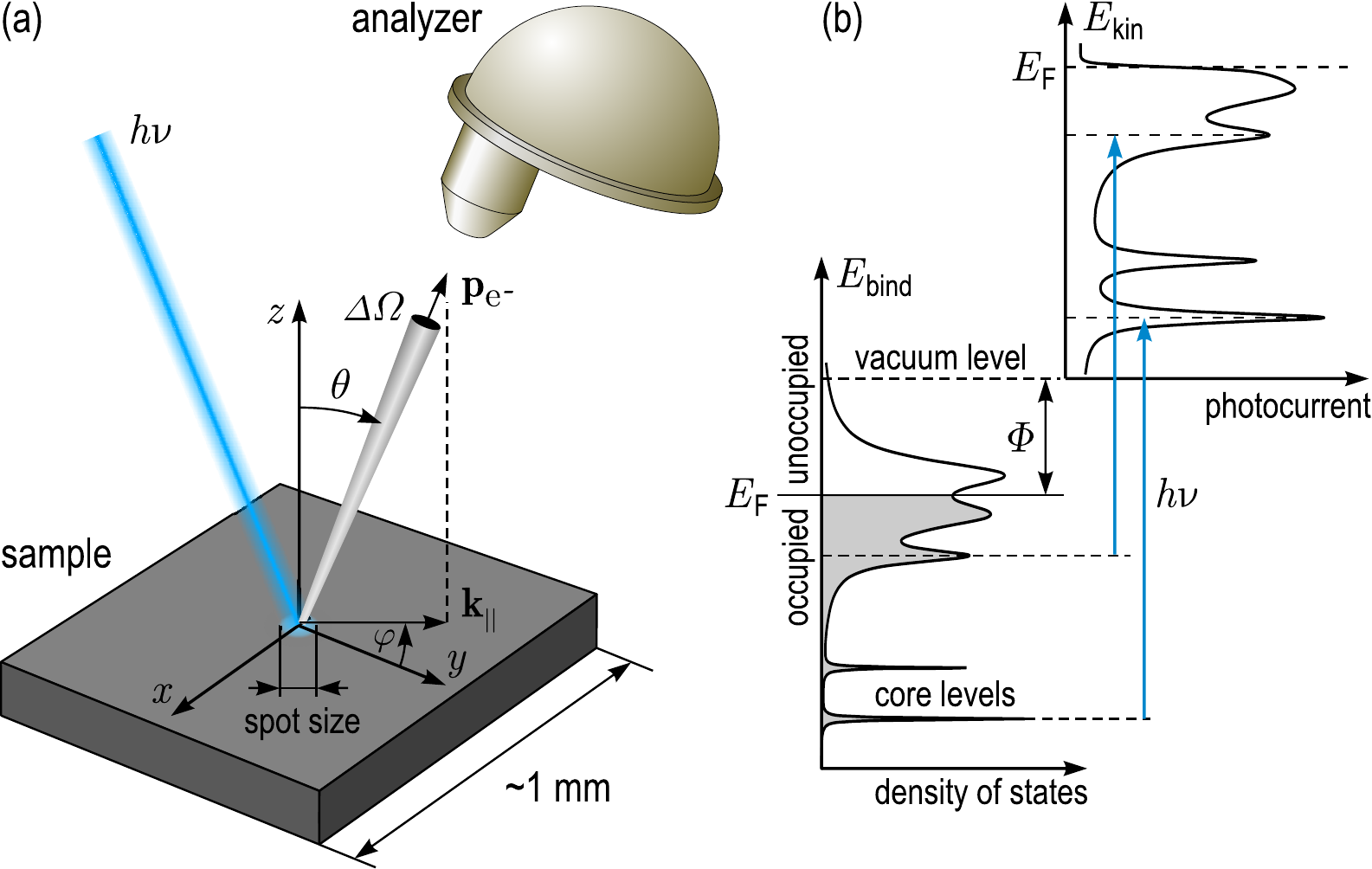}}
        {\textbf{(a)}~Schematic of a photoelectron spectroscopy experiment. An incident beam with photon energy $h\nu$
        excites photoelectrons from a finite-size spot on the surface of the sample. The number of electrons leaving
        the
        sample in the direction of the detector is registered as a function of their energy. \textbf{(b)}~Electron
        transitions in a photoelectric process (two possible transitions are indicated by blue arrows).
        $E_\text{bind}$
        and $E_\text{kin}$ stand for the binding energy of the electrons in the solid and the kinetic energy of
        photoelectrons respectively, $\mathit{\Phi}$ denotes the work function of the sample, $E_\text{F}$ is the
        Fermi
        energy (zero on the binding energy scale).} {Fig:PES}

The photoelectric effect, discovered by H. Hertz in 1887 \cite{Hertz1887} and first explained by A. Einstein in 1906
\cite{Einstein06}, underlies several spectroscopic methods combined under the general name of \textit{photoelectron
spectroscopy} (PES) \cite{Huefner95}, among them ultraviolet photoelectron spectroscopy (UPS), X-ray photoelectron
spectroscopy (XPS), and angle-resolved photoelectron spectroscopy (ARPES). A sketch of a photoelectron spectroscopy
experiment is shown in Fig.\,\ref{Fig:PES}. The electrons are excited by the incident light beam (usually from a
synchrotron source or a gas lamp) to energies above the vacuum level, so that they can escape the solid and be
detected
by the electron analyzer. The photoelectrons are characterized by their momentum $\mathbf{p}_{\text{e}^-}$ or,
equivalently, by its projection on the $x\!y$-plane $\mathbf{p}_{\parallel}$ and total kinetic energy $E_\text{kin}$.
Because the problem is invariant under the translations in the $x\!y$-plane, the $x\!y$-component of the momentum is
conserved, and assuming we can neglect the photon's momentum,\footnote{The momentum $2\piup\lambda^{-1}=E/\hslash c$
of
a 100\,eV photon is 0.05 \AA$^{-1}$\,---\kern.6ex about 3\% of the typical size of the irreducible Brillouin zone in
cuprates, which is already comparable or even above the typical experimental resolution. Therefore for excitation
energies of the order of 100\,eV or more the photon momentum may need to be taken into account.}
$\mathbf{p}_{\parallel}$ is equal to the $x\!y$-component of the electron's quasimomentum within the solid
$\mathbf{k}_{\parallel}$. On the other hand, from the energy conservation law we have
$E_\text{kin}=h\nu-E_\text{bind}-\mathit{\Phi}$, so the binding energy of the electron in a solid can be obtained from
the kinetic energy of the photoelectron. Binding energy as a function of the quasimomentum is the electron dispersion
relation, which determines many physical properties of the material.

\subsection{Photoemission from a non-interacting electron gas}

Let us consider the problem of photoemission from a non-interacting electron gas in a periodic potential. We assume
that
the sample is semi-infinite (with $z$ axis normal to the surface), and has a layered structure (layers parallel to the
$x\!y$-plane), such that the electrons within each layer can be considered independently. In such a system the
$x\!y$-component of the quasimomentum $\mathbf{k}_{\parallel}=\mathbf{k}$ together with the binding energy
$E_\text{bind}=E_\text{F}-E$ uniquely describe the electronic states within the solid.

We start from writing down the Hamiltonian for this system in the electromagnetic field of the incident light:
${\hat{\mathcal{H}\,}\kern-2pt}={\hat{\mathcal{H}\,}\kern-2pt}_0+{\hat{\mathcal{H}\,}\kern-2pt}_\text{int}$, where
${\hat{\mathcal{H}\,}\kern-2pt}_0$ is the non-interacting Hamiltonian (\ref{Eq:TightBindingHamiltonian}) and
${\hat{\mathcal{H}\,}\kern-2pt}_\text{int}$ describes the interaction with the electromagnetic field
\cite[p.\,49]{LandauLifshitz2}:
\begin{equation}
{\hat{\mathcal{H}\,}\kern-2pt}=\frac{(\hat{\mathbf{p}}-\frac{\,e\,}{\text{\raisebox{3pt}{$c$}}}\mathbf{A})^2}{2m}
={\hat{\mathcal{H}\,}\kern-2pt}_\text{0}-\frac{e}{2mc}(\mathbf{A}\cdot\hat{\mathbf{p}}+\hat{\mathbf{p}}\cdot\mathbf{A}-\frac{\,e\,}{c}\mathbf{A}^2)\text{,}
\end{equation}
where $\hat{\mathbf{p}}=-\mathrm{i}\kern.5pt\hslash\kern.5pt\nabla$ is the electronic momentum operator and
$\mathbf{A}$
is the electromagnetic vector potential. At low light intensities the quadratic term in $\mathbf{A}$ can be safely
neglected and we arrive at
\begin{equation}\label{Eq:PhotoemissionH}
{\hat{\mathcal{H}\,}\kern-2pt}_\text{int}=\frac{\mathrm{i}\hslash
e}{2mc}(\mathbf{A}\cdot\kern-.5pt\nabla+\nabla\kern-.5pt\cdot\mathbf{A})
=\frac{\mathrm{i}\hslash
e}{mc}(\mathbf{A}\cdot\kern-.5pt\nabla+{\textstyle\frac{1}{2}}\,\mathrm{div}\,\mathbf{A})\text{.}
\end{equation}
If $\mathbf{A}$ is constant over atomic dimensions, which is called the \textit{dipole approximation} and holds well
in
the ultraviolet region, the second term can be neglected for bulk photoemission. Still, it may become important at the
surface, where the electromagnetic fields may have a strong spatial dependence \cite{Damascelli03, Zabolotnyy07}, but
such surface contributions will not be considered in the current work, so we can simplify the interaction Hamiltonian
further to
\begin{equation}\label{Eq:PhotoemissionHSimplified}
{\hat{\mathcal{H}\,}\kern-2pt}_\text{int}=-\frac{e}{mc}\,\mathbf{A}\cdot\hat{\mathbf{p}}=\frac{i\hslash
e}{mc}\sum_{i,\,j}
\langle\psi_i|\mathbf{A}\cdot\kern-.5pt\nabla|\psi_j\rangle\hat{c}^\dag_i\hat{c}^{\phantom{\dag}}_j\text{.}
\end{equation}

The next logical step is to consider ${\hat{\mathcal{H}\,}\kern-2pt}_\text{int}$ as a periodic perturbation and write
down the transition probabilities between initial and final states of the electron in the first Born approximation,
which are then given by the \textit{Fermi's golden rule} \cite[p.\,153]{LandauLifshitz3}:
\begin{equation}\label{Eq:FermiGoldenRule}
w_{\text{i}\rightarrow\text{f}}=\frac{2\piup}{\hslash}\bigl|\langle\psi^N_\text{f}|{\hat{\mathcal{H}\,}\kern-2pt}_\text{int}|\psi^N_\text{i}\rangle\bigr|^2\delta(E_\text{f}^N-E_\text{i}^N-h\nu)\text{.}
\end{equation}
Here $\psi^N_\text{i}$ and $\psi^N_\text{f}$ are the $N$-particle wave functions of the initial and final states
respectively, $E_\text{i}^N$ and $E_\text{f}^N$ are the total energies of the system before and after the excitation,
and $h\nu$ is the excitation energy.

Though the problem of photoemission is complicated by the presence of the surface, so that the final and initial wave
functions can not be treated as Bloch waves in the vicinity of the solid-vacuum interface, deep within the bulk of the
crystal the Bloch wave approximation still holds. On the other hand, far outside the crystal the electrons can be
approximated as plane waves,\footnote{Even though the actual wave function outside the crystal may be not a plane
wave,
it makes sense to expand it into the plane wave basis due to the specifics of the experiment, which measures the
electron's momentum. In other words, even if before the measurement the real wave function of the photoelectron was a
mixed state of plane waves with different momenta, at the moment the electron is detected by the analyzer with a
certain
momentum $\mathbf{k}$, the wave function of the system collapses into the state that would correspond to the
plane-wave
photoelectron with momentum $\mathbf{k}$.} so that treating the solid-vacuum interface remains the main challenge.
There
are two main approaches to treating this problem: the so-called one- and three-step models \cite[p.\,245]{Huefner95}.
The \textit{one-step model} \cite{Caroli73, FeibelmanEastman74, Potthoff97, Potthoff01, Mahan70, Almbladh06,
HedinLee02}
starts from the inverse problem: propagation of the plane wave from outside the solid into the crystal. For a
particular
value of momentum, one calculates how the electron plane wave decays within the solid-vacuum interface and the
resulting
wave function (so-called \textit{reverse LEED state} \cite[p.\,270]{Huefner95}) is used as the final state, into which
an initial Bloch state is excited. In this sense, the one-step model treats photon absorption, electron removal, and
electron detection as a single coherent process \cite{Damascelli03}.

\hvFloat[floatPos=t, capWidth=1.0, capPos=r, capVPos=t, objectAngle=0]{figure}
        {\includegraphics[width=0.7\textwidth]{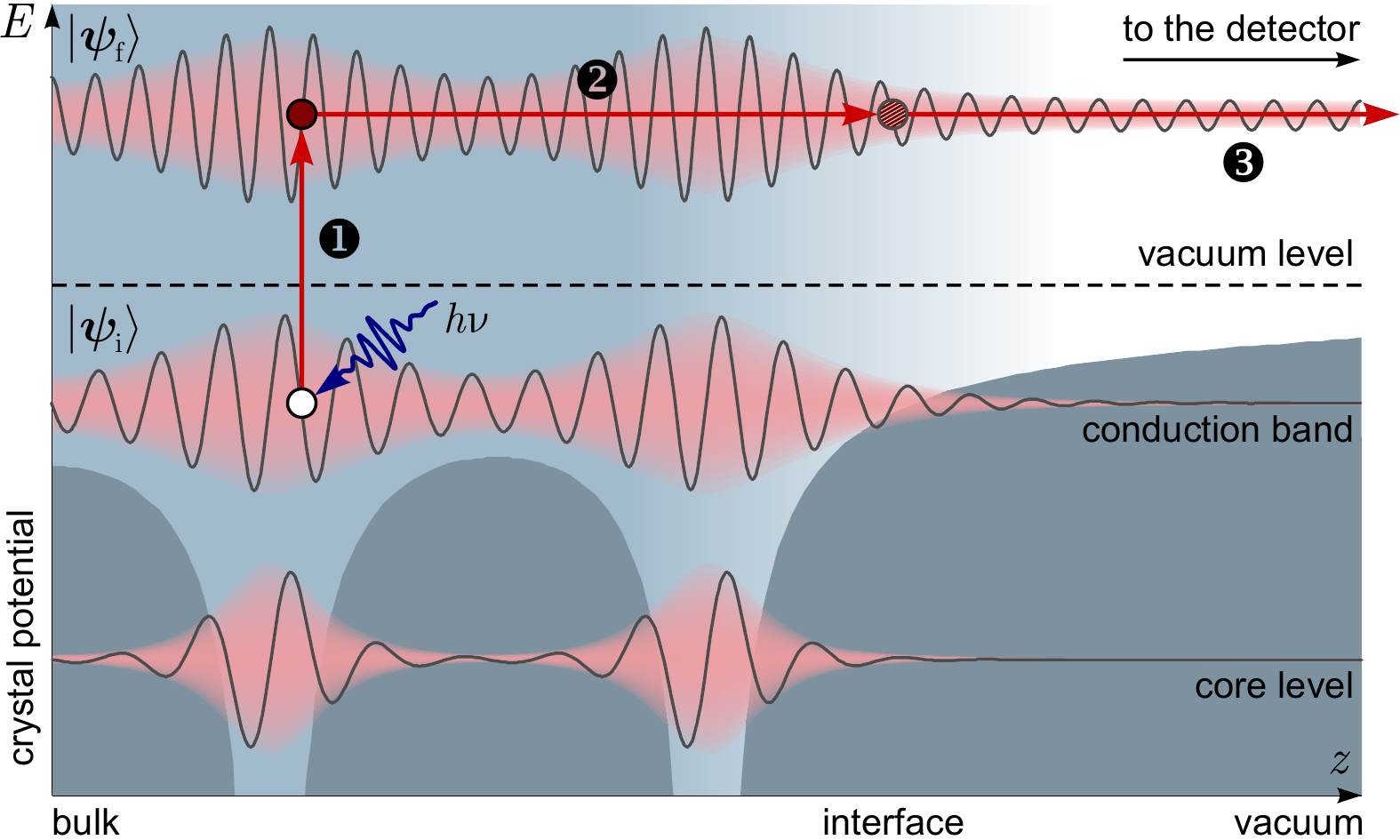}~}
        {A cartoon of the 3-step model of photoemission: (1) optical excitation of the electron in the bulk; (2)
        propagation of the
        excited electron to the surface; (3) escape of the photoelectron into vacuum.}
        {Fig:3-step}

\textbf{Three-step model.} Unfortunately, the rigorous but technically complicated one-step interpretation of
photoemission is not always practical. In many cases, a simpler \textit{three-step model} \cite[p.\,244]{Huefner95}
can
be applied (see Fig.\,\ref{Fig:3-step}). Within this approach, the photoemission process is subdivided into three
independent and sequential steps \cite{Damascelli03}: optical excitation of the electron in the bulk, propagation of
the
excited electron to the surface, and escape of the photoelectron into vacuum. The photocurrent is then determined by
the
three independent factors: optical transition probability between the initial and final states within the bulk,
scattering probability of the traveling electrons determined by the mean free path, and the transmission probability
through the surface potential barrier, depending on the energy of the excited electron and the material work function
$\mathit{\Phi}$.

To simplify things even further, one usually assumes that the net scattering\footnote{The electron-electron and
electron-phonon scattering play the major role here \cite[p.\,247]{Huefner95}.} frequency, or inverse lifetime, of the
excited electrons in the bulk is isotropic ($\mathbf{k}$-independent), so that the electron transport to the surface
affects only the absolute intensity of the photoelectrons with a given energy, leaving the $\mathbf{k}$-distribution
unchanged. The finite electron mean free path inside the crystal makes photoelectron spectroscopy a surface-sensitive
technique and becomes especially important in materials with large unit cells, comparable to the electron mean free
path, and in samples which undergo any kind of surface reconstruction, so that the surface contributions to
photocurrent
become different from the bulk contributions and can no longer be neglected.

Let us now consider the third step of the photoemission, namely the transit of the electron through the solid-vacuum
interface. Because of the broken translational symmetry along the $z$ direction, the wave function of the electron
will
have the form
$\psi_\mathbf{k}(\mathbf{r})=u_\mathbf{k}(\mathbf{r}_\parallel,z)\,\mathrm{e}^{\mathrm{i}\mathbf{k}\mathbf{r}}$, where
$u_\mathbf{k}(\mathbf{r}_\parallel,z)$ is a periodic function of $\mathbf{r}_\parallel = \{x,\,y\}$ according to the
Bloch's theorem \cite[p.\,155]{Marder00} and has a limit at $z\rightarrow\infty$, because the corresponding wave
function should approach the Bloch state of the vacuum, so the function $u_\mathbf{k}(\mathbf{r}_\parallel,z)$ must be
periodic with every period, which means it is a constant. Note that the orthogonal component of the quasimomentum
$k_z$
is defined here so that it corresponds to the $z$-component of the electron momentum at $z\rightarrow\infty$: $\hslash
k_z=p_z$, satisfying the energy conservation law. Expanding the wave function $\psi_\mathbf{k}(\mathbf{r})$ into the
Fourier series in $\mathbf{k}_\parallel$, we get:
\begin{equation}\label{Eq:MomentumConservation}
\psi_\mathbf{k}(\mathbf{r})=\sum_{\mathbf{G}_\parallel}u_{\mathbf{G}_\parallel}\text{(}z\text{)}\,\mathbf{e}^{\mathrm{i}(\mathbf{k}
+ \mathbf{G}_\parallel)\mathbf{r}}\text{.}
\end{equation}
Because the function $u_\mathbf{k}(\mathbf{r}_\parallel,z)$ has a limit at $z\rightarrow\infty$, we see that the wave
function far away from the surface of the crystal is just a sum of plane waves with the common $z$-component of the
momenta and a discrete set of the $x\!y$-projections of the momenta
$p_\parallel=\hslash\mathbf{k}_\parallel+n\hslash\mathbf{G}_\parallel$, $n\in\mathds{Z}$. This equality signifies the
conservation law of the parallel component of the momentum during the photoemission process. It can be interpreted
either as a diffraction phenomenon of bulk electrons on the crystal surface (from the ``vacuum'' point of view), or as
a
conservation law of the form $p_\parallel/\hslash=\mathbf{k}_\parallel\,\mathrm{mod}\,\mathbf{G}_\parallel$ (from the
``crystal'' point of view where momentum is defined modulo reciprocal lattice vector). The photoelectron flux
corresponding to a certain $\mathbf{G}_\parallel$ is termed \textit{cone of photoemission} \cite{Mahan70}, or
sometimes
\textit{Mahan cone} \cite[p.\,249]{Huefner95}. In the slang of the photoemission community, photoemission cones are
often called Brillouin zones by analogy with the extended zone scheme, and we will follow this loose terminology
henceforth.

Having thus shuffled off the problems of treating the second and third steps of the photoemission process in the
three-step model, we now concentrate on the first, and the most important step, which will lead us to the main results
of the theoretical part of this chapter. For this, let us turn back to Eq. (\ref{Eq:FermiGoldenRule}), which describes
the probability of a photoexcitation from the initial state of the system $\psi^N_\text{i}$ to the final state
$\psi^N_\text{f}$. In a non-interacting electron system, the matrix element of the initial and final states is
determined only by the one-particle states $|\psi_\text{i}\rangle$ and $|\psi_\text{f}\rangle$:
$\langle\psi^N_\text{f}|{\hat{\mathcal{H}\,}\kern-2pt}_\text{int}|\psi^N_\text{i}\rangle=\langle\psi_\text{f}|{\hat{\mathcal{H}\,}\kern-2pt}_\text{int}|\psi_\text{i}\rangle\stackrel{\text{def}}{=}M^\mathbf{k}_{\text{i},\,\text{f}}$.
The photocurrent as a function of energy and momentum will be therefore given by\footnote{The absolute value of the
photocurrent is difficult to estimate, because it depends on the light intensity, reflective properties of the
crystal,
density of electrons in the solid, etc. As we are interested only in the $\mathbf{k}$- and $\omega$-dependence of the
photocurrent, all coefficients that are independent of these variables are skipped, which here and henceforth is
denoted
by the proportionality sign.}
\begin{equation}
I(\mathbf{k},\omega)\propto n_\text{F}(E_\text{i})w_{\text{i}\rightarrow\text{f}}=
n_\text{F}(\epsilon_\mathbf{k})\bigl|M^\mathbf{k}_{\text{i},\,\text{f}}\bigr|^2\delta(\omega-\epsilon_\mathbf{k})
=\frac{\bigl|M^\mathbf{k}_{\text{i},\,\text{f}}\bigr|^2}{2\piup}\,n_\text{F}(\epsilon_\mathbf{k})\,A_0(\mathbf{k},\,\omega)\text{,}
\end{equation}
where $A_0(\mathbf{k},\,\omega)$ is the bare spectral function introduced in \S\ref{SubSec:SingleParticleGFint}. In
the
following we will show that a similar expression holds also in the interacting case.

\subsection{Photoemission from an interacting electron gas}\label{SubSec:SuddenApprox}

If the electron interactions are taken into account, the problem of photoemission gets more complicated. The matrix
element can no longer be factorized, so we have
\begin{equation}\label{Eq:MatrixElementInt1}
\langle\psi^N_\text{f}|{\hat{\mathcal{H}\,}\kern-2pt}_\text{int}|\psi^N_\text{i}\rangle
\propto\sum_{\nu,\,\eta}\bigl\langle\psi^N_\text{f}\big|
\underset{\displaystyle M_{\nu\eta}}{\underbrace{\langle\psi_\eta|\mathbf{A}\cdot\kern-.5pt\nabla|\psi_\nu\rangle}}
\hat{c}^\dag_\eta\hat{c}^{\phantom{\dag}}_\nu\big|\psi^N_\text{i}\bigr\rangle
=\sum_{\nu,\,\eta}M_{\nu\eta}\langle\psi^N_\text{f}
|\hat{c}^\dag_\eta\hat{c}^{\phantom{\dag}}_\nu|\psi^N_\text{i}\rangle\text{.}
\end{equation}
Say, we are interested in the spectrum of a particular single-particle final state $|\psi_\text{f}\rangle$, and we can
assume that this state is totally unoccupied, i.e. $\hat{c}_\text{f}\,|\psi^N_\text{i}\rangle=0$, then we can skip the
$\eta$-summation: $\sum_\eta M_{\nu\eta}\,\hat{c}^\dag_\eta\hat{c}^{\phantom{\dag}}_\nu|\psi^N_\text{i}\rangle
=\sum_\eta
M_{\nu\eta}\,\hat{c}^\dag_\text{f}\hat{c}^{\phantom{\dag}}_\text{f}\hat{c}^\dag_\eta\hat{c}^{\phantom{\dag}}_\nu|\psi^N_\text{i}\rangle
=M_{\nu\text{f}}\,\hat{c}^\dag_\text{f}\hat{c}^{\phantom{\dag}}_\nu|\psi^N_\text{i}\rangle$. The ($N-1$)-particle
states
$\psi^{N-1}_\text{i}=\hat{c}_\nu\psi^N_\text{i}$ and $\psi^{N-1}_\text{f}=\hat{c}_\text{f}\,\psi^N_\text{f}$ are not
eigenstates of the system, so we expand them further in the basis of ($N-1$)-particle eigenstates $\psi^{N-1}_m$:
\begin{equation}\label{Eq:MatrixElementInt2}
\kern-2.5ex\text{(\ref{Eq:MatrixElementInt1})}=\!\sum_\nu
M_{\nu\text{f}}\langle\psi^{N-1}_\text{f}|\psi^{N-1}_\text{i}\rangle
=\!\sum_\nu M_{\nu\text{f}}{\Bigl\langle\sum_m\beta^{\phantom{*}}_{\text{f}\kern.3pt
m}\psi^{N-1}_m\Big|\!\sum_m\alpha^{\phantom{*}}_{\nu m}\psi^{N-1}_m\Bigr\rangle}
\!=\!\sum_{\nu,\,m} M_{\nu\text{f}}\,\beta^*_{\text{f}\kern.3pt m}\alpha^{\phantom{*}}_{\nu m}\text{.}\kern-2ex
\end{equation}

\textbf{Sudden approximation.} We see that a single-particle transition
$|\psi_\nu\rangle\rightarrow|\psi_\text{f}\rangle$ leaves the ($N-1$)-particle system in a mixed state. Now we have to
decide how we treat the electron removal from the single-particle excited state, which is rather nontrivial. In the
most
commonly used \textit{sudden approximation} the interaction of the electron in the single-particle final state
$|\psi_\text{f}\rangle$ with the rest of the system is neglected, so the removal of the electron happens simply by the
destruction of the electron in the final state, while the rest of the system does not change.\footnote{Beyond the
sudden
approximation, one has to take into account the screening of the photoelectron by the rest of the system, which
complicates the situation drastically: the photoemission process is then described by a three-particle correlation
function (so-called \textit{generalized golden rule formulae}) \cite{Almbladh06, HedinLee02}.} As the energy of the
photoelectron is measured, the energy conservation law requires the system to collapse into one of its eigenstates
$|\psi^{N-1}_m\rangle$: $\psi^N_\text{f}=\hat{c}^\dag_\text{f}\,|\psi^{N-1}_m\rangle$. Different energies of these
eigenstates correspond to different energies of the photoelectron and, therefore, lead to the broadening of the
measured
energy spectrum. The reason for such broadening is the interaction of the initial state $|\psi_\nu\rangle$ with the
rest
of the system, which we did not neglect. If we now write down the Fermi's golden rule, rewriting the $\nu$-sum under
the
modulus squared as a double sum, we get
\begin{equation}\label{Eq:GoldenRuleSudden}
I(\mathbf{k},\,\omega)\propto
n_\text{F}\text{(}\omega\text{)}\!\sum_{\nu,\,\nu'}M^{\phantom{*}}_{\nu\text{f}}\,M^*_{\nu'\text{f}}\kern1.5pt
\underset{\displaystyle
A_{\nu'\nu}\text{(}\omega\text{)}}{\underbrace{\text{\scalebox{0.8}{$\frac{2\piup}{Zn_\text{F}\text{(}\omega\text{)}}$}}\sum_m\mathrm{e}^{-\beta
E^N_\text{i}}\langle\psi^N_\text{i}|\hat{c}^\dag_{\nu'}|\psi^{N-1}_m\rangle\langle\psi^{N-1}_m|\hat{c}^{\phantom{\dag}}_\nu|\psi^N_\text{i}\rangle\delta(\omega+E^{N-1}_m\!-E^N_\text{i})}}\text{,}
\end{equation}
where $\omega=-E_\text{bind}=E_\text{kin}+\mathit{\Phi}-h\nu$ is the energy of the photohole and
$A_{\nu'\nu}\text{(}\omega\text{)}$ is the spectral function in Lehmann representation (\ref{Eq:SpectralFunction}),
which in most applications is taken as diagonal \cite{HedinLee02}. After neglecting the off-diagonal elements
($\nu'\!\neq\nu$), we arrive at the final expression for the photocurrent:
\begin{equation}\label{Eq:PhotocurrentSudden}
I(\mathbf{k},\,\omega)\propto
n_\text{F}\text{(}\omega\text{)}\sum_\nu|M^\mathbf{k}_{\text{i},\,\text{f}}|^2\,A(\mathbf{k},\,\omega)\text{.}
\end{equation}

\subsection{Photoemission from a BCS superconductor}\label{Sec:PhotoemissionBCS}

\hvFloat[floatPos=b, capWidth=1.0, capPos=r, capVPos=t, objectAngle=0]{figure}
        {\!\includegraphics[width=0.70\textwidth]{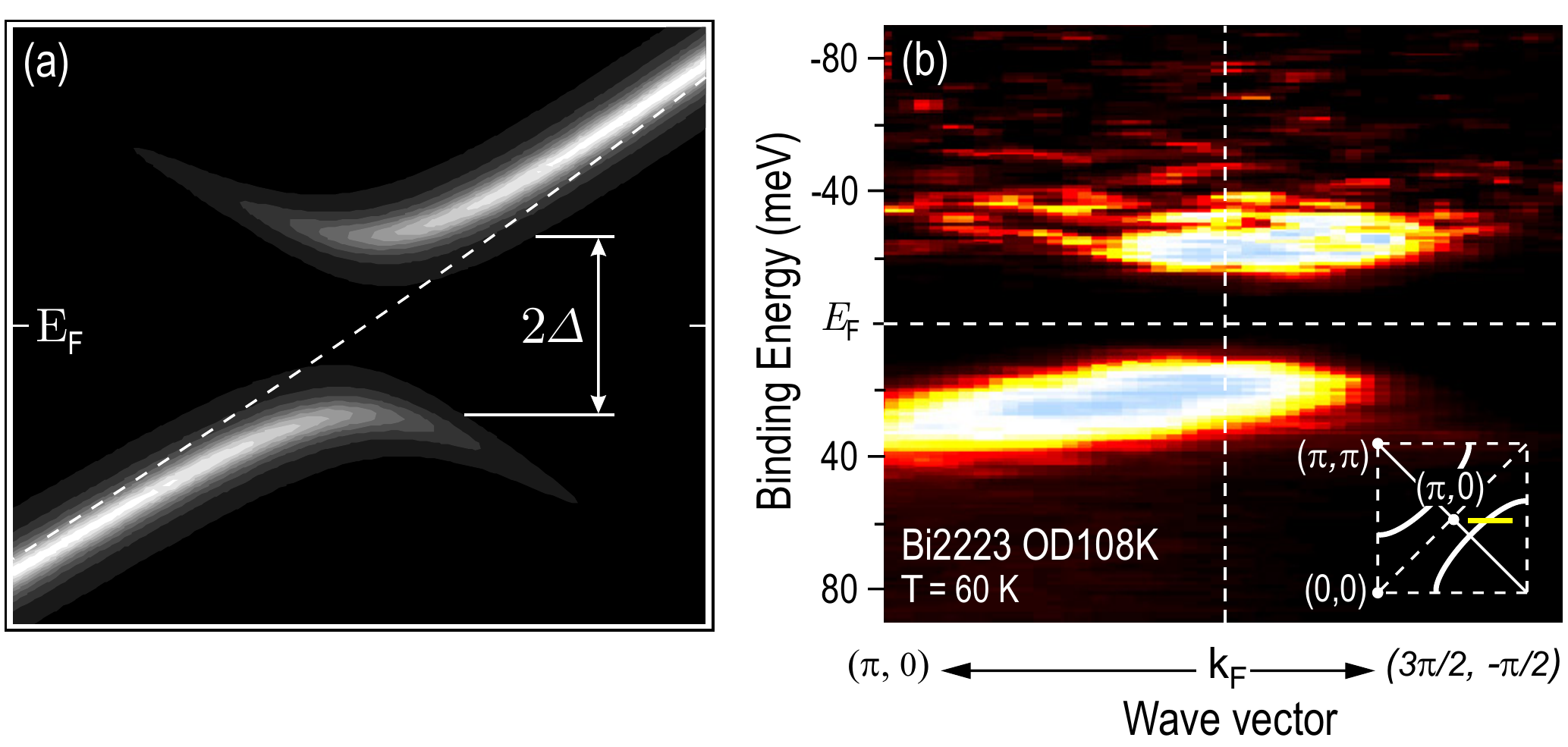}~~}
        {\textbf{(a)}~BCS spectral function near the Fermi level given by Eq. (\ref{Eq:SpectralFunctionBCS}). The bare
        dispersion $\epsilon_\mathbf{k}$ is shown by the dashed line. Delta-functions are broadened by
        Lorentzians of a fixed width. \textbf{(b)}~Experimental ARPES spectrum of a high-$T_\text{c}$ superconductor
        \cite{Matsui03}.}
        {Fig:BCS}

As was shown in \S\ref{SubSec:BCS}, the spectrum of a superconductor within BCS theory is
$E_\mathbf{k}=\pm\sqrt{\epsilon_\mathbf{k}^2+|\mathit{\Delta}_\mathbf{k}|^2}$ with the $\mathbf{k}$-dependent
superconducting order parameter $\mathit{\Delta}_\mathbf{k}$, so we expect the bare spectral function to be localized
along this dispersion. Still, to calculate the distribution of the spectral weight along the dispersion, a rigorous
calculation is required. For this we use the diagonal approximation of the spectral function
(\ref{Eq:SpectralFunction})
and rewrite it in the Bogoliubov operators (\ref{Eq:BogoliubovTransform}):
\begin{equation}
A_0(\mathbf{k},\,\sigma,\,\omega)=\frac{2\piup}{Zn_\text{F}\text{(}\omega\text{)}}\sum_{n,\,m}\mathrm{e}^{-\beta
E_n}\bigl|\langle\psi_m|u_\mathbf{k}^*\hat{\gamma}^{\phantom{\dag}}_{\mathbf{k},\,\sigma}\!+v_\mathbf{k}\hat{\gamma}^\dag_{-\mathbf{k},\,-\sigma}|\psi_n\rangle\bigr|^2\delta(\omega+E_m-E_n).
\end{equation}
The non-zero terms correspond either to $\hat{\gamma}_{\mathbf{k},\,\sigma}|\psi_m\rangle=|\psi_n\rangle$ or
$\hat{\gamma}_{-\mathbf{k},\,-\sigma}|\psi_n\rangle=|\psi_m\rangle$, so the sum can be split in two:
\begin{multline}
A_0(\mathbf{k},\,\sigma,\,\omega)=\frac{2\piup}{Zn_\text{F}\text{(}\omega\text{)}}|u_\mathbf{k}|^2\sum_{n,\,m}\mathrm{e}^{-\beta
E_n}\bigl|\langle\psi_m|\hat{\gamma}_{\mathbf{k},\,\sigma}|\psi_n\rangle\bigr|^2\delta(\omega+E_m-E_n)\\
+\frac{2\piup}{Zn_\text{F}\text{(}\omega\text{)}}|v_\mathbf{k}|^2\sum_{n,\,m}\mathrm{e}^{-\beta
E_n}\bigl|\langle\psi_n|\hat{\gamma}_{-\mathbf{k},\,-\sigma}|\psi_m\rangle\bigr|^2\delta(\omega+E_m-E_n)\\
=\frac{2\piup}{Z(1-n_\text{F}\text{(}\omega\text{)})}|u_\mathbf{k}|^2\,{\sum_m}^{\kern1pt\textstyle\prime}\mathrm{e}^{-\beta
E_m}\delta(\omega-E_\mathbf{k})+\frac{2\piup}{Zn_\text{F}\text{(}\omega\text{)}}|v_\mathbf{k}|^2\,{\sum_n}^{\kern1pt\textstyle\prime}\mathrm{e}^{-\beta
E_n}\delta(\omega+E_\mathbf{k})\text{.}
\end{multline}
In the last expression the primes over the summation signs indicate that the $m$- and $n$-summations run only over
states for which $\hat{\gamma}^\dag_{\mathbf{k},\,\sigma}|\psi_m\rangle\neq0$ and
$\hat{\gamma}^\dag_{-\mathbf{k},\,-\sigma}|\psi_n\rangle\neq0$ respectively. Recalling that
$Z=\sum_n\mathrm{e}^{-\beta
E_n}=\sum_m^{\kern1pt\textstyle\prime}\mathrm{e}^{-\beta E_m}(1+\mathrm{e}^{\pm\beta E_\mathbf{k}})$, we end up with
\begin{equation}\label{Eq:SpectralFunctionBCS}
A_0(\mathbf{k},\,\sigma,\,\omega)=2\piup\bigl(|u_\mathbf{k}|^2\delta(\omega-E_\mathbf{k})+|v_\mathbf{k}|^2\delta(\omega+E_\mathbf{k})\bigr)\text{.}
\end{equation}
As expected, the original dispersion $\epsilon_\mathbf{k}$ has split into two branches $\pm E_\mathbf{k}$ separated by
the minimal energy gap $2\mathit{\Delta}=2\mathit{\Delta}_{\mathbf{k}_\text{F}}$, with the spectral weight
distribution
determined by the gap equations (\ref{Eq:GapEquations}), as demonstrated in Fig.\,\ref{Fig:BCS} (a). The splitting of
the dispersion into the two branches was also confirmed experimentally by Matsui \textit{et al.} \cite{Matsui03}, as
shown in panel (b) of the same figure. Here the spectral function was extracted from the ARPES image of a 3-layered
cuprate Bi2223, measured at 60\,K, by dividing it by the Fermi function in order to reveal the dispersion of the
weakly
occupied states above the Fermi level.

In these examples we have considered the spectral function along a single cut in the momentum space that crosses the
Fermi surface, as shown in Fig.\,\ref{Fig:BCS}~(b, inset). Along such a cut the $\mathbf{k}$-dependence of
$\mathit{\Delta}_\mathbf{k}$ can be neglected with respect to the stronger $\mathbf{k}$-dependence of
$\epsilon_\mathbf{k}$, so the order parameter $\mathit{\Delta}_\mathbf{k}$ away from the Fermi surface crossing could
be
considered equal to the value of the energy gap at $\mathbf{k}_\text{F}$. Still, the value of the superconducting gap
may vary at different $\mathbf{k}_\text{F}$ along the normal-state Fermi surface. In what follows, by the
momentum-dependence of the superconducting energy gap, which will be discussed in detail in section
\ref{Sec:SuperconductingGap}, we will understand only the variations of $\mathit{\Delta}_\mathbf{k}$ along the Fermi
surface, while away from it the order parameter will be considered equal to the gap value at the nearest Fermi
momentum.

\subsection{Experimental effects}

As was shown above, the photocurrent within the sudden approximation is basically determined by the spectral function
of
the sample and the photoemission matrix elements. But it is also modified by several experimental effects that need to
be taken into account.

First, as suggested by Eq.\,(\ref{Eq:PhotocurrentSudden}), direct photoemission experiments can probe only occupied
states below the Fermi level and, to some extent, a small portion of states above the Fermi level that are partially
occupied due to the temperature broadening. Hence, in order to study the electronic structure in the unoccupied
region, other experimental methods, such as inverse photoelectron spectroscopy (IPES) \cite[p.\,98]{Himpsel90, LynchOlson99},
should be used.

Second, the measured spectra are broadened by the finite experimental resolution both in energy and in momentum. The
energy broadening originates both from the finite temperature effects (FWHM~$\approx\text{3.53}\, k_\text{B}T$), inherent in the
spectral function, and from the hardware (analyzer resolution, residual magnetic fields in the measurement chamber,
etc.), while the momentum resolution is mainly a hardware issue (imperfectness of the angular calibration, parasitic
magnetic fields, imperfectness of the analyzer optics, etc.). Both resolutions can be modeled with good accuracy by
the Gaussian smoothing of the ``ideal'' photocurrent intensity (\ref{Eq:PhotocurrentSudden}).

Third, there is a finite background intensity in the experimental ARPES signal. It originates mainly from the
higher-order harmonics present in the incident photon beam that excite the photoelectrons from the deep-lying and
therefore non-dispersing core levels and from the photoelectrons scattered inside the sample on their way to the
surface. The first component is equally present both above and below the Fermi level, while the second one is limited
only to the energies below the Fermi level. Because of the negligibly small momentum dependence of the background
intensity above the Fermi level, it is useful for normalization of the spectra in different momentum channels.

All the aforementioned effects can be summarized in the following formula for the experimental ARPES intensity that
will be subsequently used in section \ref{Sec:ModelingGreensFunction}:
\begin{equation}\label{Eq:PhotocurrentExperimental}
I(\mathbf{k},\,\omega)\propto\iint\!\mathrm{d}\omega'\mathrm{d}\mathbf{k}'\,n_\text{F}\text{(}\omega'\text{)}\sum_\nu|M^{\mathbf{k}'}_{\text{i},\,\text{f}}|^2\,A(\mathbf{k}',\,\omega')\,
\mathrm{e}^{-(\omega-\omega')^2/\mathit{\Delta}\omega^2-(\mathbf{k}-\mathbf{k}')^2/\mathit{\Delta}\mathbf{k}^2}\!+B(\omega)\text{,}
\end{equation}
where $\mathit{\Delta}\omega$ and $\mathit{\Delta}\mathbf{k}$ are the total hardware-related energy and momentum
resolutions, and $B\text{(}\omega\text{)}$ is the momentum-independent background.


\section{ARPES as an experimental method}

\subsection{\textit{Scienta} analyzer and experimental geometry}

XPS is known as an experimental method since 1957 after the pioneering work of Kai Siegbahn \cite{Nordling57}, for
which he received the Nobel prize in physics in 1981. The first attempts of momentum-resolved measurements were made already
in the middle of the 70s \cite{Rowe74}, and immediately after the discovery of the high-temperature superconductors in
1986 \cite{BednorzMueller86} the newly established and rapidly developing ARPES method began to be applied to study
these exotic materials \cite{StoffelChang88, StoffelMorris88, Sakisaka89, Minami89, Lindberg89}. In these early-day
experiments the momentum resolution was achieved by scanning the angular dependence of the photocurrent using
analyzers with low acceptance angle. The situation changed with the development of the first
\textit{Scienta}\footnote{\textit{Scienta Instrument AB}, founded in 1983, was then owned by \textit{SEIKO Instruments
Inc.}, but since 1997 was acquired by \textit{Gammadata Mätteknik AB}.} hemispherical analyzers \cite{BeamsonBriggs90, MartenssonBaltzer94}, which could operate in the angle-resolved mode, providing energy-momentum information not only at a single
$\mathbf{k}$ point, but along an extended cut in $\mathbf{k}$-space within a single measurement. The data used in the
present work were acquired using modern \textit{Scienta} analyzers \textit{SES\,100} \cite{SES100Booklet} and
\textit{R4000} \cite{R4000Booklet} (for comparison of technical characteristics, see table
\ref{Table:ScientaAnalyzers}).

\hvFloat[floatPos=t, capWidth=1.0, capPos=r, capVPos=t, objectAngle=0]{table}
        {\begin{tabular}[c]{l@{~~}l@{~~}l}
         \toprule
         \textit{Scienta analyzer}& \textit{SES 100} & \textit{R4000} \\
         \midrule
         Acceptance angle & $\pm 5^\circ$ & $\pm 15^\circ$ \\ \addlinespace[-0.5ex]
         Best angular resolution (FWHM) & < $0.3^\circ$ & < $0.1^\circ$ \\ \addlinespace[-0.5ex]
         Best energy resolution & $< 3$\,meV & $< 1$\,meV \\ \addlinespace[-0.5ex]
         Pass energies (angular mode) & 5\,--\,200\,eV & 1\,--\,100\,eV \\ \addlinespace[-0.5ex]
         Minimal kinetic energy & 1.0\,eV & 0.2\,eV \\ \addlinespace[-0.5ex]
         Minimal entrance slit & 0.2\,mm & 0.1\,mm \\ \addlinespace[-0.5ex]
         Maximal resolving power & 700 & 4000 \\
         \bottomrule
         \end{tabular}\quad
        }
        {Technical characteristics of the two \textit{Scienta} analyzers used in the present work.}
        {Table:ScientaAnalyzers}

\textbf{Electron analyzer.} The hemispherical electron analyzer measures the intensity of photoemitted electrons as a
function of both their kinetic energy and momentum. Its working principle is illustrated in
Fig.\,\ref{Fig:Scienta}~(a).
It is basically a hemispherical capacitor (only the inner hemisphere is shown in the figure) placed inside a
$\mu$-metal
shielded ultra-high vacuum (UHV) chamber, such as the one shown in panel (b). In the angular-resolved mode, the
photoelectrons are focused by an electron lens in such a way that they reach the entrance slit of the analyzer at
different positions depending on the direction of their initial momentum. The electron trajectories corresponding to
different momenta, represented by three different colors in the figure, end up at different places along the
$\mathbf{k}$ axis of the detector, providing the momentum resolution.\footnote{Such simplified description gives only
a basic idea about how the electron analyzer operates. Of course, a real \textit{Scienta} analyzer is a much more
complicated electro-optical system, and the electron trajectories inside the analyzer are highly complicated.} The angular range of momenta analyzed in a given measurement is called the \textit{acceptance angle} of the analyzer.

\textbf{Pass energy.} Only electrons with one particular energy, which is called the \textit{pass energy}, travel
along trajectories concentric with the hemispherical plates of the capacitor. Electrons with higher or lower energy have
trajectories with a smaller or larger curvature respectively, so that electrons with energies in some neighborhood of
the pass energy hit the detector at different places along the $E$ axis, which is orthogonal to the $\mathbf{k}$ axis.
Therefore the image on the detector contains information about both momentum- and energy-dependence of the
photoelectrons. In such a measurement, electrons with energies substantially different from the pass energy do not
reach the detector. In order to measure electrons in a different energy range, they are additionally decelerated (or sometimes accelerated) on their way to the entrance slit to match the pass energy of the analyzer. The pass energy itself can be set to a number of predefined values, if necessary, by changing the voltage between the hemispheres.

\textbf{Energy resolution.} The pass energy $E_\text{p}$ together with the width of the entrance slit $S$ are
important factors determining the energy resolution of the analyzer. The latter is limited by
$\mathit{\Delta}E=E_\text{p}\,S/D$,
where $D$ is the mean diameter of the analyzer. The resolving power of the analyzer, which is by definition the ratio
of the pass energy to the energy resolution, is therefore $R=E/\mathit{\Delta}E=D/S$. On the other hand, the intensity of
the electrons reaching the detector is inversely proportional both to $S$ and $E_\text{p}$, which makes real
measurements a compromise between the intensity and resolution. Real experimental resolution can be worse than the
theoretical limit due to the unavoidable imperfectness of the analyzer optics and calibration, residual magnetic
fields
inside the chamber, etc.

\hvFloat[floatPos=t, capWidth=1.0, capPos=b, capVPos=t, objectAngle=0]{figure}
        {\includegraphics[width=\textwidth]{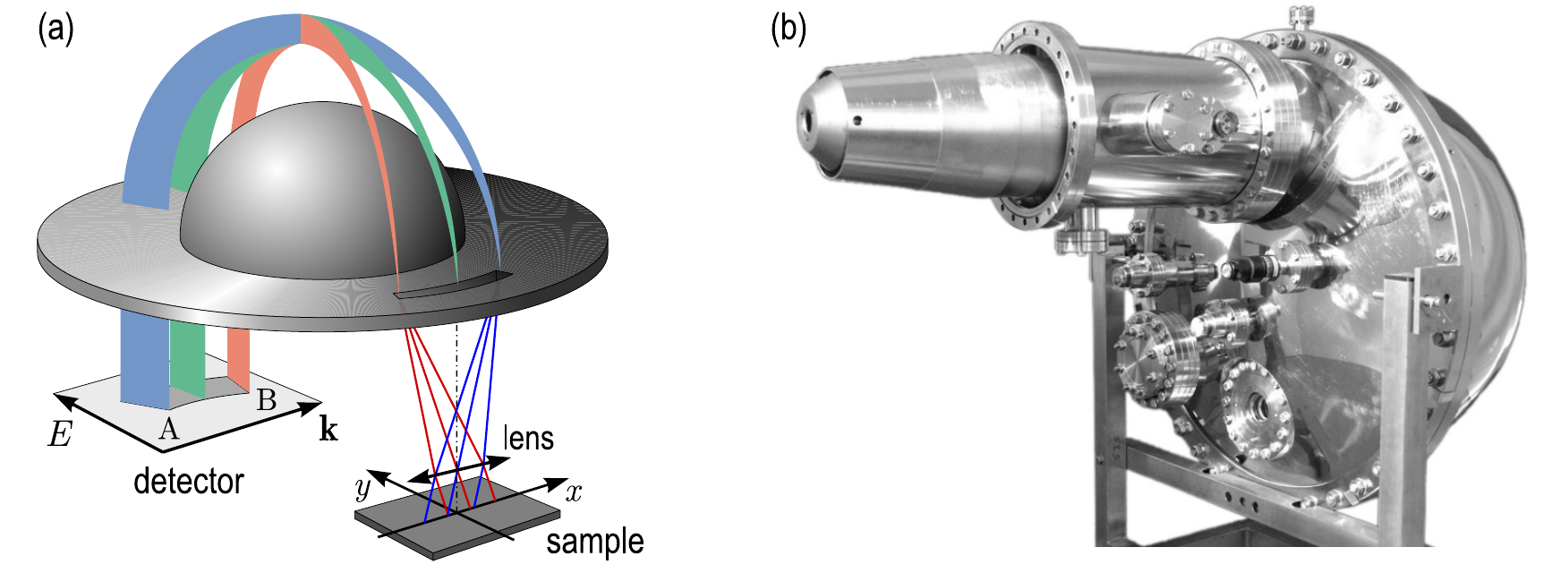}}
        {\textbf{(a)}~Schematic representation of a hemispherical electron analyzer. \textbf{(b)}~Appearance of the
        \textit{Scienta R4000} electron analyzer.}
        {Fig:Scienta}

\hvFloat[floatPos=b, capWidth=0.47, capPos=r, capVPos=t, objectAngle=0]{figure}
        {\hspace{-.7em}\includegraphics[width=0.6\textwidth]{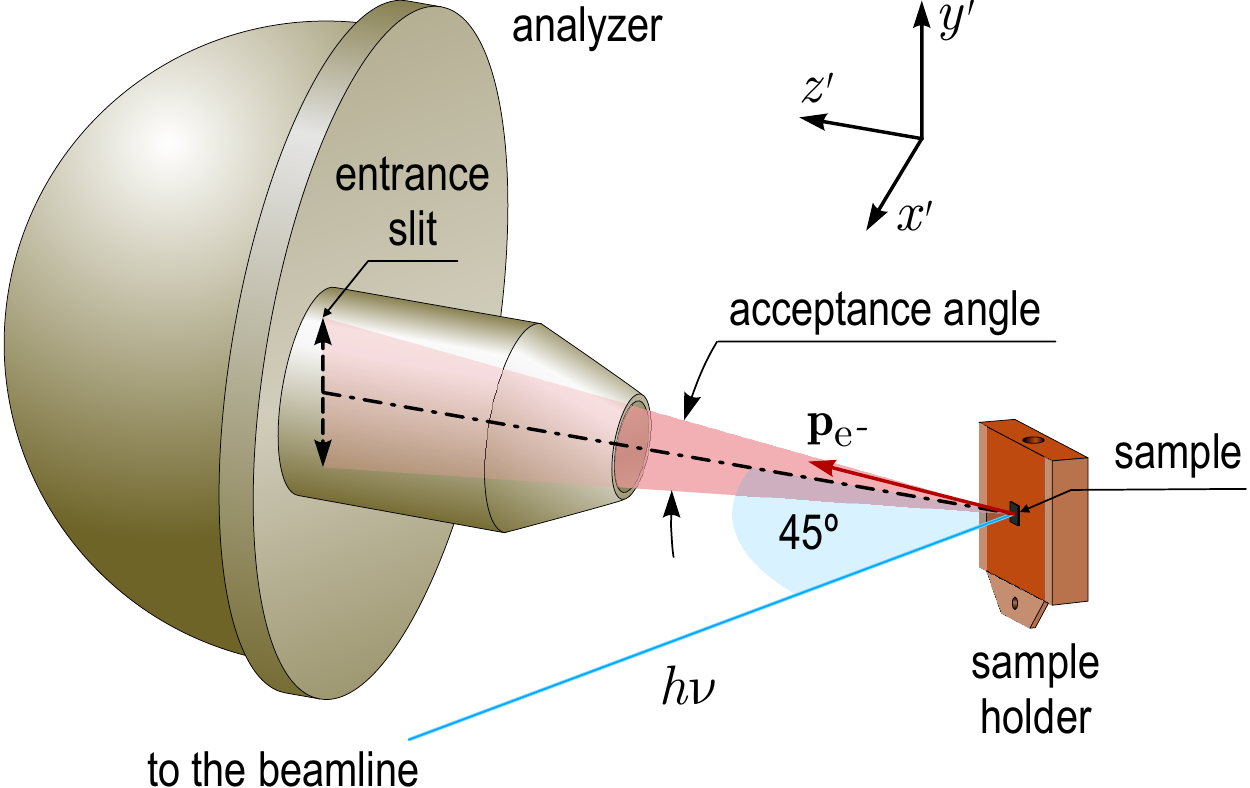}\hspace{-2.5em}}
        {Experimental geometry with the vertical orientation of the entrance slit. The
        synchrotron beam is shown by a blue line. The electron beam emitted from the sample is shown in pink.}
        {Fig:Geometry}

\textbf{Fixed and swept modes.} At each particular value of the decelerating voltage $E_\text{d}$ only electrons with
kinetic energies in the vicinity of $E_\text{d}+E_\text{p}$ can be detected. Such measurement is said to be done in
the
\textit{fixed mode}. In order to measure a larger energy window, the \textit{swept mode} is used. In the swept mode
the
decelerating voltage is scanned to cover the desired energy range, and the measured intensities corresponding to the
same kinetic energies are automatically accumulated, so that each kinetic energy is finally measured by every energy
channel of the detector. Such mode does not only allow measuring electrons in an arbitrary kinetic energy range, but
also compensates for possible inhomogeneities of the detector sensitivity by averaging over all energy channels. The
swept mode has also two drawbacks: (i) at the beginning and at the end of each scan only part of the detector is used
for data acquisition, which effectively results in longer measurement times; (ii) if the energy and angular scales of
the analyzer are imperfectly calibrated, integration may affect both energy and momentum resolutions of the
measurement.

\begin{floatingfigure}[p]{0.475\textwidth}
\noindent\includegraphics[width=0.475\textwidth]{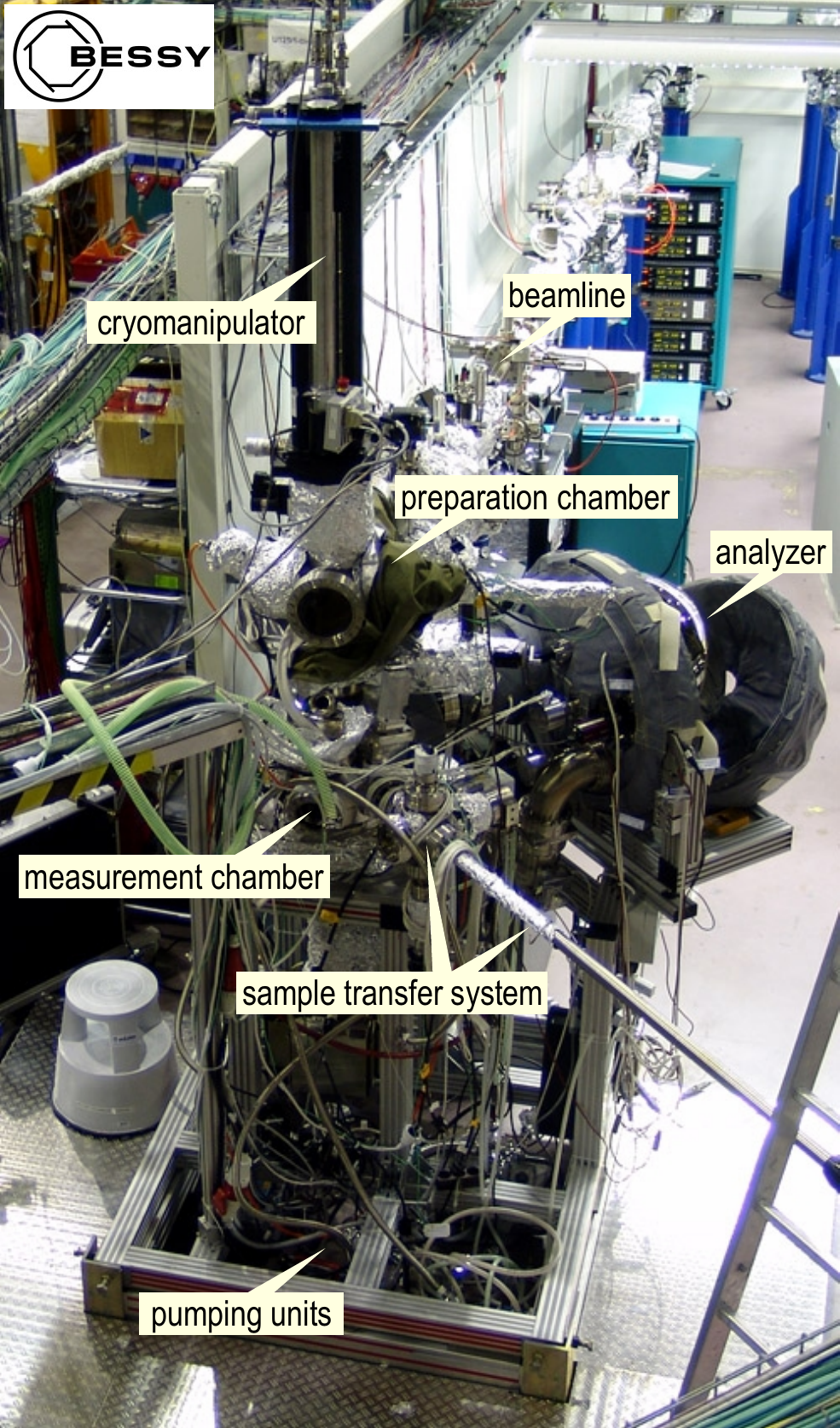}\hspace{-1em} %
\caption{Appearance of the \textit{Scienta R4000}-based experimental setup installed at the UE112-lowE PGMb beamline
of
the Berliner Elektronenspeicherring-Gesellschaft für Synchro\-tron\-strahlung (BESSY).\vspace{-1em}}
\label{Fig:ExperimentalSetup}
\end{floatingfigure}%

\textbf{Experimental geometry.} The analyzer is attached to the measurement chamber, which is connected to the
beamline
and is pumped to the ultra-high vacuum of $\sim$\,$5\cdot10^{-11}$\,mbar in order to prevent degradation of the sample
surface during the experiment. The optical axis of the analyzer and the photon beam are aligned at 45$^\circ$ to each
other in the horizontal plane in such a way that the view point of the analyzer coincides with the focus of the beam.
The geometry of both experimental setups used in the present work is such that the entrance slit of the analyzer is
placed vertically, i.e. perpendicular to the plane formed by the analyzer axis and the photon beam, as shown in
Fig.\,\ref{Fig:Geometry}. The investigated sample is fixed in the view point of the analyzer on a sample holder, which is
in turn attached to the cryo\-manipu\-lator that allows orienting the sample in three angular directions. In the
following paragraph we will stop in more detail on the description of the manipulator and its functions. Except for
the
measurement chamber, the experimental setup normally includes a sample preparation chamber, supplied with a ``fast
entry'' chamber for sample exchange. The appearance of the complete experimental setup can be seen in
Fig.\,\ref{Fig:ExperimentalSetup}.\vfill

\subsection{Cryomanipulator as a $\mathbf{k}$-space explorer}

\begin{floatingfigure}[p]{0.4\textwidth}
\noindent\includegraphics[width=0.4\textwidth]{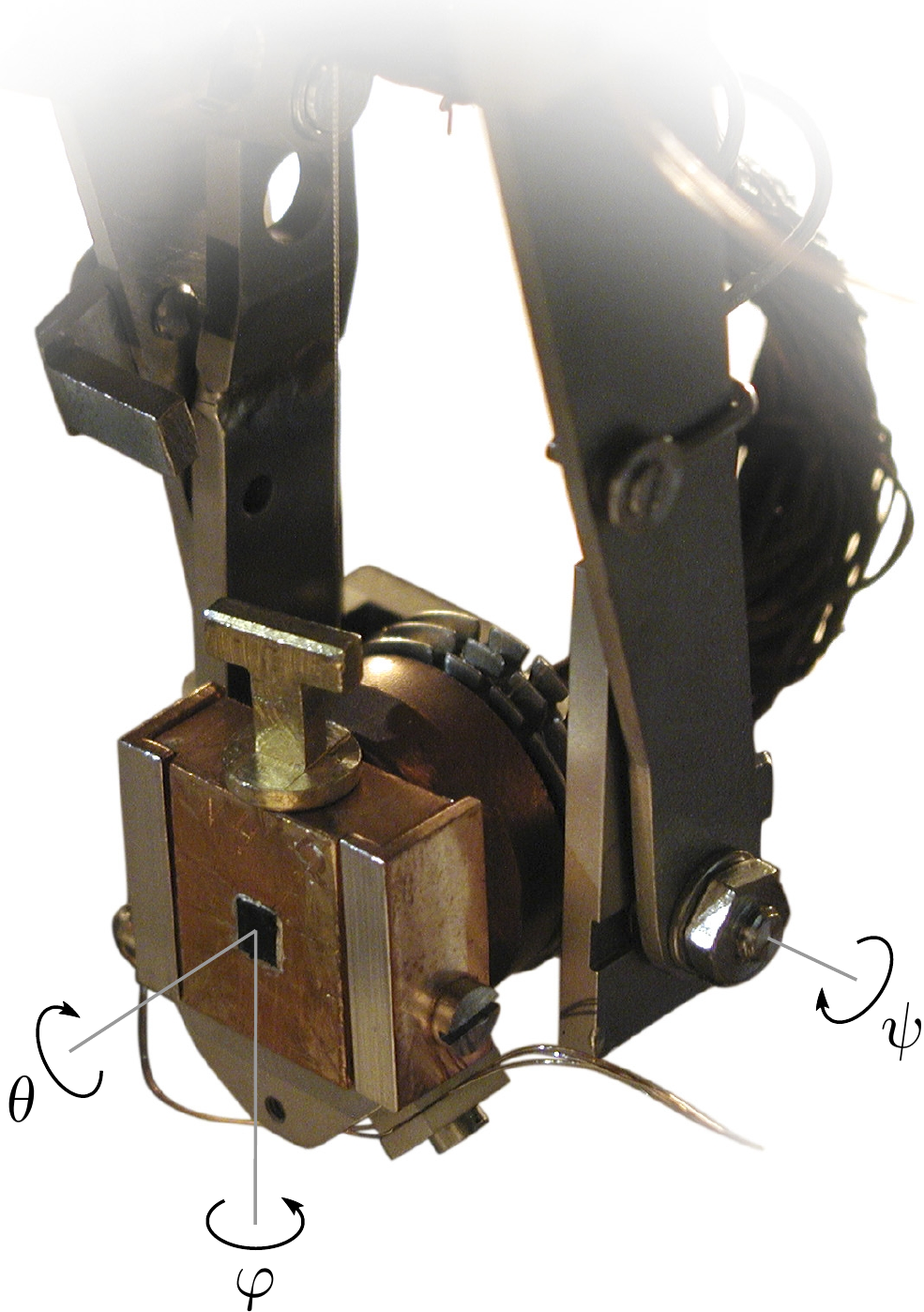}\hspace{-1em} %
\caption{The rotatable head of the \textit{Cryo\-ax~6} manipulator with a sample holder. The polar, tilt, and
azimuthal
angles are denoted by $\varphi$, $\psi$, and $\theta$ respectively. Note that only the polar and azimuthal axes pass
through the center of the sample, whereas the tilt axis is shifted, so that the vertical sample position needs to be
readjusted after every change of the tilt angle.\vspace{-1em}} \label{Fig:Manipulator}
\end{floatingfigure}%

\hvFloat[floatPos=b, capWidth=1.0, capPos=r, capVPos=t, objectAngle=0]{figure}
        {\!\includegraphics[width=0.72\textwidth]{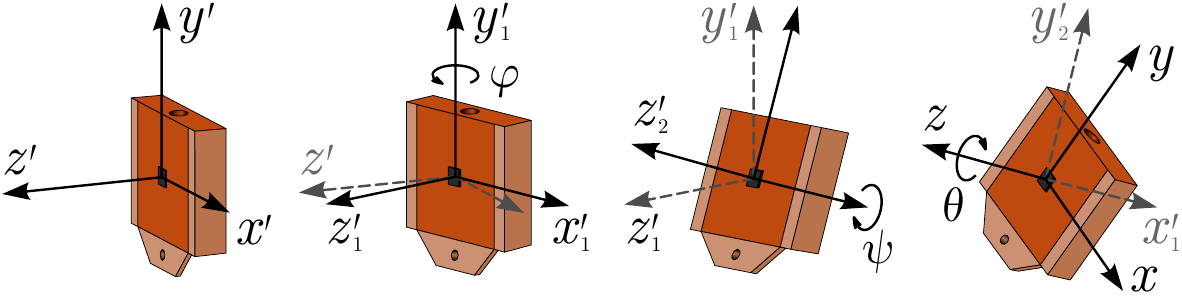}\quad}
        {Transformation between the laboratory and intrinsic coordinate systems represented as three consecutive Euler
        rotations.}
        {Fig:Euler}

The function of the cryomanipulator is to position the sample precisely in the measurement chamber at an arbitrary
position and angle while keeping it at a constant temperature between 30 and 400\,K. In the present work, the
\textit{Cryoax~6} manipulator was used (Fig.\,\ref{Fig:Manipulator}), with 3~rotational and 3~translational degrees of
freedom, and reproducibility of the angles better than 0.2$^\circ$. The temperature of the sample, cooled by liquid
He,
can be controlled by a temperature sensor attached to the sample pocket.

To understand the need for the precise sample orientation, let us see what region of the momentum space can be
measured
by the analyzer in the vertical slit geometry for each particular orientation of the sample. We will introduce two
orthonormal positively orientated coordinate systems: $\{x,\,y,\,z\}$ for the intrinsic coordinate system of the
sample
and $\{x',\,y',\,z'\}$ for the laboratory coordinate system. The first one is fixed relative to the sample as shown in
Fig.\,\ref{Fig:PES}, with the $z$ axis normal to the sample surface and $x$ and $y$ axes corresponding to the
high-symmetry directions of the sample. The second one is the laboratory coordinate system with the $z'$ axis pointing
along the optical axis of the analyzer, $y'$ directed parallel to the entrance slit of the analyzer and $x'$ lying in
the horizontal plane so that $\{x',\,y',\,z'\}$ is a right-handed orthogonal unit vector triplet (see
Fig.\,\ref{Fig:Geometry}).

The position of the intrinsic coordinate system relative to the laboratory coordinate system is uniquely determined by
three Euler angles: polar angle $\varphi$, tilt $\psi$, and azimuth $\theta$. Starting from the laboratory coordinate
system, the transformation to the intrinsic coordinate system can be represented as a composition of three consecutive
rotations (see Fig.\,\ref{Fig:Euler}): first around the $y'$ axis in the positive direction by the polar angle
$\varphi$, then around the new (rotated) $x_1'$ axis in the negative direction by the tilt angle $\psi$, and finally
around the new (twice-rotated) $z_2'$ axis in the negative direction by the azimuthal angle~$\theta$. The coordinate
transformation can be represented by the following matrix product:
\begin{equation}\label{Eq:EulerRotations}
\text{\raisebox{-0.7ex}{\scalebox{1.5}{$\Bigg($}}}\begin{matrix}x\vspace{-0.5em}\\y\vspace{-0.5em}\\z\end{matrix}\text{\raisebox{-0.7ex}{\scalebox{1.5}{$\Bigg)$}}}
=\text{\raisebox{-0.7ex}{\scalebox{1.5}{$\Bigg($}}}\kern-1ex\begin{matrix}\phantom{-}\mathrm{cos}\,\theta&\kern-.3em\mathrm{sin}\,\theta&\kern-.3em0\vspace{-0.5em}\\-\mathrm{sin}\,\theta&\kern-.3em\mathrm{cos}\,\theta&\kern-.3em0\vspace{-0.5em}\\0&\kern-.3em0&\kern-.3em1\end{matrix}\text{\raisebox{-0.7ex}{\scalebox{1.5}{$\Bigg)$}}}
\!\text{\raisebox{-0.7ex}{\scalebox{1.5}{$\Bigg($}}}\begin{matrix}1&\kern-.3em0&\kern-.3em0\vspace{-0.5em}\\0&\kern-.3em\phantom{-}\mathrm{cos}\,\psi&\kern-.3em\mathrm{sin}\,\psi\vspace{-0.5em}\\0&\kern-.3em-\mathrm{sin}\,\psi&\kern-.3em\mathrm{cos}\,\psi\end{matrix}\!\text{\raisebox{-0.7ex}{\scalebox{1.5}{$\Bigg)$}}}
\!\text{\raisebox{-0.7ex}{\scalebox{1.5}{$\Bigg($}}}\!\begin{matrix}\mathrm{cos}\,\varphi&\kern-.3em0&\kern-.3em-\mathrm{sin}\,\varphi\vspace{-0.5em}\\0&\kern-.3em1&\kern-.3em0\vspace{-0.5em}\\\mathrm{sin}\,\varphi&\kern-.3em0&\kern-.3em\phantom{-}\mathrm{cos}\,\varphi\end{matrix}\kern-.5ex\text{\raisebox{-0.7ex}{\scalebox{1.5}{$\Bigg)$}}}
\!\text{\raisebox{-0.7ex}{\scalebox{1.5}{$\Bigg($}}}\begin{matrix}x'\vspace{-0.5em}\\y'\vspace{-0.5em}\\z'\end{matrix}\!\text{\raisebox{-0.7ex}{\scalebox{1.5}{$\Bigg)$}}}
\end{equation}

The analyzer can detect only electrons with momenta $\mathbf{p}_{\text{e}^-}\!=
\{0,\,p\,\mathrm{sin}\,{\eta},\,p\,\mathrm{cos}\,{\eta}\}$ in the laboratory coordinate system, where the
\textit{analyzer angle} $\eta$ is the angle between the electron momentum and analyzer axis limited by the acceptance
angle, and $p=\sqrt{2mE_\text{kin}}\,/\,\hslash$ is the absolute value of the free electron's momentum. To find out
the
intrinsic coordinate representation of this vector, we substitute it into (\ref{Eq:EulerRotations}), which results in
the following expressions for the sought $k_x$ and $k_y$ components of the momentum:
\begin{subequations}\label{Eq:EulerMultiplicationResult}
\begin{align}
k_x=\frac{\sqrt{2\kern.2pt m\kern.2pt
E_\text{kin}}}{\hslash}\,\bigl[\mathrm{sin}\,\eta\,\mathrm{cos}\,\psi\,\mathrm{cos}\,\theta+\,\mathrm{cos}\,\eta\,(\mathrm{cos}\,\phi\,\mathrm{sin}\,\psi\,\mathrm{sin}\,\theta-\mathrm{sin}\,\phi\,\mathrm{cos}\,\theta)\bigr]\\
k_y=\frac{\sqrt{2\kern.2pt m\kern.2pt
E_\text{kin}}}{\hslash}\,\bigl[\mathrm{cos}\,\eta\,\mathrm{sin}\,\phi\,\mathrm{sin}\,\theta+\,\mathrm{cos}\,\theta\,(\mathrm{cos}\,\eta\,\mathrm{cos}\,\phi\,\mathrm{sin}\,\psi+\mathrm{sin}\,\eta\,\mathrm{cos}\,\psi)\bigr]
\end{align}
\end{subequations}
As follows from Eq.~(\ref{Eq:MomentumConservation}), the parallel components of the electron's momentum are conserved
during the photoemission process up to the multiples of the reciprocal lattice vector, so $k_x$ and $k_y$ in
(\ref{Eq:EulerMultiplicationResult}) actually represent the parallel components of the electron's quasimomentum in the
solid.

\hvFloat[floatPos=t, capWidth=1.0, capPos=r, capVPos=t, objectAngle=0]{figure}
        {\!\includegraphics[width=0.6\textwidth]{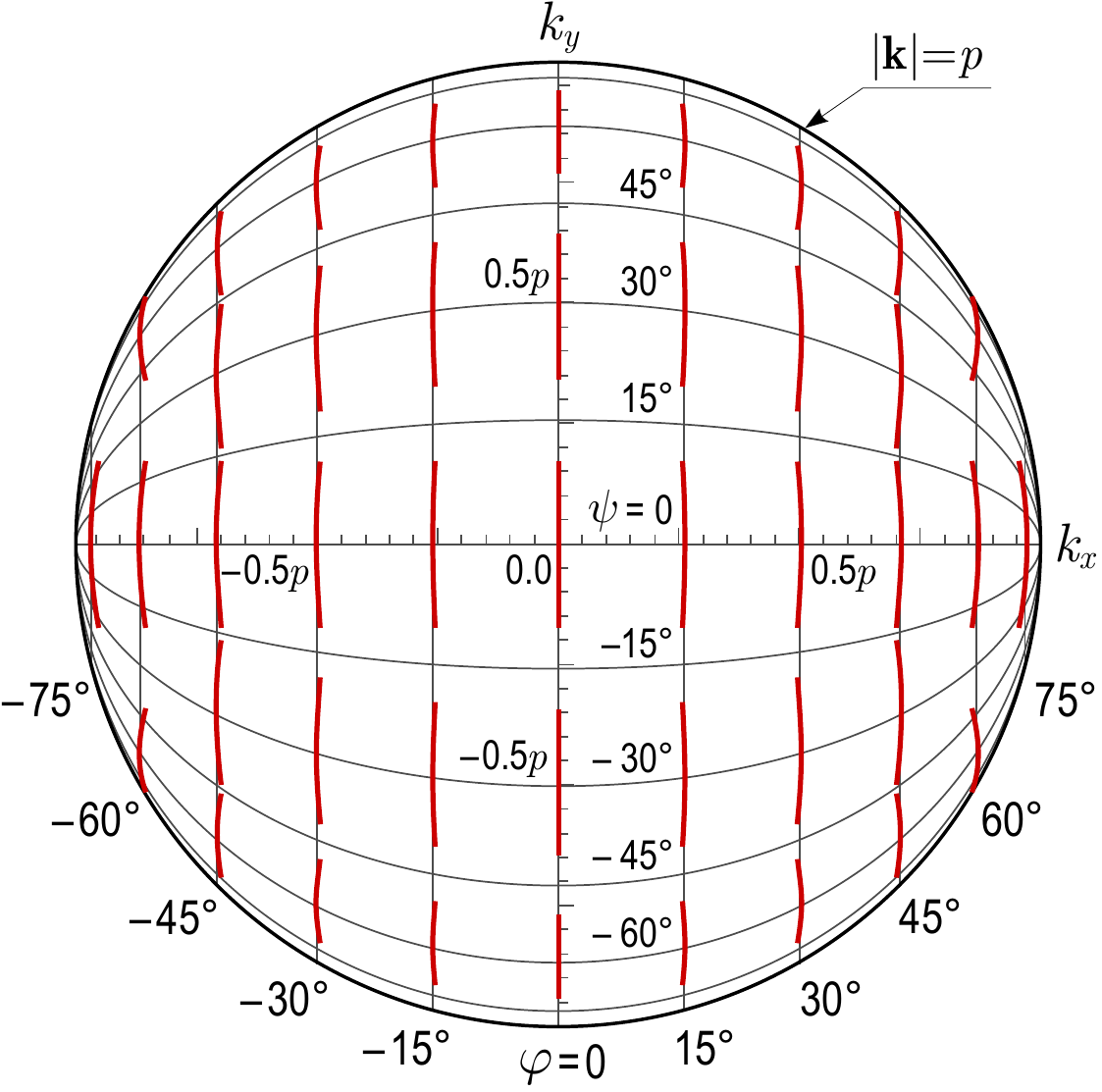}~~}
        {Position of the slit image in $\mathbf{k}$-space for different polar and tilt angles. The grid lines
        correspond
        to different values of the polar angle (vertical lines) and tilt (horizontal lines) at zero azimuth. Slit
        images
        for $\pm 10^\circ$ acceptance angle corresponding to different manipulator positions are shown in red.
        Non-zero
        azimuth would correspond to the rotation of the whole image by the azimuthal angle. Note that with the
        \textit{Cryoax~6} manipulator only tilt angles in the range from $-\text{5}^\circ$ to $\text{20}^\circ$ are
        actually accessible.}
        {Fig:Globus}

The locus of all $\mathbf{k}$ points that can be detected by the analyzer at a given manipulator position is called a
\textit{slit image}. It is a curve segment parametrically given by (\ref{Eq:EulerMultiplicationResult}) as $\eta$
varies
over the acceptance angle interval $-\eta_\text{max}<\eta<\eta_\text{max}$. Its position in $\mathbf{k}$-space can be
changed by varying the manipulator angles (Fig. \ref{Fig:Globus}). Note that all the accessible $\mathbf{k}$ vectors
lie
within a circle of radius $p$, and the length of the slit image, i.e. the distance between momenta corresponding to
the
maximal and minimal analyzer angles, is given by $2\kern.2pt p\,\mathrm{cos}\,\psi\,\mathrm{sin}\,\eta_\text{max}$.
Both
depend on the electron's kinetic energy which in turn depends on the photon energy. Therefore by increasing the photon
energy one can increase both the accessible momentum space and the momentum coverage of the slit image. On the other
hand, the energy resolution will deteriorate with increasing photon energy.

\textbf{$\mathbf{k}$-space mapping technique.} As seen from Fig.\,\ref{Fig:Globus}, with the change of the polar angle
the slit image moves orthogonally to itself, whereas with the change of the tilt angle the slit image moves parallel
to
itself. A change of azimuth is equivalent to the rotation of the slit image relative to the origin of coordinates. One
sees that by changing the polar angle in small steps one can map a certain band in $\mathbf{k}$-space of the width of
the slit image, which is called a \textit{polar map}. Measuring several polar maps at different values of the tilt one
can extend the width of the measured band of the $\mathbf{k}$-space. The direction, in which the polar mapping is
performed, is given by the azimuthal angle. Sometimes it is convenient to perform mapping along different directions,
e.g. along all high-symmetry directions of the sample. The movement of the manipulator responsible for the mapping can
be performed automatically, resulting in a single three-dimensional data set of the photocurrent as a function of
kinetic energy, analyzer angle, and manipulator angle. After rescaling the data according to
(\ref{Eq:EulerMultiplicationResult}) one ends up with a spectrum in the coordinates of binding energy and two
projections of the momentum. As was shown in the previous section, such spectrum can be described within a number of
approximations by Eq.\,(\ref{Eq:PhotocurrentExperimental}) and can therefore provide important information about the
single-particle spectral function of the sample.\vspace{1em}

\subsection{Synchrotron light sources}\vspace{-1em}

\hvFloat[floatPos=t, capWidth=1.0, capPos=b, capVPos=t, objectAngle=0]{table}
        {\begin{tabular}{lccc}
         \toprule
         Synchrotron: & SLS & \multicolumn{2}{c}{BESSY II}   \\
         \midrule
         Electron energy: &  2.4\,GeV & \multicolumn{2}{c}{1.7/1.9\,GeV}   \\
         Circumference: & 288\,m & \multicolumn{2}{c}{240\,m} \\
         \midrule
         Beamline: & SIS X9L & UE112-PGM1 & UE112-LowE PGMb\\
         \midrule
         Energy range\\
         \quad linear horizontal: & 10\,--\,800\,eV  & 15\,--\,600\,eV & 5\,--\,250\,eV\\
         \quad linear vertical:   & 100\,--\,800\,eV & --- & ---\\
         \quad circular:          & 50\,--\,800\,eV  & --- & in project\\
         Beam spot size:          & $50\times100\,\mu$m & --- & slitsize $\times 100\mu$m\\
         Resolving power:         & $10^4$ & $10^4$ & $>10^5$ \\
         Undulator: & UE212 & UE112 & UE112 \\
         \bottomrule
         \end{tabular}\quad}
        {Comparative technical characteristics of the synchrotrons and beamlines used in the present work.}
        {Table:Beamlines}

\begin{figure}[b]
\includegraphics[width=\textwidth]{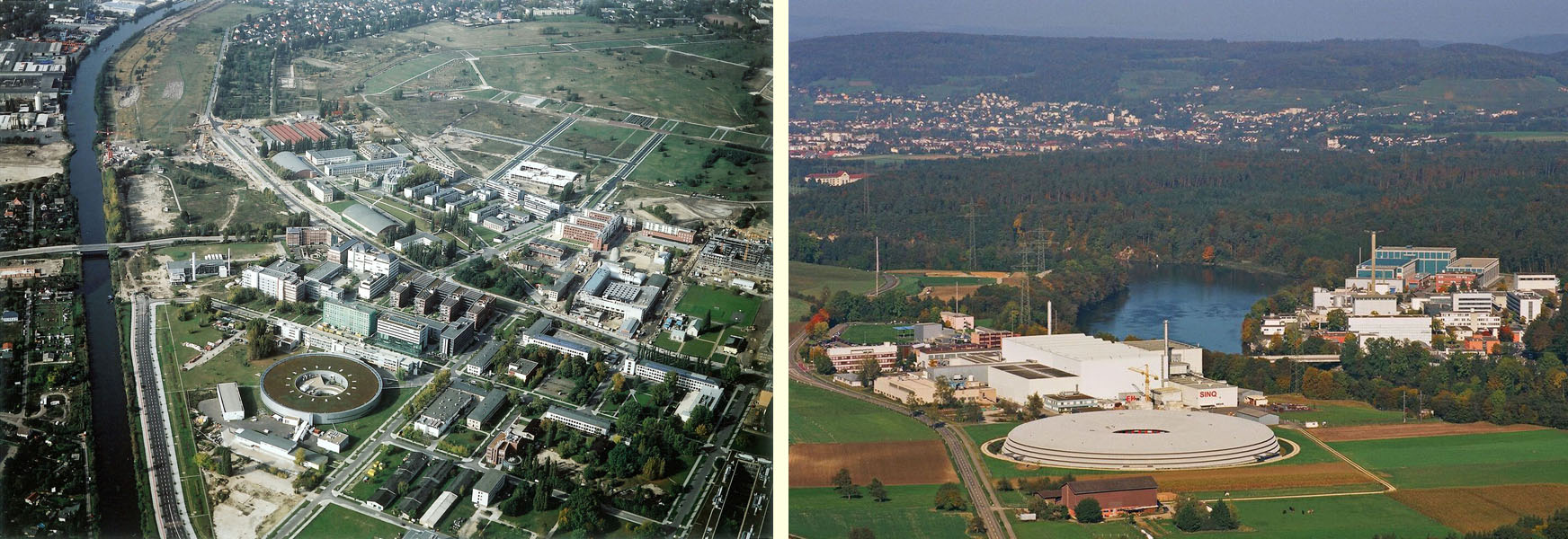}
\caption{BESSY II, Berlin-Adlershof, Germany (left), and SLS, Villigen, Switzerland (right), synchrotron facilities.}
\label{Fig:Synchrotrons}
\end{figure}

Modern ARPES experiments require high-quality light sources with high energy resolution (narrow bandwidth), variable
photon energy and polarization, small spot size, high intensity, and high stability of the beam. All these conditions
can be met only with synchrotron radiation at the built-on-purpose beamlines specially optimized for this class of
experiments. The working principle of synchrotron light sources is based on the fact that high-energy charged
particles
(such as electrons, protons, or positrons) emit electromagnetic radiation in the ultra-violet to X-ray range when
subjected to large accelerations orthogonal to their velocity \cite{IwanenkoPomeranchuk44, ElderLangmuir48, Ternov95}.
The electrons are accelerated in a linear accelerator (linac) and then in the booster to energies of several GeV and
then injected into the storage ring (see Fig.\,\ref{Fig:Synchrotron}). The closed trajectory of the electrons is
controlled by a set of bending magnets, and the energy losses at each revolution are compensated by the radio
frequency
cavity. To produce an intense photon beam, the electrons are led through undulators\,---\,special devices with a
periodic pattern of the magnetic field. The undulators can be tuned to have a maximum of photon emission at a given
energy (\textit{fundamental mode}). Due to the relativistic effects the directional pattern of photon emission is
peaked
in the forward direction of the electron beam and is further enhanced by interference effects between photons emitted
at
different undulator wiggles. The photon beam generated in the undulator is then fed into one of several beamlines,
where
it is monochromatized and focused before it reaches the measurement chamber.

The data for the present work has been acquired at two electron synchrotrons (Fig.\,\ref{Fig:Synchrotrons}): at the
SIS-9L beamline of the Swiss Light Source (SLS) in Villigen, Switzerland \cite{SLSWebSite}, and at the Berliner
Elektronenspeicherring-Gesellschaft für Synchrotronstrahlung m.\,b.\,H. (BESSY) in Germany, beamlines UE112-PGM1 and
UE112-LowE PGMb \cite{BESSYWebSite, KraemerBESSY}. All three beamlines are fitted with the plane grating
monochromators
(PGM). The summary of technical characteristics of the synchrotrons and the beamlines can be found in table
\ref{Table:Beamlines}.

\hvFloat[floatPos=t, capWidth=1.0, capPos=b, capVPos=t, objectAngle=0]{figure}
        {\!\includegraphics[width=\textwidth]{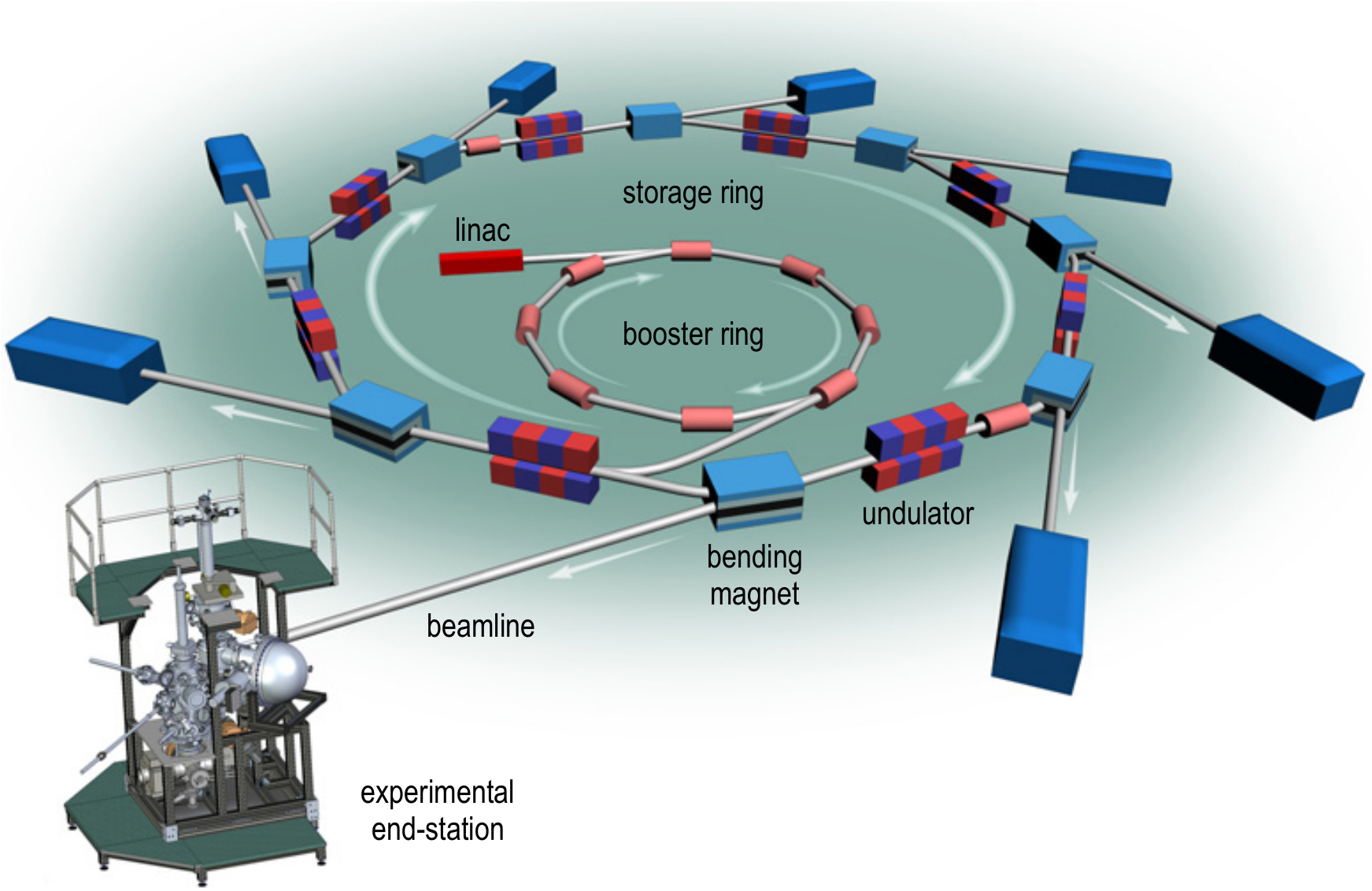}}
        {Sketch of a synchrotron facility with an ARPES end-station.}
        {Fig:Synchrotron}

\subsection{Sample preparation}

In the present work, cuprates of the Bi2212 and Y123 families were studied (for the crystal structure, see section
\ref{Sec:CrystalStructure}). The single crystals of the slightly overdoped (Bi,\,Pb)$_2$Sr$_2$CaCu$_2$O$_{8+\delta}$
($T_\text{c}=\text{71}$\,K), slightly underdoped (Bi,\,Pb)$_2$Sr$_2$Ca$_{1-x}$Tb$_x$Cu$_2$O$_8$ and optimally doped
Bi$_2$Sr$_2$CaCu$_2$O$_{8+\delta}$ were grown by the self-flux method in the group of Dr.~H.~Berger.\footnote{Institut
de Physique de la Matière Complexe, Lausanne, Switzerland.} The nearly optimally doped untwinned
YBa$_2$Cu$_3$O$_{6.85}$
($T_\text{c}=\text{92}$\,K) samples were synthesized by the solution-growth method in the group of
Prof.~B.~Keimer.\footnote{Max-Planck-Institut für Festkörperforschung, 70569 Stuttgart, Germany.} Detwinning was
achieved by applying a uniaxial mechanical stress to the sample at elevated temperatures.

\textbf{Sample cleavage}. Because ARPES is a surface-sensitive technique, the bulk samples need to be prepared by
cleaving \textit{in situ} in UHV conditions ($P\approx \text{5}\cdot\text{10}^{-11}$\,mbar) immediately before the
measurement. At
such vacuum conditions, the samples can usually be measured within approximately 24 hours after cleavage without any
significant degradation of the surface. In order to transfer a sample into the UHV chamber and cleave it, it has to be
glued to the sample holder as seen in Fig.\,\ref{Fig:Manipulator} using a conducting silver epoxy glue.

\hvFloat[floatPos=t, capWidth=1.0, capPos=r, capVPos=t, objectAngle=0]{figure}
        {\hspace{-0.5ex}\includegraphics[width=0.7\textwidth]{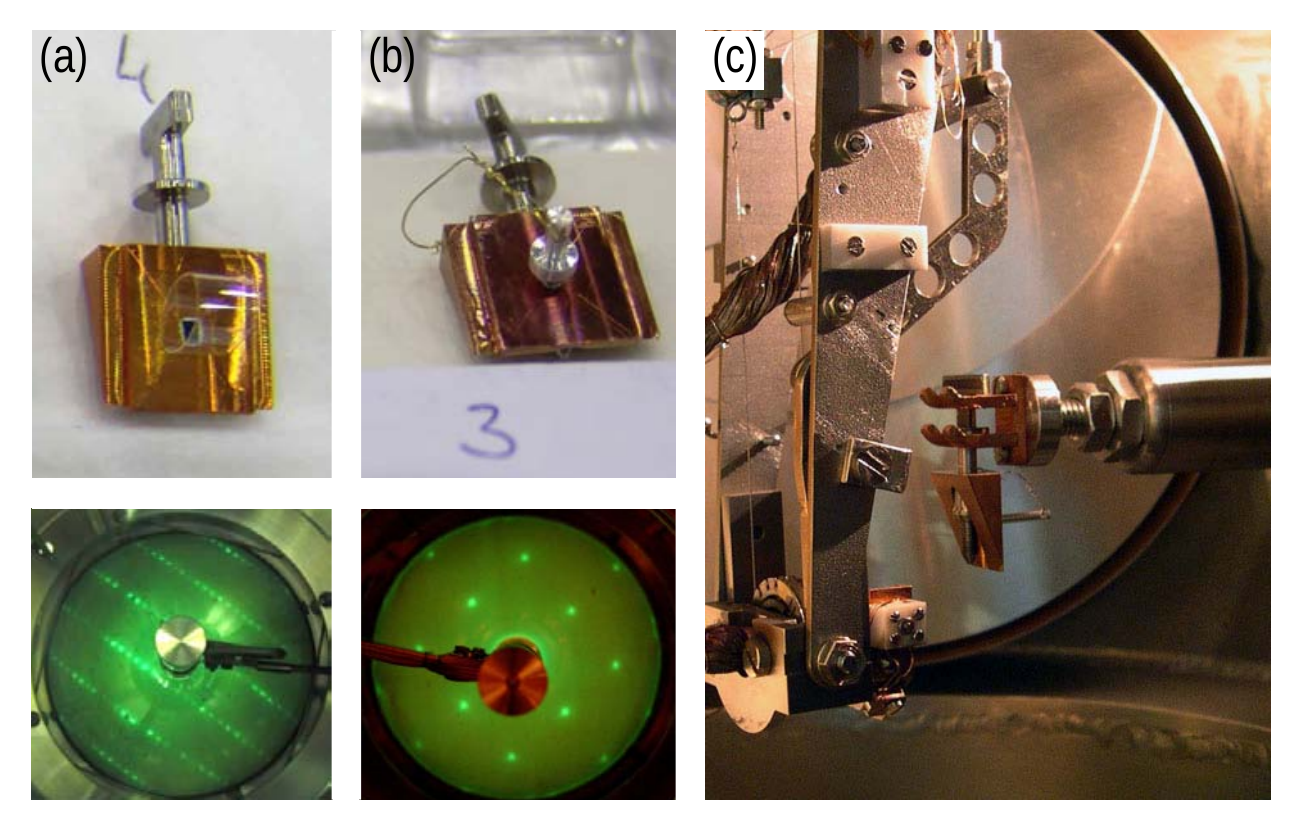}~}
        {Cleavage methods of BSCCO and YBCO samples and the typical LEED images taken from the cleaved surfaces.
        \textbf{(a)}~BSCCO samples can be cleaved with an O-shaped stripe of a sticky tape. \textbf{(b)}~YBCO samples
        have to be cleaved with an aluminum top-post glued to the top. \textbf{(c)}~Transfer of a sample into the
        preparation chamber for cleavage.}
        {Fig:Cleavage}

In case of a BSCCO sample, weak Van-der-Waals bonds between the BiO layers make cleavage especially easy. It can be
done
simply with an O-shaped stripe of a sticky tape glued to the free surface of the sample as shown in
Fig.\,\ref{Fig:Cleavage}~(a). In the UHV chamber, the stripe is then removed by the ``fork'' of the sample transfer
system together with the upper layers of the crystal, which usually results in an atomically clean, mirror-like
surface.
The lower image in panel (a) shows a typical low-energy electron diffraction (LEED) picture taken from the surface of
a
Pb-free Bi2212 sample after cleavage.

Cleavage of YBCO samples requires a little more effort, but is also straightforward. As a rule, an aluminum top-post
is
glued to the free surface of the sample using the same silver epoxy glue, as shown in Fig.\,\ref{Fig:Cleavage}~(b).
The
top-post is tied to the sample holder with a thin golden wire in order to keep it from falling after cleavage. After
the
sample is transferred into the preparation chamber [see panel (c) of the same figure], the top-post is kicked off with
a
screwdriver, which usually results in at least part of the sample to have a mirror-like atomically clean surface
suitable for measurements. Such method of cleavage allows one to apply a substantial cleaving force, which makes it
possible
to cleave even harder samples, such as LSCO, with a significantly non-zero chance of success.

\subsection{Experimental data preprocessing}

\textbf{Intensity normalization}. Possible inhomogeneities of the detector sensitivity in different momentum channels
can lead to a slight $\mathbf{k}$-dependent intensity modulation in the raw ARPES spectra. Moreover, when measuring an
angular map, the intensity may vary from one momentum cut to another due to various reasons. To eliminate these
effects
and to bring all spectra to a single intensity scale, the spectra are usually normalized to the background intensity
integrated above the Fermi level. This spectral weight comes from the non-dispersing core levels excited by the higher
harmonics of the synchrotron radiation, and is therefore expected to have no significant dependence on momentum. The
normalized ARPES intensity is therefore given by
\begin{equation}
I_\text{norm}(E_\text{kin},\,\eta)=I_\text{raw}(E_\text{kin},\,\eta)\,\Bigg/\!\int_{E_1}^{E_2}\!I_\text{raw}(E_\text{kin},\,\eta)\,\mathrm{d}E_\text{kin}\text{,}
\end{equation}
where $[E_1,\,E_2]$ is a kinetic energy window above the Fermi level: $E_\text{F}<E_1<E_2$.

\textbf{Energy scale corrections}. The energy scale of a raw ARPES spectrum represents the kinetic energy of
photoelectrons, which still has to be transformed to the binding energy scale. Naively, one could expect to be able to
do the transformation using the energy conservation law $E_\text{bind}=h\nu-E_\text{kin}-\mathit{\Phi}$.
Unfortunately, this is not possible, because the work function is not known with the required precision.
Instead, a reference spectrum of a freshly evaporated silver or gold film has to be measured, which is known to
produce a perfect Fermi step. Because of the electrical contact between the sample, silver film, and analyzer, the Fermi
levels in both spectra should coincide. $E_\text{F}$ can be therefore determined from the reference spectrum in every
momentum channel independently (usually with subsequent smoothing\footnote{For more details, see Ref.~\citenum{ZabolotnyyThesis},
p.~38\,--\,39.}), and the energy scale transformation takes the form \begin{equation}
\omega(\eta)=-E_\text{bind}(\eta)=E_\text{kin}-E_\text{F}(\eta)\text{.}
\end{equation}
The momentum-dependence of the Fermi energy determined in such a way comes mainly from the imperfectness of the energy calibration curves and the form of the analyzer slit. By the described procedure these deviations can be eliminated, which results in the accuracy of the energy scale $\sim$\,1\,meV. For the best accuracy, it is important that the reference spectrum is taken with exactly the same experimental settings (analyzer slit, beamline settings, pass energy, etc.) as the spectra that are to be calibrated. For example, changing the photon energy of the beamline to a different value and back may result in an apparent shift of the Fermi level due to the imperfect reproducibility of the monochromator, so the reference spectrum will have to be remeasured.


\chapter{Electronic structure and renormalization effects in Bi$_2$Sr$_2$CaCu$_2$O$_{8+\delta}$}

\section{Crystal structure of layered cuprates}\label{Sec:CrystalStructure}

\subsection{Basic cuprate families}

After the high-temperature superconductivity had been discovered in Ba$_x$La$_{5-x}$Cu$_5$O$_{5(3-y)}$
\mbox{\cite{BednorzMueller86}, the\,high-$T_\text{c}$\,cuprates\,became\,one\,of\,the\,most\,studied\,materials\,in\,all
of\,science,\,yielding} in the research activity perhaps only to silicon. Even though the physics of superconductivity
in these materials is not yet fully understood, their studies have been regularly producing new fundamental results in
physics for more than 20 years since their discovery in 1986.

\hvFloat[floatPos=t, capWidth=1.0, capPos=l, capVPos=t, objectAngle=0]{figure}
        {\hspace{-27.7em}\includegraphics[width=\textwidth]{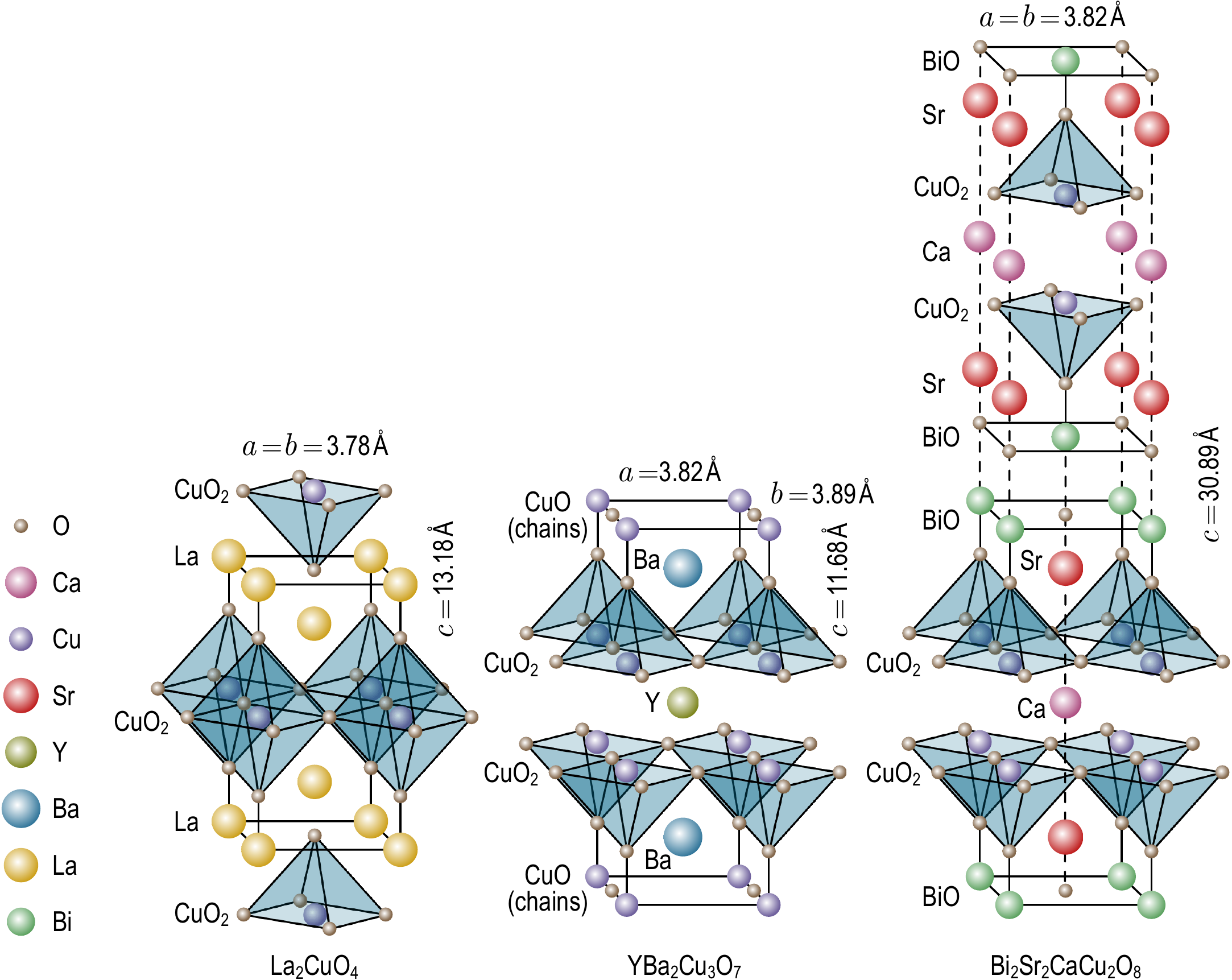}\!}
        {Crystal structure of the parent compounds for the three most studied families of high-$T_\text{c}$ cuprates:
        LSCO, YBCO, and BSCCO, shown in chronological order of their discovery. The size of the atoms represents their
        atomic radii. The perovskite-like oxygen coordination of Cu atoms is marked by the blue polyhedra. The lattice
        parameters $a$, $b$, and $c$ are shown beside each drawing. Note that LSCO has a single CuO$_2$ layer per
        formula unit, while YBCO and BSCCO have a CuO$_2$ bilayer per formula unit. The unit cells of LSCO and BSCCO
        consist of two formula units shifted by $\text{(}\mathbf{a}+\mathbf{b}\text{)}/2$, while the unit cell of YBCO
        consists of a single formula unit.}
        {Fig:CrystalStructure}

The crystal structure of the parent compounds for the three most actively studied families of cuprates is shown in
Fig.\,\ref{Fig:CrystalStructure}. Among them, the La-based family, including La$_{2-x}$Ba$_x$CuO$_4$ (`Zürich oxide',
LBCO) and La$_{2-x}$Sr$_x$CuO$_4$ (LSCO), was the first to be discovered \cite{BednorzMueller86, Kitazawa87}. These are
the hardest of the three materials, which makes the growth of larger single crystals ($\sim$\,1\,cm) possible. On the
other hand, the stronger chemical bonds complicate the cleavage, making it difficult to obtain atomically flat surfaces
for surface-sensitive measurements.

\begin{figure}[b]
\vspace{1em}\center\includegraphics[width=1.02\textwidth]{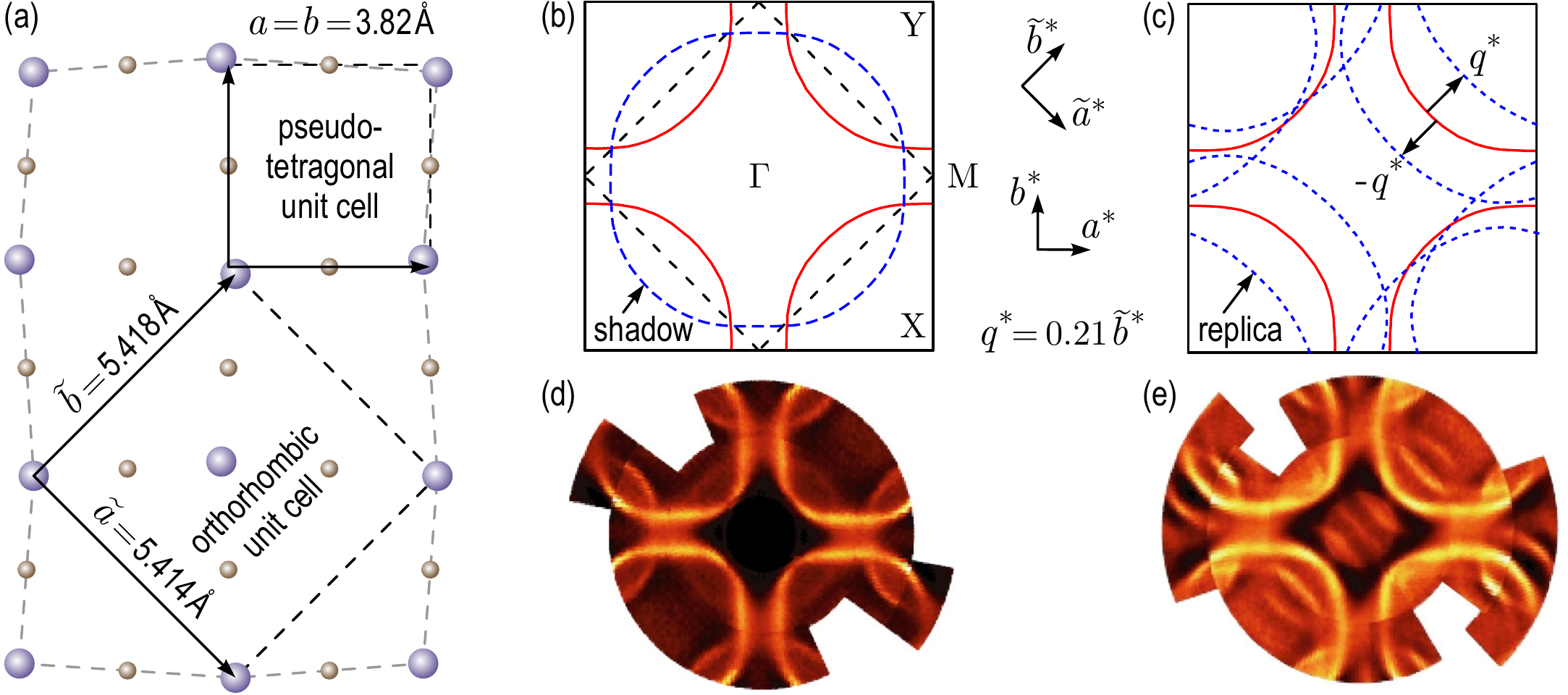}%
\caption{\textbf{(a)}~2\,$\times$\,2 periodic lattice modulation in the CuO$_2$ plane with orthorhombic and
pseudotetragonal unit cells. \textbf{(b)}~Appearance of the shadow Fermi surface (blue dashed line) when changing from
the tetragonal (solid square) to the orthorhombic (dashed square) unit cell due to the folding of the original Fermi
surface (red lines). \textbf{(c)}~Appearance of the superstructure replicas due to the incommensurate periodic lattice
modulation with the period of $\text{4.75}\,\tilde{b\,}$\!. \textbf{(d}\,--\,\textbf{e)}~Manifestation of the shadow
Fermi surface and superstructure replicas in ARPES spectra, experimental data after
Ref.~\citenum{BorisenkoNature04}.\vspace{-1.5em}}
\label{Fig:PLD}
\end{figure}\afterpage{\clearpage}

The second family of compounds, YBa$_2$Cu$_3$O$_{7-\delta}$ (YBCO, or Y123), was also discovered within a year after
the
discovery of high-$T_\text{c}$ superconductivity \cite{HorGao87, WuAshburn87}. Unlike LSCO, it was a bilayer cuprate,
and was the first superconductor to break the 77\,K (liquid nitrogen) temperature limit. YBCO can be prepared in
relatively large, clean, highly ordered single crystals, which makes it the best cuprate for studies by bulk-sensitive
techniques such as INS. YBCO crystals are easily cleavable along the CuO chain layer. Unfortunately, such cleavage
leads to the charge redistribution at the surface and, as a result, to significant overdoping of the upper CuO$_2$ bilayer
\cite{ZabolotnyyOD07}, precluding the studies of its bulk properties by surface-sensitive techniques such as ARPES.
Note the slight orthorhombic distortion (1.8\%) in the CuO$_2$ planes of YBCO due to the anisotropy caused by the presence
of the one-dimensional CuO chains.

The third compound, Bi$_2$Sr$_2$CaCu$_2$O$_{8+\delta}$ (BSCCO, or Bi2212), was independently discovered in 1988 by
several groups \cite{MaedaTanaka88, HazenPrewitt88, ShawShivashankar88, SunshineSiegrist88, TarasconPage88} to become
the first high-temperature superconductor not containing a rare earth element. It is usually available in thin, small
single crystals, but is easily cleavable and is therefore the favorite material for STM and ARPES. Unlike YBCO, BSCCO
has no chain layers, which makes its spectra easier to analyze. The cleavage happens symmetrically along the weakest
bonds between the BiO layers and does not therefore distort the nearest CuO$_2$ bilayers, so that they can be
considered representative of the crystal's bulk properties. On the other hand, because of the small volume of the samples, the
bulk-sensitive measurements of BSCCO are rather difficult.

\hvFloat[floatPos=b, capWidth=1.0, capPos=l, capVPos=t, objectAngle=0]{figure}
        {\includegraphics[width=0.59\textwidth]{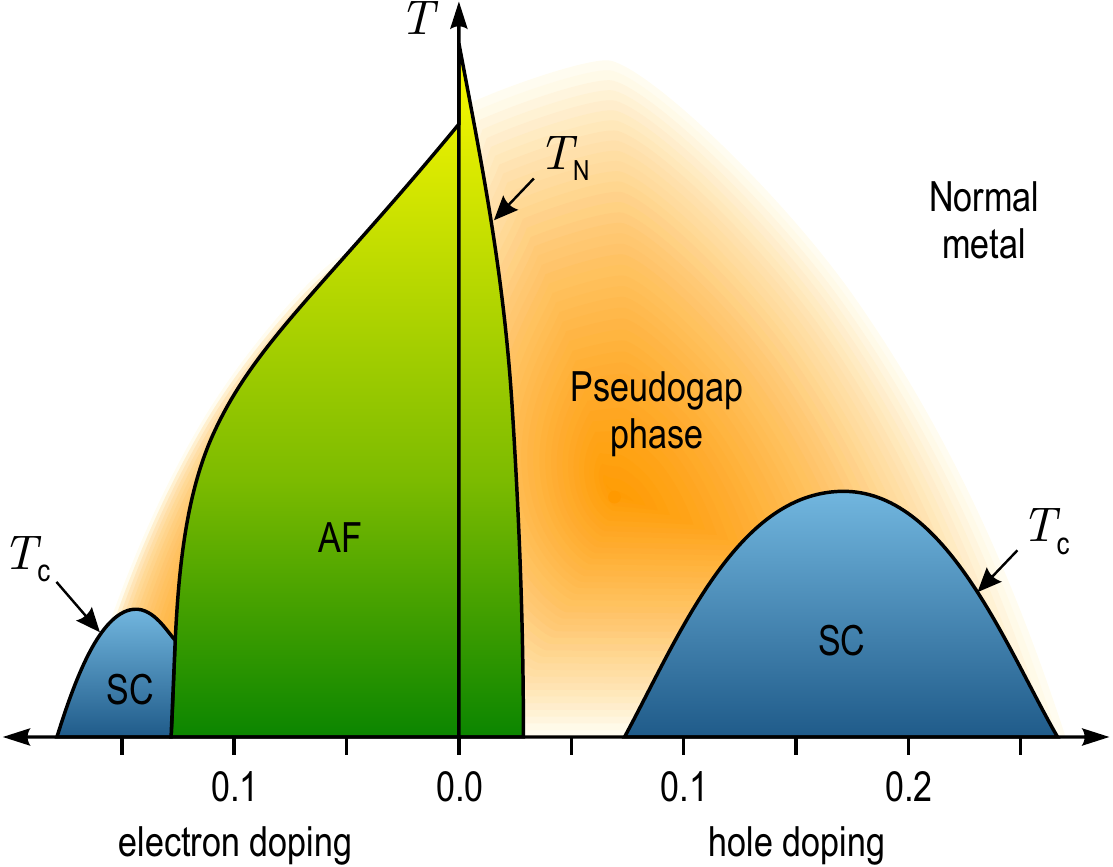}}
        {Schematic doping phase diagram of electron- and hole-doped high-$T_\text{c}$ superconductors
         showing, in particular, the superconducting (SC) and antiferromagnetic (AF) phases
         \cite{Fischer07}.\vspace{-1em}}
        {Fig:PhaseDiagram}

\subsection{Periodic lattice modulations}\label{SubSec:PLD}

The unit cell of BSCCO, shown in Fig.\,\ref{Fig:CrystalStructure}, is not exactly tetragonal. Within the
$x\!y$-plane, there is a slight periodic distortion of the lattice. First, there is a commensurate lattice
modulation in the $\tilde{\mathbf{a}}$ direction as shown in Fig.\,\ref{Fig:PLD}\,(a) that changes the original
tetragonal unit cell to a larger orthorhombic unit cell \cite{KoitzschBorisenko04, MansSantoso06, ArpiainenLindroos06,
NakayamaSato06} with lattice parameters $\tilde{a}=$\,5.414\,\AA, $\tilde{b}=$\,5.418\,\AA, and $c=$\,30.89\,\AA. If
one nevertheless considers the original (pseudotetragonal) unit cell, in the reciprocal space this periodic modulation will manifest itself
as shadows of the main bands shifted by ($\piup,\,\piup$), as shown in Fig.\,\ref{Fig:PLD}\,(b) and (d). Moreover,
there is an incommensurate lattice modulation in the $\tilde{\mathbf{b}}$ direction with a period of 4.75\,$\tilde{b\,}\!$
\cite{MilesKennedy06, Gladyshevskii96, CalestaniRizzoli89}, clearly observed both in surface- \cite{HazenPrewitt88,
ShawShivashankar88, PunHudson98, PunONeal01} and bulk-sensitive \cite{CalestaniRizzoli89} experiments. This modulation
leads to the appearance of superstructure replicas in the ARPES spectra as shown in Fig.\,\ref{Fig:PLD}\,(c) and (e),
which are an additional obstruction for data analysis \cite{DingBellman96, ChuangGromko99, BorisenkoGolden00,
AsensioAvila03}. To minimize their effect, BSCCO is often doped by Pb, which partially replaces the Bi atoms in the
lattice, enhancing the $T_\text{c}$ \cite{SunshineSiegrist88} and removing the superstructure distortion
\cite{RameshHegde89, Eibl91, XianhuiYitai92, XianhuiCheng93}. The Fermi surfaces in Fig.\,\ref{Fig:PLD}\,(d) and (e)
were measured from BSCCO crystals with and without Pb doping respectively.

\begin{table}[t]
\center\includegraphics[width=0.9\textwidth]{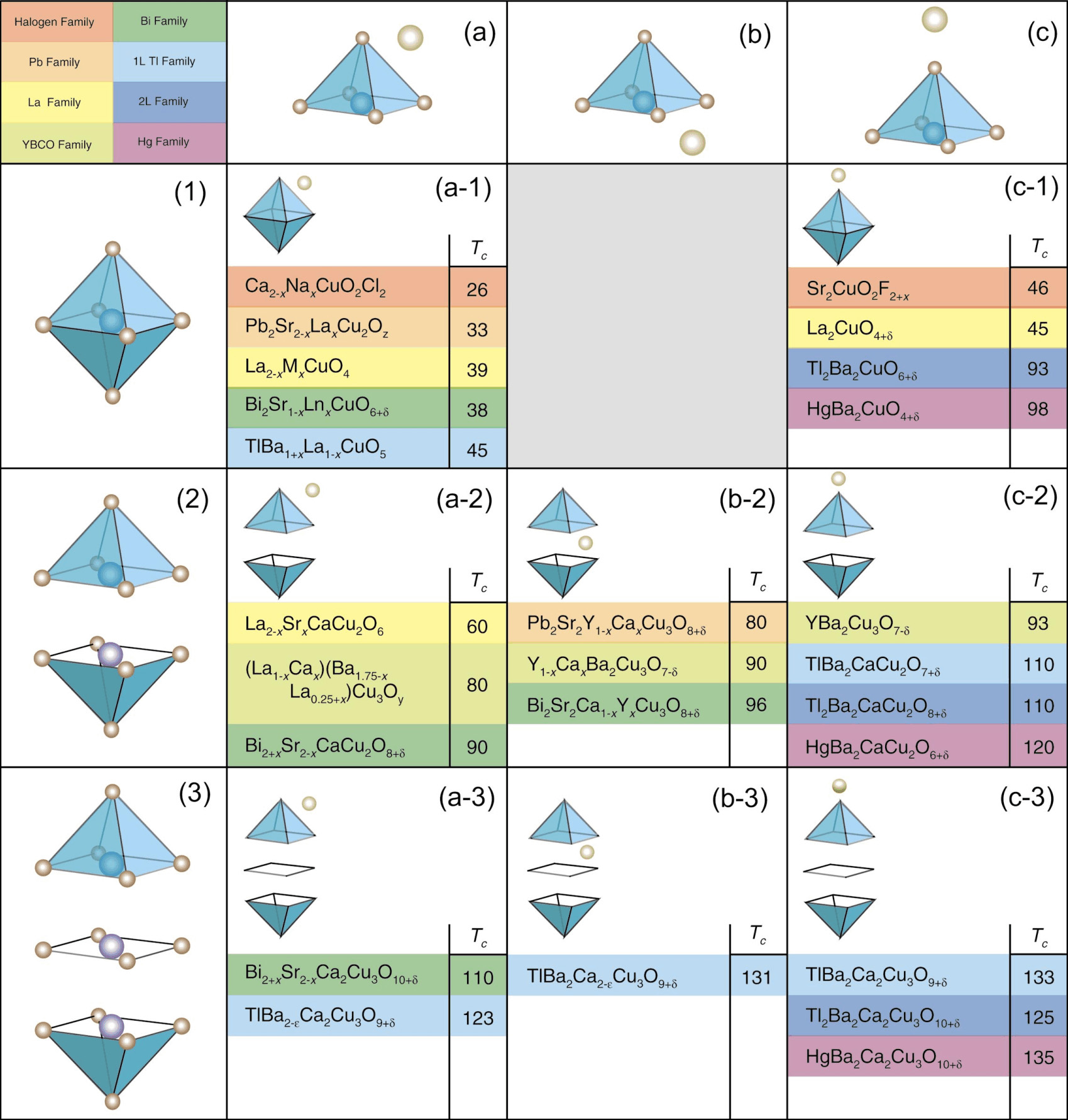} \caption{Classification of hole-doped high-$T_\text{c}$
cuprates in terms of the disorder site and the number of CuO$_2$ layers and their critical temperatures at optimal
doping taken from Ref.~\citenum{EisakiKaneko04}.} \label{Table:CupratesZoo}
\end{table}

\subsection{The phase diagram}

The layered structure of cuprates makes their electronic structure quasi-two-dimensional. Their parent compounds at
zero
doping are antiferromagnetic Mott insulators at half-filling (one hole per Cu site). The superconductivity and
metallic
behavior set in when the crystals are doped either by holes or electrons, as can be seen in the schematic phase
diagram
in Fig.\,\ref{Fig:PhaseDiagram} \cite{Fischer07, LeeNagaosa06, Dagotto94}. As more holes are added into the CuO$_2$
planes, the O\,2p$^5$ and Cu\,3d$^9$ orbitals hybridize into the so-called \textit{Zhang-Rice singlet} state
\cite{ZhangRice88}, which leads to the antiferromagnetic superexchange between O and Cu holes. In the metallic and
superconducting regimes the conduction bands are formed exclusively by the states within the CuO$_2$ planes (in case
of
YBCO also by the CuO chains). The superconductivity occurs only within the CuO$_2$ planes, whereas the rest of the
crystal structure serves as a charge reservoir to control the doping of the planes.

\subsection{Variety of high-$T_\text{c}$ cuprates}

In Ref.~\citenum{EisakiKaneko04}, H.~Eisaki \textit{et al.} provide a classification of the hole-doped cuprates in
terms
of the disorder site and the number of CuO$_2$ layers, which is shown in table~\ref{Table:CupratesZoo}, together with
the corresponding critical temperatures. Note that $T_\text{c}$ increases both across the rows and down the columns of
the chart for each single family of materials, each denoted by a different color. By changing the number of layers,
the
disorder atom and its position, dopants, etc. the macroscopic properties of the high-$T_\text{c}$ cuprates, including
$T_\text{c}$, can be significantly optimized. In most recent attempts to increase $T_\text{c}$, crystals with the
so-called planar weight disparity\footnote{Planar weight disparity exists whenever planes within the layered
perovskites
are alternated light-heavy.} (PWD) \cite{BarreraSarmiento07, Eck175K} were prepared, some of which are claimed to
exhibit signatures of superconductivity at temperatures as high as 175\,K at ambient pressure \cite{Eck175K}.

\section{Electronic band dispersion and self-energy}\label{Sec:DispersionSE}

\subsection{Electronic structure of Bi2212 as seen by ARPES}

\hvFloat[floatPos=t, capWidth=1.0, capPos=r, capVPos=t, objectAngle=0]{figure}
        {\includegraphics[width=0.68\textwidth]{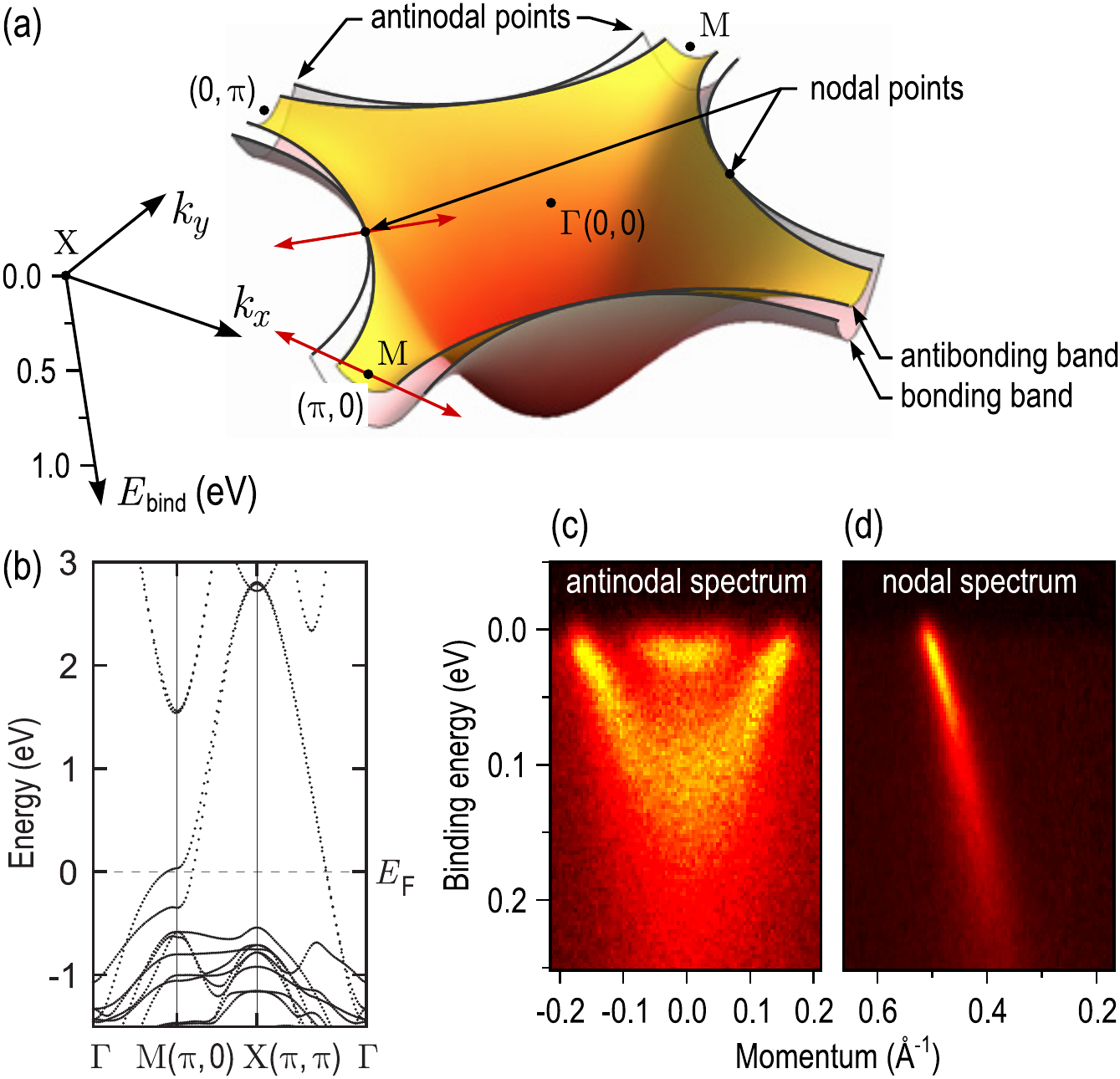}\quad}
        {\textbf{(a)}~Schematic dispersion of the conductance band of optimally doped Bi2212 in a tight-binding model
        (parameters from Ref.\,\citenum{KordyukBorisenko03}). The bonding and antibonding sub-bands are shown in pink
        and yellow color respectively. \textbf{(b)}~LDA band structure of Bi$_2$Sr$_2$CaCu$_2$O$_{8.3}$ along the
        high-symmetry directions from Ref.\,\citenum{LinSahrakorpi06}. \textbf{(c)}~and \textbf{(d)}~Typical ARPES
        images measured in the vicinity of the antinodal and nodal points in the directions shown in panel (a) by the
        red arrows. The color scale represents photoemission intensity.\vspace{-1em}}
        {Fig:ElectronicStructure}

Due to the layered crystal structure with the conductance electrons confined to the CuO$_2$ planes, the electronic
structure of the high-$T_\text{c}$ cuprates is quasi-two-dimensional, which significantly simplifies the ARPES
measurements and data treatment. Though a small but finite $k_z$-dispersion has been reported \cite{SterneWang88,
MarkiewiczSahrakorpi05, BansilLindroos05, LindroosSahrakorpi06}, much of the existing ARPES work on the cuprates
implicitly assumes the system to be perfectly two-dimensional \cite{Damascelli03,CampuzanoNorman04}, ignoring the
effects of interlayer hopping.

As mentioned earlier, the conductance band of a doped CuO$_2$ plane is formed by the hybridized O\,2p$^5$ and
Cu\,3d$^9$ states\,---\,the \textit{Zhang-Rice singlet} \cite{ZhangRice88}. In a solitary CuO$_2$ layer, this state corresponds
to a single band in the momentum space, whereas in the double-layer compounds it is additionally split into the
bonding
and antibonding sub-bands \cite{KordyukBorisenko02, BorisenkoKordyuk02, KordyukBorisenko02prl, KordyukBorisenko04} due
to the non-zero interaction between the layers. The dispersion of the conductance band in Bi2212 at optimal doping is
schematically presented in Fig.\,\ref{Fig:ElectronicStructure}~(a) (only the occupied states below the Fermi energy
are
shown). As illustrated in the figure, the Fermi surface consists of the two hole barrels formed by the bonding and
antibonding bands. The bilayer splitting is evident both from the local-density approximation (LDA) based band
structure
calculations [panel (b)] and from the experimental ARPES data [panel (c)]. As was already discussed in
\S\ref{SubSec:PLD}, the situation is additionally complicated by the shadow Fermi surface and superstructure replicas,
possibly due to the periodic lattice modulations in these materials. The typical experimental Fermi surfaces of
Bi(Pb)-2212 and Bi2212 can be seen in Fig.\,\ref{Fig:PLD}~(d) and (e).

\subsection{Self-energy effects in the ARPES spectra of Bi2212}

\hvFloat[floatPos=t, capWidth=1.0, capPos=r, capVPos=t, objectAngle=0]{figure}
        {\includegraphics[width=0.4\textwidth]{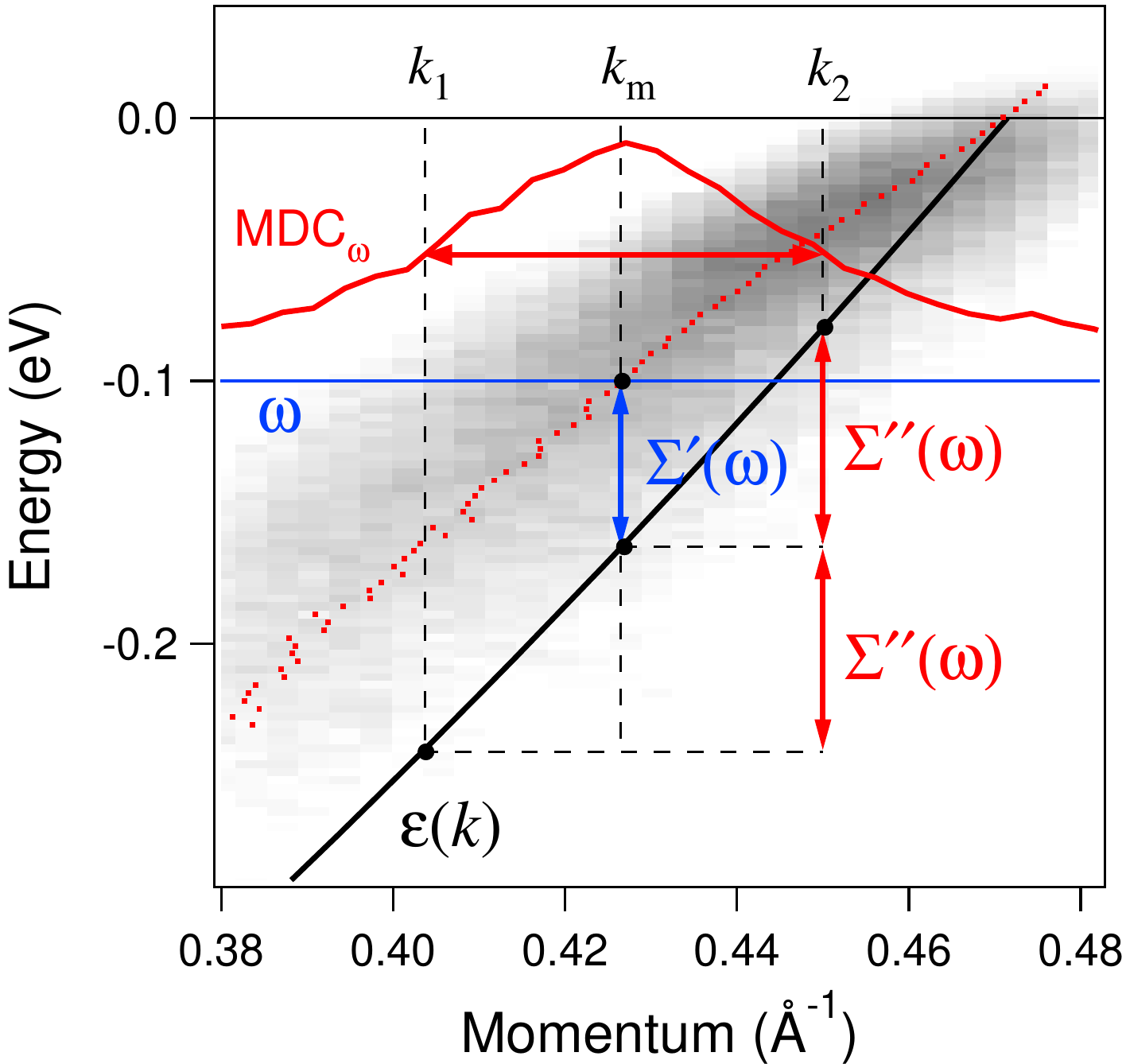}\quad}
        {Real and imaginary parts of the self-energy (shown by blue and red double headed arrows respectively) on a
        photoemission image. Bare band dispersion (black solid line) and renormalized dispersion (red points). Red
        solid
        line represents the momentum distribution curve (MDC) taken at $\omega$. The position of MDC maximum
        determines
        $k_\text{m}$, and the momenta of its half-maximum level determine $k_1$ and $k_2$. The figure is reproduced
        from
        Ref.~\citenum{KordyukBorisenko05condmat}.\vspace{-1em}}
        {Fig:NodalSE}

Except for the electron dispersion, the spectral function measured by ARPES also contains information about the
self-energy, which is reflected in the broadening of the spectrum and the deviation of the experimental dispersion
from
the bare band. Possibility to extract the self-energy from experimental data is of great importance, as it gives
access
to the understanding of interactions within the many-body system. The method of simultaneous self-consistent
evaluation
of the bare band dispersion and self-energy from ARPES spectra in the nodal direction has been developed in
Ref.~\citenum{KordyukBorisenko03, KoitzschBorisenko04a, KordyukBorisenko05, KordyukBorisenko05condmat} and will be
discussed in detail in the next paragraph. It is remarkable that the $k_z$ dispersion vanishes at the nodal point
\cite{LindroosSahrakorpi06}, therefore there is no additional broadening introduced to the nodal spectrum due to the
three-dimensionality, which is not the case at other points along the Fermi surface. This fact, as well as the maximal
steepness of the dispersion, makes the self-energy analysis easiest at the node.

Neglecting for the moment the effects of the energy and momentum resolutions as well as the influence of matrix
elements, one can consider the ARPES signal to be simply proportional to the spectral function, as follows from
Eq.~(\ref{Eq:PhotocurrentExperimental}). The self-energy effect on the ARPES spectrum is then given by
Eq.~(\ref{Eq:RenormG}), which we can rewrite by taking the imaginary part and writing explicitly the real and
imaginary
parts of the self-energy as
\begin{equation}\label{Eq:RenormA}
   A(\mathbf{k},\,\omega)=-\frac{1}{\piup}\,\frac{\mathit{\Sigma}''\text{(}\omega\text{)}}{[\omega-\epsilon_\mathbf{k}-\mathit{\Sigma}'\text{(}\omega\text{)}]\mathstrut^2+\mathit{\Sigma}''\text{(}\omega\text{)}\mathstrut^2}\text{.}
\end{equation}

Formula (\ref{Eq:RenormA}) is illustrated in Fig.\,\ref{Fig:NodalSE}. The background image (in gray scale) represents
a
typical nodal spectrum in the vicinity of the Fermi level. The red curve shows a constant energy cut, or momentum
distribution curve (MDC) for a particular energy $\omega$. The red dots mark the experimental dispersion, i.e.
positions
of MDC maxima $k_\text{m}\text{(}\omega\text{)}$ taken at different $\omega$. The solid black line is the bare band
dispersion $\epsilon_k$. Points $k_1\text{(}\omega\text{)}$ and $k_2\text{(}\omega\text{)}$ are the momenta at which
MDC$_\omega$ reaches its half-maximum level (note that $k_2-k_\text{m}\neq k_\text{m}\!-k_1$). Then the real part of
the
self-energy $\mathit{\Sigma}'\text{(}\omega\text{)}=\omega-\epsilon_{k_\text{m}}$ is the deviation of the experimental
dispersion from the bare band at $k_\text{m}\text{(}\omega\text{)}$, while the imaginary part is given by
\cite{KordyukBorisenko05condmat}
\begin{equation}\label{Eq:SEIm}
|\mathit{\Sigma}''\text{(}\omega\text{)}|=\epsilon_{k_2}\!-\epsilon_{k_\text{m}}=\epsilon_{k_\text{m}}\!-\epsilon_{k_1}=(\epsilon_{k_2}\!-\epsilon_{k_1})/2\text{.}
\end{equation}

\hvFloat[floatPos=t, capWidth=1.0, capPos=r, capVPos=t, objectAngle=0]{figure}
        {\includegraphics[width=0.55\textwidth]{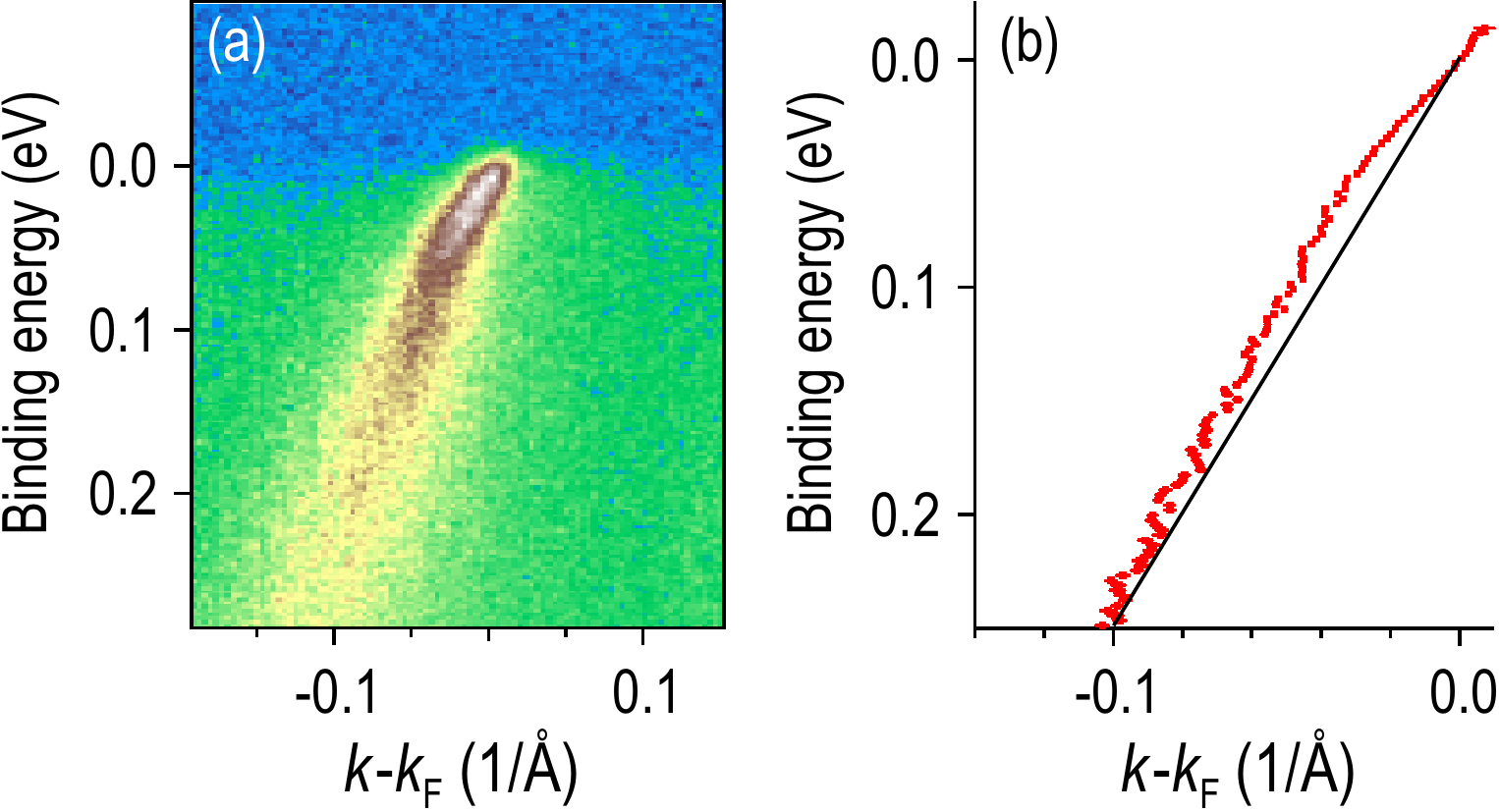}\quad}
        {``Kink'' in the experimental dispersion along the nodal direction in pure Bi2212. \textbf{(a)}~Photoemission
        intensity
        distribution along the nodal direction. \textbf{(b)}~Experimental dispersion (red dots). The solid black line
        is a guide
        to the eye. The figure is reproduced from Ref.~\citenum{ZabolotnyyBorisenko06prl}.\vspace{-1em}}
        {Fig:Kink}

There are two major self-energy effects that can be observed experimentally along the nodal direction. First, the MDC
width increases with $|\omega|$ at low energies, which is a typical behavior both for normal and marginal Fermi
liquids
(see \S\ref{SubSec:SelfEnergyFL}). Then, there is an anomaly in the experimental dispersion, the so-called ``kink'',
at
about 0.05\,--\,0.10~eV binding energy \cite{VallaFedorov99, KaminskiRanderia01, BogdanovLanzara00, LanzaraBogdanov01,
JohnsonValla01, KordyukBorisenko04prl, ZabolotnyyBorisenko06prl}, as shown in Fig.\,\ref{Fig:Kink}. According to the
most common understanding, this energy scale is a result of the bosonic mode coupling term in the self-energy (see
\S\ref{SubSec:SelfEnergyMode}) that originates from coupling either to a phonon \cite{BogdanovLanzara00,
LanzaraBogdanov01} or a collective mode of electronic origin \cite{KaminskiRanderia01, JohnsonValla01,
KordyukBorisenko04prl, ZabolotnyyBorisenko06prl}. In the following, we will consider the two constituents of the
self-energy in more detail.

\hvFloat[floatPos=b, capWidth=1.0, capPos=b, capVPos=t, objectAngle=0]{figure}
        {\includegraphics[width=0.63\textwidth]{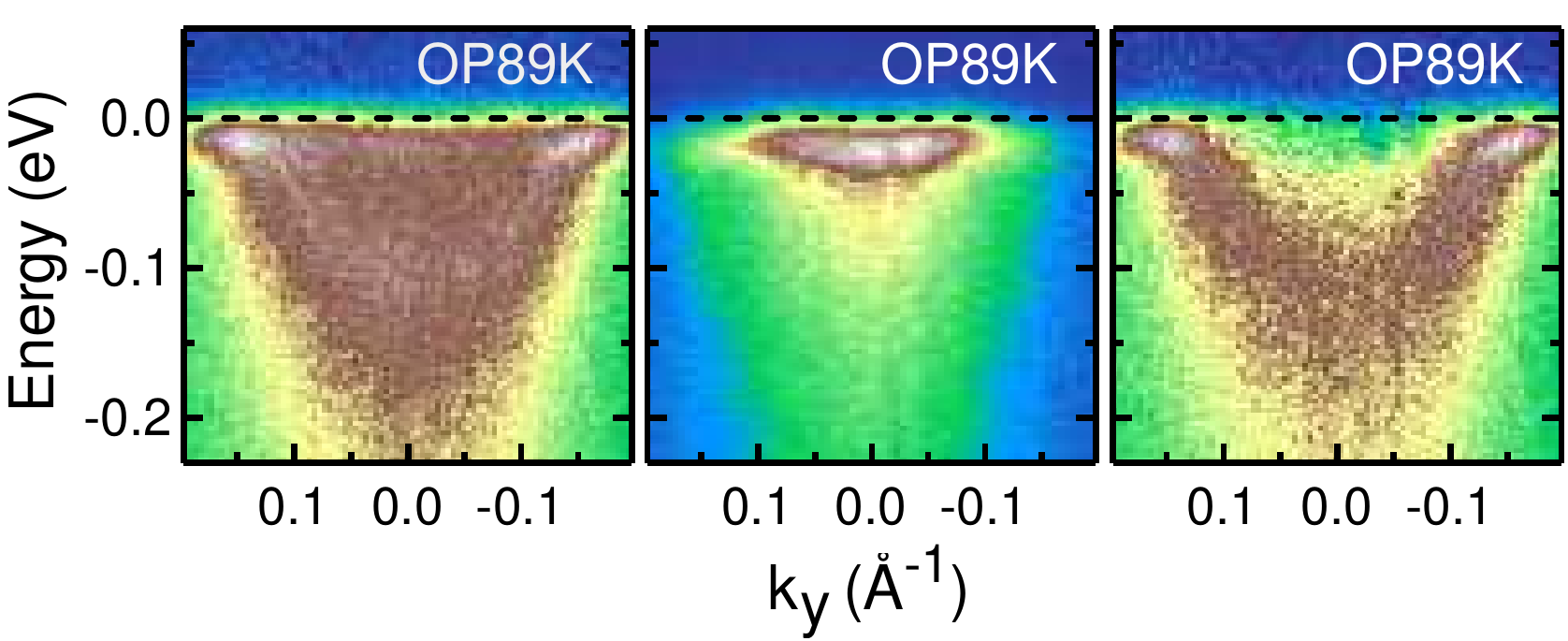}\quad}
        {Renormalization in the antinodal region of $k$-space. Left: data taken with a photon energy of
        $h\nu=$~38\,eV, at which the signal from the bonding band is maximal. Middle: data taken at $h\nu=$~50\,eV,
        where the signal from the antibonding band is dominant. Right: subtraction of the latter from the former
        yielding the spectral weight of the bonding band. The figure is reproduced from
        Ref.~\citenum{KimKordyuk03}.\vspace{-1em}}
        {Fig:AntinodalSpectra}

As one moves away from the nodal point towards the antinode \cite{KaminskiRanderia01, KimKordyuk03}, the situation is
complicated by the opening of the superconducting gap, smaller band width, stronger renormalization, and non-zero
$k_z$-dispersion, which leads to an additional broadening of the spectrum, precluding exact quantitative self-energy
analysis in this region of the momentum space. Nevertheless, the signatures of the renormalization similar to those at
the nodal point can be clearly observed, as seen from Fig.\,\ref{Fig:AntinodalSpectra}. As the renormalization becomes
stronger towards the antinode, the spectral weight becomes ``squeezed'' between the superconducting gap energy and the
energy of the bosonic mode, significantly reducing the band width. For comparison, see the model spectra from
Fig.\,\ref{Fig:SE_Model}.

\subsection{Self-consistent evaluation of the bare dispersion and self-energy}\label{SubSec:SelfConsistentSE}

Let us now return to the analysis of nodal spectra and scrutinize the procedure for extraction of the self-energy from
photoemission data \cite{KordyukBorisenko03, KoitzschBorisenko04a, KordyukBorisenko05, KordyukBorisenko05condmat,
EvtushinskyKordyuk06}. As follows from Eq.\,(\ref{Eq:SEIm}), the extraction of the imaginary part is straightforward
if
one knows the dispersion of the bare band. One has then simply to fit each MDC$_\omega$ with some, generally speaking,
asymmetric lineshape, determine momenta $k_{1,2}\text{(}\omega\text{)}$ at which the fitting curve reaches its
half-maximum level, and substitute them into (\ref{Eq:SEIm}).

Unfortunately, the bare band dispersion is usually not known with the required accuracy. Though band structure
calculations can provide us with the bare electronic structure, it is \textit{a priori} unknown whether these results can
be relied upon. It is therefore desirable to be able to extract the bare band dispersion from the experimental data
together with the self-energy. This becomes possible, if we make some assumptions about the functional form of the
dispersion. As soon as we analyze the electronic structure in the vicinity of the Fermi level, it is a good
approximation to assume that the dispersion has a parabolic shape given by
\begin{equation}
\epsilon_k=\omega_0\bigl[(k/k_\text{F})^2-1\bigr]\text{.}
\end{equation}
The Fermi momentum $k_\text{F}$ is simply determined as the crossing point of the experimental dispersion with the
Fermi
level. We are therefore left with a single free parameter $\omega_0>0$.

\textbf{Imaginary part ($\mathit{\Sigma}''$).} For a quadratic dispersion, formula (\ref{Eq:SEIm}) can be rewritten as
\begin{equation}\label{Eq:SEImQuadr}
|\mathit{\Sigma}''|=\frac{\omega_0}{2k_\text{F}^2}\,(k_2^2-k_1^2)=\frac{\omega_0}{k_\text{F}^2}\,W(k_2+k_1)=\frac{2\omega_0}{k_\text{F}^2}W\sqrt{k_\text{m}^2-W^2}\text{,}
\end{equation}
where $W=(k_2-k_1)/2$ is the half-width at half-maximum (HWHM) of the MDC peak. The last equality follows from the
following property of a parabolic function: if $2\,\epsilon(k_\text{m})=\epsilon(k_2)+\epsilon(k_1)$, then
$4k_\text{m}^2=(k_2+k_1)^2+(k_2-k_1)^2$.

There is another useful consequence of this property. It is interesting that the ratio $L=\sqrt{k_\text{m}^2-\langle
k\rangle^2}/W$, where $\langle k\rangle=(k_1+k_2)/2$, is identical to unity in case of a parabolic dispersion, i.e.
$k_\text{m}^2-\langle k\rangle^2=W^2$. We can rewrite this equality as
\begin{equation}\label{Eq:MDCAsymmetry}
\frac{k_\text{m}-\langle k\rangle}{W}=\frac{W}{k_\text{m}+\langle k\rangle}\text{.}
\end{equation}
The left part of (\ref{Eq:MDCAsymmetry}) characterizes the asymmetry of the MDC peak. Indeed, the position of the peak
maximum $k_\text{m}$ deviates from the center of the interval $[k_1,\,k_2]$ by $k_\text{m}-\langle k\rangle$. The
ratio
$(k_\text{m}-\langle k\rangle)/W$ then gives an estimate of the relative error that one introduces by neglecting the
asymmetry of the peak. On the other hand, in the vicinity of the Fermi level the right part of the same equation
(\ref{Eq:MDCAsymmetry}) is of the order of $W/2\,k_\text{F}\ll1$. This observation leads us to a useful conclusion: it
is often sufficient to use a symmetric line shape in the fitting procedure (e.g. a Lorentzian or Voigt profile), as
long
as $W\text{(}\omega\text{)}\ll2\,k_\text{m}$ (which can be called \textit{low-energy region}). As a rule of thumb,
this
approximation holds approximately half-way down to the bottom of the band, i.e. down to $\sim$\,0.5\,eV binding
energy.
At energies that high, the parabolic approximation for the bare band dispersion probably breaks down as well, so
taking
the asymmetry of the peak into account is not expected to improve the accuracy of the self-energy analysis procedure,
at
least in this simple model. Moreover, our analysis is essentially based on the assumption that the self-energy is a
function of one variable, which is justified only if the spectral function is well localized in momentum (see
\S\ref{SubSec:NotionFL}). It should be therefore emphasized that the described procedure is only capable of studying
the
behavior of the self-energy in the low-energy region, but not in the vicinity of the band bottom.

\textbf{Relation to the real part ($\mathit{\Sigma}'$).} Formula (\ref{Eq:SEImQuadr}) gives the imaginary part of the
self-energy in the assumption of the parabolic bare band. We can replace the unknown constant $\omega_0$ by the Fermi
velocity of the bare band $v_\text{F}=2\omega_0/k_\text{F}$. A similar expression can be written down for the real
part
of the self-energy in the same approximations, which leads to
\begin{subequations}\label{Eq:ExperimSE}
\begin{align}
\mathit{\Sigma}'\text{(}\omega\text{)}&=\omega+\frac{v_\text{F}}{2k_\text{F}}[k_\text{F}^2-k_\text{m}^2\text{(}\omega\text{)}]\text{,}\\
\mathit{\Sigma}''\text{(}\omega\text{)}&=-\frac{v_\text{F}}{k_\text{F}}\,W\text{(}\omega\text{)}\sqrt{k_\text{m}^2\text{(}\omega\text{)}-W^2\text{(}\omega\text{)}}\text{.}
\end{align}
\end{subequations}
The real and imaginary parts of the self-energy are related by the Kramers-Kronig transformation (\ref{Eq:KK}), which
will let us determine the unknown factor $v_\text{F}$ in formulae (\ref{Eq:ExperimSE}). In order to perform the
Kramers-Kronig transformation, the experimentally extracted self-energy curves have to be symmetrized to the
unoccupied
part of the spectrum ($\omega>0$) and extrapolated to the high-energy region using some empirical model of the
`tails',
for example:
\begin{equation}\label{Eq:SelfEnergyTails}
\mathit{\Sigma}''_\text{mod}\text{(}\omega\text{)}=-\frac{\alpha\omega^2+C}{1+|\omega/\omega_\text{c}|^n}\text{,
$n\geq2$.}
\end{equation}
The constants $C$ and $\alpha$ are determined in a way such that the `tails' are smoothly joined to the experimental
data. The value of $n=3$ usually provides a reasonable result, while the cut-off energy $\omega_\text{c}$ can be
either
treated as an independent fitting parameter, or chosen from the following rule of thumb:
$\omega_\text{c}\approx\omega_0/2$ \cite{KordyukBorisenko05}. This energy approximately corresponds to the point where
$|\mathit{\Sigma}''\text{(}\omega\text{)}|$ reaches its maximum and starts to decrease, while
$|\mathit{\Sigma}'\text{(}\omega\text{)}|$ crosses zero.

The fitting procedure is done in three steps. In the first two steps, the real part of the self-energy, for given
$\omega_0$ and $\omega_\text{c}$, is calculated in two independent ways: (i) $\mathit{\Sigma}_\text{disp}'$ determined
directly from Eq.\,\ref{Eq:ExperimSE}a; (ii) $\mathit{\Sigma}_\text{KK}'$ calculated indirectly from
Eq.\,\ref{Eq:ExperimSE}b with subsequent Kramers-Kronig transform. Then, in step (iii), the fitting parameters are
adjusted to minimize the difference
$\mathit{\Delta}\mathit{\Sigma}'=\mathit{\Sigma}_\text{KK}'-\mathit{\Sigma}_\text{disp}'$. The real and imaginary
parts
are considered to be Kramers-Kronig consistent, if this can be done with a reasonable accuracy determined by the
experimental resolution.

\hvFloat[floatPos=t, capWidth=1.0, capPos=r, capVPos=t, objectAngle=0]{figure}
        {\includegraphics[width=0.45\textwidth]{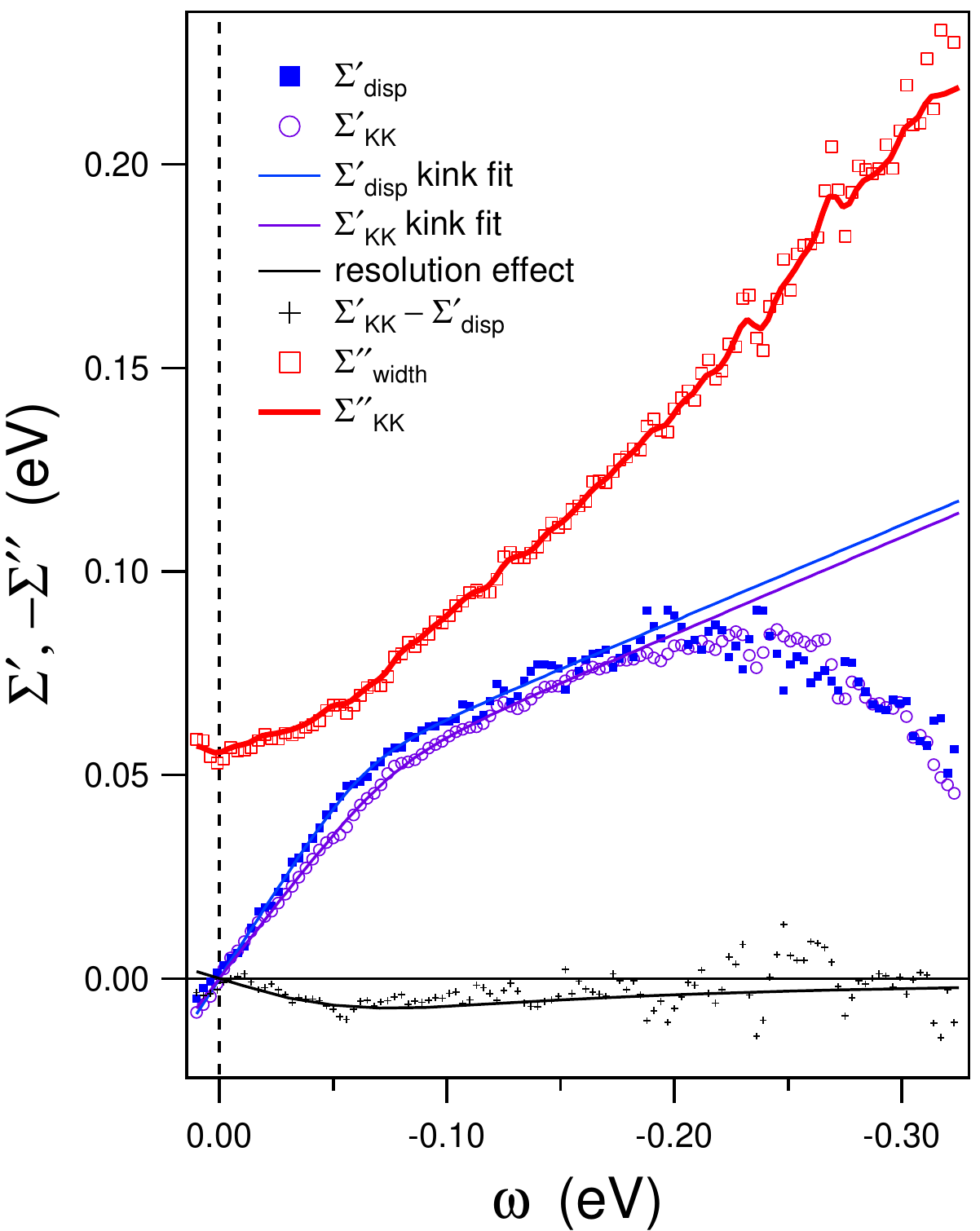}\quad}
        {Real and imaginary parts of the self-energy extracted from the experiment with the described procedure.
        Figure
        reproduced from Ref.\,\citenum{KordyukBorisenko05}.\vspace{-1em}}
        {Fig:ExperimentalSE}

\textbf{Influence of the experimental resolution.} It should be mentioned that the imaginary part of the self-energy
determined from experimental data is affected by the experimental resolution. This influence can be approximated by
the
following formula:
\begin{equation}
\mathit{\Sigma}_\text{width}''\text{(}\omega\text{)}=\sqrt{\mathit{\Sigma}''\text{(}\omega\text{)}\mathstrut^2+R\mathstrut^2}\text{,}
\end{equation}
where the empirical parameter $R$ characterizes the overall resolution. The frequency-dependent contribution of the
resolution to the imaginary part of $\mathit{\Sigma}\text{(}\omega\text{)}$, which we will call here the
\textit{resolution function}, is then given by
\begin{equation}
R''\text{(}\omega\text{)}=\sqrt{\mathit{\Sigma}''\text{(}\omega\text{)}\mathstrut^2+R\mathstrut^2}-\mathit{\Sigma}''\text{(}\omega\text{)}\text{,}
\end{equation}
Due to the linearity of the Kramers-Kronig transform, the contribution of the resolution function to the real part of
the self-energy $R'\text{(}\omega\text{)}$ will be simply given by the Kramers-Kronig transform of
$R''\text{(}\omega\text{)}$.

Returning to the fitting procedure described above, note that $\mathit{\Sigma}_\text{disp}'$ is not influenced by the
resolution function, therefore the difference $\mathit{\Delta}\mathit{\Sigma}'$ has to be fitted not to zero, but to
the
real part of the resolution function $R'\text{(}\omega\text{)}$ (Fig.\,\ref{Fig:ExperimentalSE}). For details, see
Ref.\,\citenum{KordyukBorisenko05}.

A more subtle way to account for the experimental resolution is to use the Voigt profile for MDC fitting
\cite{EvtushinskyKordyuk06}. As the intrinsic MDC line shape is a Lorentzian, and the resolution effect is equivalent
to
Gaussian broadening, such procedure allows one to separate the intrinsic line width from the resolution effect.

\textbf{Tight-binding model.} The above-described procedure allows one to determine the energy scale of the band
structure,
i.e. either $\omega_0$ or the nodal Fermi velocity $v_\text{F}$. In Pb-Bi2212, their values are 0.86$\pm$0.03\,eV and
2.46$\pm$0.07\,\AA$\cdot$eV respectively \cite{KordyukBorisenko05}. With this information at hand, the whole band
structure can be fitted by a tight-binding model to the experimentally measured Fermi surface map. The Fermi surface
shape is reproduced by fitting the dimensionless tight-binding parameters to the Fermi surface map, and the energy
scale
is taken from the self-consistent self-energy analysis. This procedure is justified, as the real part of the
self-energy
is zero at the Fermi level, so the photocurrent is localized along the Fermi surface given by the bare band structure.

We will use the tight-binding model limited to the three nearest neighbor intraplane hopping terms and one interlayer
hopping term that is responsible for the bilayer splitting. Rewriting the expression (\ref{Eq:TightBindingDispersion})
for the pseudo-tetragonal crystal lattice of Bi2212, one gets the following model for the dispersion
\cite{KordyukBorisenko03}:
\begin{multline}
\epsilon_\mathbf{k}=\mathit{\Delta}\epsilon-2t(\mathrm{cos}\,k_x+\mathrm{cos}\,k_y)+4t'\mathrm{cos}\,k_x\mathrm{cos}\,k_y\\-2t''(\mathrm{cos}\,2k_x+\mathrm{cos}\,2k_y)\pm
t_\perp(\mathrm{cos}\,k_x-\mathrm{cos}\,k_y)^2/4\text{.}
\end{multline}
In Ref.~\cite{KordyukBorisenko03}, this model has been fitted to the experimental ARPES data from two Bi2212 samples
with different doping: overdoped ($T_\text{c}$=69\,K) and underdoped ($T_\text{c}$=77\,K). The corresponding
tight-binding parameters are given in table~\ref{Table:TightBinding}. In the following, we will use linear
interpolation
between these two samples to find the tight-binding parameters for the intermediate doping levels.

\hvFloat[floatPos=t, capWidth=1.0, capPos=r, capVPos=t, objectAngle=0]{table}
        {\begin{tabular}[c]{l@{~~~}c@{~~}c@{~~}c@{~~}c@{~~}c}
         \toprule
         Sample & $t$\,(eV) & $t'$\,(eV) & $t''$\,(eV) & $t_\perp$\,(eV) & $\mathit{\Delta}\epsilon$\,(eV)\\
         \midrule
         OD\,69\,K & 0.40 & 0.090 & 0.045 & 0.082 & 0.43\\
         UD\,77\,K & 0.39 & 0.078 & 0.039 & 0.082 & 0.29\\
         \bottomrule
         \end{tabular}\quad
        }
        {Tight-binding parameters of the Bi2212 conductance band from Ref.~\cite{KordyukBorisenko03}.}
        {Table:TightBinding}

\subsection{Models of the self-energy} \label{SubSec:SelfEnergyModel}

Understanding the nature of the self-energy provides an insight into the many-particle interaction processes in the
system. It is therefore essential not only to be able to evaluate the self-energy curves from experiment, but to
understand the origin of its constituents. It turns out that the self-energy in cuprates can we well reproduced by a
relatively simple model which involves two kinds of interactions: Fermi-liquid like component originating from the
electron-electron interaction (see \S\ref{SubSec:SelfEnergyFL}) and the second term which originates from the coupling
to a bosonic mode (see \S\ref{SubSec:SelfEnergyMode}), which we will denote here by $\mathit{\Sigma}_\textup{el}$ and
$\mathit{\Sigma}_\textup{bos}$ respectively \cite{InosovBorisenko07}. The self-energy can be assumed to be the same
for
the bonding and antibonding bands, although strictly speaking, one should also account for the difference in
scattering
rates between the two bands \cite{BorisenkoKordyuk06}, which is however small enough to be neglected in our case.

To reproduce the experimental data, we model the imaginary part of the self-energy, and calculate the real part by
means
of the Kramers-Kronig transformation. This is done independently in the nodal and antinodal directions, so that the
self-energy at an arbitrary $\mathbf{k}$ point can be found by a d-wave interpolation between the node and antinode:
\begin{equation}
\mathit{\Sigma}''(\textbf{k},\omega) = \mathit{\Sigma}''_\textup{n}\text{(}\omega\text{)} +
\text{\scalebox{0.8}{$\,\frac{1}{4}\,$}}
[\mathit{\Sigma}''_\textup{a}\text{(}\omega\text{)} -
\mathit{\Sigma}''_\textup{n}\text{(}\omega\text{)}](\mathrm{cos}\,k_x-\mathrm{cos}\,k_y)^2
\end{equation}

\textbf{Nodal direction.} In the nodal direction, where the superconducting gap is zero and the partial density of
states can be considered constant in the vicinity of the Fermi level, the electron-electron scattering rate
$\mathit{\Sigma}''_\textup{el}$ is assumed to have a quadratic energy dependence typical for a normal Fermi liquid:
\begin{equation}
\mathit{\Sigma}''_\textup{el} = -\alpha\kern1pt\omega^2\text{.}
\end{equation}
The electron-boson part is modeled as
\begin{equation}\label{Eq:NodalBosonicSE}
\mathit{\Sigma}''_\textup{bos} =
-\beta_\textup{n}\biggl[1\kern-2pt+\mathrm{exp}\biggl(\frac{-|\omega|\,
+\,\mathit{\Omega}_\textup{n}}{\delta\omega_\textup{n}}\biggr)\biggr]^{-1}
\end{equation}
This function is analogous to (\ref{Eq:SelfEnergyT0}), but also takes into account the finite broadening of the
bosonic
mode, which is given here by the empirical parameter $\delta\omega_\textup{n}$. In the limit
$\delta\omega_\textup{n}\rightarrow0$, this corresponds to a step in the nodal scattering rate at the energy
$\mathit{\Omega}_\textup{n}$ with amplitude $\beta_\textup{n}$.

The broadened bosonic mode, as introduced in (\ref{Eq:NodalBosonicSE}), is well localized in energy, i.e. it has
exponentially decaying tails. This formula is purely phenomenological, as in practice the exact shape of the bosonic
spectrum is difficult to account for. Another possible way to introduce the broadening of the mode would be to offset
the pole in the real part of the self-energy (\ref{Eq:SelfEnergyT0}) \cite{EngelsbergSchrieffer63}:
\begin{equation}
\mathit{\Sigma}_\text{bos}'(\omega)=
-\frac{g}{\piup}\,\mathrm{ln}\biggl|\frac{\omega+\mathit{\Omega}+\mathrm{i}\mathit{\Gamma}}{\omega-\mathit{\Omega}+\mathrm{i}\mathit{\Gamma}}\biggr|\text{,}
\end{equation}
where $\mathit{\Gamma}$ is a small broadening parameter. Though this might seem to be a more natural way, it is not
very
convenient in practical situations because of the slowly decaying tails, which produce nonzero scattering rate near
the
Fermi level.

The total scattering rate is calculated from the two constituents by the following formula:
\begin{equation}
\mathit{\Sigma}'' = \frac{\mathit{\Sigma}''_\textup{el} +
\mathit{\Sigma}''_\textup{bos}}{1+(|\omega|/\omega_\text{c})^3}
\end{equation}
The denominator is necessary to guarantee convergence of the Kramers-Kronig transformation and represents a special
case
of the `tails' model (\ref{Eq:SelfEnergyTails}) for $n=3$. In Ref.~\cite{InosovBorisenko07} the cutoff energy
$\omega_\text{c}$ was chosen to be one third of the bare band bottom energy
$\omega_0=\epsilon_\mathbf{k}\big|_{\mathbf{k}=0}$.

\textbf{Antinodal direction}. In the antinodal direction, the situation is more complicated, because the partial
density
of states can no longer be considered constant in the superconducting state. Due to the nonzero energy gap, the
density
of states develops a pile-up peak, which can be directly observed in scanning tunneling spectroscopy experiments
\cite{Fischer07, HoogenboomBerthod03, SacksCren06}. Ref.\,\citenum{SacksCren06} provides a convenient model for the
partial density of states, which is
\begin{equation}
P(\omega,\delta,\mathit{\Delta}) = \Biggl|\,\textup{Re}\,\frac{\omega}{\kern-2pt\sqrt{(\omega\kern1pt-\kern1pt
i\,\delta)^2\kern1pt-\kern1pt\mathit{\Delta}^2}}\,\Biggr|\text{,}
\end{equation}
where $\mathit{\Delta}$ is the superconducting gap and $\delta$ is a broadening parameter (see
Ref.\,\citenum{SacksCren06} and references therein).

\hvFloat[floatPos=b, capWidth=1.0, capPos=r, capVPos=t, objectAngle=0]{figure}
        {\includegraphics[width=0.63\textwidth]{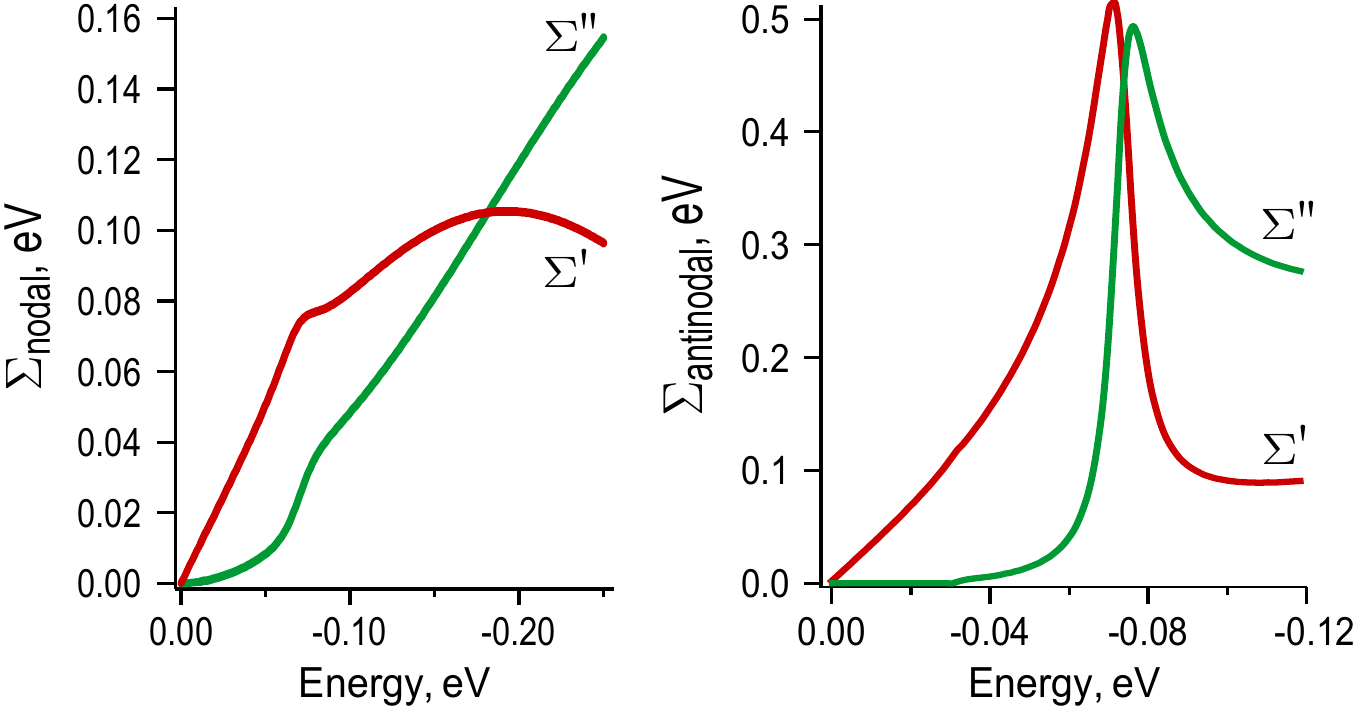}\quad}
        {The Kramers-Kronig consistent real and imaginary parts of the nodal and antinodal self-energies. The figure
        is
        reproduced from Ref.\,\citenum{InosovBorisenko07}.\vspace{-1em}}
        {Fig:SelfEnergy}

The electron-electron scattering rate will be then proportional to the number of states above the energy $\omega$
squared:
\begin{equation}
\mathit{\Sigma}''_\textup{el} = -\alpha\kern1pt\Biggl[\,\smallint_0^\omega
P(\omega,\delta\omega_\textup{a},\mathit{\Delta}_0)\,\textup{d}\omega\Biggr]^2\!\!\!\text{,}
\end{equation}
where $\mathit{\Delta}_0$ is the maximal superconducting gap at the ($\piup,0$) point. The bosonic part can be
approximated by
\begin{equation}
\mathit{\Sigma}''_\textup{bos} =
-\beta_\textup{a}\,P(\omega,\delta\omega_\textup{a},\mathit{\Delta}_0\kern1pt+\kern1pt\mathit{\Omega}_\textup{a})
=\kern-1pt -\beta_\textup{a}\,\Biggl|\,\textup{Re}\,
\frac{\omega}{\kern-2pt\sqrt{(\omega\kern1pt-\kern1pt
i\,\delta\omega_\textup{a})^2\kern1pt-\kern1pt(\mathit{\Delta}_0\kern1pt+\kern1pt\mathit{\Omega}_\textup{a})^2}}\,\Biggr|\text{.}
\end{equation}
Here $\mathit{\Omega}_\text{a}$ is the energy of the mode and $\delta\omega_\textup{a}$ is the broadening parameter at
antinodal point.

We assume the particle-hole symmetry in $\mathit{\Sigma}''$ to evaluate the self-energy in the unoccupied part of the
spectrum ($\omega>0$), so that the real and imaginary parts of the self-energy are odd and even functions of energy
respectively.

\textbf{Fitting procedure.} In the nodal direction, the self-energy model can be directly fitted to the experimentally
measured self-energy. At the antinode, this is not possible, so one has to calculate the spectral function from the
self-energy (see section \ref{Sec:ModelingGreensFunction}) and compare it to the experimentally observed one to achieve
a good correspondence between the two. This was done in Ref.\,\citenum{InosovBorisenko07}, where all the free parameters
were adjusted during comparison with a set of ARPES spectra of Bi2212 to achieve the best correspondence. The resulting
self-energies are plotted in Fig.\,\ref{Fig:SelfEnergy}. The best-fit parameters of the model are listed in the
following table:

\begin{center}\noindent\footnotesize\begin{tabular}{llll}\toprule\vspace{-1.85em}\\\label{Table:BestFit}
$\alpha = \text{3.0}$\,eV$^{-1}$ & $\beta_\textup{n} = \text{30}$\,meV & $\beta_\textup{a} = \text{200}$\,meV &
$\delta\omega_\textup{n} = \text{10}$\,meV \\
\smallskip
$\delta\omega_\textup{a} = \text{0.08}\,\mathit{\Delta}_0$ & $\mathit{\Omega}_\textup{n} = \text{60}$\,meV &
$\mathit{\Omega}_\textup{a} = \text{42}$\,meV & $\mathit{\Delta}_0 = \text{35}$\,meV  \\
\bottomrule
\end{tabular}\end{center}\normalsize\vspace{-0.2em}


\section{Superconducting energy gap}\label{Sec:SuperconductingGap}

\subsection{Momentum-dependence of the gap}

Before proceeding to the description of the Green's function model, which will be the main result of this chapter, we
need to pause for a more detailed discussion of the superconducting energy gap and its evaluation from the ARPES data.
As we know from BCS theory (\S\ref{SubSec:BCS}), the gap function $\mathit{\Delta}_\mathbf{k}$ is the order parameter of
the superconducting phase transition. The question of the symmetry of the superconducting gap has therefore become a hot
topic in high-$T_\text{c}$ superconductivity research and is crucial to understand the nature of superconductivity in
cuprates \cite{Annett90, Harlingen95, TsueiKirtley00}. Though there are still hot debates about the exact behavior of
the gap as a function of momentum \cite{ShenDessau93, DingCampuzano95, HaasBalatsky97, MatsuiTerashima05,
KanigelNorman06, KanigelChatterjee07, KondoTakeuchi07, LeeVishik07}, temperature \cite{LeeVishik07, DingYokoya96,
NormanDing98, NormanPines05, BoyerWise07}, doping \cite{LoeserShen96, MesotNorman99, TaconSacuto06, HuefnerHossain08},
and its space variations \cite{PunONeal01, McElroySimmonds03}, it is now well established that the order parameter, and
therefore the pairing interaction, have a predominantly $d_{x^2-y^2}$ symmetry of the gap in the superconducting state
\cite{WengerOestlund93, TsueiKirtley00, SatoKamiyama01, DingNorman96, NewnsTsuei07}:
\begin{equation}\label{Eq:GapDwave}
\mathit{\Delta}(\textbf{k}) = \frac{1}{2}\,\mathit{\Delta}_0\,(\mathrm{cos}\,k_x-\mathrm{cos}\,k_y)\text{,}
\end{equation}
possibly with a minor s-wave component \cite{KirtleyTsuei06}. Such momentum-dependent order parameter is one of the
major distinctions of the high-$T_\text{c}$ cuprates from conventional superconductors, where the gap has an
s-symmetry
and is therefore $\mathbf{k}$-independent. Though some data suggest that (\ref{Eq:GapDwave}) might be incomplete and
that higher order terms might also be present in the momentum-dependence of the gap function at low doping
\cite{MesotNorman99, Ghosh99}, for the purposes of the present work it is a good approximation to consider the gap as
having a perfect d-wave shape.

\subsection{Evaluating the gap from ARPES data}

One has to draw a clear distinction between the gap function $\mathit{\Delta}_\mathbf{k}$, which is defined for all
values of $\mathbf{k}$ and enters the Green's function of the superconducting state (\ref{Eq:GreensFuncTensor}), and
the
energy gap extracted in a photoemission experiment \cite{ManzkeBuslaps89, OlsonLiu89, DingCampuzano95, DingNorman96,
MarshallDessau96, DingNorman97, BorisenkoKordyuk02, KordyukBorisenko03}, usually defined as the leading edge gap (LEG)
\cite{MarshallDessau96, DingNorman97, BorisenkoKordyuk02, KordyukBorisenko03} and measured along the Fermi surface
contours. The leading edge gap is determined as the lowest binding energy at which the energy distribution curve (EDC)
reaches half of its maximum. It is understandable, and usually admitted, that the LEG should depend on any parameters
that determine the EDC line shape and, unless the relation between the LEG and the real gap is known, can be
considered
only as a qualitative representation of the real gap. So, analyzing the experimental data it is very important to
distinguish between the artificial variations of the LEG and changes caused by the real gap in the electronic density
of
states. We will not stop here on the details of the LEG behavior as a function of experimental parameters, which can
be
found in Ref.\,\cite{KordyukBorisenko03}. Instead, we will consider a couple of examples in which the momentum
dependence of the LEG is revealed from ARPES data analysis.\enlargethispage{1.5em}

\hvFloat[floatPos=b, capWidth=1.0, capPos=r, capVPos=t, objectAngle=0]{figure}
        {\raisebox{13.1em}{\footnotesize(a)}\hspace{-0.8em}\includegraphics[height=0.354\textwidth]{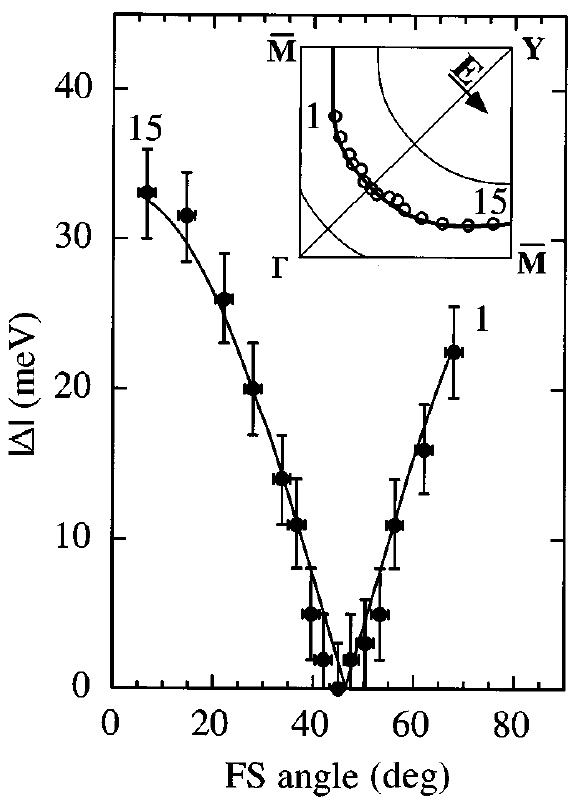}~
         \raisebox{13.1em}{\footnotesize(b)}\hspace{-0.8em}\includegraphics[height=0.35\textwidth]{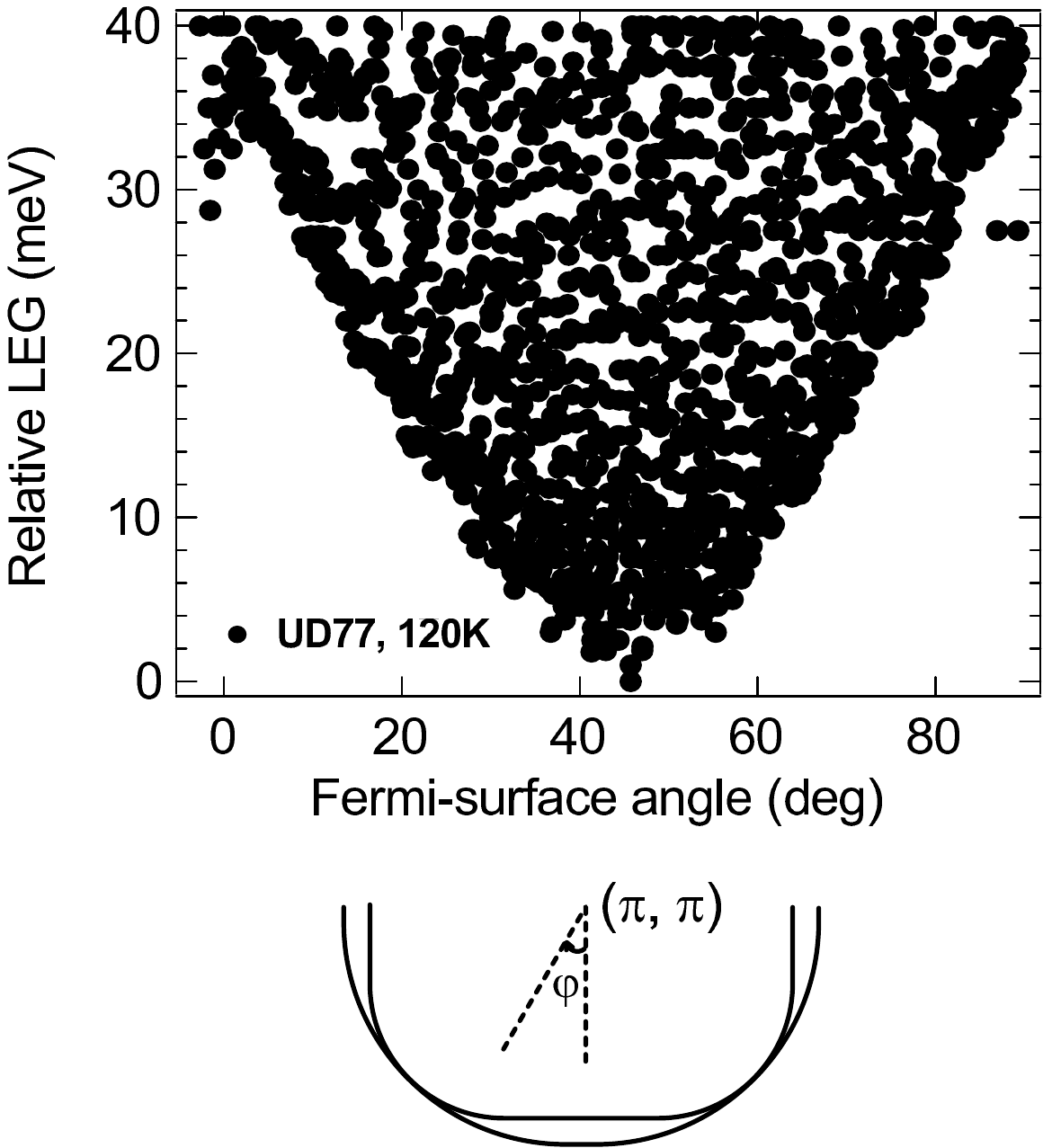}\quad}
        {Plots of the LEG anisotropy as a function of the Fermi surface angle $\varphi$ within the quadrant of the
        Brillouin zone [see inset at the bottom of panel (b)]: \textbf{(a)}~for a nearly optimally doped Bi2212 sample
        measured at 13\,K \cite{DingNorman96}; \textbf{(b)}~for an underdoped Bi2212 sample measured at 120\,K (above
        the transition temperature) in the pseudogap phase \cite{EcklHanke04}.}
        {Fig:Gap}

It is clear that in case of a non-zero energy gap, the spectral weight at the Fermi level vanishes and the Fermi
surface
is therefore no longer defined. One can however redefine it as the locus of $\mathbf{k}$ points where the LEG reaches
its minimum \cite{KordyukBorisenko03, EcklHanke04, BorisenkoNature04}. This permits plotting the gap as a function of
the
Fermi surface angle, measured from the center of the barrel. In Fig.\,\ref{Fig:Gap} two such dependencies are shown
for
a nearly optimally doped Bi2212 sample at low temperature [panel (a)] and for underdoped Bi2212 sample above the
transition temperature in the pseudogap phase [panel (b)]. In the first panel, only points measured along the minimal
gap locus are shown. In the second case, all momenta are plotted, so that the actual $k$-dependence of the pseudogap
is
represented by the low-gap envelope curve of all data points. In panel (a), the angular variation of the gap is in
excellent agreement with that expected from a d-wave order parameter of the $\mathrm{cos}\,k_x-\mathrm{cos}\,k_y$
form.
In the pseudogap case (b), significant deviations from this shape are observed.


\section{Modeling of the Green's function}\label{Sec:ModelingGreensFunction}

\subsection{Normal Green's function in the superconducting state}

The self-energy obtained in \S\ref{SubSec:SelfEnergyModel} can be used to model the Green's functions\,---\,a
fundamental result that can be used in a variety of different calculations and enables comparisons of ARPES with other
experimental methods, as will be shown in chapter \ref{Chap:Relation}. It is tempting to use the raw ARPES data for
such
calculations, which is however unreliable. The problem is that the measurements are strongly affected by matrix
element
effects, as will be demonstrated in the next chapter, by the experimental resolution, and possibly some experimental
artifacts \cite{BorisenkoKordyuk01}, such as intensity variations during measurements, normalization procedure applied
to the data, etc. Experimental spectra include contributions from both bonding and antibonding bands that are
difficult
to separate in a wide range of momenta. The spectral function originally measured by ARPES also lacks the absolute
intensity scale, which precludes quantitative calculations.

Instead, in Ref.\,\citenum{InosovBorisenko07} we have employed a model of the Green's function based on the bare
electron dispersion studied in Ref.\,\citenum{KordyukBorisenko03} and a model for the imaginary part of the
self-energy
described in \S\ref{SubSec:SelfEnergyModel}. The Green's function was then calculated according to
\cite{ChubukovNorman04}:
\begin{equation}\label{Eq:NormalG}
G(\textbf{k},\omega)\kern-1pt=\kern-1pt\frac{1}{2\piup}\frac{\omega-\mathit{\Sigma}(\textbf{k},\omega)+\epsilon_\textbf{k}}
{\bigl[\omega\kern-1pt-\kern-1pt\mathit{\Sigma}(\textbf{k},\omega)\bigr]^2\kern-2.5pt-\mathit{\Delta}^2(\textbf{k})\Bigl[1\kern-1pt-\frac{\mathit{\Sigma}(\textbf{k},\omega)}{\omega}\Bigr]^2
\kern-3pt-\epsilon_\textbf{k}^2}
\end{equation}
\noindent where $\mathit{\Delta}(\textbf{k})$ is the superconducting d-wave gap (\ref{Eq:GapDwave}) changing from zero
along the Brillouin zone diagonals to the maximal value of $\pm\mathit{\Delta}_0$ along the antinodal directions. The
term
$\mathit{\Phi}(\mathbf{k},\,\omega)=\mathit{\Delta}(\textbf{k})\Bigl[1-\frac{\mathit{\Sigma}(\textbf{k},\omega)}{\omega}\Bigr]$
in the denominator of (\ref{Eq:NormalG}) is the anomalous vertex function,\footnote{The superconducting gap in BCS
theory is the solution of $\mathit{\Sigma}(\mathit{\Delta})=\mathit{\Phi}(\mathit{\Delta})$ \cite{AbanovChubukov01}.
Alternatively to $\mathit{\Sigma}\text{(}\omega\text{)}$ and $\mathit{\Phi}\text{(}\omega\text{)}$, one can introduce
the complex effective mass function, $Z\text{(}\omega\text{)}=\mathit{\Sigma}\text{(}\omega\text{)}/\omega$, and the
complex effective gap function $\mathit{\Delta}\text{(}\omega\text{)}=
\mathit{\Phi}\text{(}\omega\text{)}/Z\text{(}\omega\text{)}$ \cite{Eliashberg60, ScalapinoSchrieffer66}.} which is
related to the pairing gap and is generally energy-dependent \cite{ChubukovNorman04, HaslingerChubukov03,
AbanovChubukov01, EschrigNorman00, Carbotte90, ScalapinoSchrieffer66, Eliashberg60}, though its energy dependence is
often neglected \cite{ChubukovNorman04}.

To achieve the best reproduction of the experimental data, all the free parameters of the self-energy were adjusted
during comparison with a set of ARPES spectra of Bi2212 to achieve the best correspondence (see Fig.\,\ref{Fig:Model}
for the comparison of the model with experimental spectra and the table on page~\pageref{Table:BestFit} for the
fitting
parameters). It is worth stressing that a simple self-energy model that includes coupling only to a single bosonic
mode
can accurately reproduce the state of the art ARPES spectra of Bi2212, as Fig.\,\ref{Fig:Model} clearly shows.

\hvFloat[floatPos=t, capWidth=1.0, capPos=b, capVPos=t, objectAngle=0]{figure}
        {\includegraphics[width=1.01\textwidth]{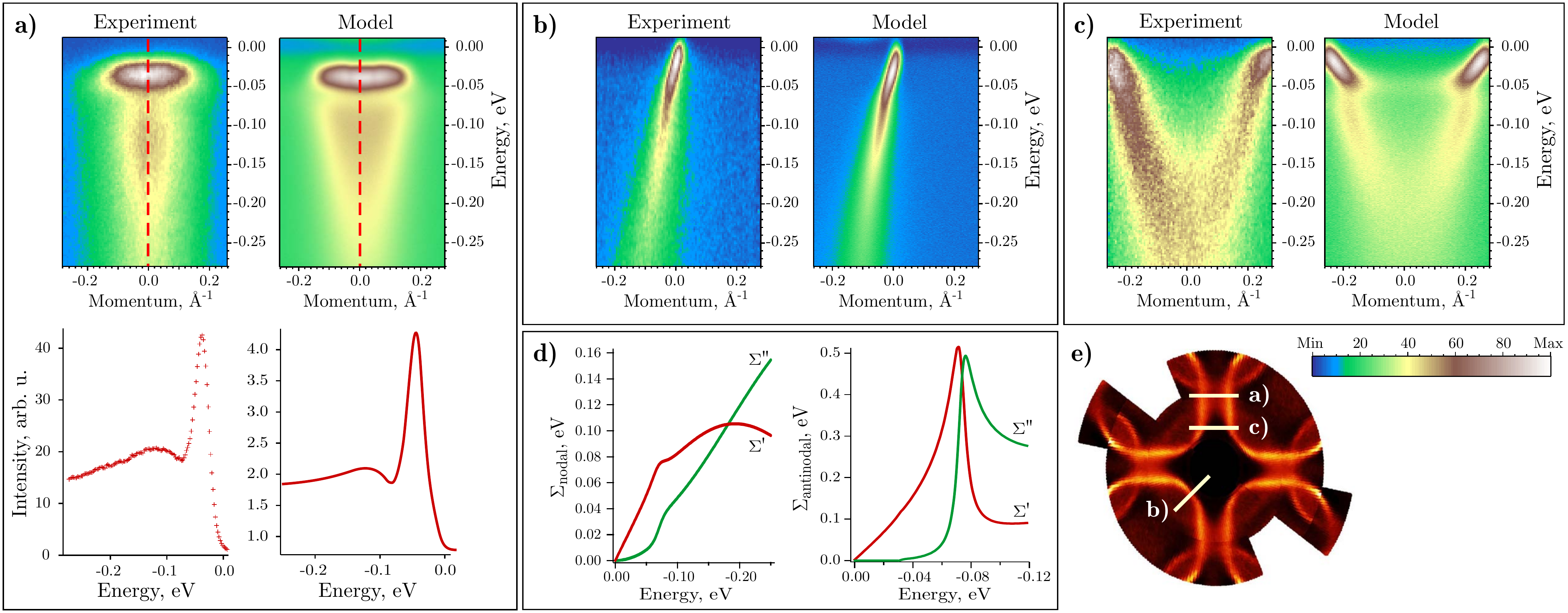}\quad}
        {Comparison of the model with experimental ARPES spectra of optimally doped Bi2212 at 30~K taken with 50~eV
        (a,~b) and 38~eV
        (c) photon energies. The model spectra are smoothed with a Gaussian to account for 20\,meV energy resolution
        and
        0.025\,\AA$^{-1}$ angular resolution. \textbf{(a)}~Spectra at the $(\piup,\,0)$ point with the corresponding
        energy
        distribution curves (below) taken along the dashed lines. \textbf{(b)}~Nodal spectra along the
        $(\piup,\,\piup)$
        direction. \textbf{(c)}~Comparison of the experimental and model spectra taken at an intermediate position in
        $k$-space to check the validity of the interpolation of the self-energy between the nodal and antinodal
        directions. \textbf{(d)}~The Kramers-Kronig consistent real and imaginary parts of the nodal and antinodal
        self-energies. \textbf{(e)}~Positions of the cuts (a\,--\,c) in $\mathbf{k}$-space. The figure is
        reproduced from Ref.\,\citenum{InosovBorisenko07}.}
        {Fig:Model}

\subsection{Anomalous Green's function}\label{SubSec:AnomalousG}

As was shown in \S\ref{SubSec:SupercondG}, the Green's function of the superconducting state is actually a 2$\times$2
spinor, which includes an anomalous off-diagonal element. The anomalous Green's function can not be directly measured
in
an ARPES experiment, but with the self-energy $\mathit{\Sigma}(\textbf{k},\omega)$ and the pairing vertex
$\mathit{\Phi}(\mathbf{k},\,\omega)$ present in the model (\ref{Eq:NormalG}), it becomes possible to calculate the
anomalous Green's function $F(\textbf{k},\omega)$ analytically \cite{HaslingerChubukov03}:
\begin{equation}\label{Eq:AnomalousG}
F(\textbf{k},\omega)\kern-1pt=\kern-1pt\frac{1}{2\piup}\frac{\mathit{\Delta}(\textbf{k})\Bigl[1\kern-1pt-\frac{\mathit{\Sigma}(\textbf{k},\omega)}{\omega}\Bigr]}
{\bigl[\omega\kern-1pt-\kern-1pt\mathit{\Sigma}(\textbf{k},\omega)\bigr]^2\kern-2.5pt-\mathit{\Delta}^2(\textbf{k})\Bigl[1\kern-1pt-\frac{\mathit{\Sigma}(\textbf{k},\omega)}{\omega}\Bigr]^2
\kern-3pt-\epsilon_\textbf{k}^2}
\end{equation}

The described model of the Green's functions is free of the disadvantages inherent in the raw ARPES data. Only in such
a
way can one completely separate the bonding and antibonding bands, which is impossible to achieve in the experiment.
Besides the already mentioned absence of matrix element effects and experimental resolution, the formulae
(\ref{Eq:NormalG}) and (\ref{Eq:AnomalousG}) also allow one to obtain both real and imaginary parts of the Green's
functions
for all $\textbf{k}$ and $\omega$ values including those above the Fermi level. It automatically implies the
particle-hole symmetry ($\epsilon_{\textbf{k}_\textup{F}-\textbf{k}} = -\epsilon_{\textbf{k}_\textup{F}+\textbf{k}}$)
in
the vicinity of the Fermi level, which in case of the raw data would require a complicated symmetrization procedure
based on Fermi surface fitting, being a source of additional errors. Finally, it provides the Green's function in
absolute units, allowing for quantitative comparison with other experiments and theory, even though the spectral
function originally measured by ARPES lacks the absolute intensity scale. Thereupon, we find the proposed analytical
expressions to be better estimates for the self-energy and both Green's functions and therefore helpful in
calculations
where comparison to the experimentally measured spectral function is desirable.


\chapter{Matrix element effects in photoemission spectra of cuprates}

\section{Manifestation of the photoemission matrix elements in an ARPES experiment}

\subsection{Excitation energy dependence of the matrix elements}

\hvFloat[floatPos=t, capWidth=1.0, capPos=r, capVPos=t, objectAngle=0]{figure}
        {\includegraphics[width=0.5\textwidth]{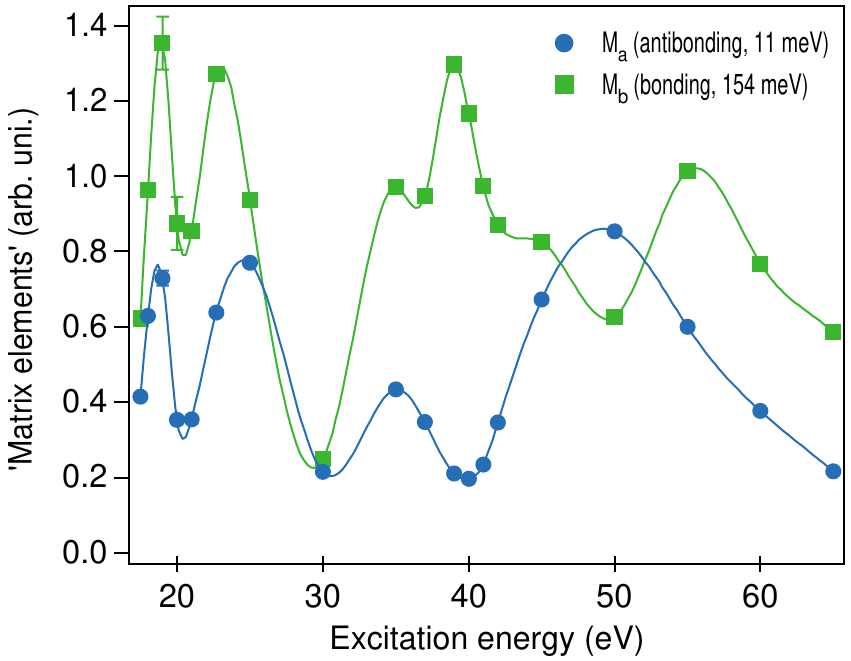}\quad}
        {The intensities of bonding and antibonding bands in an antinodal spectrum of Bi2212 as a function of
        excitation
        energy extracted from experimentally measured ARPES spectra. The figure is reproduced from
        Ref.\,\citenum{KordyukBorisenko02prl}.\vspace{-1em}}
        {Fig:MEexp}

In section \ref{Sec:TheoryOfPhotoemission} it was shown that ARPES essentially measures the one-particle spectral
function of the initial states. Much of the data on the cuprates have been analyzed within this approximation. While
this simple approach yields insights into the underlying physics, a satisfactory description of the spectra must
necessarily model the photoexcitation process properly by taking into account the matrix element involved, the complex
modifications of the wave functions resulting from a specific surface termination, and the effects of multiple
scattering and of finite lifetimes of the initial and final states \cite{BansilLindroos99}. It is now well recognized
\cite{InosovFink07, InosovKordyuk07, BorisenkoKordyuk01, KordyukBorisenko02prl, BorisenkoKordyuk03, BansilLindroos99,
BansilLindroos98, LeeFujimori02, LindroosSahrakorpi02, BansilMarkiewicz04, ChuangGromko04, ArpiainenZalobotnyy08} that
this interplay between the bulk and surface phenomena\,---\,the so called ``matrix element effect'', is essential for
a
satisfactory interpretation of the ARPES data.

In a photoemission experiment, matrix elements manifest themselves as an additional modulation of the photocurrent
with
momentum, excitation energy, and polarization of the incident photons. On the one hand, such modulation is harmful, as
it complicates data analysis and distorts the underlying electronic structure. On the other hand, it can be used to
enhance or mask the ARPES signal from individual bands, for example bonding and antibonding, by tuning the excitation
energy so that only one of the sub-bands is enhanced (see Fig.\,\ref{Fig:AntinodalSpectra} as an example). It is
therefore essential to know the behavior of the matrix elements depending on the experimental parameters in order to
uncover the underlying electronic structure.

The relative contributions of the bonding and antibonding bands to the spectrum of Bi2212 taken near the antinodal
point
are known as a function of photon energy both from experiment \cite{BorisenkoKordyuk01, KordyukBorisenko02prl,
BorisenkoKordyuk03} and theoretical calculations \cite{BansilLindroos98, BansilLindroos99, LeeFujimori02,
LindroosSahrakorpi02, BansilMarkiewicz04, ChuangGromko04, ArpiainenZalobotnyy08}. The experimentally measured
intensity
prefactors for the two sub-bands are shown in Fig.\,\ref{Fig:MEexp}. One can see that 38\,eV photons effectively
excite
electrons only from the bonding band, while the antibonding band is enhanced by 50\,eV photons.

In the nodal direction, a similar quantitative measurement is complicated in Bi2212 because of the very small bilayer
splitting \cite{KordyukBorisenko04}. But the modulations of the relative bonding/antibonding intensity have
nevertheless
been observed both in Bi2212 \cite[Fig.\,1\,(d)\,--\,(g)]{KordyukBorisenko04} and in Y123
\cite{BorisenkoKordyuk06kinks}, though not quantified.

\subsection{Momentum and polarization dependence of the matrix elements}

The dependence of the photoemission matrix elements on momentum has been studied mostly at the Fermi level, i.e. as
the
variation of photocurrent intensity along the Fermi surface contours. Unfortunately, the published results of
numerical
calculations covering an extended region of $\mathbf{k}$-space are available only for a couple of excitation energies:
21.2\,eV \cite{AsensioAvila03} and 22\,eV \cite{BansilLindroos98, BansilLindroos99}, and there are no systematic
studies
of the matrix element effects as a function of both photon energy and momentum, as well as at higher binding energies.

\hvFloat[floatPos=t, capWidth=1.0, capPos=r, capVPos=t, objectAngle=0]{figure}
        {\includegraphics[width=0.35\textwidth]{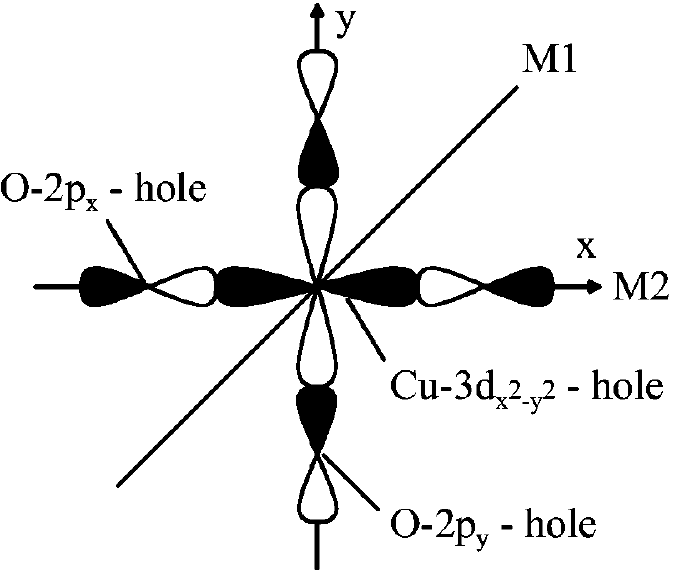}\quad}
        {Schematic of the Zhang-Rice singlet state in the one-particle representation. The Zhang-Rice
        singlet has the same symmetry as a $\mathrm{d}_{x^2-y^2}$ orbital, implying the existence of two mirror planes
        perpendicular to the CuO$_2$ plane, labeled M1 and M2. The black and white colors represent opposite signs of
        the
        orbital lobes. The figure is reproduced from Ref.~\citenum{DuerrLegner01}.\vspace{-1em}}
        {Fig:ZhangRice}

Some conclusions about the behavior of the matrix elements can be made, though, even from simple symmetry and parity
considerations. In the one-band Hubbard model, the $t$-$J$ model and its extensions, the Zhang-Rice singlet state is a
one-particle state that has the same symmetry as a Cu\,3d$_{x^2-y^2}$ orbital. Fig.\,\ref{Fig:ZhangRice} shows a
schematic representation of the Zhang-Rice singlet state, including the two mirror planes M1 and M2 which are
perpendicular to the CuO$_2$ plane and which are relevant for photoemission along
$\text{(}0,\,0\text{)}$\,--\,$\text{(}\piup,\,\piup\text{)}$ and
$\text{(}0,\,0\text{)}$\,--\,$\text{(}\piup,\,0\text{)}$ respectively \cite{DuerrLegner01}. The wave function of the
photoelectron to be detected by the analyzer must always be even with respect to the detector plane, otherwise it
would
be zero at the detector. The initial state (ground state) is a one-hole state with $\mathrm{d}_{x^2-y^2}$ symmetry,
and
therefore has even parity with respect to M2 and odd parity with respect to M1. In the assumption of the Zhang-Rice
singlet construction applicability to the first electron-removal final state, this state is totally symmetric. Let's
consider two possible polarizations of the incident photons: (i) polarization plane coincides with the emission plane
(\textit{parallel} polarization); (ii) polarization plane is orthogonal to the emission plane (\textit{perpendicular}
polarization). The photoemission operator $\mathbf{A}\cdot\mathbf{p}$ has even ($+$) parity in a parallel and odd
($-$)
parity in a perpendicular experimental geometry. The matrix element thus formally vanishes for the two cases:
\begin{equation}
M^\mathbf{k}_{\text{i},\,\text{f}}=\langle\psi_\text{f}|\,\mathbf{A}\cdot\mathbf{p}\,|\psi_\text{i}\rangle
=\Biggl\{\kern-5pt\begin{array}{l}\langle+|+|-\rangle\text{, i.e. parallel geometry for
$\text{(}0,\,0\text{)}$\,--\,$\text{(}\piup,\,\piup\text{)}$;}\\
\text{\raisebox{5pt}{$\langle+|-|+\rangle$, i.e. perpendicular geometry for
$\text{(}0,\,0\text{)}$\,--\,$\text{(}\piup,\,0\text{)}$.}}\end{array}
\end{equation}
Therefore along the $\text{(}0,\,0\text{)}$\,--\,$\text{(}\piup,\,\piup\text{)}$ direction the first electron-removal
state feature is expected to show the highest photoemission intensity with perpendicular geometry, whereas along
$\text{(}0,\,0\text{)}$\,--\,$\text{(}\piup,\,0\text{)}$ maximal intensities are to be observed in the parallel
geometry. This simple consideration explains the strong momentum and polarization dependence of the matrix elements
and
demonstrates that in particular geometries the ARPES signal may completely vanish due to the symmetry selection
rules.\vfill


\section{Anomalous high-energy dispersion as a matrix element effect}

\subsection{``Waterfalls'' in cuprates: first observations}

\nocite{GrafGweon07, VallaKidd07, HwangNicol07, ZhangLiu08, XieYang07, PanRichard06, GrafLanzara07, MeevasanaZhou07,
MeevasanaBaumberger08}

Appearance of the new generation of electron spectrometers with the wide acceptance angle ($\pm$\,15$^\circ$ for
\textit{Scienta~R4000}) has opened up the possibility of viewing the electronic structure of cuprates over a broad
momentum range covering more than one Brillouin zone\footnote{Unlike the standard definition of the $n^\text{th}$
Brillouin zone in solid state physics, here by different Brillouin zones we mean primary and secondary cones of
photoemission (Mahan cones), as introduced in Ref.\,\citenum{Mahan70}.} in a single measurement. This has triggered a
series of publications evidencing anomalous high-energy dispersion in the renormalized band structure of
Bi$_2$Sr$_2$CaCu$_2$O$_{8+\delta}$ \cite{GrafGweon07, VallaKidd07, HwangNicol07, ZhangLiu08, XieYang07, PanRichard06,
GrafLanzara07, MeevasanaZhou07}, Bi$_2$Sr$_2$CuO$_{6+\delta}$ \cite{XieYang07, PanRichard06, GrafLanzara07,
MeevasanaZhou07, MeevasanaBaumberger08}, La$_{2-x}$Sr$_x$CuO$_4$ \cite{MeevasanaZhou07, GrafLanzara07, ChangPailhes07},
La$_{2-x}$Ba$_x$CuO$_4$
\cite{VallaKidd07}, Pr$_{1-x}$LaCe$_x$CuO$_4$ \cite{PanRichard06},
Ba$_2$Ca$_3$Cu$_4$O$_8$(O$_\delta$F$_{1-\delta}$)$_2$
\cite{MeevasanaZhou07}, and Ca$_2$CuO$_2$Cl$_2$ \cite{RonningShen05} at the binding energies higher than
$\sim$\,0.3\,--\,0.5~eV\,---\,a region that has previously been scarcely explored. All of these reports seem to agree
on
the qualitative appearence of the spectra: (i) in the $(0,0)$\,--\,$(\piup,\piup)$ (nodal) direction the
``\textit{high-energy kink}\kern.8pt'' (``\textit{giant kink}\kern.8pt'') at $\sim$\,0.4~eV is followed by a nearly
vertical dispersion (``\textit{waterfall}\kern.8pt'') that ends up below 1\,eV with a barely detectable band bottom
approaching that of the bare band (see Fig.\,\ref{Fig:WaterfallsOld}); (ii) the band bifurcates near the high-energy
kink, forming another branch with a bottom at $\sim$\,0.5~eV \cite{GrafGweon07,PanRichard06,MeevasanaZhou07}; (iii)
some
authors claim that the actual band width is even larger than that of the bare band, reaching $\sim$\,1.6~eV
\cite{XieYang07, MeevasanaZhou07}; (iv) as one moves away from the nodal direction, the ``\textit{vertical
dispersion}\kern.4pt'' persists surprisingly up to the $(\piup,\,0)$ (antinodal) point, forming a
``\kern-.5pt\textit{diamond}\kern.8pt'' shape in momentum space at $\sim$\,0.5~eV \cite{GrafLanzara07, VallaKidd07,
ChangPailhes07} [Fig.\,\ref{Fig:WaterfallsOld} (b) and (c)].

These observations suggested the presence of new energy scales in the electronic structure. Except for the well-known
``kink'' at $\sim$\,0.05\,--\,0.10~eV, the second energy scale corresponding to the ``high-energy kink'' at
$\sim$\,0.4~eV, and even a third energy scale at $\sim$\,0.9~eV \cite{HwangNicol07} have been reported. It is widely
believed that understanding the interplay between different energy scales in any physical system, and in
high-$T_\text{c}$ superconductors in particular \cite{Dagotto94, Damascelli03, LeeNagaosa06}, holds the key to
mastering
their physical properties. The existence (or non-existence) of the new high energy scales is of fundamental importance
for the dressing of the charge carriers in high-$T_\text{c}$ superconductors and may be related to the strange normal
state properties of these materials and possibly even to the mechanism of high-$T_\text{c}$ superconductivity.
Furthermore, ``waterfalls'' have been detected for the first time in the ARPES spectra of cuprates, but never in any
other correlated or uncorrelated material. Hence, the clarification of this phenomenon is also of great importance for
the ARPES method itself, which has now developed into one of the most powerful experimental methods in solid state
physics. This explains the heated interest to the new observations, and gives us the motivation to study the
``waterfalls'' phenomenon in greater detail.

As one works with the spectra taken in a momentum window as wide as the Brillouin zone, and several electron-volts
deep
in momentum, the simplified notion of the ``experimental dispersion'' breaks down and can no longer be used, because
the
observed spectral features are not localized in momentum. The experimental dispersions extracted from momentum and
energy distribution curves can be drastically different \cite{RonningShen05, PanRichard06, ZhangLiu08,
MeevasanaBaumberger08}.
This renders many simplified procedures commonly used for ARPES data treatment, such as self-energy analysis or direct
comparison with band structure calculations, virtually impossible. The very notion of an electronic band with
``vertical'' dispersion (a whole energy range corresponding to the same quasimomentum) or an S-shaped dispersion, where
$\epsilon_\mathbf{k}$ becomes an ambiguous function of momentum as observed in MDC-derived experimental dispersions
\cite{HwangNicol07, XieYang07, MeevasanaZhou07}, is unphysical. The quantum number $\mathbf{k}$ enumerates different
electronic states within each electronic band, therefore each $\mathbf{k}$ value for a particular band corresponds to a
unique eigenstate with a well-defined eigenvalue, leaving no place for ambiguity. This means that the experimental
dispersion can not represent the underlying bare band structure, but is a more complex many-body effect.

\hvFloat[floatPos=t, capWidth=1.0, capPos=b, capVPos=t, objectAngle=0]{figure}
        {\includegraphics[width=\textwidth]{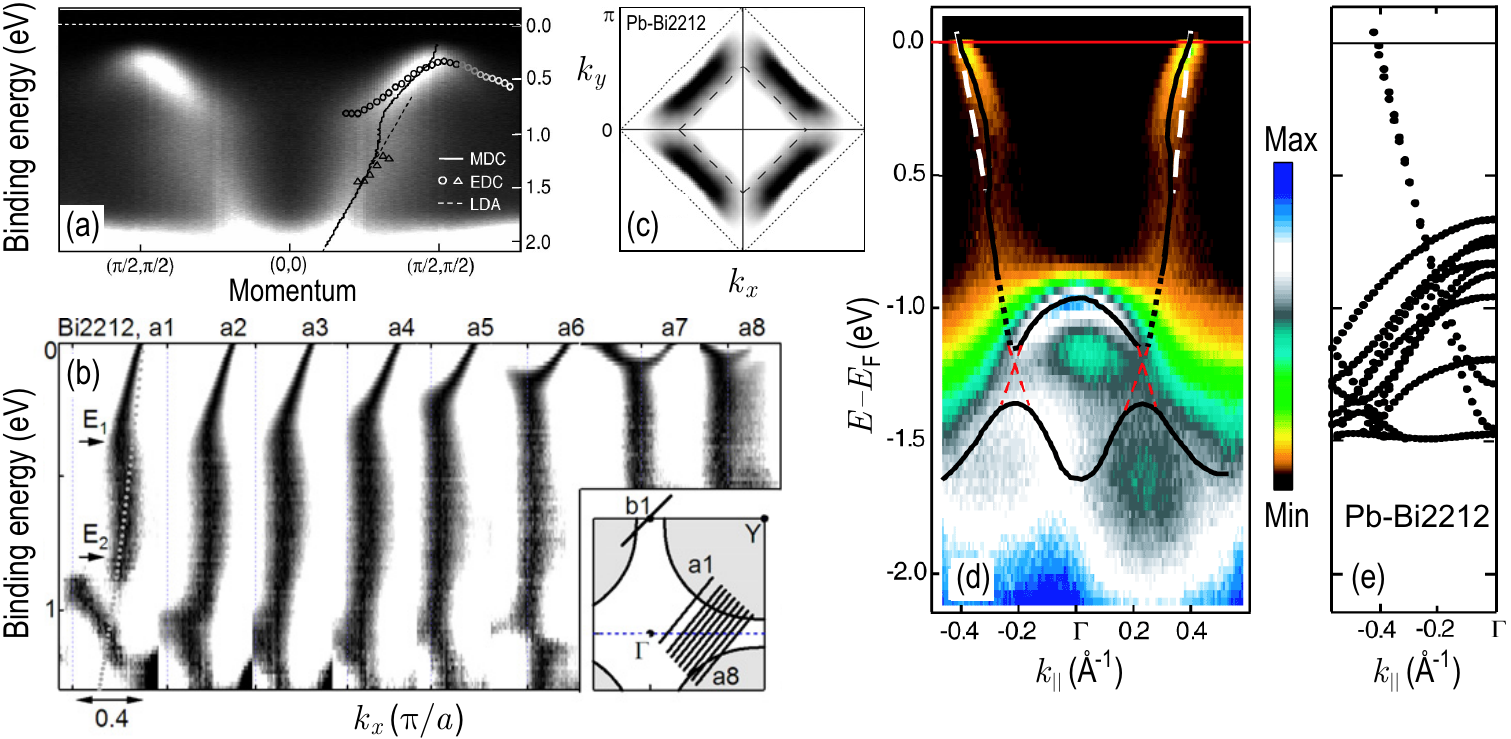}\vspace{-1em}}
        {Some of the first experimental observations of the high-energy anomaly. \textbf{(a)}~The first published
        evidence of the anomalous dispersion in Ca$_2$CuO$_2$Cl$_2$ \cite{RonningShen05}. \textbf{(b)}~Evolution of
        the
        ``waterfalls'' in Bi2212 with momentum as one proceeds from the node towards the antinode as shown in the
        inset
        \cite{GrafLanzara07}. \textbf{(c)}~Photoemission intensity integrated from $E_\text{F}$ to 0.8\,eV binding
        energy revealing a diamond-like shape \cite{GrafLanzara07}. \textbf{(d)}~ARPES intensity map along the nodal
        direction of Bi2201 from Ref.\,\citenum{XieYang07} compared to the LDA band structure of Bi2212 from
        Ref.\,\citenum{LinSahrakorpi06} [panel (e)].}{Fig:WaterfallsOld}

\subsection{Possible origins of the ``waterfalls''-resembling phenomena}

Unfortunately there is still no consensus on the physics responsible for the high-energy anomaly. In principle any
strong coupling to a bosonic mode would lead to the appearance of the incoherent spectral weight below the energy of
the
mode \cite{EngelsbergSchrieffer63}, which can resemble the ``vertical dispersion'', so distinguishing between
different
mechanisms is impossible without accurate quantitative comparison between theory and experiment. Up to now, a number
of
qualitative explanations have been proposed for the high-energy anomaly, including a disintegration of the
quasiparticles into a spinon and holon branch \cite{GrafGweon07}, coherence-incoherence crossover \cite{PanRichard06,
ChangPailhes07}, disorder-localized band-tailing \cite{AlexandrovReynolds07}, polarons \cite{MeevasanaZhou07},
rotationally symmetric charge modulations \cite{ZhouWang07}, familiar $t$-$J$ model with \cite{Manousakis07} or
without
\cite{TanWan07} string excitations, Hubbard model \cite{ByczukKollar07, SrivastavaGhosh07, LeighPhillips07}, as well
as
the self-energy approach with strong local spin correlations \cite{XieYang07}, itinerant spin fluctuations
\cite{VallaKidd07, MeevasanaZhou07, MacridinJarrell07, MarkiewiczSahrakorpi07}, quantum criticality \cite{ZhuAji08},
plasmons \cite{MarkiewiczBansil07}, or any arbitrary bosonic mode pairing \cite{CojocaruCitro07}. The reported
``diamond''-like momentum distribution of the ``waterfalls'' in \hbox{Bi-2212} with its sides pinned at
$(\pm\piup/4,\pm\piup/4)$ around the BZ center could be a sign of BZ folding due to some form of antiferromagnetism
\cite{GrafGweon07, PanRichard06}. However, such picture is violated in other cuprates and does not appear to be
universal \cite{VallaKidd07}. Let us briefly summarize some of the above-mentioned alternative explanations to get an
impression about the obvious difficulties that one runs into, attempting to find a theoretical background for the
anomalous behavior of the dispersion. For brevity, we will dwell here only on theories that have provided us with
numerical results that can be directly compared with experiment.

\hvFloat[floatPos=t, capWidth=1.0, capPos=b, capVPos=t, objectAngle=0]{figure}
        {\includegraphics[width=\textwidth]{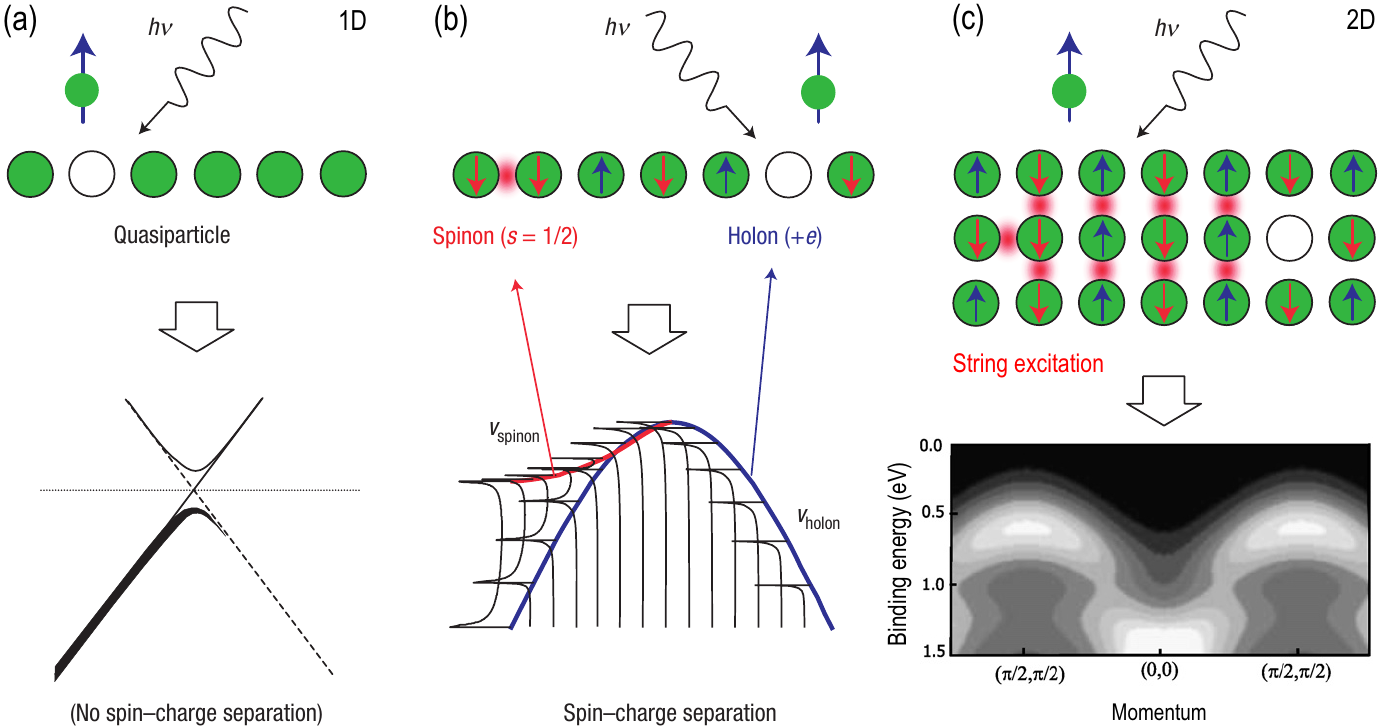}\vspace{-0.5em}}
        {Cartoon comparison of the spin-charge separation and electron-removal excitation spectrum in one- and
        two-dimensional materials. \textbf{(a)}~Within the band theory, a single branch in the excitation spectrum
        exists. \textbf{(b)}~If interactions are properly taken into account (e.g. in a $t$-$J$ model), in a
        one-dimensional antiferromagnet spin-charge separation leads to the appearence of two branches, as shown in
        red
        and blue at the bottom of the panel. \textbf{(c)}~In two dimensions, charge separation is not possible, as the
        holon and spinon are always bound by a string of ``disturbed'' spins. Taking into account these string
        excitations within the $t$-$J$ model, E.~Manousakis \cite{Manousakis07} has obtained the excitation spectrum
        shown
        at the bottom of the panel. The figure is reproduced (with modifications) from Ref.\,\citenum{KimKoh06}, the
        bottom image in panel (c) is taken from Ref.\,\citenum{Manousakis07}.}{Fig:ChargeSeparation}

\textbf{Spin-charge separation.} The disintegration of the low energy quasiparticles into a spinon and holon branch at
the energy of the ``giant kink'' was suggested as a possible mechanism underlying the high energy scale, because of
the
apparent similarity to the quasi-one-dimensional materials with antiferromagnetic correlations, such as SrCuO$_2$
\cite{GrafGweon07}, where spin-charge separation and, consequently, appearence of the two bands (\textit{spinons} and
\textit{holons}) scaling with $J$ and $t$ respectively in a $t$-$J$ model, are an established phenomenon
\cite{Dagotto94, KimKoh06}. But a closer look shows that such simple analogy with two-dimensional cuprates does not
hold. First, SrCuO$_2$ is a one-dimensional insulator, and the bifurcation of the highest occupied electronic band
into
the spinon and holon branch happens at about 1.0~eV, whereas in cuprates it is observed at $\sim$\,0.4~eV in the
conductance band, so even the visual similarity between the two spectra in not complete. Second, it is well known that
in two dimensions spin-charge separation does not occur because of the formation of the so-called \textit{strings}
(see
Fig.\,\ref{Fig:ChargeSeparation}). As the photohole moves in a N\'{e}el background of antiferromagnetically ordered
spins, it creates a disturbance that raises the energy of the system \cite{Dagotto94, KaneLee89, DagottoJoynt90},
creating a one-dimensional \textit{string excitation} that binds the photohole to its original position. If in the
one-dimensional case the spinon and holon are fully independent, each possessing its own energy scale, in two
dimensions
they are bound by an effective potential growing linearly with the distance between the two ``particles'', so the
string
excitation as a whole is characterized by a single energy scale of the order of $J$. The excitation spectrum of such a
system was calculated within the $t$-$J$ model (see Fig.\,\ref{Fig:ChargeSeparation}) \cite{Manousakis07}. Though it
bears a slight resemblance to the ``waterfalls'' kind of spectrum, it no longer possesses the two energy scales and
fails to reproduce other experimental details, such as long vertically dispersing ``waterfalls'' of almost constant
width.

\hvFloat[floatPos=t, capWidth=1.0, capPos=r, capVPos=t, objectAngle=0]{figure}
        {\!\includegraphics[width=0.40\textwidth]{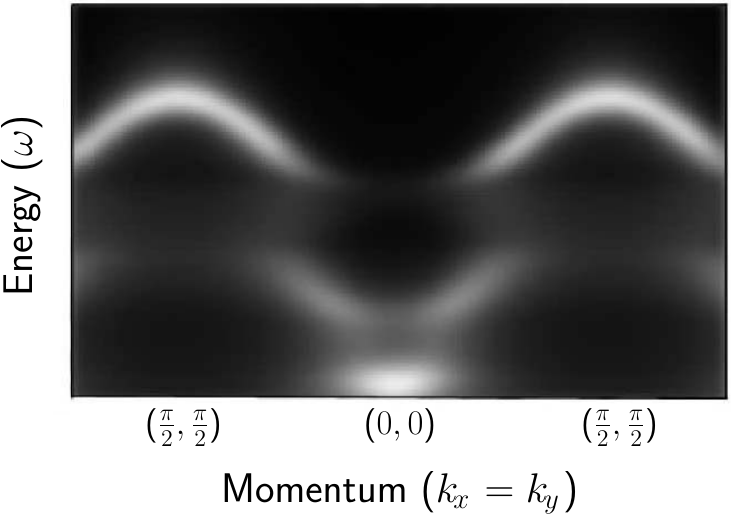}\quad}
        {Result of the Hubbard model calculation including finite-$U$ double-occupancy effects on magnetic
        excitations, reproduced from Ref.~\citenum{SrivastavaGhosh07}. The figure is to be compared with panel (a) in
        Fig.\,\ref{Fig:WaterfallsOld}.}
        {Fig:Srivastava}

It is worth mentioning that spin-charge separation is not the only mechanism that may lead to a bifurcation of the
low-energy band, as similar behavior has been reproduced within a Hubbard model calculation for the two-dimensional
doped Mott insulator \cite{LeighPhillips07, ChoyLeigh08}. There the inner branch of the dispersion corresponds to the
standard lower Hubbard band of the $t$-$J$ model, while the outer branch represents a new charge $e$ state that arises
from a binding of the hole with the charge 2$e$ boson.

\textbf{Strong correlations.} Several other theoretical calculations employing the Hubbard or $t$-$J$ models have been
made in the attempt to reproduce the experimentally observed spectra in the high-energy range. Calculations by
K.\,Byczuk \textit{et al.} \cite{ByczukKollar07} within the Hubbard model solved by many-body dynamical mean-field
theory (DMFT) showed the possibility to reproduce the high-energy kinks, but again, no vertically dispersing features
similar to ``waterfalls'' can be observed in the calculated spectra. In the work by P.~Srivastava \textit{et al.}
\cite{SrivastavaGhosh07}, is was shown that the high-energy kink in the calculated hole dispersion is strongly
enhanced
by the finite-$U$ double-occupancy effects within the Hubbard model, exhibiting qualitative agreement with the
experiment on Ca$_2$CuO$_2$Cl$_2$ \cite{RonningShen05} (see Fig.\,\ref{Fig:Srivastava}). Another work by Y.~Wan
\textit{et al.} \cite{TanWan07} provides a $t$-$J$ model calculation, where the double-occupancy effects are
implicitly
excluded, and reports a ``waterfall''-like feature that again lacks the extent in energy and is much weaker than the
experimentally observable spectral weight.

\hvFloat[floatPos=b, capWidth=1.0, capPos=r, capVPos=t, objectAngle=0]{figure}
        {\!\includegraphics[width=0.50\textwidth]{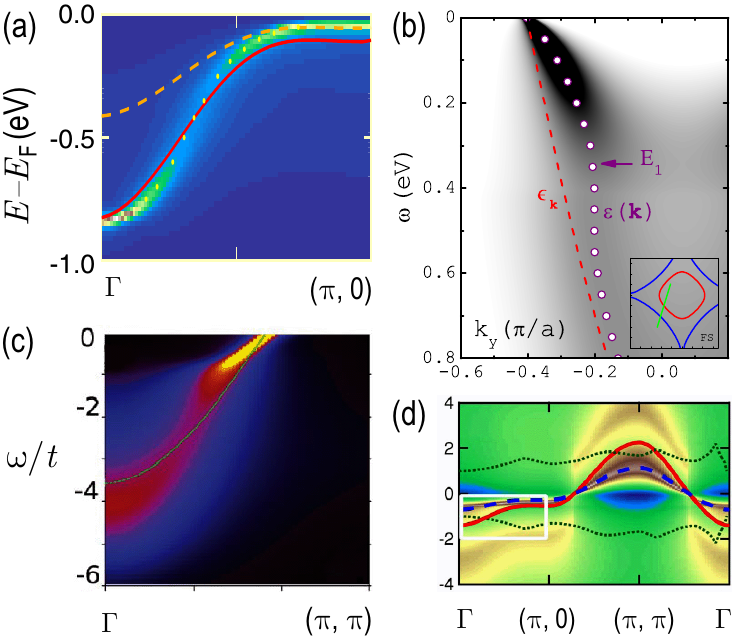}\quad}
        {Summary of the self-energy based models that lead to the appearence of high-energy anomalies in the spectral
        density: \textbf{(a)} coupling to magnons \cite{MarkiewiczSahrakorpi07}, bare and renormalized dispersions are
        shown by the solid red and dashed orange lines respectively, MDC dispersion is shown by dots; \textbf{(b)}
        quantum critical fluctuations \cite{ZhuAji08}, the position in momentum space is shown in the inset;
        \textbf{(c)} spin fluctuations \cite{MacridinJarrell07}, thin line indicates bare dispersion; \textbf{(d)}
        plasmons
        \cite{MarkiewiczBansil07}, bare and renormalized dispersions shown by red solid and blue dashed lines
        respectively.} {Fig:WaterfallsSE}

\textbf{Self-energy approach.} Several self-energy models have been proposed that at least partially reproduce the
high-energy anomaly. These include coupling to bosonic modes \cite{CojocaruCitro07}, including spin fluctuations
\cite{VallaKidd07, MeevasanaZhou07, MacridinJarrell07}, magnons\cite{MarkiewiczSahrakorpi07}, plasmons
\cite{MarkiewiczBansil07}, or quantum-critical fluctuations of a loop-current phase \cite{ZhuAji08}. Some of these
results are summarized in Fig.~\ref{Fig:WaterfallsSE}.

\hvFloat[floatPos=t, capWidth=1.0, capPos=r, capVPos=t, objectAngle=0]{figure}
        {\!\includegraphics[width=0.56\textwidth]{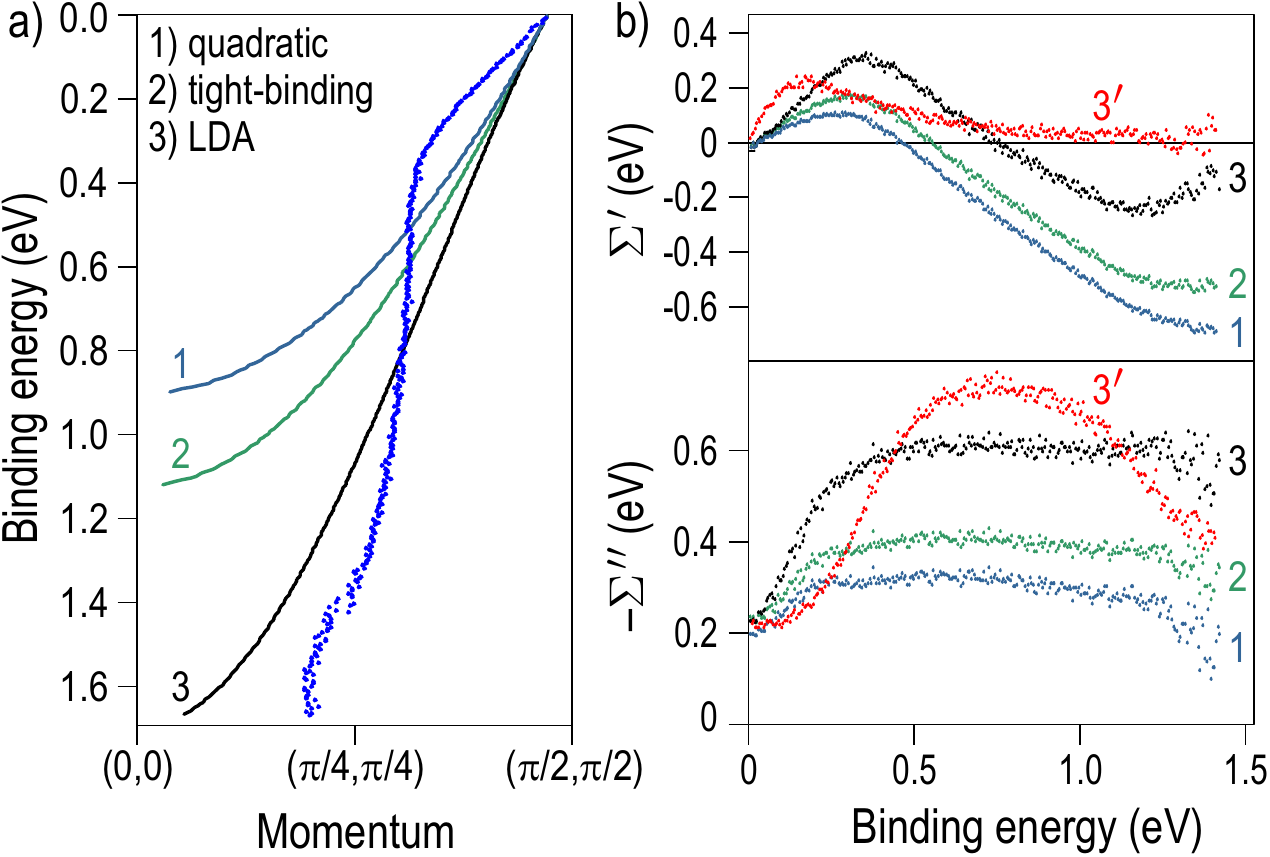}\quad}
        {Kramers-Kronig inconsistency of the experimental self-energy. \textbf{(a)}
        Experimental dispersion (points) shown together with different bare band models: parabolic (1)
        \cite{KordyukBorisenko05}, tight binding (2) \cite{KordyukBorisenko03}, and LDA (3)
        \cite{LinSahrakorpi06}. \textbf{(b)} Real (top) and imaginary (bottom) parts of the self-energy extracted from
        the experimental dispersion and MDC width, respectively. Different curves (1\,--\,3)
        correspond to different bare band models. The red curves ($\text{3}'$) in both panels represent the
        KK-consistent
        counterparts for the LDA bare bands.}{Fig:WaterfallsKK}

Other evidence, however, lets us exclude the self-energy as the sole reason for the ``waterfalls'' behavior
\cite{InosovKordyuk07, InosovFink07, InosovSchuster08, MeevasanaBaumberger08}. Before we consider more arguments below,
let us first see what information one can obtain directly from the self-energy analysis of the photoemission data. As
we know, in the common
assumption of momentum-independent self-energy (i.e. when the self-energy is considered a function of one variable
$\omega$), the real and imaginary parts are to be Kramers-Kronig consistent. Let us check if the consistency holds for
the experimental self-energy extracted from a ``waterfall'' spectrum. To do this, we extract the self-energy
$\mathit{\Sigma}$ from the nodal spectrum of Bi2212, such as the one shown in Fig.\,\ref{Fig:WaterfallsMomentum}\,(a),
assuming three different forms of the bare band with the same Fermi momentum and velocity: parabolic
\cite{KordyukBorisenko05}, tight-binding model \cite{KordyukBorisenko03}, and LDA \cite{LinSahrakorpi06}. In
Fig.\,\ref{Fig:WaterfallsKK}\,(a) these bare bands are shown together with the experimental dispersion obtained by
fitting the momentum distribution curves with Lorentzians. By the procedure described in
\S\ref{SubSec:SelfConsistentSE}
we extract the real and imaginary parts of the self-energy from the experimental dispersion and half-width at
half-maximum (HWHM) of the MDC:
\begin{equation}\label{Eq:SigmaWaterfalls}
\begin{split}
\mathit{\Sigma}^{\prime}(\omega)&=\omega-\varepsilon(k_\omega),\\
\mathit{\Sigma}^{\prime\prime}(\omega)&=\varepsilon(k_\omega+\delta k_\omega)-\varepsilon(k_\omega-\delta k_\omega)/2.
\end{split}
\end{equation}
Here $k_\omega$ is the experimental MDC dispersion, $\delta k_\omega$ is HWHM of the MDC at energy $\omega$, and
$\varepsilon(k)$ is the bare dispersion.\footnote{The formulae (\ref{Eq:SigmaWaterfalls}) are exact, except that we
neglect here the asymmetry of the actual MDC curve, which turns out to be a very good approximation whenever $k_\omega
\gg \delta k_\omega$ (so that the bare band can be well approximated with a linear function within the momentum window
defined by the MDC width) and starts to break down only in the close vicinity of the bare band bottom (at about
$\sim$\,70\% of the band width).}

The results for all three different bare band models are shown in Fig.\,\ref{Fig:WaterfallsKK}\,(b). As one can see,
for
all three models the scattering rate $\mathit{\Sigma}^{\prime\prime}$ appears constant at high energies (along the
``waterfall'') with an onset at about 0.15\,eV. One would expect from Kramers-Kronig relations that the maximum in
$\mathit{\Sigma}^\prime$ would coincide with this onset. Instead, the maximum appears to be shifted to
0.3\,--\,0.35\,eV. Hence, we can conclude that the spurious early onset of the constant scattering rate is a result of
the spectral function being distorted by photoemission matrix element effects. To quantify this result,
Fig.\,\ref{Fig:WaterfallsKK}\,(b) also shows the Kramers-Kronig consistent counterparts of the experimental
$\mathit{\Sigma}^{\prime}$ and $\mathit{\Sigma}^{\prime\prime}$. The curves $\text{3}^{\prime}$ in both upper and lower
panels
are the Kramers-Kronig transforms of the curves 3 in the lower and upper panels respectively. As one can see from
comparison of the curves 3 and $\text{3}^{\prime}$ in both panels, the real and imaginary parts of $\mathit{\Sigma}$
dramatically disagree, which means that the experimental self-energy obtained without proper treatment of matrix
element
effects is not Kramers-Kronig consistent and is therefore unphysical. Hence the extensive ``waterfalls'' are present
only in the raw ARPES spectra, but not in the spectral function, which explains the difficulties in their theoretical
interpretation. In order not to overload the figure, the Kramers-Kronig transforms are shown only for the LDA-based
curves, but the other two self-energies (curves 1 and 2) result in an equally bad match.

\hvFloat[floatPos=t, capWidth=1.0, capPos=b, capVPos=t, objectAngle=0]{figure}
        {\includegraphics[width=\textwidth]{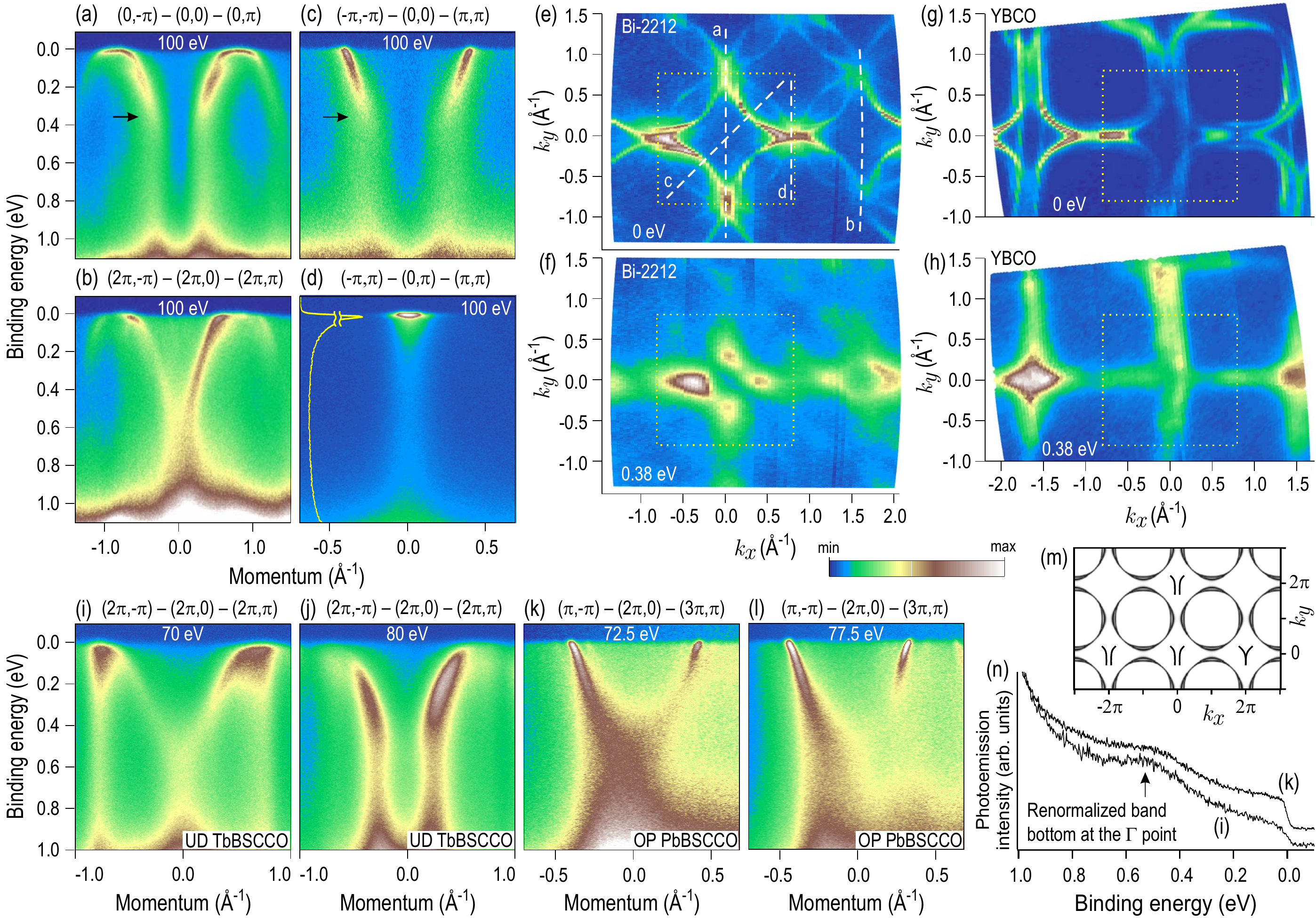}\afterpage{\clearpage}}
        {(\textbf{a})\,--\,(\textbf{h}) Typical snapshots of
         the one-particle excitation spectra of Bi-2212 (\textbf{a}\,--\,\textbf{f}) and Y-123
         (\textbf{g},\,\textbf{h}) measured
         by angle-resolved photoemission with 100~eV photon energy. The spectra (\textbf{a})\,--\,(\textbf{d}) are
         measured along
         the high-symmetry directions marked by the dashed lines on the Fermi surface map (\textbf{e}). Spectra
         (\textbf{a}) and
         (\textbf{c}) from the first Brillouin zone exhibit strong high-energy kinks (black arrows) and ``waterfalls'',
         while the equivalent
         spectrum (\textbf{b}) from the second Brillouin zone exhibits no pronounced high-energy scales. Additional
         spectral weight is clearly
         seen in panel (\textbf{d}), where the curve at the left of the panel shows the energy distribution curve at
         $(\piup,0)$.
         (\textbf{f}) Constant-energy cut at 0.38\,eV below the Fermi level in Bi-2212 showing spectral weight
         depletion along
         the first Brillouin zone diagonals. (\textbf{g}),\,(\textbf{h}) The respective constant-energy maps of Y-123.
         The first Brillouin zone on the
         constant-energy maps is confined by the dotted squares. (\textbf{i}),\,(\textbf{j}) and
         (\textbf{k}),\,(\textbf{l})
         Pairs of equivalent spectra of Tb- and Pb-doped Bi2212 taken in the second Brillouin zone along the
         $(\piup,0)$ and $(\piup,\piup)$
         directions with two different excitation energies as indicated on top of each panel. In both directions, the
         onset of
         the ``waterfalls'' behavior suddenly occurs at about 75\,eV photon energy. The color scale in all panels
         represents
         photoelectron intensity. The spectra are normalized to the background above the Fermi level. The spectra in
         panels
         (\textbf{k}) and (\textbf{l}) are in addition multiplied by a linear function of momentum to enhance the
         right-hand part
         of the spectrum, which otherwise has much lower intensity than the left-hand part due to the experimental
         geometry.
         (\textbf{m}) Schematic representation of the experimentally accessible regions of momentum space showing
         different
         behaviors of the high-energy dispersion in different Brillouin zones immediately below 75\,eV photon energy.
         Positive $k_x$ values
         correspond to the experimental geometry approaching normal incidence. (\textbf{n}) Energy distribution curves
         taken at
         the $\mathrm{\Gamma}$ point from spectra shown in panels (\textbf{i}) and (\textbf{k}), showing a distinct
         bottom of the
         renormalized band at about 0.5 eV. The figure is reproduced from
         Ref.\,\citenum{InosovFink07}.}{Fig:WaterfallsMomentum}

\enlargethispage{-1.3em}A careful reader might remark here that the self-energy analysis can be affected by several
factors which we did not take into account. Namely, (i) the $\textbf{k}$-dependence of the self-energy, (ii)
sensitivity
of the Kramers-Kronig transform to the assumed behavior of the ``tails'' \cite{KordyukBorisenko05}, and (iii) bilayer
splitting effects. Let us comment briefly on these three points. (i) The $\textbf{k}$-dependence of the self energy is
estimated to be much less than 20\% across the whole Brillouin zone \cite{ZhuAji08}, which by itself is too weak an
effect to account for the observed inconsistency of the experimental self-energies. Moreover, the experimental
self-energy tracks the renormalization effects only along a particular curve in $(\textbf{k},\,\omega)$ space, so even
if there was some $\textbf{k}$-dependence of $\mathit{\Sigma}$, it would be equally present both in
$\mathit{\Sigma}^\prime$ and $\mathit{\Sigma}^{\prime\prime}$, not affecting their Kramers-Kronig consistency. Only
presence of a notable $\textbf{k}$-dependence on the scale of the MDC width could influence our arguments, which is
definitely not the case. (ii) We have checked how different forms of the ``tails'' terminating the experimental
self-energy at $+\infty$ influence the result. If one reasonably assumes that the self-energy decreases monotonically
after the cut-off energy ($\sim$1.5~eV), the result of the Kramers-Kronig transform will not significantly depend on
the
actual form of the ``tail''. We have checked this for ``tails'' of different steepness, from abrupt zeroing at the
cut-off energy up to a constant non-zero value at high energies. It slightly influenced the high-energy part of the
resulting curve, while all the energy scales essential for our argument remained practically unchanged. (iii) The
bilayer splitting can, indeed, affect the self-energy analysis, because the intensity ratio of the bonding and
antibonding bands should generally depend both on energy and momentum. If this effect is significant, it only confirms
our conclusion that within the broad energy and momentum ranges under consideration matrix element effects can not be
neglected. In \S\ref{SubSec:WaterfallsExcitEn}, however, we will present additional evidence for the bilayer splitting
effects not being responsible for the anomalous dispersion.

\subsection{Momentum dependence and line shape analysis}

Now we will consider the momentum dependence of the ``waterfalls''. We observe strong differences in the shape of the
single-particle excitation spectrum between different Brillouin zones and its strong dependence on the excitation
energy. This indicates that photoemission matrix elements strongly influence the recorded spectral weight and that the
reported values for a high energy scale, as well as the respective physical models, may be incorrect.

Here we present several counter-examples which show that the ``waterfalls'' do not necessarily reveal a ``new energy
scale''. Fig.\,\ref{Fig:WaterfallsMomentum} \cite{InosovFink07} shows several typical photoemission spectra of Bi2212
along high-symmetry directions [panels (a)\,--\,(d)] and the constant-energy maps at the Fermi level [panels (e) and
(g)] and at 0.38~eV below it for Pb-Bi2212 and Y123 [panels (f) and (h)]. As can be seen from comparison of panels (a)
and (b), presenting the spectra taken along equivalent cuts in momentum space in the first and second Brillouin zones,
the high-energy kinks and \,\hbox{\waterfalls-}shaped ``waterfalls'' appear in the first Brillouin zone, while in the
second Brillouin zone neither of these features is observed. Since both the electronic band structure and many body
effects remain invariant under any translation by a reciprocal lattice vector, the difference between these two images
can come only from the photoemission matrix elements which, as a rule, strongly depend on momentum and excitation
energy
\cite{BansilLindroos99, BorisenkoKordyuk01, AsensioAvila03}.

Panels (i)\,--\,(l) in the same figure show the energy dependence of the spectra along the $(\piup,\,0)$ and
$(\piup,\,\piup)$ directions in the second Brillouin zone taken with the photon energies close to the binding energy
of
the Cu\,3p level (75.1~eV), where the photoionization cross section is modified by interchannel coupling of the direct
photoemission process with an Auger decay of the photoexcited Cu\,3p core hole \cite{DavisFeldkamp81}. Here we also
observe an abrupt transition from a~\hbox{\waterfalls-}shaped to a \hbox{\champagne-}shaped dispersion at
$\text{75}\pm\text{1}$~eV photon energy. Panels (i) and (k) show spectra below the transition that are to be compared
with the equivalent spectra shown in panels (j) and (l) above the transition energy. Using different experimental
geometries, i.\,e. different sample positions relative to the analyzer, we can access the second Brillouin zone both
at
$k_x > \text{0}$ (experimental geometry approaching normal incidence) and $k_x < \text{0}$ (experimental geometry
approaching grazing incidence). It is remarkable that we do not see a distinct transition neither in the first, nor in
the second Brillouin zone at $k_x < \text{0}$, as shown schematically in panel (m). At the $\mathrm{\Gamma}$ points
marked by ``\kern1pt\waterfalls\kern1.5pt'', the \hbox{\waterfalls-}shaped dispersion persists at all energies, while
at
the $\mathrm{\Gamma}$ point marked by ``\kern1pt\champagne\kern2.2pt'', a sharp transition from the~\hbox{\champagne-}
to the \hbox{\waterfalls-}like behavior is observed at 75~eV photon energy. In panel (n) we show energy distribution
curves at the $\mathrm{\Gamma}$ point extracted from spectra (i) and (k), where a distinct band bottom at about 0.5~eV
is observed.

However, the matrix elements can not explain all of the high-energy effects. As can be clearly seen both in Bi2212
[panels (b), (d), (f), and (i)] and Y123 [panel (h)], additional incoherent spectral weight is aggregated along the
bonding directions in the momentum space, i.\,e. $(2\piup\kern.3pt n,\,k_y)$ and $(k_x,\,2\piup\kern.3pt n)$, $n\in
\mathbb{Z}$, persisting deeply below the saddle-point of the conductance band [panel (d)] and forming a grid-like
structure in the momentum space [panel (h)]. At the center of the first Brillouin zone this incoherent component is
suppressed by matrix elements together with the coherent part of the spectrum, forming the ``waterfalls''\,---\,two
long
vertically dispersing tails seen in panels (a) and (c), and high-energy kinks.

In addition, we should mention that in our studies we have not detected any significant dependence of the high-energy
dispersion neither on doping nor on temperature.

Both photon energy dependence near 75~eV and the dependence on the Brillouin zone may be related to the resonant
enhancement of the Cu photoionization cross section at this energy \cite{HenkeGullikson93}. In the Auger process which
resonates at $h\nu = \text{75}$~eV with the normal photoemission process, near the threshold the core hole is produced
by a Cu~3p\,$\rightarrow$\,3d$_{x^2-y^2}$ transition. According to the dipole selection rules \cite{FinkNuecker94}
this
transition is allowed for $\vec{E}$ vectors parallel to the surface of the sample (or to the Cu~3d$_{x^2-y^2}$
orbital)
and forbidden for $\vec{E}$ vectors perpendicular to the surface. Thus we would expect an enhancement of the Cu
ionization cross section above the resonance for those momenta (sample positions), for which the component of
$\vec{E}$
parallel to the surface is significant, i.\,e., for $k_x > \text{0}$. This would mean that changing the photon energy
from below to above the resonance, small changes in the spectral weight should be expected for $k_x < \text{0}$ and
large changes should be expected for $k_x > \text{0}$, which is in good agreement with our experimental findings.

This suggests that in those Brillouin zones where we see for $h\nu < \text{75}$ eV a \hbox{\champagne-}shaped
dispersion, we directly probe the spectral function of the renormalized band without significant distortion, while in
the other Brillouin zones and at higher excitation energies the \hbox{\waterfalls\kern.3pt-\kern.3pt}shaped
``waterfalls'' are produced by a strong suppression of the spectral weight near the $\mathrm{\Gamma}$ point due to
matrix element effects \cite{DuerrLegner01, MesotRanderia01}. According to such interpretation, the peak seen near
0.5~eV in the energy distribution curves extracted from the \hbox{\champagne-}shaped spectra (Fig. 1(n)) would be the
bottom of the renormalized conduction band. It is interesting to compare the observed band width with recent
calculations of the band renormalization due to a coupling to the charge carrier plasmon \cite{MarkiewiczBansil07}.
There the renormalization factor $Z=\text{0.5}$ has been derived, corresponding to the bandwidth narrowing by a factor
of two, which is in reasonable agreement with the observed bottom of the conduction band at 0.5~eV.

\hvFloat[floatPos=t, capWidth=1.0, capPos=b, capVPos=t, objectAngle=0]{figure}
        {\includegraphics[width=0.8\textwidth]{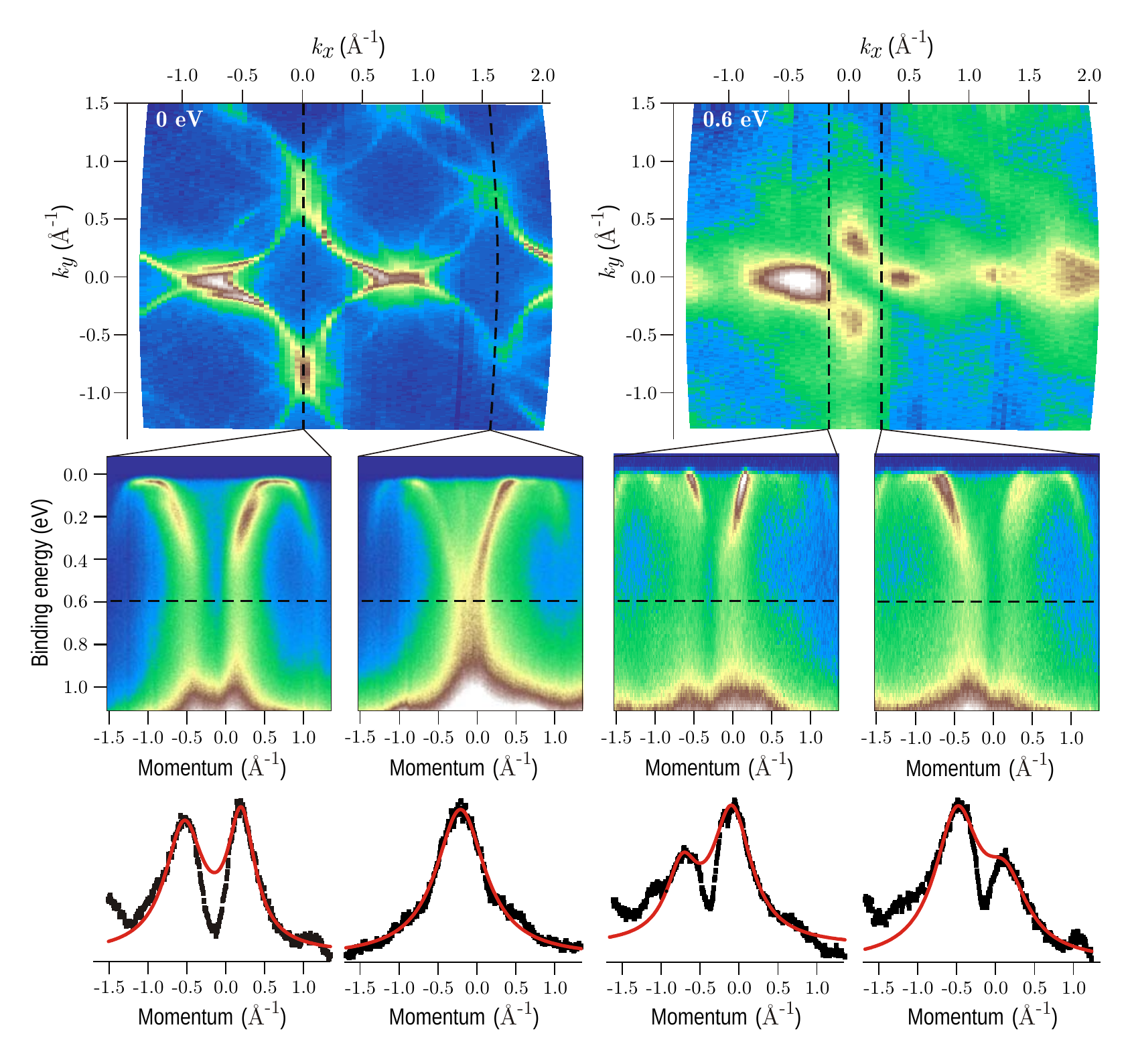}}
        {Momentum dependence of the high-energy dispersion. \textbf{Top row:} constant energy maps of Bi2212 at the
        Fermi level and at 0.6~eV binding energy, where the depletion of the spectral weight along the Brillouin zone
        diagonals is seen. \textbf{Second row:} ARPES images taken along the momentum cuts marked on the maps by the
        dashed lines. The first and second images correspond to equivalent cuts in $\mathbf{k}$-space measured in the
        first and second Brillouin zones respectively. \textbf{Bottom row:} corresponding MDC profiles at 0.6~eV
        compared to the sum of two Lorentzian peaks (red curves) to emphasize the depletion of the spectral
        weight.}{Fig:LineShape}

Our reasoning is additionally supported by Fig.~\ref{Fig:LineShape}, which presents the line shape analysis of the two
kinds of spectra. One sees that the MDC of a \hbox{\champagne-}shaped spectrum taken at 0.6~eV binding energy has a
Lorentzian line shape, while the corresponding MDC of a \hbox{\waterfalls-}shaped spectrum can not be fitted with two
Lorentzians due to the suppressed spectral weight at the $\mathrm{\Gamma}$ point. As one moves away from the
high-symmetry direction, the intensity of the peaks becomes asymmetric, as the suppression of spectral weight moves
along the Brillouin zone diagonal and is stronger along one diagonal than along the other.

Finally, we discuss the vertical feature close to $(\piup,\,0)$ which extends from $\sim$\,0.1 to 1.0~eV. Along the
cut
(d) in Fig.~\ref{Fig:WaterfallsMomentum}, it has almost constant intensity below the saddle point of the conductance
band, visible at a variety of excitation energies. Its intensity is approximately twice as large as could be explained
simply by a sum of the tails of the conductance and valence bands. It is interesting that its distribution in momentum
space is localized along the bonding directions [Fig.~\ref{Fig:WaterfallsMomentum}\,(h)]. In agreement with this,
along
the $(0,\,0)$\,--\,$(2\piup,\,0)$ cut (or equivalent) no feature is observed at $(\piup,\,0)$ neither at low
($h\nu=\text{50}$~eV) nor at high ($h\nu=\text{100}$~eV) photon energies [see e.\,g.
Fig.~\ref{Fig:WaterfallsMomentum}\,(b)]. But surprisingly, in a similar cut taken with $h\nu=\text{70}$~eV [see Fig.
1(i)] we also see vertical features. They might possibly stem from the shadow bands, caused by the orthorhombic
lattice
distortions of Bi-2212, which are seen cutting the line ``a'' in Fig.~\ref{Fig:WaterfallsMomentum}\,(e) near the
$(\piup,\,0)$ point at an angle of 90$^{\circ}$. Evidently there is a strong enhancement of the shadow bands near the
photon energy $h\nu=\text{70}$~eV, which is natural, because strong matrix element effects of the spectral weight of
these bands have been previously detected \cite{MansSantoso06}.

\subsection{Excitation energy dependence}\label{SubSec:WaterfallsExcitEn}

Let us now have a closer look at the photon energy dependence of the high-energy dispersion. As already mentioned,
there
are two distinctive types of behavior observed near the $\mathrm{\Gamma}$ point in the second Brillouin zone in the
binding energy range between 0.4 and 0.8~eV. Fig.\,\ref{Fig:WaterfallsMDC} gives an example of two equivalent ARPES
spectra of slightly overdoped Pb-Bi2212 taken along the $(2\piup,-\piup)$\,--\,$(2\piup,\,0)$\,--\,$(2\piup,\,\piup)$
direction in the momentum space at two different excitation energies: 64~eV (a) and 81~eV (b). The first image shows a
\hbox{\champagne-}shaped (``champagne glass'') type of dispersion with a single vertical stem in the high energy
region,
while the second image exhibits the \hbox{\waterfalls-}like (``waterfalls'') behavior with two vertically dispersing
features in the same energy range. In panel (c), the momentum distribution curves (MDC) of the photocurrent integrated
in a small binding energy window around 0.6~eV are plotted for several excitation energies, showing a smooth crossover
between the two types of spectra at about 72~eV. It is remarkable that such behavior is universal for different
families
of cuprates \cite{InosovFink07}.

\hvFloat[floatPos=t, capWidth=1.0, capPos=b, capVPos=t, objectAngle=0]{figure}
        {\includegraphics[width=0.85\textwidth]{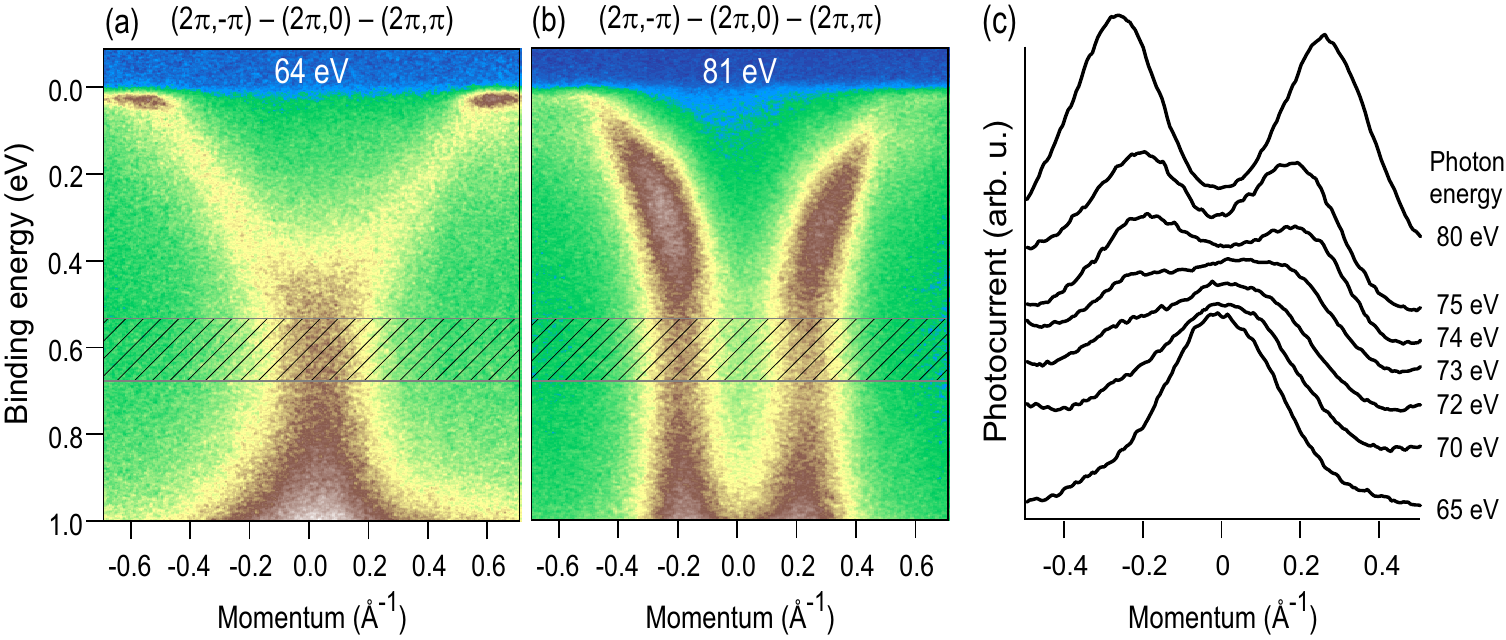}}
        {Photon energy dependence of the high-energy anomaly in Pb-Bi2212 \cite{InosovSchuster08}. A pair of equivalent
        spectra taken in the second Brillouin zone along the
        $(2\piup,\,-\piup)$\,--\,$(2\piup,\,0)$\,--\,$(2\piup,\,\piup)$ direction with excitation energies 64 and 81~eV
        are shown in panels (a) and (b) respectively. The spectrum (a) is an example of the ``champagne glass''
        dispersion, while spectrum (b) represents the ``waterfalls'' behavior. The momentum distribution curves
        integrated in a small energy window around 0.6~eV binding energy (hatched area) are shown in panel (c) for a
        number of excitation energies, showing a transition between the two types of behavior at about
        70~eV.}{Fig:WaterfallsMDC}

Two seemingly reasonable explanations for these chan\-ges in behavior could be related to (i) bilayer splitting, i.e.
modulation of the relative intensity of the bonding and antibonding bands due to the photoemission matrix elements;
(ii)
effects of the $k_z$ dispersion that cause periodic changes of the ARPES signal with varying excitation energy. In the
following, we will show that both these hypotheses are inconsistent with the experimental observations.

Fig.\,\ref{Fig:WaterfallsEdep} shows an excitation energy map along the same cut
$(2\piup,-\piup)$\,--\,$(2\piup,\,0)$\,--\,$(2\piup,\,\piup)$ in momentum space \cite{InosovSchuster08}. The color
scale
represents photoemission intensity integrated in a small binding energy window around 0.6~eV. Each vertical cut
corresponds to an MDC similar to those shown in Fig.\,\ref{Fig:WaterfallsMDC} (c), measured with a 1~eV step in
excitation energy (plotted along the horizontal axis). The intensity of each MDC is normalized by its average value. A
single MDC maximum at the $\mathrm{\Gamma}$ point corresponds to the ``champagne glass'' behavior, while the two split
maxima represent the ``waterfalls''. Except for the already known transition at $\sim$\,70~eV, there are two more
transitions observed around 50 and 90~eV.

One can see that the distance between the MDC maxima changes continuously within each transition. The two maxima in
the
``waterfalls'' region do not lose intensity, giving place to the central peak, as one would possibly expect in the
case
of bilayer splitting; they rather change their position in momentum gradually, merging into a single peak. This lets
us
rule out the bilayer splitting hypothesis.

\hvFloat[floatPos=b, capWidth=1.0, capPos=r, capVPos=t, objectAngle=0]{figure}
        {\includegraphics[width=0.6\textwidth]{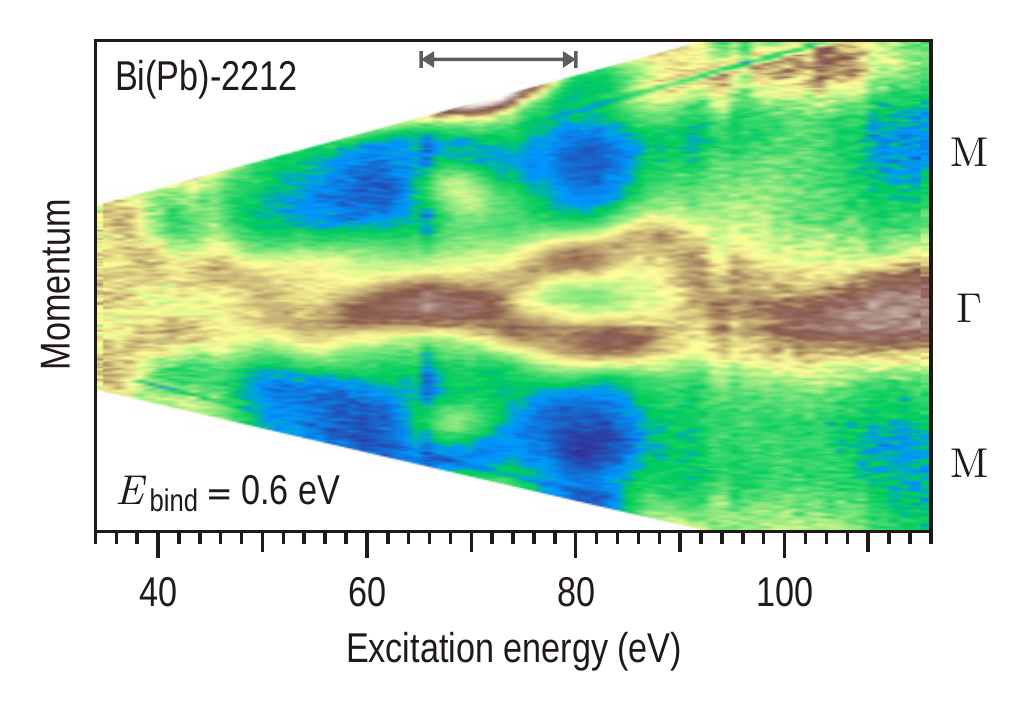}\quad}
        {Momentum distribution of the photocurrent along the M\,--\,$\mathrm{\Gamma}$\,--\,M direction measured in the
        second Brillouin zone as a function of excitation energy \cite{InosovSchuster08}, showing several alterations
        of the high energy dispersion behavior. The color scale represents photoemission intensity integrated in a
        small binding energy window around 0.6~eV and normalized by the average intensity along each cut. The
        double-headed arrow marks the energy range covered by Fig.\,\ref{Fig:WaterfallsMDC}~(c).}{Fig:WaterfallsEdep}

It is also illustrative to compare Fig.\,\ref{Fig:WaterfallsEdep} to the experimentally measured photon energy
dependence curves for the matrix elements of the bonding and antibonding bands (see Fig.\,\ref{Fig:MEexp}). The
relative
intensity of the bonding band near the Fermi level is known to reach maxima at 38 and 56~eV, while the antibonding
band
is enhanced by 50~eV photons. On the other hand, the transitions seen in Fig.\,\ref{Fig:WaterfallsEdep} do not follow
this pattern. Comparison to the theoretical photoemission intensity curves available for the bonding and antibonding
bands in an even wider photon energy window \cite{LeeFujimori02, BansilMarkiewicz04} will lead us to the same
conclusion.

Let us now turn to the consideration of the possible role of the $k_z$ dispersion. It is well known that by varying
the
excitation energy in a photoemission experiment, one can probe different $k_z$ points \cite{Huefner95}. As the Bi2212
crystals are known to be not perfectly two-dimensional \cite{MarkiewiczSahrakorpi05, LindroosSahrakorpi06}, this might
lead to periodic variations of the observed electronic structure as a function of photon energy. The easiest way to
estimate the period of such variations is to use the three-step model in the free electron approximation
\cite{Huefner95}. The kinetic energy of the photoelectron is given by
\begin{equation}
E_\text{kin}=(p_\perp^2+p_\parallel^2)/2m=h\nu-E_\text{bind}-\mathit{\Phi}\text{,}
\end{equation}
where $p_\perp$ and $p_\parallel$ are the normal and parallel components of the electron's momentum in vacuum, $h\nu$
is
the photon energy, $E_\text{bind}$ is the binding energy of the electron in the solid, and $\mathit{\Phi}$ is the work
function. The component of the wave vector perpendicular to the surface is
\begin{multline}\label{Eq:Kperp}
k_\perp\!+n_\perp\!G_\perp\!=\sqrt{\frac{2m}{\hslash^2}\,(E_\text{kin}+V_0)-(k_\parallel+n_\parallel G_\parallel)^2}\\
=\sqrt{\text{0.262}\,\frac{\,\text{\AA}^{-2\kern-5pt}}{\text{eV}}\,\,(h\nu-E_\text{bind}+V_0-\mathit{\Phi})-(k_\parallel+n_\parallel
G_\parallel)^2}\text{,}
\end{multline}
where $V_0>0$ is the inner potential of the crystal, $G_\parallel$ is the reciprocal lattice vector;
$n_\perp,\,n_\parallel\in\mathbb{Z}$. At the $\mathrm{\Gamma}$ point, $k_\parallel=0$. The periodicity in $k_\perp$
should correspond to $G_\perp=2\piup/c=2\piup/\text{30.89}\text{\AA}\approx\text{0.2}\,\text{\AA}^{-1}$, where $c$ is
the lattice constant along $z$ direction. If the periodic changes in Fig.\,\ref{Fig:WaterfallsEdep} originated from
the
$k_z$ dispersion, one period in $k_\perp$ would fit approximately between $h\nu_1=\text{50}$\,eV and
$h\nu_2=\text{90}$\,eV. Using formula (\ref{Eq:Kperp}), we find that this is not possible to achieve for any
reasonable
value of $V_0-\mathit{\Phi}$. Indeed, solving the equation
$k_\perp(h\nu_2)-k_\perp(h\nu_1)=\text{0.2}\,\text{\AA}^{-1}$
yields an unphysically large minimal value of $V_0-\mathit{\Phi}=\text{2550}$\,eV that corresponds to $n_\parallel=0$,
which lets us also reject the $k_z$ dispersion as a possible reason for the observed changes in behavior.

\subsection{Summary and outlook}

The anomalous high-energy dispersion in the electronic structure of cuprates remains a hot topic in the
high-temperature
superconductivity research. After multiple attempts to explain this phenomenon as an intrinsic property of the
spectral
function, we have finally shown that the experimentally observed dispersion significantly depends on the experimental
conditions, such as photon energy and the experimental geometry, which suggested that the influence of photoemission
matrix elements distorts the real behavior of the conductance band at high binding energies. This distortion can be
explained by the photoemission matrix element effect that suppresses the total photoemission signal near the
$\mathrm{\Gamma}$ point at particular excitation energies. Such conclusion agrees with the $\mathbf{k}$-dependent
matrix element with a minimum at the $\mathrm{\Gamma}$ point recently found by W.~Meevasana~\textit{et al.}
\cite{MeevasanaBaumberger08} after a complicated two-dimensional fitting of the spectral function to the ARPES data.

The photon energy dependence of the high-energy dispersion lets us rule out both the bilayer splitting and the
$k_z$-dispersion as possible causes of the changes in behavior. We conclude that it can only be caused by the
suppression of the total photoemission signal near the $\mathrm{\Gamma}$ point due to the matrix element effect, which
would mean that the real underlying electronic structure is closer to the ``champagne glass'' type, rather than to the
''waterfalls''. Further theoretical work still needs to be done in order to understand all the details of the
high-energy anomaly behavior as a function of photon energy and gain more insight into the underlying electronic
structure.

On the other hand, up to now there is no clear understanding of the source of the additional spectral weight observed
along the bonding directions, which manifests itself as the vertical feature at $(\piup,\,0)$ below the saddle point
[Fig.~\ref{Fig:WaterfallsMomentum}\,(d) and (i)] and as the ``waterfalls'' at the $\mathrm{\Gamma}$ point extending
below the bottom of the conductance band in the energy range between 0.5 and 1.0~eV
[Fig.~\ref{Fig:WaterfallsMomentum}\,(i) and (k)]. Here we mention only that such an additional component is supported
by
recent optical experiments \cite{HwangNicol07}. Evidently, this component, either incoherent or extrinsic, represents
a
new phenomenon which deserves more systematic studies as a function of photon energy and momentum. Possible
explanations
can be related with the disorder-localized in-gap states \cite{AlexandrovReynolds07}. The inelastic scattering of
photoelectrons \cite{TougaardSigmund82} can be another option. On the other hand, the grid-like momentum distribution
of
this additional spectral weight may hint at the presence of a one-dimensional structure \cite{ZhouBogdanov99}. If so,
then the photoemission spectra consist of two components: one from the well studied two-dimensional metallic phase and
another from an underdoped one-dimensional phase. Such a scenario would be consistent with the ``checkerboard''
structure observed by scanning tunneling spectroscopy in lightly hole-doped cuprates \cite{HanaguriLupien04}.


\chapter{Relation between the single-particle spectral function and the two-particle correlation
functions}\label{Chap:Relation}

\section{Spin excitations in cuprates probed by the inelastic neutron scattering}

\subsection{Inelastic neutron scattering spectroscopy as an experimental method}

\begin{wrapfigure}[14]{p}{0.263\textwidth}
\vspace{-1.1em}~\includegraphics[width=0.263\textwidth]{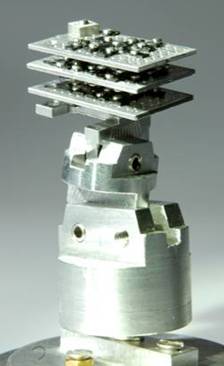}~ \caption{Array of 180 untwinned co-oriented
YBCO
single crystals for an INS measurement \cite{KeimerWebSite}.} \label{Fig:YBCOarray}
\end{wrapfigure}

Inelastic neutron scattering is an experimental technique capable of probing magnetic and crystal field excitations of
a
solid by measuring the change in kinetic energy and momentum of the scattered neutron beam \cite{INS_Notes07}. It has
played a major role in characterizing the nature and strength of antiferromagnetic interactions and spin fluctuations
in
cuprates \cite{Tranquada05, SidisPailhes04}. What INS actually measures is the imaginary part of the dynamic spin
susceptibility $\chi''(\mathbf{q},\,\mathit{\Omega})$ of the sample, where the scattering vector $\mathbf{q}$ is the
difference between the incoming and outgoing wave vectors, and  $\mathit{\Omega}$ is the energy change experienced by
the sample (negative that of the scattered neutron). Different contributions to the susceptibility may come from all
possible kinds of magnetic excitations in the system, therefore an important part of the INS data analysis is to
distinguish between these contributions.

\begin{wrapfigure}[15]{l}{0.37\textwidth}
\includegraphics[width=0.37\textwidth]{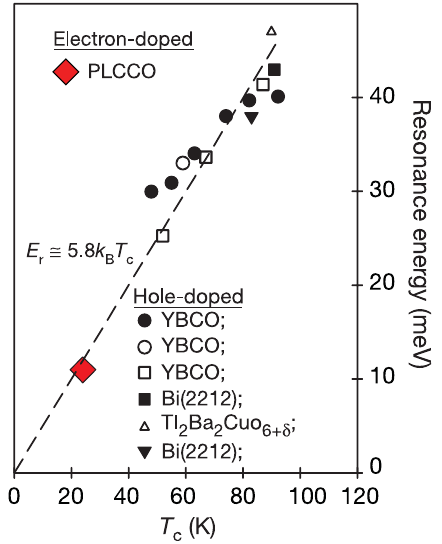} \caption{Resonance energy as a function of $T_\text{c}$ for hole-
and electron-doped cuprates. The figure is reproduced from Ref.~\citenum{WilsonDai06}.} \label{Fig:58kBTc}
\end{wrapfigure}

ARPES and INS are complementary momentum-resolved techniques that probe charge and spin excitation spectra,
respectively. Comparison of ARPES and INS data is therefore expected to shed light on the interactions between spin
and charge excitations, which according to many models are at the root of the mechanism of high-$T_\text{c}$
superconductivity \cite{Eschrig06}. Neutron scattering (in contrast to ARPES) is a bulk sensitive technique, and its
intensity is determined in particular by the volume of the single crystals available for measurements. This is why INS
measurements were prohibited for a long time by the small size of the single crystals. One often has to prepare huge
arrays of co-oriented single crystals (see Fig.\,\ref{Fig:YBCOarray}) in order to reach reasonable intensity and
resolution of an INS measurement \cite{HinkovPailhes04}.

\hvFloat[floatPos=b, capWidth=1.0, capPos=b, capVPos=t, objectAngle=0]{figure}
        {\includegraphics[width=0.8\textwidth]{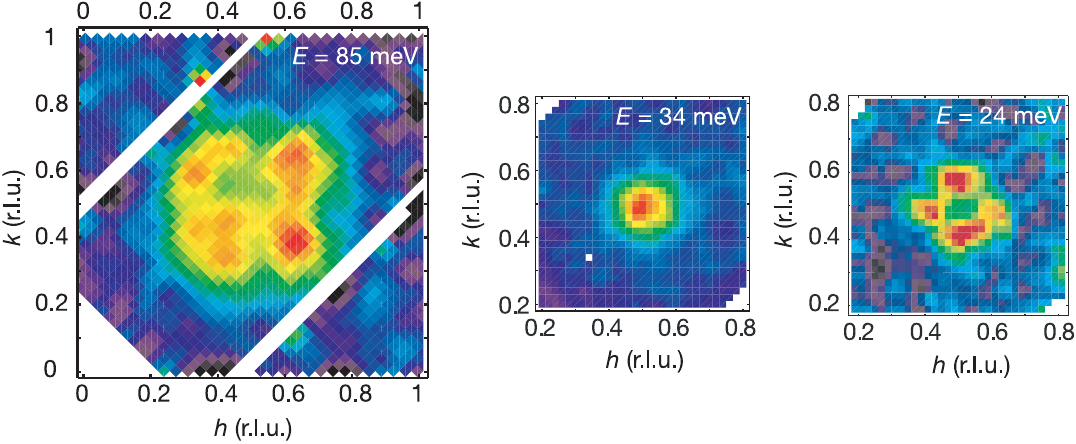}\vspace{-1em}}
        {Images of magnetic excitations in YBa$_2$Cu$_3$O$_{6.6}$ above (left), below (right), and at the resonance
        energy (middle). The figure is reproduced from Ref.\,\citenum{HaydenMook04}.}{Fig:Incommensurate}

Most of the neutron scattering studies of cuprate superconductors \cite{RossatMignod91, FongBourges00, DaiMook01,
PailhesSidis03, HaydenMook04, HinkovPailhes04, PailhesSidis04, PailhesUlrich06, WooDai06, TranquadaWoo04,
VignolleHayden07} have focused on two families: LSCO and YBCO for the simple reason that these are the only compounds
for which large crystals have been available. For a comparison with ARPES this is rather disappointing, as
surface-sensitive techniques perform much better with BSCCO samples, which are easily cleavable to obtain atomically
clean surfaces representative of the bulk. Unfortunately, ARPES measurements of YBCO and LSCO are complicated by the
surface effects \cite{ZabolotnyyOD07}, while INS measurements of BSCCO have long been prohibited because of the small
size of the single crystals. Because of the weak bonding along the $\mathbf{c}$ direction in BSCCO, the crystals
typically grow as thin plates with volumes much too small for INS. This problem has recently been overcome
\cite{FongBourges99, HeSidis01, CapognaFauque07, FauqueSidis07}, but the resolution of neutron scattering measurements
of BSCCO is still far behind that of LSCO and YBCO.

\subsection{Magnetic resonance structure observed in high-$T_\text{c}$ cuprates}

\hvFloat[floatPos=t, capWidth=1.0, capPos=b, capVPos=t, objectAngle=0]{figure}
        {\includegraphics[width=\textwidth]{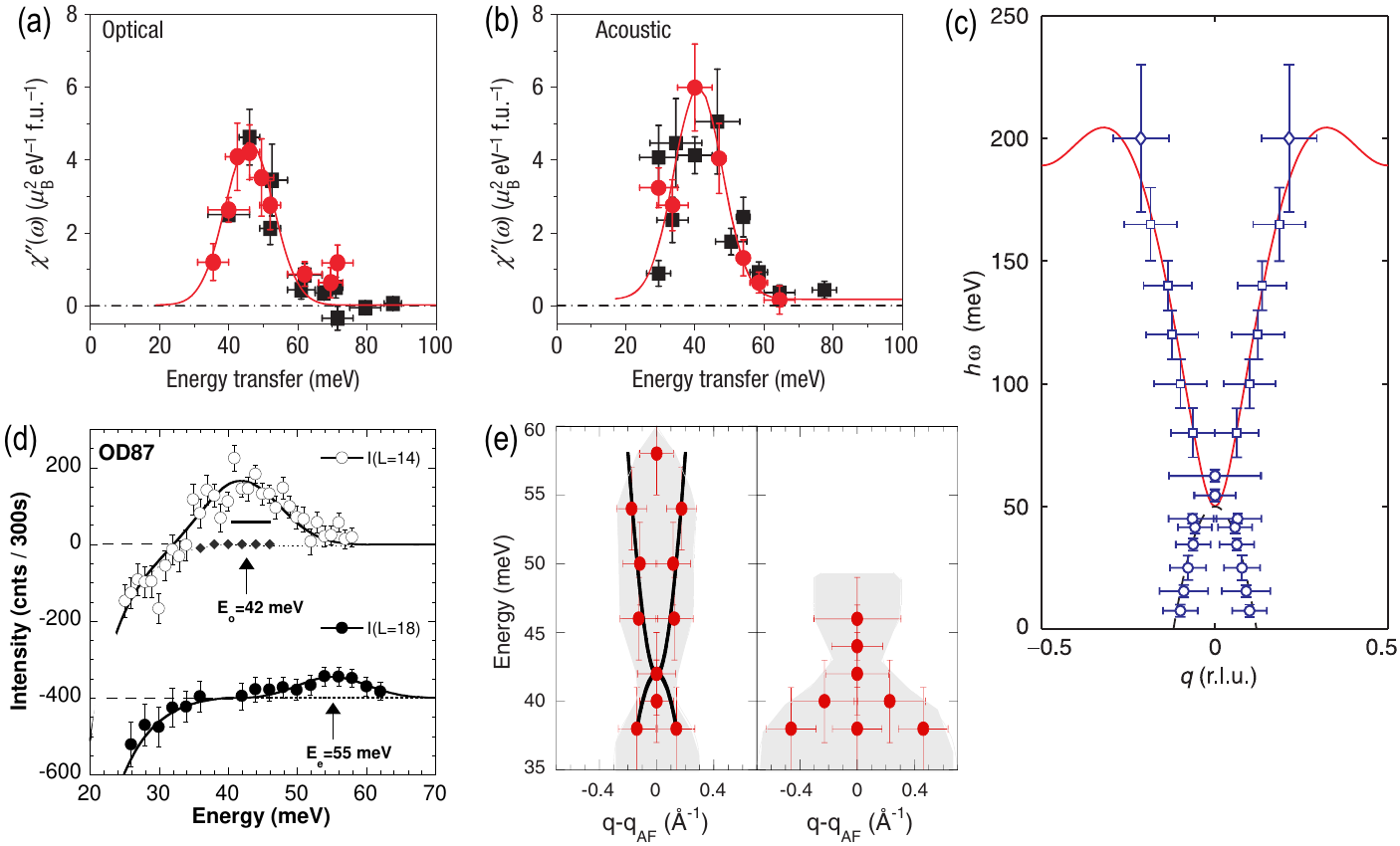}}
        {Summary of the $\mathbf{q}$- and $\omega$-dependent susceptibility of YBCO and BSCCO. \textbf{(a)} Local
        susceptibility of YBa$_2$Cu$_3$O$_{6.95}$ at $(\piup,\,\piup)$ in the optical (even) channel in absolute units
        \cite{WooDai06}. \textbf{(b)} Same for the acoustic (odd) mode. \textbf{(c)} Incommensurability of spin
        excitations in YBCO as a function of energy \cite{TranquadaWoo04}. \textbf{(d)} Acoustic (top) and optical
        (bottom) resonances in Bi2212 \cite{CapognaFauque07}.  \textbf{(e)} Dispersion of the spin excitations in
        Bi2212
        along the (130) and (110) directions \cite{FauqueSidis07}.}{Fig:NeutronResonances}

Inelastic neutron scattering measurements have brought to light the existence of unusual antiferromagnetic excitations
in cuprates that develop below $T_\text{c}$, but their origin is still highly controversial \cite{SidisPailhes04}. The
spin excitation spectrum is dominated by a sharp magnetic excitation at the planar antiferromagnetic wave vector
$\mathbf{q}_\text{AF}=(\piup,\,\piup)$\,---\,the so-called magnetic resonance mode \cite{RossatMignod91,
MookYethiraj93,
FongKeimer95, FongKeimer96, BourgesRegnault96}. The energy of the mode depends on the particular material and was
found
to scale approximately linearly with $T_\text{c}$ as $\mathit{\Omega}_0\approx\text{5.8}k_\text{B}T_\text{c}$ for a
wide
range of both hole- and electron-doped cuprates \cite{WilsonDai06}, as shown in Fig.\,\ref{Fig:58kBTc}. It is
therefore
a fundamental property of the superconducting state that might play an important role in the mechanism of
high-$T_\text{c}$ superconductivity.

The resonance has been originally discovered in YBCO \cite{RossatMignod91} at the energy of $\sim$\,40\,meV. In
optimally doped BSCCO a similar magnetic resonance peak has been observed at $\sim$\,43\,meV \cite{FongBourges99}.
Further, it has been seen in Tl$_2$Ba$_2$CuO$_{6+\delta}$ as high as at 47\,meV \cite{HeBourges02}. Except for the
main
resonance, INS provides evidence for incommensurate spin fluctuations both above and below the resonance
\cite{HaydenMook04, TranquadaWoo04}, as shown in Fig.\,\ref{Fig:Incommensurate}. The incommensurability increases
below
$T_\text{c}$ with decreasing temperature and decreases upon approaching the resonance energy in the superconducting
state, forming an hourglass-like shape [Fig.\,\ref{Fig:NeutronResonances}\,(c)]. The symmetry of the incommensurate
resonances also changes across $\mathit{\Omega}_0$: below the resonance the intensity is concentrated along the
$(q,\,0)$ and $(0,\,q)$ directions, while above the resonance it prevails along the diagonal directions $(q,\pm q)$.

In bilayer cuprates, the interlayer exchange coupling leads to the formation of two non-degenerate modes of magnetic
excitations characterized by odd (o) and even (e) symmetries with respect to exchange of the layers
\cite{PailhesSidis03, PailhesSidis04}. These are often called acoustic and optical modes respectively, by analogy with
the acoustic and optical magnons in the insulating parent compounds \cite{SidisPailhes04}. The odd and even channels
can
be separated in an INS experiment by their differences in $q_z$ dependence, as the total magnetic response
$\chi''(\mathbf{q},\,\mathit{\Omega})$ is given by
\begin{equation}
   \chi''(q_x,\,q_y,\,q_z,\,\omega)=\chi''_\text{o}(q_x,\,q_y,\,\omega)\,\mathrm{sin}^2(q_zd/2)+\chi''_\text{e}(q_x,\,q_y,\,\omega)\,\mathrm{cos}^2(q_zd/2)\text{,}
\end{equation}
where $d$ is the distance between the nearest CuO$_2$ planes along the $c$ axis \cite{WooDai06}.

For a long time, the magnetic resonance peak was observed only in the odd channel, where it is stronger. However, the
resonance in the even channel was subsequently found both in Y123 \cite{PailhesSidis03, PailhesSidis04, WooDai06,
PailhesUlrich06} at $\sim$\,43\,meV and in Bi2212 \cite{CapognaFauque07} at $\sim$\,55\,meV (both values correspond to
nearly optimal doping). These results are summarized in Fig.\,\ref{Fig:NeutronResonances}.


\section{Dynamic spin susceptibility of Bi$_2$Sr$_2$CaCu$_2$O$_{8+\delta}$ in the random phase
approximation}\label{Sec:BSCCO_RPA}

\subsection{Localized versus itinerant electron models}\label{SubSec:LocalizedItinerant}

The localized and itinerant electron models \cite[p.\,4]{Moriya85} have diametrically opposed starting points. The
former starts with the electronic states localized in the real space, while the latter starts with those localized in
the reciprocal or wave-vector space. Since early theories of magnetism based on these two mutually opposite models had
complementary merits and demerits, famous controversies over these two models have lasted for quite a long time.

The \textit{localized electron approach}, described by the Heisenberg (\ref{Eq:HeisenbergHamiltonian}) or $t$-$J$
(\ref{Eq:tJHamiltonian}) Hamiltonians, is justified from the microscopic point of view when well-defined local atomic
moments exist. This is established to be the case in magnetic insulator compounds \cite{Anderson59}, where the
electrons
are localized owing to the Mott mechanism of strong correlations \cite{Mott49}, and in the majority of rare-earth
materials. The magnetic susceptibility within the Heisenberg model is given by
\begin{equation}
\chi=Ng^2\mu_\text{B}^2\,S(S+1)/3k_\text{B}(T-T_\text{C})\text{,}
\end{equation}
where $N$ is the number of atoms in the crystal, $g$ is the gyromagnetic ratio ($g$-factor), $S$ is spin,
$T_\text{C}=2\sum_jJ_{ij}\,S(S+1)/3k_\text{B}$ is the Curie temperature, and $J_{ij}$ are the interatomic exchange
interaction constants.

On the other hand, the \textit{itinerant electron approach} has been motivated by the progress in the band
calculations
and by the experimental observation of d-electron Fermi surfaces in magnetic transition metals. The itinerant electron
theory of ferromagnetism has first been developed in the works of Bloch \cite{Bloch29}, Wigner \cite{Wigner34}, Slater
\cite{Slater36}, and Stoner \cite{Stoner38}. The Stoner model considers spin flip excitations of the electrons across
the Fermi surface, or equivalently the excitations of electron-hole pairs with opposite spins (so-called \textit{spin
fluctuations}). These are exactly the excitations described by RPA theory (see section
\ref{Sec:TwoParticleCorrelation}). The itinerant susceptibility is therefore given by (\ref{Eq:HiItinerant}).

It should be emphasized that although itinerant and localized models approach the electron subsystem from two
different
points of view, originating in the reciprocal and real space respectively, they are originally equivalent and simply
correspond to a change of basis (e.g. from Bloch waves to Wannier functions). If the series (\ref{Eq:FeynSeriesHi})
could be summed up exactly, it would provide an exact expression for the correlation function of the electron
subsystem
independently of the nature of the electronic states, in contrast to a popular belief that the itinerant approach is
not
applicable to electronic states localized in real space. In fact, for a non-interacting electron system the
correlation
function is exactly given by the Lindhard function, independently of the localization of electrons in real space. As
the
interactions are included, the other terms in (\ref{Eq:FeynSeriesHi}) become important, and for a strongly interacting
system the series is expected to converge more slowly, which ultimately makes RPA a bad approximation, though it is
difficult to estimate rigorously how strong the interactions should be for this to happen.

In this section we will calculate the itinerant magnetic spectrum of nearly optimally doped
Bi$_2$Sr$_2$CaCu$_2$O$_{8+\delta}$ in the RPA approximation. Even though high-$T_\text{c}$ superconductors are
generally
considered to have highly correlated electronic subsystems, the spin susceptibility estimated in the itinerant model
is
in reasonable agreement with the INS data, which might mean that the RPA approximation still holds in the vicinity of
the optimal doping in these materials.

\subsection{Dilemma of the magnetic resonance origin in cuprates}

The origin of the magnetic resonance structure observed in the superconducting state of Y123 \cite{RossatMignod91,
FongBourges00, DaiMook01, PailhesSidis03, HaydenMook04, HinkovPailhes04, PailhesSidis04, PailhesUlrich06, WooDai06},
Bi2212 \cite{FongBourges99, HeSidis01, CapognaFauque07, FauqueSidis07}, and other families of cuprates
\cite{HeBourges02, TranquadaWoo04, VignolleHayden07} is one of the most controversial topics in today's
high-$T_\text{c}$ superconductor physics. Existing theories waver between the itinerant magnetism resulting from the
fermiology \cite{FongKeimer95, LiuZhaLevin95, BrinckmannLee99, AbanovChubukov99, KaoSi00, Norman00, ManskeEremin01,
Norman01, ChubukovJanko01, OnufrievaPfeuty02, EreminMorr05, EreminManske05, EreminMorr07} and the local spins pictures
(such as static
and fluctuating ``stripes'', coupled spin ladders, or spiral spin phase models) \cite{TranquadaWoo04, VojtaUlbricht04,
UhrigSchmidt04, KrugerScheidl04, Tranquada05, SeiboldLorenzana05, VojtaVojta06, ReznikIsmer06}, as it appears that
both
approaches can qualitatively reproduce the main features of the magnetic spectra in the neighborhood of the optimal
doping. It is a long standing question, which one of these two components (itinerant or localized) predominantly forms
the integral intensity and the momentum-dependence of the magnetic resonances. It is therefore essential to estimate
their contribution quantitatively, carefully taking into account all the information about the electronic structure
available from experiment. However, such a comparison, which could shed light on the dilemma, is complicated, as it
requires high-quality INS data and the extensive knowledge of the electronic structure for the same family of
cuprates.
On the other hand, APRES data for Y-based compounds, for which the best INS spectra are available, are complicated by
the surface effects \cite{ZabolotnyyOD07}, while for the Bi-based cuprates, most easily measured by surface-sensitive
techniques such as ARPES, the INS measurements show much lower resolution due to small crystal sizes. Only thanks to
the
recent progress in INS on Bi2212 discussed above, the direct comparison of theory and experiment finally became
possible.

Before we turn to our calculation, let us briefly compare some of the available results from both ``local'' and
``itinerant'' camps to pinpoint the differences between the two approaches. The localized theories have been
successful
in explaining the data on LSCO and LBCO utilizing the models of static stripes and coupled spin ladders
\cite{TranquadaWoo04, UhrigSchmidt04, VojtaUlbricht04, SeiboldLorenzana05}, reproducing the universal ``hourglass''
dispersion of magnetic excitations [Fig.\,\ref{Fig:LocalModels}\,(a)\,--\,(c)]. To explain the neutron scattering data
on YBCO within the stripe picture, where no static order has been observed, one has to introduce the notion of
\textit{fluctuating stripes} \cite{Sachdev03, KivelsonBindloss03, Hasselmann99, VojtaVojta06}, i.e. locally
one-dimensional magnetic order that fluctuates both in space and time [Fig.\,\ref{Fig:LocalModels}\,(d)]. On the one
hand, this model leads to spin excitations very similar to those observed in experiments. On the other hand, the
RPA-type (itinerant) models have been equally successful in reproducing the same data \cite{ManskeEremin01,
EreminMorr05} [Fig.\,\ref{Fig:LocalModels}\,(e)\,--\,(g)]. Although RPA-based models tend to produce dispersions that
are closer to W-shape than to the ``hourglass'', the distinction between the two shapes in the experimental data is
still ambiguous. At first, it seemed that the ``smoking gun'' experiment to distinguish between the two models would be
the
observation of one-dimensionality in the INS signal from an array of detwinned YBCO crystals \cite{HinkovPailhes04},
but
it turned out that it can be explained within itinerant models as well, if one accounts for orthorhombicity
\cite{EreminManske05,
EreminManske06, SchnyderManske06, ZhaoLi07, EreminManske07}. Recently, the itinerant models have been developed to the
point where they can also qualitatively reproduce the effects of bilayer splitting (acoustic and optical resonances)
\cite{EreminMorr07}, and resonances in the electron-doped cuprates \cite{IsmerEremin07}.

\hvFloat[floatPos=t, capWidth=1.0, capPos=r, capVPos=t, objectAngle=0]{figure}
        {\includegraphics[width=0.65\textwidth]{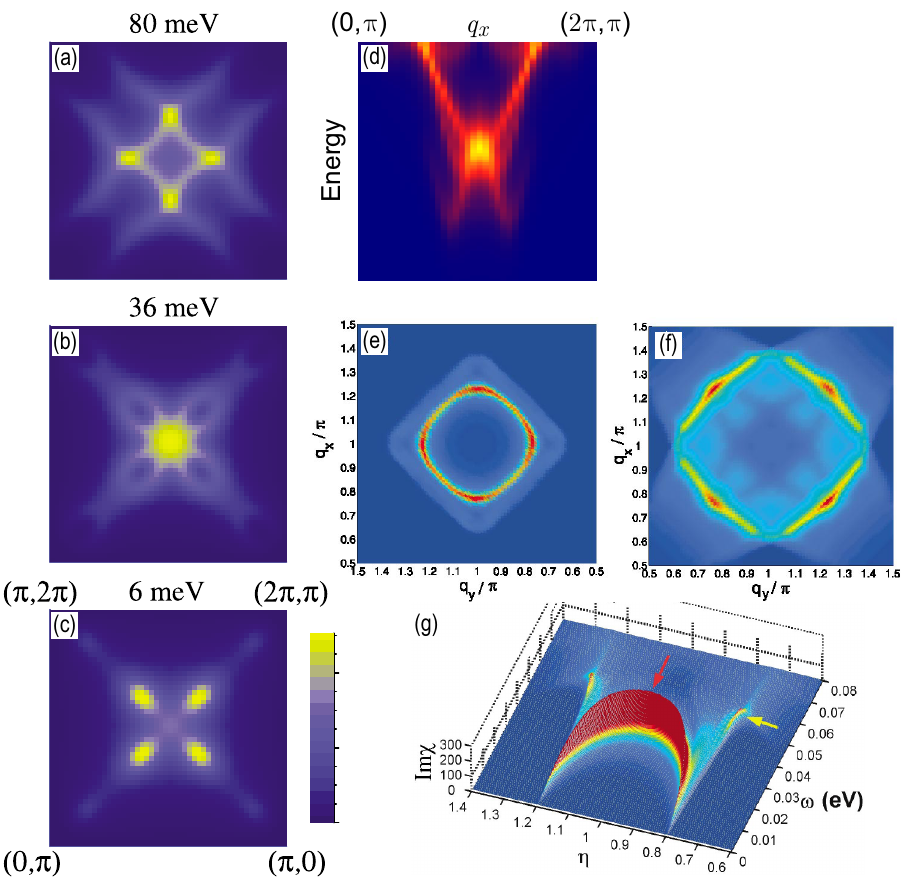}\hspace{-7.5em}}{Local and itinerant models of spin
        fluctuations. \textbf{(a)\,--\,(c)}~Constant energy cuts in the bond-centered stripes model
        \cite{VojtaUlbricht04} (cf.\,Fig.\,\ref{Fig:Incommensurate}).\,\textbf{(d)} Dispersion along $(q_x,\piup)$ in
        the fluctuating stripes model \cite{VojtaVojta06}. \textbf{(e)\,--\,(g)} Constant energy cuts and dispersion
        in
        an RPA (itinerant) model \cite{EreminMorr05}.}{Fig:LocalModels}

At this point, only an accurate \textit{quantitative} comparison of the local and itinerant models can prove helpful in
defining the applicability ranges of the two opposed alternatives. Unfortunately, this is complicated, as the
above-mentioned theories are to a great extent phenomenological and neither of them can provide quantitative results
that can be directly compared to experiment. In the RPA, the problem is that the results of the calculations are
sensitive to the fine details of the electronic structure and renormalization effects, which up to now have not been
fully accounted for. In the following paragraph, we will address these problems in detail and will show how ARPES can
provide the necessary data for the quantitative comparison. Starting from the experimentally measured single-particle
spectra, we will calculate the RPA susceptibility that accurately accounts for the renormalization effects and bilayer
splitting, which we will compare to the recent INS measurements.

\subsection{Establishing the relation between ARPES and INS experiments}

Here we will establish an important relation between the single-particle spectral function measured by ARPES and the
INS
response. The dynamic spin susceptibility in the odd (o) and even (e) channels can be estimated within the random
phase
approximation from the single-particle spectral function, including many-body effects \cite{InosovBorisenko07}. As
already shown in section \ref{Sec:TwoParticleCorrelation}, the two-particle correlation function measured by INS
spectroscopy can be calculated within RPA from the single-particle Green's function. In the normal state, the
irreducible part of the susceptibility function is given by Eq.\,(\ref{Eq:RenormHi0}), while in the superconducting
state an additional term appears due to the anomalous Green's function, resulting in Eq.\,(\ref{Eq:RenormHi0Nambu}).
The
bilayer splitting leads to an additional summation over the four possible pairs of sub-bands, of which bonding-bonding
and antibonding-antibonding terms contribute only to the odd susceptibility component, while the two
bonding-antibonding
terms contribute only to the even component \cite{EschrigNorman02}. The normal-state Lindhard function can be
therefore
related to the quasiparticle Green's function via the following Matsubara frequency summation \cite{BruusFlensberg81,
MonthouxPines94}:
\begin{equation}\label{Eq:Hi0MatsubaraBilayer}
\chi_0(\textbf{Q},\mathrm{i}\mathit{\Omega}_n)=\frac{\text{1}}{2\piup^{2}}\sum_{\lower1em\hbox{$\stackrel{\scriptstyle
i=j\,\textup{(o)}}{\scriptstyle i\neq j\,\textup{(e)}}$}}\sum_m\,\int\!
G_i(\textbf{k},\mathrm{i}\omega_m)\,G_j(\textbf{k}+\textbf{Q},\mathrm{i}\omega_m+\mathrm{i}\mathit{\Omega}_n)\,\mathrm{d}\textbf{k}\text{,}
\vspace{-0.5em}
\end{equation}
where indices $i$ and $j$ numerate the bonding and antibonding bands. Rewriting the Matsubara summation as a double
integral along the real energy axis \cite{DahmTewordt95, EschrigNorman03}, we obtain:
\begin{equation}\label{Eq:Hi0RenormBilayer}
\chi_0^{\textup{o},\kern.5pt\textup{e}}(\textbf{Q},\mathit{\Omega})=\sum_{\lower1em\hbox{$\stackrel{\scriptstyle
i=j\,\textup{(o)}}{\scriptstyle i\neq j\,\textup{(e)}}$}}\iint_{-\infty}^{+\infty}\kern-3pt
C_{ij}(\textbf{Q},\epsilon,\nu)~
\frac{n_\textup{F}\text{(}\nu\text{)}-n_\textup{F}\text{(}\epsilon\text{)}}
{\mathit{\Omega}+\nu-\epsilon+\mathrm{i}\mathit{\Gamma}}~\mathrm{d}\nu\kern.5pt\mathrm{d}\epsilon\text{,}
\vspace{-0.5em}
\end{equation}
where
$C_{ij}(\textbf{Q},\epsilon,\nu)=\frac{1}{2\piup^2}\textstyle{\int}A_i(\textbf{k},\epsilon)\,A_j(\textbf{k}+\textbf{Q},\nu)\,\mathrm{d}\textbf{k}$
is the cross-correlation of every pair of constant-energy cuts of the spectral function over the Brillouin zone that
can
be efficiently calculated in the Fourier domain by means of the cross-correlation theorem \cite{Papoulis62}.

In the superconducting state, the anomalous Green's function $F(\textbf{k},\epsilon)$ additionally contributes to
$\chi_0$ \cite{AbanovChubukov99}:
\begin{equation}
C_{ij}(\textbf{Q},\epsilon,\nu)=\frac{1}{2\piup^4}
\kern-2pt\int\bigl[\mathrm{Im}\kern.2pt G_i(\textbf{k},\epsilon)\,\mathrm{Im}\kern.2pt G_j(\textbf{k}+\textbf{Q},\nu)
+\mathrm{Im}\kern.2pt F_i(\textbf{k},\epsilon)\,\mathrm{Im}\kern.2pt
F_j(\textbf{k}+\textbf{Q},\nu)\bigr]\mathrm{d}\textbf{k}
\end{equation}
Although $\mathrm{Im}\kern.2pt F$ is not directly measured by ARPES, it can be modeled using the self-energy as
described earlier in \S\ref{SubSec:AnomalousG}.

After one knows the Lindhard function $\chi_0$ (frequently referred to as the bare spin susceptibility), one can
finally
obtain from RPA the dynamic spin susceptibility $\chi$ \cite{LiuZhaLevin95}, the imaginary part of which is directly
proportional to the measured INS intensity \cite{Tranquada05}:
\begin{equation}\label{Eq:RPAoe}
\chi^{\text{o},\kern.5pt\text{e}}(\textbf{Q},\mathit{\Omega}) =
{\chi_0^{\text{o},\text{e}}(\textbf{Q},\mathit{\Omega})}\Big/{\bigl[1-J_Q^{\,\text{o},\text{e}}\,\chi_0^{\text{o},\text{e}}(\textbf{Q},\mathit{\Omega})\bigr]}
\end{equation}

The coefficient $J_Q^{\,\text{o},\text{e}}$ in the denominator of (\ref{Eq:RPAoe}) describes the effective Hubbard
interaction. In our calculations we employed the model for $J_Q^{\,\textup{o},\kern.5pt\textup{e}}$ discussed in
Ref.~\citenum{EreminMorr07} and \citenum{BrinckmannLee01}, namely:\vspace{-0.5em}
\begin{equation}
J_Q^{\,\textup{o},\kern.5pt\textup{e}}=-J_\|(\mathrm{cos}\,{Q_x}+\mathrm{cos}\,{Q_y})\pm J_\perp,
\end{equation}
where the first term accounts for the momentum dependence due to the in-plane nearest-neighbor superexchange, and the
second term arises from the out-of-plane exchange interaction.

Thus, knowing the single-particle Green's function leads us to a comparison of ARPES results with the INS data. The
previous calculations based on this idea \cite{AbanovChubukov99, EreminMorr05, EreminMorr07} were performed for the
bare
band structure only, disregarding the renormalization effects, which makes the conclusions based on comparison with
the
INS data rather uncertain. The recent work by U.\,Chatterjee \textit{et~al.} \cite{ChatterjeeMorr07} is the only
available paper that includes the many-body effects from experimental data (in a procedure different from ours), but
it
does not account for the bilayer splitting (necessary for reproducing the odd and even INS channels), provides the
results in arbitrary units only, rather than on an absolute scale, and gives only an estimate for the anomalous
contribution to $\chi\kern-.5pt$. So we will address these issues in more detail below.

In section \ref{Sec:ModelingGreensFunction} we have introduced an analytical model that can reproduce the ARPES
measurements within a wide energy range and all over the Brillouin zone. As in a single experiment it is practically
impossible to obtain a complete data set of ARPES spectra, such a model allows making use of all the available data
measured from a particular sample and calculating the full 3D data set afterwards. In such a way the effect of matrix
elements and experimental resolution is also excluded. We employed this model of the Green's function based on the
bare
electron dispersion and self-energy extracted from ARPES data \cite{InosovBorisenko07} to compute both normal and
anomalous Green's functions and consequently calculate the dynamic spin susceptibility (\ref{Eq:RPAoe}).

\hvFloat[floatPos=t, capWidth=1.0, capPos=b, capVPos=t, objectAngle=0]{figure}
        {\includegraphics[width=0.7\textwidth]{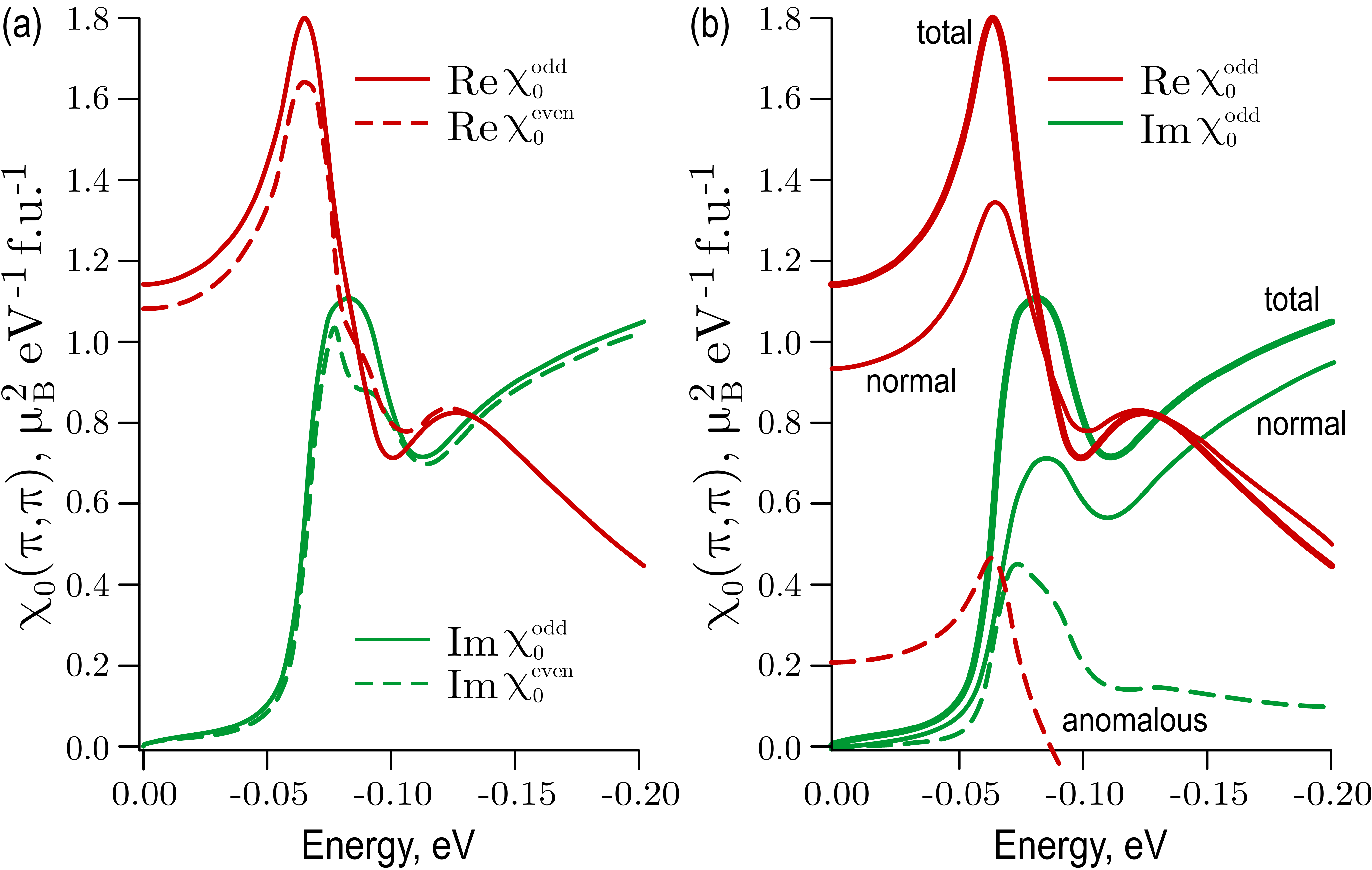}}
        {\textbf{(a)}~Energy dependence of the real and imaginary parts of the Lindhard function $\chi_0$ at the
        $(\piup,\piup)$ point for the odd and even channels. \textbf{(b)}~Contributions of the normal (thin solid
        curves) and anomalous (dashed curves) components to the real and imaginary parts of $\chi_0^\textup{odd}$ in
        the
        superconducting state. The sum of two components is shown as thicker curves. In our calculations we used the
        $\mathit{\Gamma}$ value in (\ref{Eq:Hi0RenormBilayer}) of 5~meV, which could introduce insignificant
        additional
        broadening of $\chi_0$ as compared to the bare band calculations. The energy integration range in
        (\ref{Eq:Hi0RenormBilayer}) was chosen to be $\pm$0.25~eV. The figure is reproduced from
        Ref.~\citenum{InosovBorisenko07}.}{Fig:Chi0}

Starting from the model data set built for optimally doped Bi2212 at 30\,K, with the maximal superconducting gap of
35~meV, we have calculated the Lindhard function [Eq.\,(\ref{Eq:Hi0RenormBilayer})] in the energy range of
$\pm$0.25~eV
in the whole Brillouin zone for the odd and even channels of the spin response [see Fig.\,\ref{Fig:Chi0}\,(a)]. To
demonstrate that the contribution of the anomalous Green's function is not negligibly small, in
Fig.\,\ref{Fig:Chi0}\,(b) the normal and anomalous components of $\chi_0^\textup{odd}$ are also shown.

\hvFloat[floatPos=t, capWidth=1.0, capPos=b, capVPos=t, objectAngle=0]{figure}
        {\includegraphics[width=0.7\textwidth]{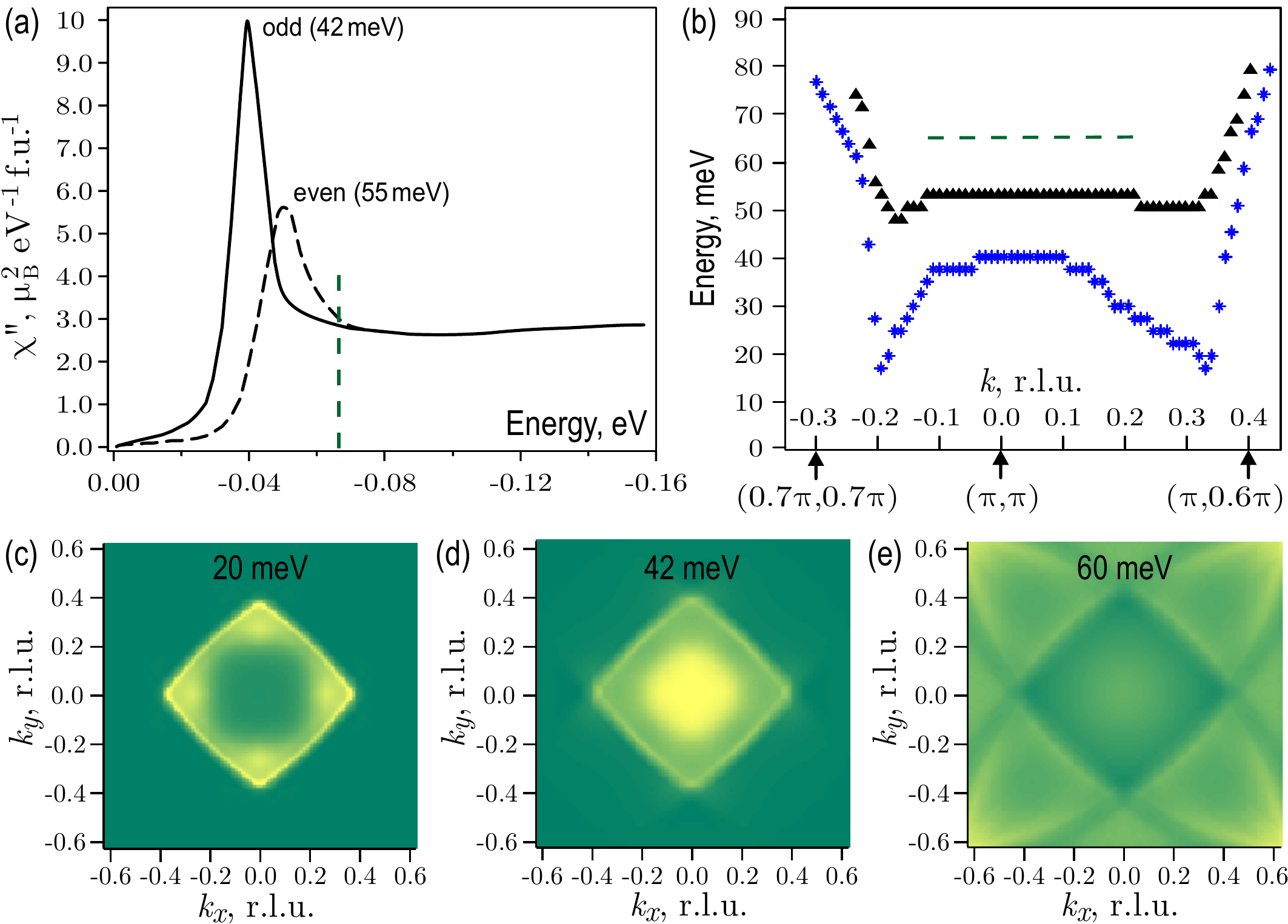}}
        {\textbf{(a)} $\textbf{k}$-integrated $\chi''\kern-0.3em=\kern-0.1em\mathrm{Im}\text{(}\chi\text{)}$ in the
        odd
        (solid curve) and even (dashed curve) channels. \textbf{(b)}~momentum dependence of the resonance energies in
        odd (\textcolor{DarkBlue}{$\times\kern-1.645ex$+}) and even (\raise0.1em\hbox{$\scriptstyle\blacktriangle$})
        channels along the high-symmetry directions $(0,0)$\,--\,$(\piup,\piup)$\,--\,$(\piup,0)$. The dashed lines
        mark
        the onset of the particle-hole continuum (position of the ``step'' in $\chi''_0$). \textbf{Second
        row:}~Constant
        energy cuts of $\chi''$ in the odd channel below the resonance (c), at the resonance energy (d), and above the
        resonance (e). The center of each Brillouin zone image corresponds to the $(\piup,\piup)$ point. The figure is
        reproduced from Ref.~\citenum{InosovBorisenko07}.}{Fig:Chi}

After that we calculated $\chi\kern-.5pt$ (Eq.\,\ref{Eq:RPAoe}) by adjusting the $J_\|$ and $J_\perp$ parameters to
obtain correct resonance energies at $(\piup,\piup)$ in the odd and even channels (42 and 55~meV respectively), as
seen
by INS in Bi2212 \cite{FongBourges99, HeSidis01, CapognaFauque07, FauqueSidis07}. The resulting
$\chi\kern-.5pt^\textup{o,\kern.5pt e}$ are qualitatively similar to those obtained for the bare Green's function
\cite{EreminMorr07}. The intensity of the resonance in the even channel is approximately two times lower than in the
odd
channel, which agrees with the experimental data \cite{PailhesUlrich06, CapognaFauque07} shown in
Fig.~\ref{Fig:NeutronResonances} (d). On the other hand, for $J_\perp=0$ the splitting between odd and even resonances
does not exceed 5\,--\,6~meV, which is two times less than the experimental value. This means that the out-of-plane
exchange interaction (in our case $J_\perp/J_\|\approx\text{0.09}$) is significant for the splitting and the
difference in $\chi_0$ alone between the two channels cannot fully account for the effect.

In Fig.\,\ref{Fig:Chi}\,(a) we show both resonances, momentum-integrated all over the BZ. Here we should pay attention to the absolute intensities of the resonances. A good estimate for the integral intensity in this case is the product of the peak amplitude and the full width at half maximum, which for the odd resonance results in 0.12~$\mu_\textup{B}^2/$f.u. in our case. This is in good agreement with the corresponding intensity in latest experimental spectra on YBCO ($\sim$\,0.11~$\mu_\textup{B}^2/$f.u.) \cite{WooDai06}.

As for the momentum dependence of $\chi\kern-.5pt$, Fig.\,\ref{Fig:Chi}\,(b) shows the dispersions of incommensurate
resonance peaks in both channels along the high-symmetry directions, calculated from the Green's function model with
the
self-energy derived from the ARPES data. We see the W-shaped dispersion similar to that seen by INS on Y123
\cite{HaydenMook04, Tranquada05} and to the one calculated previously by RPA for the bare Green's function
\cite{EreminMorr05, EreminMorr07}. At $(\piup,\,\piup)$ both resonances are well below the onset of the particle-hole
continuum at $\sim\,$65\,meV (dashed line), which also agrees with previous observations \cite{PailhesSidis04,
EreminMorr05, EreminMorr07}. At higher energies magnetic excitations are overdamped, so the upper branch of the
``hourglass'' near the resonance at $(\piup,\,\piup)$ suggested by some INS measurements \cite{HaydenMook04,
PailhesSidis04, Tranquada05} is too weak to be observed in the itinerant part of $\chi$ and is either not present in
Bi2212 or should originate from the localized spins.

In Fig.\,\ref{Fig:Chi} we additionally show three constant-energy cuts of $\chi$ in the odd channel below the
resonance,
at the resonance energy, and above the resonance. As one can see, besides the main resonance at $(\piup,\,\piup)$ the
calculated $\chi$ reproduces an additional incommensurate resonance structure, qualitatively similar to that observed
in
INS experiments \cite{HaydenMook04}. Below the resonance the intensity is concentrated along the $(k,\,0)$ and
$(0,\,k)$
directions, while above the resonance it prevails along the diagonal directions $(k,\pm k)$.


\section{Relating the FT-STS data to ARPES measurements}\label{Sec:FT-STS}

\subsection{Spatial modulations of the electronic density of states in the direct and reciprocal spaces.}

In this and the following sections, we will consider two more examples of how the model of the Green's function derived
from ARPES data can prove useful in bridging different experimental techniques and theories. We will start with the
problem of spatial inhomogeneities seen in the local density of states (LDOS) of high-$T_\text{c}$ cuprates by scanning
tunneling spectroscopy (STS) \cite{PanHudson00, PunONeal01, HudsonPan99, HudsonLang01, LangMadhavan02}, which has
attracted much attention of the scientific community because of its clear relation to the central problem of
high-$T_\text{c}$ superconductivity\,---\,the evolution of an antiferromagnetic insulator to a superconductor with
doping. Recent breakthrough in the development of STS technique to a level where the Fourier transformed (FT) STS
images have been crystallized into well-defined symmetric patterns \cite{HoffmanMcElroy02, McElroySimmonds03,
HowaldEisaki03, HanaguriLupien04, VershininMisra04} has revealed the existence of regular inhomogeneities which could
relate the high temperature superconductivity problem to self-ordering phenomena in correlated electron systems.

\hvFloat[floatPos=b, capWidth=1.0, capPos=r, capVPos=t, objectAngle=0]{figure}
        {\includegraphics[width=0.58\textwidth]{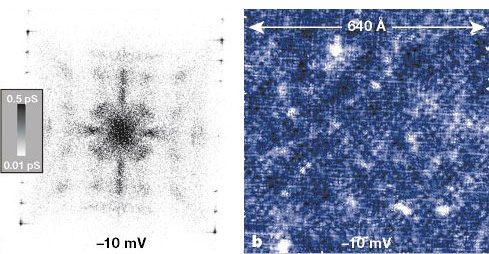}\quad}
        {An FT-STS image at 10\,meV bias voltage (left) obtained from the spatially resolved $\mathrm{d}I/\mathrm{d}V$
        map of Bi2212 (right). The figure is adapted from Ref.~\citenum{McElroySimmonds03}.}{Fig:FT-STS}

The local density of states $g_\omega\text{(}\mathbf{r}\text{)}$ is obtained in STS by spatial mapping of the
tip-sample differential tunneling conductance ($\mathrm{d}I/\mathrm{d}V$) as a function of real-space variable
$\mathbf{r}$ at each bias voltage $V=\omega/e$. An FT-STS image for a particular energy $\omega$ is then obtained by
the Fourier transform of the corresponding LDOS:
$\mathcal{F}_\mathbf{r}[g_\omega\text{(}\mathbf{r}\text{)}]\text{(}\mathbf{q}\text{)}$, where $\mathcal{F}_\mathbf{r}$
is the Fourier transform operator with the integration variable $\mathbf{r}$. Fig.\,\ref{Fig:FT-STS}
\cite{McElroySimmonds03} gives an example of the 10\,meV FT-STS image and the corresponding LDOS map measured from a
Bi2212 sample. As one sees, the seemingly random inhomogeneities of the LDOS in real space reveal a regular pattern of
peaks in the spatial frequency domain.

The most common explanation for the regular patterns seen in the FT-STS images is given by the so-called
\textit{impurity scattering hypothesis}, which identifies the inhomogeneities with the quasiparticle interference
mediated by the impurity scattering \cite{HoffmanMcElroy02, McElroySimmonds03, HowaldEisaki03, HanaguriLupien04,
VershininMisra04, Markiewicz04, WangLee03}. In the framework of this hypothesis, the FT-STS function,
$\mathcal{F}_\mathbf{r}[g_\omega\text{(}\mathbf{r}\text{)}]\text{(}\mathbf{q}\text{)}$, is proportional to the joint
density of states $C_\omega\text{(}\mathbf{q}\text{)}$ defined as the autocorrelation of the single-particle spectral
function $A_\omega\text{(}\mathbf{k}\text{)}$:
\begin{equation}\label{Eq:Autocorrelation}
\mathcal{F}_\mathbf{r}[g_\omega\text{(}\mathbf{r}\text{)}]\text{(}\mathbf{q}\text{)} =
C_\omega\text{(}\mathbf{q}\text{)} = [\star_\mathbf{k}\,
A_\omega\text{(}\mathbf{k}\text{)}]\text{(}\mathbf{q}\text{)}\text{.}
\end{equation}
Here $[\star_\mathbf{k}\, A_\omega\text{(}\mathbf{k}\text{)}]\text{(}\mathbf{q}\text{)}\stackrel{\text{def}}{=}\int\!
A_\omega\text{(}\mathbf{k+q}\text{)}A_\omega\text{(}\mathbf{k}\text{)}\,\mathrm{d}\mathbf{k}$ denotes the
autocorrelation of the spectral function $A_\omega\text{(}\mathbf{k}\text{)}\equiv
A\text{(}\mathbf{k},\,\omega\text{)}$ taken over the momentum variable $\mathbf{k}$ at some fixed value of $\omega$.

\subsection{Can one see the Fermi surface with STS?}

The\,impurity\,scattering\,hypothesis\,provides\,a\,link\,between\,the\,joint $\mathbf{q}$-space\,of\,FT-STS\,and the
reciprocal $\mathbf{k}$-space where the quasiparticle spectral function $A_\omega\text{(}\mathbf{k}\text{)}$ is known
from ARPES with a tremendous accuracy \cite{Damascelli03}. Formula (\ref{Eq:Autocorrelation}) is an attractive
possibility to compare the FT-STS images to ARPES data. Here two approaches are possible. First, knowing the spectral
function $A\text{(}\mathbf{k}\text{)}$, one can straightforwardly calculate the left part of (\ref{Eq:Autocorrelation})
and compare the result with the STS measurements. Second, one can try to solve the inverse problem\,---\,to restore the
spectral function from its autocorrelation measured experimentally by STS. In Ref.\,\citenum{KordyukZabolotnyy07}, we
make the first step in this direction and propose a procedure to uniquely recover $A_\omega\text{(}\mathbf{k}\text{)}$
from $g_\omega\text{(}\mathbf{r}\text{)}$, therefore shifting the problem of FT-STS to the ARPES domain.

For a better understanding of the problem, let us rewrite Eq.\,(\ref{Eq:Autocorrelation}) using the Wiener-Khinchin
theorem \cite{WienerKhinchin}, which states that the autocorrelation is simply given by the inverse Fourier transform
of the function's Fourier amplitude squared:
\begin{equation}\label{Eq:WienerKhinchin}
[\star_\mathbf{k}\,A_\omega\text{(}\mathbf{k}\text{)}]\text{(}\mathbf{q}\text{)}=\Bigl[\!\mathcal{F}^{-1}_\mathbf{r}\bigl|[\mathcal{F}_\mathbf{k}\,A_\omega\text{(}\mathbf{k}\text{)}]\text{(}\mathbf{r}\text{)}\bigr|^2\Bigr]\!\text{(}\mathbf{q}\text{)}
\end{equation}
In other words, the autocorrelation operator preserves information about the function's Fourier amplitude, but loses
information about its phase. The Wiener-Khinchin theorem provides a convenient way to calculate the autocorrelation
using fast Fourier transform, which is much more efficient than straightforward integration. More important, it
demonstrates the irreversibility of the autocorrelation procedure in the general case, as all complex functions that
have the same Fourier amplitude (but different phase) will yield the same autocorrelation. Luckily, this
irreversibility can be overcome if one knows additional information about the function. For example, if the function is
real-valued and possesses particular symmetries, the Fourier phase can be restored from the amplitude using iterative
numerical algorithms. This problem is well known from applied optics, where a number of \textit{phase retrieval
algorithms} were developed \cite{Fienup82, IdellFienup87, MiaoSayre98}. These algorithms involve iterative Fourier
transformation back and forth between the object and Fourier domains with application of the known constraints.

\begin{figure}[t]
\center\includegraphics[width=\textwidth]{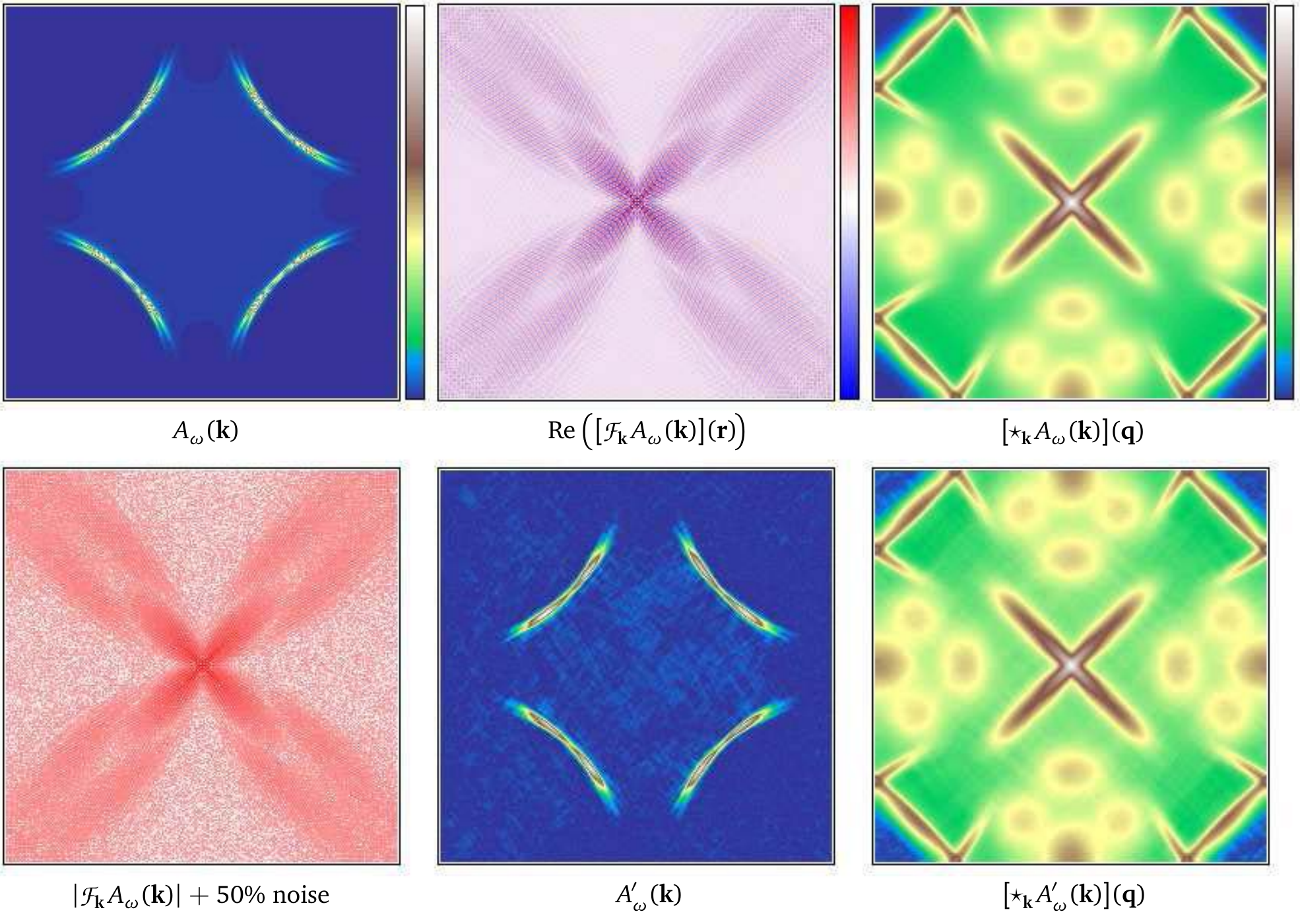}%
\caption{\textbf{Top row:} modeled spectral function $A_\omega\text{(}\mathbf{k}\text{)}$ of an optimally doped
superconducting CuO-bilayer at $\omega=\text{20}$\,meV binding energy (left), the real part of its Fourier image
$\mathrm{Re}\,\bigl([\mathcal{F}_\mathbf{k}\,A_\omega\text{(}\mathbf{k}\text{)}]\text{(}\mathbf{r}\text{)}\!\bigr)$
(middle), and the result of its autocorrelation $[\star_\mathbf{k}\,
A_\omega\text{(}\mathbf{k}\text{)}]\text{(}\mathbf{q}\text{)}$ (right).
\textbf{Bottom row:} to demonstrate the stability of the phase retrieval algorithm to the experimental uncertainty, a
noise has been added to $|\mathcal{F}_\mathbf{k}\,A_\omega\text{(}\mathbf{k}\text{)}|$ (left). The restored spectral
function $A'_\omega\text{(}\mathbf{k}\text{)}$ (middle) and its autocorrelation (right) are also
shown.}\label{Fig:Autocorrelation}
\end{figure}

Fig.\,\ref{Fig:Autocorrelation} (top row) shows a model of the spectral function $A_\omega\text{(}\mathbf{k}\text{)}$
of an optimally doped superconducting CuO-bilayer at $\omega=\text{20}$\,meV binding energy (left), the real part of
its Fourier image,
$\mathrm{Re}\,\bigl([\mathcal{F}_\mathbf{k}\,A_\omega\text{(}\mathbf{k}\text{)}]\text{(}\mathbf{r}\text{)}\bigr)$
(middle), and the result of its autocorrelation, $[\star_\mathbf{k}\,
A_\omega\text{(}\mathbf{k}\text{)}]\text{(}\mathbf{q}\text{)}$ (right). $A_\omega\text{(}\mathbf{k}\text{)}$ is
simulated from the experimentally determined tight-binding parameters of the bare band dispersion for an optimally
doped Bi2212 compound taking into account a d-wave superconducting gap $\mathit{\Delta} =
\mathit{\Delta}_0(\mathrm{cos}\,k_x - \mathrm{cos}\,k_y)/\text{2}$ with $\mathit{\Delta}_0 = \text{40}$\,meV. We note
that $\mathrm{Im}[\mathcal{F_\mathbf{k}}\,A_\omega(\mathbf{k})] = 0$ due to the even symmetry of the spectral function,
$A_\omega\text{(}\mathbf{k}\text{)}=A_\omega\text{(}\mathbf{-k}\text{)}$, so the inverse problem is reduced to
recovering a symmetric real-valued function from its experimentally measured Fourier amplitude
$R\text{(}\mathbf{r}\text{)}$. Such a procedure can be realized, in principle, but it can be potentially sensitive to
the finite resolution and unavoidable uncertainty of the experiment. The bottom row of Fig.\,\ref{Fig:Autocorrelation}
illustrates the robustness of the phase retrieval algorithm used in our work to experimental uncertainty:
$A'_\omega\text{(}\mathbf{k}\text{)}$ is recovered from a noisy
$|\mathcal{F}_\mathbf{k}\,A_\omega\text{(}\mathbf{k}\text{)}|$ (the additive noise was simulated from a Gaussian
distribution with the standard deviation equal to half of the average value of
$|\mathcal{F}_\mathbf{k}\,A_\omega\text{(}\mathbf{k}\text{)}|$).

The phase retrieval algorithms are essentially discrete. Their convergence to a unique solution, in case of positive
and spatially confined object, has been proved theoretically, while the stability to noise and speed of convergence are
the issues of continuous development \cite{Fienup82, IdellFienup87, MiaoSayre98}. The recovering of
$A_\omega\text{(}\mathbf{k}\text{)}$ from $|\mathcal{F}_\mathbf{k}\,A_\omega\text{(}\mathbf{k}\text{)}|$ by means of a
phase retrieval algorithm without noise and with the 50\% Gaussian noise is illustrated by two movies available online
\cite{AutocorrelationMovies}. We used a modified ``input-output'' algorithm \cite{Fienup82, IdellFienup87,
MiaoSayre98}, the $n$-th iteration of which can be formulated as follows:
\begin{subequations}
\begin{eqnarray}
& R_n\text{(}\mathbf{r}\text{)} &\kern-1.3ex=
[\mathcal{F}_\mathbf{k}\,A_{n}\text{(}\mathbf{k}\text{)}]\text{(}\mathbf{r}\text{)},\\
& \tilde{R}_n\text{(}\mathbf{r}\text{)} &\kern-1.3ex=
R\text{(}\mathbf{r}\text{)}\,\mathrm{exp}\bigl\{\mathrm{i}\cdot\mathrm{arg}[R_{n}\text{(}\mathbf{r}\text{)}]\bigr\},\\
& \tilde{A}_n\text{(}\mathbf{k}\text{)} &\kern-1.3ex=
[\mathcal{F}_\mathbf{r}^{-1}\tilde{R}_n\text{(}\mathbf{r}\text{)}]\text{(}\mathbf{k}\text{)},\\
& A_{n+1} &\kern-1.3ex= \Biggl\{\!\!
    \begin{array}{l}
    \mathrm{Re}\,[\kern.3pt\tilde{A}_n\text{(}\mathbf{k}\text{)}] \text{,\hspace{4.3em} if $\mathrm{Re}\,[\kern.3pt
    \tilde{A}_{n}\text{(}\mathbf{k}\text{)}] \ge 0$}, \\
    \raisebox{5pt}{\text{$\mathrm{Re}\,[\kern.3pt A_n\text{(}\mathbf{k}\text{)}-\beta
    \tilde{A}_n\text{(}\mathbf{k}\text{)}] \text{, if $\mathrm{Re}\,[\kern.3pt \tilde{A}_{n}\text{(}\mathbf{k}\text{)}]
    < 0$}$,}}
    \end{array}
\end{eqnarray}
\end{subequations}
where $R\text{(}\mathbf{r}\text{)}$ is the ``source'' function and $\beta$ is a constant which we choose between 1 and
2 to compromise between speed of convergence and stability of the algorithm. As an initial guess, we used a Gaussian
distribution with random noise: $A_0\text{(}\mathbf{k}\text{)} = \mathrm{exp}(|\mathbf{k}|^2/w^2)+\text{noise}$.

Finally, we discuss the existent attempts to compare the STS and ARPES data in the $\mathbf{q}$-domain
\cite{Markiewicz04, McElroyGweon06, ChatterjeeShi06}. It has been shown that the intensity maps measured by ARPES, when
autocorrelated, do not result in such distinct spots as those observed in FT-STS images \cite{Markiewicz04}. There is
still hope for better correspondence assuming better energy resolution in ARPES \cite{Markiewicz04}, or
$\mathbf{k}$-dependent matrix elements in STS \cite{McElroyGweon06}. We believe that both effects should be taken into
account together with the gapped and highly anisotropic quasiparticle self-energy, which could be a topic of future
research.
We are not aware of any earlier publications that would attempt to recover the spectral function from FT-STS data. If
the momentum resolution of FT-STS ever allows this, the direct comparison with the spectral function directly measured
by ARPES will possibly clarify such long standing problems as tunneling matrix elements \cite{HoogenboomBerthod03} and
inconsistency in the values of the quasiparticle self-energies determined from STS and ARPES experiments on high-$T_c$
cuprates \cite{HoogenboomBerthod03}.


\section{Probing the Peierls instability conditions by ARPES}

\subsection{Charge density waves and conditions for their formation.}

In 1955 Peierls \cite{Peierls55} suggested that in one-dimensional metals a spontaneous formation of periodic lattice
distortions (PLD) and charge density waves (CDW) can be energetically favorable under certain conditions. Since then,
CDW formation has been experimentally observed in many anisotropic compounds, such as transition metal chalcogenides
\cite{WilsonDiSalvo75, MonctonAxe77, GabovichVoitenko00, GabovichVoitenko01, GabovichVoitenko02}. This kind of symmetry
breaking, which happens upon cooling at a certain transition temperature $T_\text{CDW}$, is known as a Peierls phase
transition. The physical mechanisms of CDW formation are now well understood and are generally known to be determined in
particular by the Fermi surface geometry \cite{Peierls55, ChanHeine73, RiceScott75, Gruener94}, though some aspects of
CDW formation in two-dimensional metals, including its possible relation to the problem of high-$T_\text{c}$
superconductivity, are still actively discussed \cite{RiceScott75, McMillan77, EreminEremin97, Klemm00, Klemm00notes,
CastroNeto01, GabovichVoitenko00, GabovichVoitenko01, GabovichVoitenko02, CebulaZielinski01, BarnettPolkovnikov06,
VallaFedorov04, VallaFedorov06, JohannesMazin06, BorisenkoKordyuk08, EvtushinskyKordyuk08}.

In its simplest form, the instability condition for the formation of CDW/PLD in an electronic system can be written as
\cite{ChanHeine73}\vspace{-0.2em}
\begin{equation}\label{Eq:InstabilityCDW}
4\bar{\eta}_\mathbf{q}^2/\hbar\omega_\mathbf{q}-\text{2}\bar{U}_\mathbf{q}+\bar{V}_\mathbf{q}\geq
1/\chi_\mathbf{q}\text{,}\vspace{-0.3em}
\end{equation}
\mbox{where
$\bar{U}_\mathbf{q}\!=\!\langle\mathbf{k}+\mathbf{q}~\mathbf{k}'|\hat{U}|\mathbf{k}'\!+\mathbf{q}~\mathbf{k}\rangle$ and
$\bar{V}_\mathbf{q}\!=\!\langle\mathbf{k}+\mathbf{q}~\mathbf{k}'|\hat{V}|\mathbf{k}~\mathbf{k}'\!+\mathbf{q}\rangle$ are
the direct and exchange}
\mbox{Coulomb\,interactions\,in\,the\,local\,approximation\,\cite[p.\,115\,--\,192]{HalperinRice68},
$\bar{\eta}_\mathbf{q}$ is\,the\,local\,electron-} phonon interaction, and
$\chi_\mathbf{q}\!=\!\sum_\mathbf{k}[n_\text{F}(\epsilon_\mathbf{k})-n_\text{F}(\epsilon_{\mathbf{k}+\mathbf{q}})]/
(\epsilon_\mathbf{k}-\epsilon_{\mathbf{k}+\mathbf{q}})$ is the real part of the bare spin susceptibility (Lindhard
function) at $\omega\rightarrow0$, which can be successfully evaluated from the ARPES data \cite{InosovBorisenko07}.
Here by $n_\text{F}(\epsilon)=1/[\mathrm{exp}(\epsilon/k_\text{B}T)+1]$ we denote the Fermi-Dirac distribution function.
Note that the imaginary part of $\chi_\mathbf{q}$ vanishes in the static limit.

From Eq.\,(\ref{Eq:InstabilityCDW}) one sees that if the electron-phonon interaction is strong enough for the left part
of the inequality to be positive, a divergence or a strong peak in $\chi_\mathbf{q}$ at a particular wave vector
$\mathbf{q}$ would lead to the CDW/PLD instability. The phase transition would be then preceded by the softening of a
phonon mode, until it ``freezes'' at $T_\text{CDW}$, giving rise to the PLD with the same (or similar) wave vector
$\mathbf{q}$. Appearance of such a divergence in the static susceptibility we will call \textit{nesting}. In the
simplest scenario, such sharp peak will arise if the Fermi surface possesses parallel fragments such that many pairs of
electronic states can be connected by the same wave vector $\mathbf{q}$, which results in an enhancement of the
susceptibility at this vector.

\hvFloat[floatPos=b, capWidth=1.0, capPos=r, capVPos=t, objectAngle=0]{figure}
        {\raisebox{15.9em}{$\mathstrut$}\includegraphics[width=0.525\textwidth]{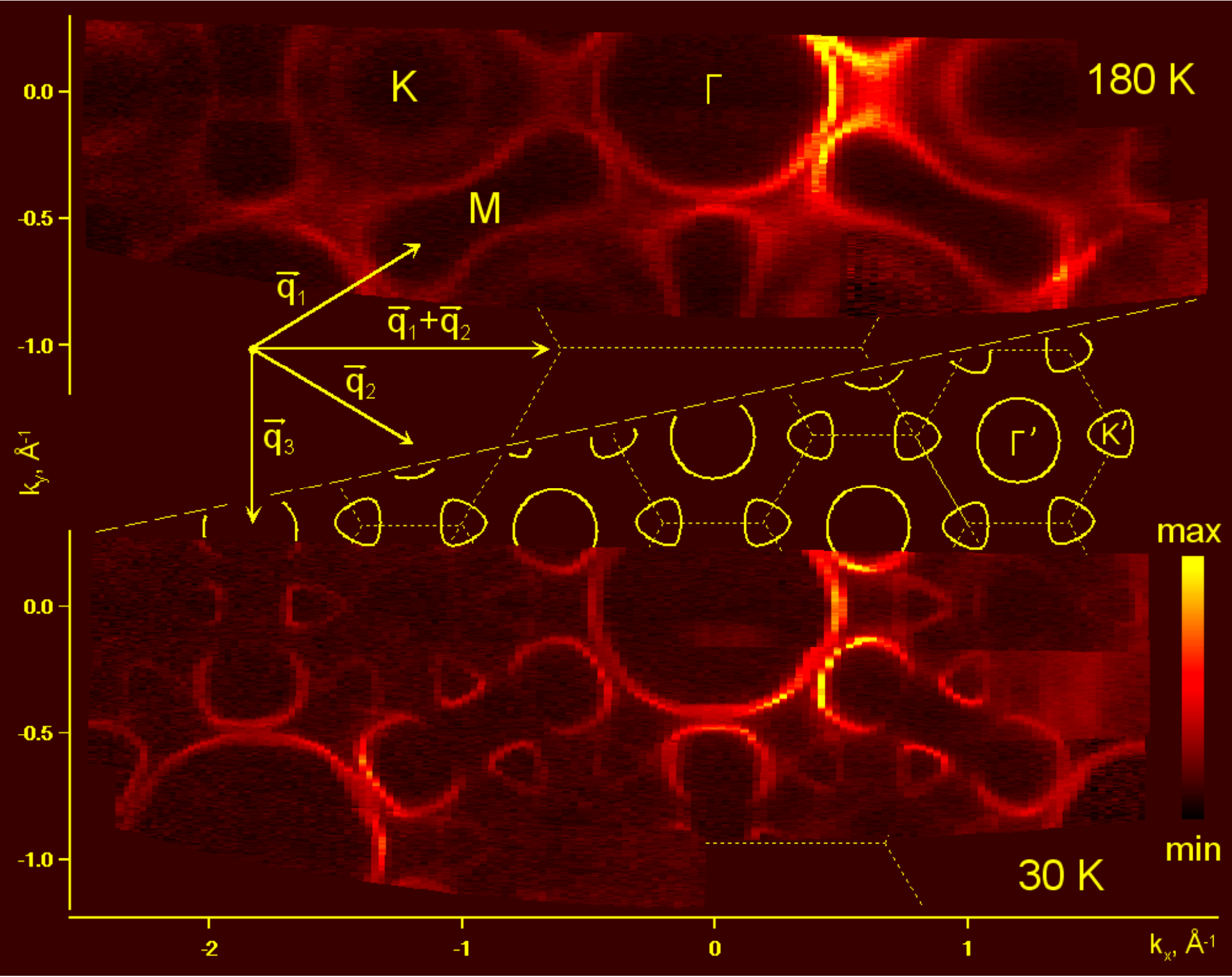}\quad}
        {Fermi surface maps of 2H-TaSe$_2$ at 180\,K (top) and 30\,K (bottom). Thin dotted lines are the Brillouin zone
        boundaries. At high temperatures the Fermi surface consists of two kinds of hole-like barrels centered at the
        $\mathrm{\Gamma}$ and K points and electron-like ``dogbones'' around the M-point. In the commensurate CDW
        state, the Fermi surface reconstructs as schematically shown in the middle of the figure, consisting of nearly
        circles around new $\mathrm{\Gamma}'$ points and rounded triangles around new $\mathrm{K}'$ points in the
        folded Brillouin zone. The figure is reproduced from Ref.~\citenum{BorisenkoKordyuk08}.}{Fig:TaSe2Maps}

It is a long-standing argument, however, whether such simple mechanism of Fermi surface instabilities underlies the CDW
formation in transition metal dichalcogenides, such as TaSe$_2$ and NbSe$_2$, which are the subject of this letter. In
some of the earlier studies the existence of necessary nesting conditions in transition metal chalcogenides was
questioned \cite{SmithKevan85, WithersWilson86, RossnagelRotenberg05, JohannesMazin06}, and some alternative mechanisms
of CDW instability were proposed \cite{RiceScott75, McMillan77, CastroNeto01, BarnettPolkovnikov06}. We find several
instability scenarios proposed in the literature: (i) simple Fermi surface nesting \cite{StraubFinteis99}, which in some
studies was considered too weak to be responsible for the instability \cite{RossnagelSeifarth01}, (ii) nesting of the
van Hove singularities (saddle points) \cite{RiceScott75, LiuOlson98, KissYokoya07}, and (iii) combination of the two:
partial nesting of the FS with the saddle band \cite{TonjesGreanya01}.

To clarify the role of simple nesting as the driving force of the CDW instabilities, we performed high resolution
measurements of several transition metal dichalcogenides using modern angle-resolved photoelectron spectroscopy, which
let us accurately determine the Fermi surface geometries and assess their nesting properties and their variations with
temperature. As will be shown in the following, our results not only support the Fermi surface nesting scenario of CDW
formation, but also reveal new aspects of nesting geometry: incommensurability of the nesting vector and its
universality among several transition metal dichalcogenides.

\enlargethispage{1em} As already demonstrated in section \ref{Sec:BSCCO_RPA}, the spin susceptibility can be calculated
from ARPES data, which we successfully did for the Bi2212 cuprate. Then we were interested in both real and imaginary
parts of the Lindhard function as functions of energy, and were accounting for the many-body effects. Here we will use a
much simpler version of the same technique to calculate the static susceptibility at $\omega=0$, neglecting the
renormalization effects, for a different layered compound\,---\,transition metal dichalcogenide 2H-TaSe$_2$.

\subsection{Nesting properties of 2H-TaSe$_2$ as a function of temperature.}

2H-TaSe$_2$ (trigonal prismatic tantalum diselenide) is a quasi-two-dimensional CDW-bearing material with two phase
transitions at accessible temperatures: a second-order incommensurate CDW transition at 122\,K and a first-order
commensurate 3$\times$3 CDW lock-in transition at 90\,K \cite{MonctonAxe77}. The temperature evolution of its Fermi
surface is presented in Fig.\,\ref{Fig:TaSe2Maps} \cite{BorisenkoKordyuk08}. The Fermi surface sheets originate from two
bands: one is responsible for the $\mathrm{\Gamma}$ and K barrels with a saddle point in between, the other one supports
the ``dogbone'' with another saddle point at M. The dispersions in the normal and incommensurate CDW states are
qualitatively similar \cite{LiuTonjes00, VallaFedorov00, RossnagelRotenberg05, BorisenkoKordyuk08}. In contrast, the
lock-in transition to the commensurate CDW state at 90\,K is clearly pronounced, resulting in a new folded Fermi surface
consisting of a set of nearly circles around new $\mathrm{\Gamma}$ points and rounded triangles around new $\mathrm{K}$
points \cite{BorisenkoKordyuk08}. The commensurate CDW vectors $q_n=\frac{2}{3}\kern.3pt\mathrm{\Gamma M}$ are well
known from experiments \cite{WilsonDiSalvo75, MonctonAxe77}.

\hvFloat[floatPos=t, capWidth=1.0, capPos=r, capVPos=t, objectAngle=0]{figure}
        {\includegraphics[width=\textwidth]{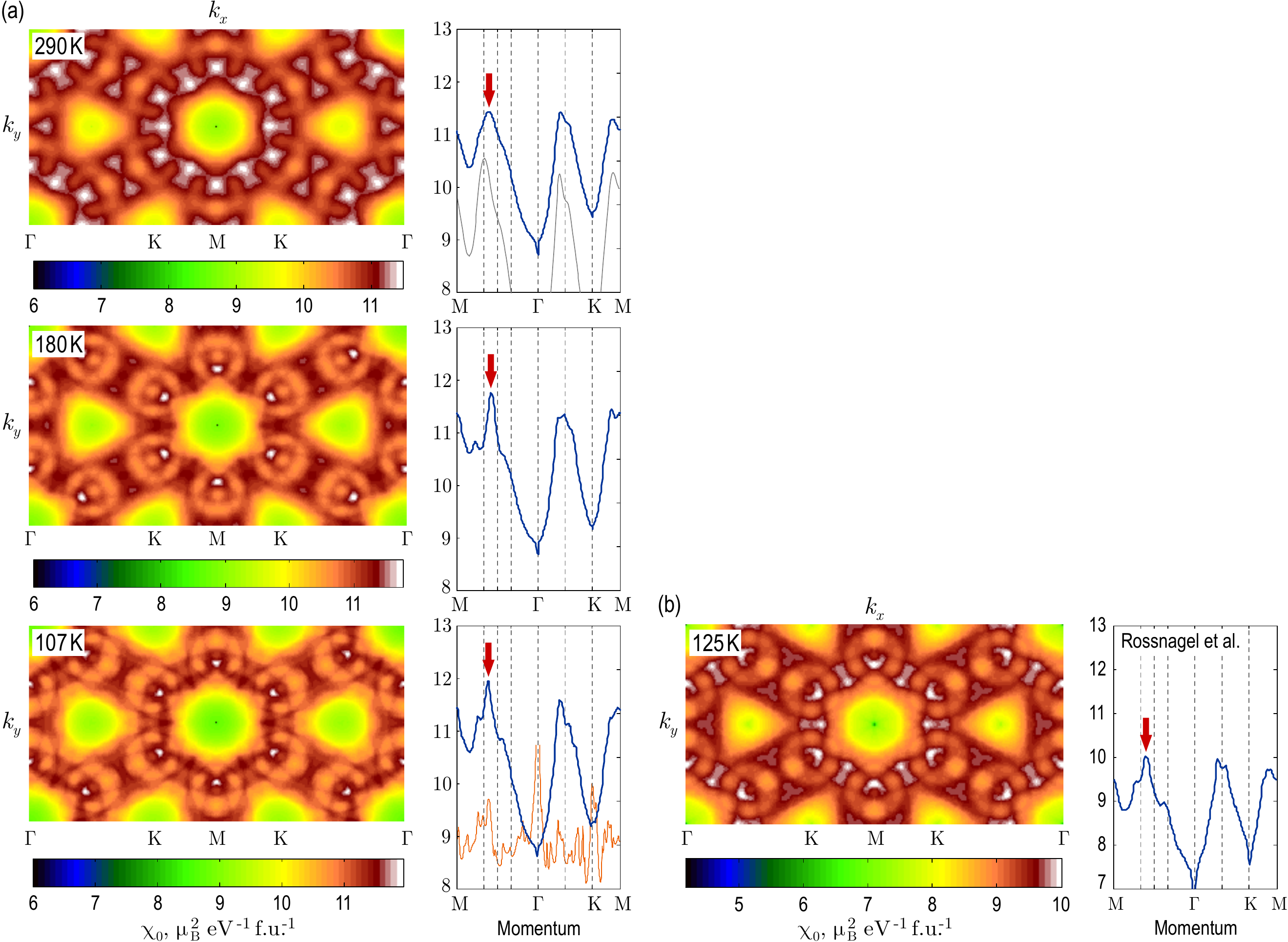}\hspace{-0.4795\textwidth}}
        {\textbf{(a)} Nesting properties of the 2H-TaSe$_2$ Fermi surface at different temperatures. Color plots show
        the real part of the Lindhard function in the static limit ($\omega\rightarrow0$) as a function of momentum.
        Corresponding profiles along high-symmetry directions are shown to the right of each panel. Red arrows mark the
        position of the peak in susceptibility close to the CDW wave vector. The thin orange curve in the bottom panel
        shows the autocorrelation of the Fermi surfaces (off scale). \textbf{(b)} The same plots for the tight-binding
        model of Rossnagel~\textit{et~al.}\,at 125\,K \cite{RossnagelRotenberg05} are in good agreement with our
        calculations.}{Fig:NestingTaSe2}

The question that we want to address here is whether the Fermi surface geometry observed in the normal state by ARPES
possesses the nesting properties that could explain the transition to the CDW state upon cooling. It is also interesting
to which extent the Fermi surface varies with temperature in the neighborhood of the incommensurate CDW transition. To
our knowledge, in earlier studies such minor variations of the Fermi surface could not be detected, and the dispersion
was considered unchanged down to 90\,K \cite{LiuTonjes00, VallaFedorov00, RossnagelRotenberg05}, until the Fermi surface
reconstruction due to the commensurate CDW lock-in transition finally occurred. On the other hand, we have recently
reported a noticeable variation of the distance between the M- and K-barrels (see Fig.\,4a in
Ref.\,\citenum{BorisenkoKordyuk08}), which gave us the motivation to study temperature variations of the Fermi surface
all over the momentum space in order to estimate their effect on the nesting properties.

We have fitted the experimental dispersion of 2H-TaSe$_2$ measured with high resolution at three different temperatures
using the following tight-binding expansion (momenta $k_x$ and $k_y$ enter the formula in dimensionless units):
\begin{multline}
\epsilon_\mathbf{k} = t_0 +
t_1\Bigl[2\,\mathrm{cos}\,\frac{k_x}{2}\,\mathrm{cos}\,\frac{\sqrt{3}\,k_y}{2}+\mathrm{cos}\,k_x\Bigr] +
t_2\Bigl[2\,\mathrm{cos}\,\frac{3k_x}{2}\,\mathrm{cos}\,\frac{\sqrt{3}\,k_y}{2}+\mathrm{cos}\,\sqrt{3}\,k_y\Bigr] \\+
t_3\, [\,2\,\mathrm{cos}\,k_x\,\mathrm{cos}\,\sqrt{3}\,k_y+\mathrm{cos}\,2k_x] +
t_4\,[\,2\,\mathrm{cos}\,3k_x\,\mathrm{cos}\,\sqrt{3}\,k_y+\mathrm{cos}\,2\sqrt{3}\,k_y].
\end{multline}
The tight-binding parameters were independent for the two bands, which resulted in the total of 10 fitting parameters.
The fitting was done by precisely measuring the relative Fermi momenta (distances between Fermi surface contours) and
Fermi velocities along several high-symmetry directions that were exactly determined from Fermi surface maps at
different temperatures. The tight-binding parameters were then found by solving an overdetermined system of 15 equations
relating the Fermi momenta and velocities of the model to the experimentally measured ones, so that the resulting
tight-binding model best reproduces both the Fermi surface contours and the experimental dispersion in the vicinity of
the Fermi level. Such fitting procedure has been applied to the Fermi surface maps of 2H-TaSe$_2$ independently at three
temperatures: 107\,K (in the incommensurate CDW state), 180\,K, and 290\,K (both in the normal state). The corresponding
tight-binding parameters are given in Table~\ref{Table:TaSe2TB}.

\hvFloat[floatPos=b, capWidth=1.0, capPos=b, capVPos=t, objectAngle=0]{table}
        {\begin{tabular}[c]{l|@{~~}c@{~~}c@{~~}c@{~~}c@{~~}c|@{~}c@{~~}c@{~~}c@{~~}c@{~~}c}
         \toprule
         &&\multicolumn{3}{c}{$\mathrm{\Gamma}$- and K-barrels} &&& \multicolumn{3}{c}{M-barrels
         (``dogbones'')}&\\
         $T$\,(K)& $t_0^\text{a}$ & $t_1^\text{a}$ & $t_2^\text{a}$ & $t_3^\text{a}$ & $t_4^\text{a}$ & $t_0^\text{b}$ & $t_1^\text{b}$ & $t_2^\text{b}$ & $t_3^\text{b}$ & $t_4^\text{b}$\\
         \midrule
         290 & --0.027 & 0.199 & 0.221 & 0.028 & 0.013~ & ~0.407 & \phantom{--}0.114 & 0.444 & --0.033 & 0.011\\
         180 & --0.051 & 0.172 & 0.248 & 0.005 & 0.011~ & ~0.355 & --0.015 & 0.406 & --0.069 & 0.013\\
         107 & --0.064 & 0.167 & 0.211 & 0.005 & 0.003~ & ~0.369 & \phantom{--}0.074 & 0.425 & --0.049 & 0.018\\
         \bottomrule
         \end{tabular}\quad
        }
        {Experimental tight-binding parameters of 2H-TaSe$_2$ independently determined for three different temperatures. All values are given in eV.}
        {Table:TaSe2TB}

We then calculated the Lindhard functions at $\omega\rightarrow0$ as
\begin{multline}\label{Eq:Susceptibility}
\chi_\mathbf{q}=\sum_\mathbf{k}\,\frac{n_\text{F}(\epsilon^\text{a}_\mathbf{k})-n_\text{F}(\epsilon^\text{a}_{\mathbf{k}+\mathbf{q}})}{\epsilon^\text{a}_\mathbf{k}-\epsilon^\text{a}_{\mathbf{k}+\mathbf{q}}}
+\sum_\mathbf{k}\,\frac{n_\text{F}(\epsilon^\text{a}_\mathbf{k})-n_\text{F}(\epsilon^\text{b}_{\mathbf{k}+\mathbf{q}})}{\epsilon^\text{a}_\mathbf{k}-\epsilon^\text{b}_{\mathbf{k}+\mathbf{q}}}\\
+\sum_\mathbf{k}\,\frac{n_\text{F}(\epsilon^\text{b}_\mathbf{k})-n_\text{F}(\epsilon^\text{a}_{\mathbf{k}+\mathbf{q}})}{\epsilon^\text{b}_\mathbf{k}-\epsilon^\text{a}_{\mathbf{k}+\mathbf{q}}}
+\sum_\mathbf{k}\,\frac{n_\text{F}(\epsilon^\text{b}_\mathbf{k})-n_\text{F}(\epsilon^\text{b}_{\mathbf{k}+\mathbf{q}})}{\epsilon^\text{b}_\mathbf{k}-\epsilon^\text{b}_{\mathbf{k}+\mathbf{q}}}\text{,}
\end{multline}
where indices a and b indicate the two bands forming the $\mathrm{\Gamma}$- and K-centered hole barrels and M-centered
electron ``dogbones'' respectively. The results of the calculation are shown in Fig.\,\ref{Fig:NestingTaSe2}\,(a). The
sharp peak seen near the $\frac{2}{3}\kern.3pt\mathrm{\Gamma M}$ wave vector (red arrows) is a clear evidence of
nesting. Surprisingly, it does not exactly coincide with the CDW vector, but appears at
$\sim$\,0.58\,--\,0.60\,$\mathrm{\Gamma M}$. The same calculation performed for the tight-binding model of
Rossnagel~\textit{et al.}\,\cite{RossnagelRotenberg05}, as shown in Fig.\,\ref{Fig:NestingTaSe2}\,(c), yields the same
pattern of somewhat weaker peaks at remarkably similar positions. As will be shown later, similar incommensurate nesting
peak appears to be universal between different transition metal dichalcogenides.

The temperature behavior of the nesting vector observed in TaSe$_2$ agrees with our previous observations
\cite{BorisenkoKordyuk08}. Upon lowering the temperature towards the incommensurate CDW transition, the nesting vector
moves away from the commensurate position, which means that the system feels the instability and starts to avoid it
already above the transition. In the incommensurate state, the nesting peak seems to be slightly driven in the opposite
direction upon cooling, which finally drives the commensurate transition at 90\,K.

As also seen from Fig.\,\ref{Fig:NestingTaSe2}, the absolute intensity of the dominant nesting peak slightly decreases
with temperature due to natural temperature broadening, which finally leads to the phase transition as soon as the
instability criterion (\ref{Eq:InstabilityCDW}) is satisfied. This natural scenario is confirmed by the observation of a
Kohn-like anomaly in the $\mathrm{\Sigma}_1$ phonon branch already at 300\,K, which softens even more as the transition
is approached \cite{Gruener94}. Such a mutual response can signify a strong electron-phonon interaction in 2H-TaSe$_2$.
It is interesting that at first the system does not develop a static commensurate CDW order. Instead, it opens up a
pseudogap \cite{BorisenkoKordyuk08} and falls into an incommensurate CDW state, which shifts the nesting vector closer
to the commensurate position [see the 107\,K curve in Fig.\,\ref{Fig:NestingTaSe2}\,(a)] preserving its strength. This
new nesting peak in the incommensurate state may finally drive the commensurate CDW transition at lower temperatures.

We note here that the effect of temperature on the absolute value of the susceptibility may be even higher due to the
renormalization effects, as the self-energy is usually temperature-dependent. Therefore many-body effects, which we
neglect in our calculations, may lead to additional temperature broadening of the spectral function and consequently of
the susceptibility.

Our results are at variance with the previous observations \cite{WithersWilson86, JohannesMazin06,
RossnagelRotenberg05}, which have found susceptibility for 2H polytypes to take a broadly humped form without strong
signatures of nesting, and with the earlier band structure calculations \cite{SmithKevan85}, which fail to reproduce the
Fermi surface topology and therefore its nesting properties.

It is worth mentioning that in our previous work \cite{BorisenkoKordyuk08} the calculation of the Lindhard function as
described above was replaced by the autocorrelation of the Fermi surface maps, which is easier to calculate. This
procedure is partially justified, as the peaks that are present in the Lindhard function are also to be found in the
autocorrelation. One has to be careful, however, as the latter may additionally include many ``false'' peaks that do not
represent relevant nesting vectors [compare two curves in the bottom panel of Fig.\,\ref{Fig:NestingTaSe2}\,(a)]. The
rigorous calculation of the bare susceptibility should be therefore preferred whenever allowed by the computational
capability.

\subsection{Universality of the nesting vector.}

\afterpage{
\begin{table}[t]
        \begin{tabular}[c]{l|@{~~}c@{~~}c@{~~}c@{~~}c@{~~}c|@{~}c@{~~}c@{~~}c@{~~}c@{~~}c}
         \toprule
         &&\multicolumn{3}{c}{inner barrels} &&& \multicolumn{3}{c}{outer barrels}&\\
         & $t_0^\text{a}$ & $t_1^\text{a}$ & $t_2^\text{a}$ & $t_3^\text{a}$ & $t_4^\text{a}$ & $t_0^\text{b}$ & $t_1^\text{b}$ & $t_2^\text{b}$ & $t_3^\text{b}$ & $t_4^\text{b}$\\
         \midrule
         2H-NbSe$_2$ & 0.000 & 0.082 & 0.167 & 0.044 & 0.016~ & ~0.173 & 0.101 & 0.227 & 0.037 & --0.005\\
         2H-Cu$_{0.2}$NbS$_2$ & --0.029 & 0.191 & 0.235 & 0.108 & 0.000~ & ~0.011 & 0.196 & 0.230 & 0.098 & \phantom{--}0.000 \\
         \bottomrule
         \end{tabular}
         \caption{Experimental tight-binding parameters of 2H-NbSe$_2$ and 2H-Cu$_{0.2}$NbS$_2$. All values are given in eV.}
         \label{Table:OtherTB}
\end{table}

\hvFloat[floatPos=b, capWidth=1, capPos=b, capVPos=t, objectAngle=0]{figure}
        {\includegraphics[width=\textwidth]{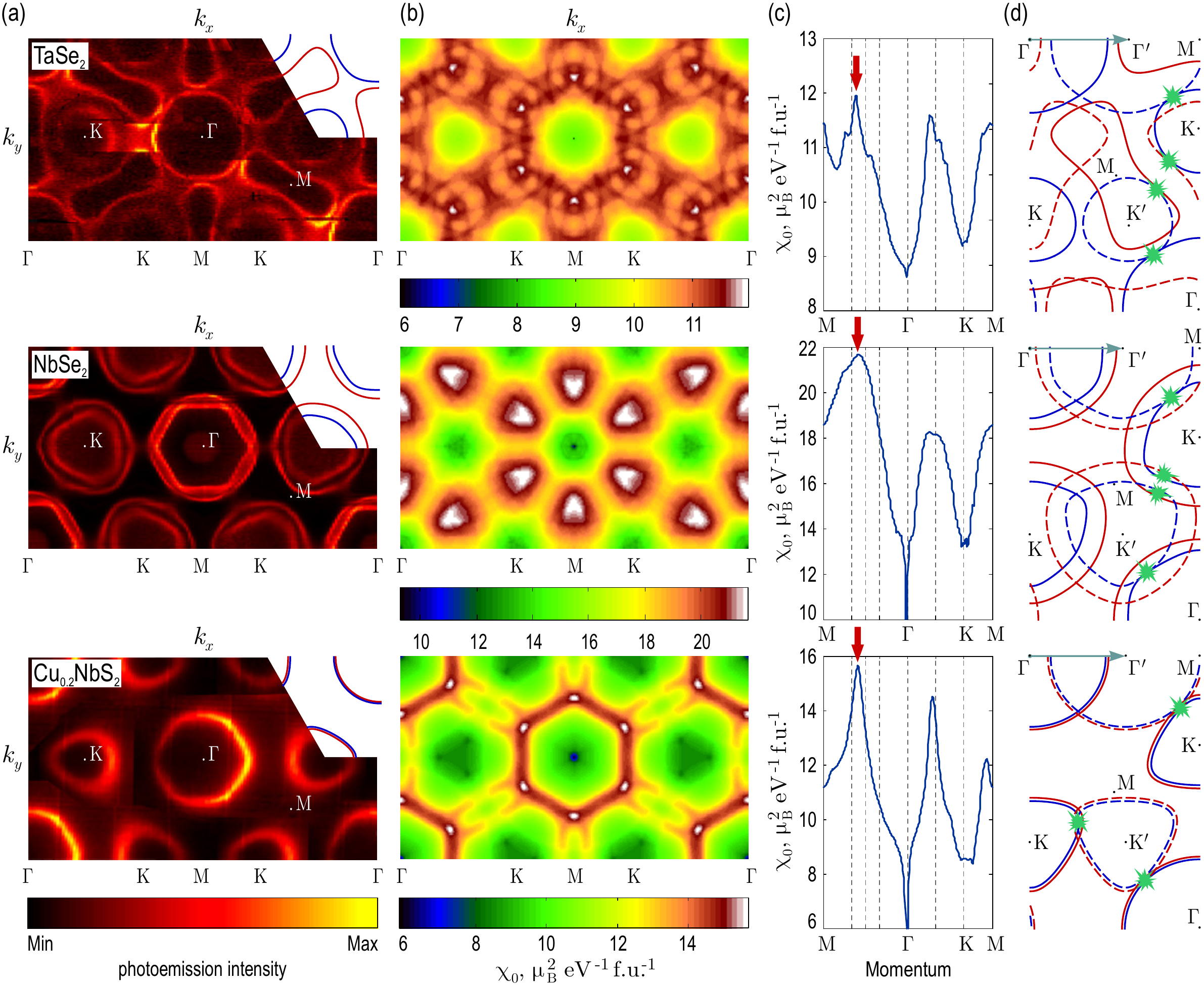}}
        {Nesting properties of three transition metal dichalcogenides: 2H-TaSe$_2$ (top row), 2H-NbSe$_2$ (middle row), and
        2H-Cu$_{0.2}$NbS$_2$ (bottom row). \textbf{(a)}~Experimental normal-state Fermi surfaces as seen by ARPES. The maps
        are symmetrized so that they fully cover the rectangular unitary cell in the momentum space. \textbf{(b)}~Real
        part of the Lindhard function at $\omega\rightarrow0$ as a function of momentum. \textbf{(c)}~Corresponding
        profiles along high-symmetry directions, with the dominant nesting vector marked by the red arrow. The same
        vectors can be seen in panel (b) as white spots. \textbf{(d)}~Fermi surface contours (solid lines) are shown
        together with their replicas shifted by the nesting vector (dashed lines) to demonstrate the nesting geometry. The
        nesting vector is shown on top of each image. The parts of the Fermi surface that give most contribution to the
        susceptibility are marked by green ``sparks''.}{Fig:Nesting3in1}
\clearpage}

For a comparison with TaSe$_2$, we have chosen two other transition metal dichalcogenides of the same 2H polytype,
namely NbSe$_2$ (niobium diselenide) and Cu$_{0.2}$NbS$_2$ (copper intercalated niobium disulphide). Both compounds
possess different Fermi surface topology, as in contrast to 2H-TaSe$_2$, both saddle points along the $\mathrm{\Gamma}$K
line are located below the Fermi level, therefore the Fermi surface consists of double hole barrels around the
$\mathrm{\Gamma}$ and K points, rather than single hole barrels and single electron ``dogbones'' as in case of TaSe$_2$.
The experimental Fermi surfaces of all three compounds are presented in Fig.\,\ref{Fig:Nesting3in1}\,(a), measured at
180\,K, 20\,K, and 30\,K respectively. Both TaSe$_2$ and NbSe$_2$ are half-filled band metals, while copper
intercalation in Cu$_{0.2}$NbS$_2$ is responsible for the 20\% electron doping. These doping levels are in good
agreement with Fermi surface areas determined from ARPES datasets.

The ordering temperatures of the three materials are also different. 2H-TaSe$_2$ experiences an incommensurate CDW phase
transition at 122\,K, commensurate CDW phase transition at 90\,K, and a superconducting transition at 0.2\,K
\cite{VanMaarenSchaeffer66}. In 2H-NbSe$_2$ incommensurate CDW transition happens at much lower temperature of 33.5\,K
\cite{WilsonDiSalvo75, MonctonAxe77, HarperGeballe75}, the commensurate CDW transition does not occur, whereas the
superconducting critical temperature is enhanced to 7.2\,K \cite{GabovichVoitenko01}. 2H-NbS$_2$ shows no CDW
transitions (neither incommensurate nor commensurate) \cite{NaitoTanaka82}, but to our knowledge it has not been studied
whether CDW ordering exists in Cu-intercalated samples around 20\% doping. Superconductivity is still present below
6.3\,K in pure NbS$_2$ \cite{VanMaarenSchaeffer66}, but is suppressed to 2.5\,K in Cu$_{0.2}$NbS$_2$.

One would expect the nesting properties in these three materials to be different, the absolute value of the
susceptibility at the dominant nesting vector being in direct correspondence to the CDW transition temperature.
Surprisingly, the results of our calculations shown in Fig.\,2\,(b) reveal a totally different picture. First, the
maximal value of the susceptibility is not correlated with the CDW transition temperature. It is the lowest among the
three compounds in TaSe$_2$, where $T_\text{CDW}$ is the highest, is higher in Cu$_{0.2}$NbS$_2$, where no CDW order has
been observed, and is maximal in NbSe$_2$, which has an intermediate transition temperature. This probably could be
explained by the difference in the phonon spectra and electron-phonon interaction in these materials or by the
differences in $k_z$-dispersion, which we neglect in our calculations. More surprising is that the dominant nesting
vector in both compounds coincides with that of TaSe$_2$ and is located at 0.60$\pm$0.05\,$\mathrm{\Gamma M}$ [see panel
(c) of the same figure]. Such coincidence can hardly be accidental, as we know that the nesting properties are extremely
sensitive to the Fermi surface geometry. In fact, a rigid band shift of the TaSe$_2$ band structure by 20\,meV is enough
to displace the nesting peak in the room temperature susceptibility from 0.6\,$\mathrm{\Gamma M}$ to the commensurate
position at $\frac{2}{3}\kern.3pt\mathrm{\Gamma M}$. Such shift corresponds to the electron doping of about 4\%, which
is much smaller than the 20\% doping of Cu$_{0.2}$NbS$_2$. Similarly, an arbitrary distortion of the Fermi surface in
any of these compounds was shown to destroy the universal nesting vector or shift it to a different position.

The tight binding parameters that were fitted to experimental ARPES datasets and used for susceptibility calculations
are given in Table~\ref{Table:OtherTB}. The corresponding Fermi surfaces for all three materials are shown in
Fig.\,\ref{Fig:Nesting3in1}\,(d) together with their replicas shifted by the nesting vector (dashed lines) to show the
different nesting geometry. In TaSe$_2$ the dominant contribution to the nesting peak comes from simultaneous tangency
of the M and K barrels, K and $\mathrm{\Gamma}$ barrels, and the K barrel with itself. In NbSe$_2$ the broad nesting
peak is a sum of several peaks that originate from the pairwise tangency of the two K-centered barrels with the
$\mathrm{\Gamma}$ barrels and with themselves, while in Cu$_{0.2}$NbS$_2$ the splitting between the two bands is too
small to be resolved, so they can be thought of as a single degenerate band that produces the sharp nesting peak by the
external contact of the K-centered barrel with the $\mathrm{\Gamma}$ barrel and with itself, as shown in the figure.
Such different nesting geometries make the coincidence of their nesting vectors even more puzzling.

\subsection{Conclusions.}

We have found that the Fermi surfaces of TaSe$_2$, NbSe$_2$, and Cu$_{0.2}$NbS$_2$ possess a strong nesting vector in
the vicinity of the CDW wave vector, which supports the Peierls instability scenario for the formation of CDW/PLD in
these materials. We observe however an offset of the nesting vector from the commensurate position that is persistent
over Fermi surfaces of all three materials and over different temperatures in TaSe$_2$, and note that the absolute
maximal value of the susceptibility does not correlate with the ordering temperature.

Here several questions can be posed. First, how to explain the inconsistency of the maximal susceptibility values with
the transition temperature. As one would suspect that the differences in the electron-phonon coupling might be a
possible answer, a comparative study of the phonon spectrum in these materials seems to be necessary. Second, why is the
nesting vector shifted from the commensurate position, if we know that even in the incommensurate CDW state the
incommensurability of the CDW wave vector does not exceed 2\% \cite{MonctonAxe77}? A small shift of the CDW wave vector
relative to the nesting vector is natural, if one recalls that in the parent compound the frequency of the phonon mode
is nonzero, which means that it will couple to the susceptibility at a finite energy of the phonon mode (which is of the
order of 10\,meV), rather than at the Fermi level. Evidently, the CDW does not necessarily form exactly at the same wave
vector at which the instability criterion (\ref{Eq:InstabilityCDW}) is first satisfied (a similar discrepancy in pure Cr
and its possible mechanisms are discussed for example in Ref.\,\citenum{FishmanJiang98}). In fact,
Eq.\,(\ref{Eq:InstabilityCDW}) provides the conditions at which the CDW transition becomes energetically favorable, but
does not specify the exact lattice configuration, at which the new energy minimum is reached. As the phonon frequency
softens towards zero, it will couple to the electronic susceptibility at different energies, which correspond to
slightly different positions of the nesting peak in momentum space. Moreover, the electronic system itself may react to
the ongoing transition, changing its nesting vector. Slight deviations of the CDW wave vector from the peak in
susceptibility would also be possible if the electron-phonon interaction is strongly momentum-dependent, which is
however an unlikely explanation, as the nesting peaks, at least in TaSe$_2$ and Cu$_{0.2}$NbS$_2$, are very sharp.

Finally, we note that the universality of the nesting vector among different compounds is in line with the neutron
scattering measurements performed on the same materials \cite{MonctonAxe77}, which show surprisingly identical
incommensurate wave vectors in 2H-TaSe$_2$ and 2H-NbSe$_2$. The fact that the nesting vector in Cu$_{0.2}$NbS$_2$ is the
same possibly means that Cu-intercalation enhances the CDW instability is this material, even though no CDW order is
observed in the pure compound. The presence of CDW order in Cu$_{0.2}$NbS$_2$ would not be surprising, being analogous
to the strong changes in the magnetic transition temperatures observed in NbS$_2$ upon intercalation by the first row
transition metals (Mn, Fe, Co, Ni) \cite{FriendBeal77, YamamuraaMoriyamaa04} to comparable doping levels. Alternatively,
our findings might suggest that chemical intercalation does not result in a simple rigid band shift of the bands, which
would immediately destroy the nesting vector, but leads to more complex changes in the dispersion that pin some parts of
the Fermi surface relevant for the nesting to their original position. A similar effect has been observed by
C.\,Battaglia \textit{et al.} \cite{BattagliaCercellier07}, who have shown that the $\mathrm{\Gamma}$-centered barrel in
Ni- and Mn-intercalated NbS$_2$ remains practically unaffected by the presence of intercalant species in violation of
the rigid band approximation. To distinguish between these two possibilities, more systematic studies of the nesting
properties as a function of doping and in other transition metal dichalcogenides might be helpful.


\section{Summary and outlook}

In this chapter, some important connections between different experimental techniques were drawn. In particular, we
have demonstrated the basic relationships between the ARPES, INS, and FT-STS data and have shown that the
single-particle Green's function extracted from ARPES can be useful in predicting physical properties of the solids.

Our comparison of ARPES and INS data from section \ref{Sec:BSCCO_RPA} supports the idea that the magnetic response
below $T_\textup{c}$ (or at least its major constituent) can be explained by the itinerant magnetism. Namely, the
itinerant component of $\chi$, at least near optimal doping, has enough intensity to account for the experimentally
observed magnetic resonance both in the acoustic and optic INS channels. The energy difference between the acoustic and
optic resonances seen in the experiments on both BSCCO and YBCO, cannot be explained purely by the difference in
$\chi_0$ between the two channels, but requires the out-of-plane exchange interaction to be additionally considered. In
this latter case the experimental intensity ratio of the two resonances agrees very well with our RPA results. Also the
calculated incommensurate resonance structure is similar to that observed in the INS experiment. Such quantitative
comparison becomes possible only if the many-body effects and bi-layer splitting are accurately accounted for. A
possible way to do that is to use the analytical expressions for the normal and anomalous Green's functions proposed in
this work. We point out that such method is universal and can be applied also to other systems with electronic
structure describable within the self-energy approach.

In section \ref{Sec:FT-STS}, we have sketched a way that could make the direct comparison between FT-STS and ARPES data
principally possible. One can approach this problem from two directions: (i) starting from the model of the Green's
function described in section \ref{Sec:ModelingGreensFunction}, calculate the autocorrelation functions given by
Eq.\,(\ref{Eq:Autocorrelation}), and compare them with FT-STS images; (ii) apply the phase retrieval algorithms to the
FT-STS data and attempt to restore the spectral function that is to be compared with the ARPES spectra. Both approaches
still meet more or less significant obstacles on their way to real application, as discussed above, but have been shown
to work in principle.

Finally, we have used the same relation between the single-particle spectral function measured in ARPES and the
two-particle response functions to investigate the nesting properties of the 2H-TaSe$_2$ Fermi surface. A rigorous
calculation that starts with a high-quality tight-binding model of the ARPES data confirms that conventional Fermi
surface nesting scenario is most probably responsible for the CDW/PLD instability in transition metal chalcogenides.
This seemingly irrelevant example was chosen to demonstrate universality of the approach: for a wide variety of
materials, ARPES can be a starting point for intriguing connections between different experiments and theories.

In outlook, it would be interesting to develop accurate tight-binding models and study the nesting properties of other
CDW-bearing materials. As soon as the quality of ARPES spectra on YBCO and LSCO cuprates allow, repeating the
susceptibility calculations for these materials is another challenge. One could attempt to apply similar techniques to
calculate the transport properties of the high-$T_\text{c}$ cuprates and transition metal chalcogenides, where the
resistivity and Hall coefficient show non-trivial behavior \cite{LeeGarzia70, NaitoTanaka82, LeBoeuf07}. The first
successful steps in this direction have already been made \cite{EvtushinskyKordyuk08}.


\appendix\chapter{Kramers-Kronig relations and transforms}\label{Appendix:KK}

The retarded Green's function, self-energy, Lindhard function, and correlation functions introduced in chapter
\ref{Chap:Theory} are causal functions of $\omega$. Causality \cite{Dienstfrey01} is the property of any transfer function of a linear
time-invariant physical system complying with the cause-and-effect relationship. According to the
Paley-Wiener theorem \cite{DymMcKean72}, causality in the time domain is equivalent in frequency domain to the
analyticity in the upper half-plane together with the following integrability property:
\begin{equation}\label{Eq:HardyCondition}
\underset{\omega''>0\,}{\mathrm{sup}}\,\int_{-\infty}^{\infty}\bigl|f(\omega'+\mathrm{i}\omega'')\bigr|^2\,\mathrm{d}\omega'=
\int_{-\infty}^{\infty}\bigl|f(\omega')\bigr|^2\,\mathrm{d}\omega'<\infty\text{.}
\end{equation}
The functions satisfying these two conditions are called \textit{Hardy functions}. Any Hardy function can be obtained
as
the Fourier transform of a function supported on $(0,\infty)$, and vice versa. For any Hardy function
$f\text{(}\omega\text{)}$ the following Cauchy integral formula can be written:
\begin{equation}\label{Eq:CauchyS}
   \fint_{-\infty}^{\infty}\frac{f(\omega')\,\mathrm{d}\omega'}{\omega'-\omega}=\mathrm{i}\piup\,f\text{(}\omega\text{)}\text{.}
\end{equation}
By taking the real and imaginary parts of (\ref{Eq:CauchyS}) we obtain the \textit{Kramers-Kronig relations}, which
where mentioned in \S\ref{SubSec:SelfEnergy} for the self-energy in formulae (\ref{Eq:KK}):
\begin{subequations}\label{Eq:KKAppend}
\begin{equation}
   \mathrm{Re}\,f\text{(}\omega\text{)}=\phantom{-}\frac{1}{\piup}\,\fint_{-\infty}^{\infty}\frac{\mathrm{Im}\,f(\omega')\,\mathrm{d}\omega'}{\omega'-\omega}\text{;}
\end{equation}
\begin{equation}
   \mathrm{Im}\,f\text{(}\omega\text{)}=-\frac{1}{\piup}\,\fint_{-\infty}^{\infty}\frac{\mathrm{Re}\,f(\omega')\,\mathrm{d}\omega'}{\omega'-\omega}\text{.}
\end{equation}
\end{subequations}
These relations suggest that the real and imaginary parts of Hardy functions are not independent, but can be obtained
from one another. This procedure, called \textit{Kramers-Kronig transformation}, is important in processing ARPES
data,
as demonstrated in section \ref{Sec:DispersionSE}. Below an effective algorithm which was used to perform such
calculations will be given.

Two other useful properties can be derived from the analytical properties of Hardy functions. (i) The Hardy
integrability condition (\ref{Eq:HardyCondition}) implies that the function has to vanish at
$|\omega|\rightarrow\infty$. (ii) If in the time domain the causal function is real-valued, then in the frequency
domain
its real and imaginary parts are respectively even and odd. Incorporating these symmetries into (\ref{Eq:KKAppend})
gives
\begin{subequations}\label{Eq:KKSym}
\begin{equation}
   \mathrm{Re}\,f\text{(}\omega\text{)}=\phantom{-}\frac{2}{\piup}\,\fint_{0}^{\infty}\frac{\omega'\,\mathrm{Im}\,f(\omega')\,\mathrm{d}\omega'}{\omega'\,^2-\omega^2}\text{;}
\end{equation}
\begin{equation}
   \mathrm{Im}\,f\text{(}\omega\text{)}=-\frac{2\omega}{\piup}\,\fint_{0}^{\infty}\frac{\mathrm{Re}\,f(\omega')\,\mathrm{d}\omega'}{\omega'\,^2-\omega^2}\text{.}
\end{equation}
\end{subequations}

In applications, the universal problem is that it is possible to take frequency measurements only within a bounded
interval of frequencies, at a finite number of points, and with limited accuracy. This imposes some restrictions upon
the practical use of Kramers-Kronig transformations. Even though the analytical continuation of a function given
within
a bounded interval is unique, the restriction of Hardy functions to any finite interval is dense in $L^2$ according to
the Riesz theorem \cite{Partington97}. This suggests that for any experimentally measured pair of functions
$\bigl\{f'\text{(}\omega\text{)},\,f''\text{(}\omega\text{)}\bigr\}$ their Kramers-Kronig consistency cannot be
checked
without imposing some additional constraints on their smoothness and behavior outside of the experimentally accessible
frequency window (`tail regions'). Neither is it possible to perform the Kramers-Kronig transformation of an
experimentally measured function without imposing such external constraints. In both cases a necessary condition is
that
the tail regions \textit{a priori} contain a negligibly small amount of their $L^2$-weight compared to that inside the
interval. Under this assumption the Kramers-Kronig transformations become well-defined and practically independent of
the shape of `tails' outside the interval.

\textbf{Kramers-Kronig transformations in Fourier domain.} As seen from (\ref{Eq:KKAppend}), the real part of a Hardy
function is the convolution of its imaginary part with $h\text{(}\omega\text{)}=-1/\piup\omega$, while the inverse
transform is equivalent to the convolution with $-h\text{(}\omega\text{)}$. Rewriting the convolution as a product in
Fourier domain, we get
\begin{subequations}\label{Eq:KKFourier}
\begin{equation}
   \mathrm{Re}\,f\text{(}\omega\text{)}=\mathcal{F}^{-1}\bigl\{-\mathrm{i}\kern1.5pt\mathrm{sign}\text{(}t\text{)}\,\mathcal{F}\bigl\{\mathrm{Im}\,f\text{(}\omega\text{)}\bigr\}\text{(}t\text{)}\bigr\}\text{(}\omega\text{)}\text{;}
\end{equation}
\begin{equation}
   \mathrm{Im}\,f\text{(}\omega\text{)}=\mathcal{F}^{-1}\bigl\{\mathrm{i}\kern1.5pt\mathrm{sign}\text{(}t\text{)}\,\mathcal{F}\bigl\{\mathrm{Re}\,f\text{(}\omega\text{)}\bigr\}\text{(}t\text{)}\bigr\}\text{(}\omega\text{)}\text{.}
\end{equation}
\end{subequations}
Here $\mathcal{F}$ is the Fourier transform defined as
$\mathcal{F}\bigl\{f\text{(}t\text{)}\bigr\}\text{(}\omega\text{)}=\int\kern-1.5pt\mathrm{d}t\,f\text{(}t\text{)}\,\mathrm{e}^{\mathrm{i}\omega
t}$.

In practice, either real or imaginary part of a causal, real-valued in time domain physical quantity is usually known
only within a limited interval $[-\mathit{\Omega}\kern.3pt_\text{max};~\mathit{\Omega}\kern.3pt_\text{max}]$, outside
of
which it can be safely neglected. Within this interval it is usually sampled in $N$ equally spaced points:
$f(\omega_n)$. Under these conditions equations (\ref{Eq:KKFourier}) can still be applied upon replacing the operators
$\mathcal{F}$ and $\mathcal{F}^{-1}$ by the discrete fast Fourier transform (FFT) and inverse fast Fourier transform
(IFFT) respectively:
\begin{subequations}\label{Eq:KKFFT}
\begin{equation}\label{Eq:KKFFTRe}
   \mathrm{Re}\,f(\omega_n)=-\mathrm{Re}~\mathrm{IFFT}\bigl\{-\mathrm{i}\kern1.5pt\mathrm{sign}(t_n)\,\text{FFT}\bigl\{\mathrm{Im}\,f(\omega_n)\bigr\}(t_n)\bigr\}\text{;}
\end{equation}
\begin{equation}
   \mathrm{Im}\,f(\omega_n)=-\mathrm{Re}~\mathrm{IFFT}\bigl\{\mathrm{i}\kern1.5pt\mathrm{sign}(t_n)\,\text{FFT}\bigl\{\mathrm{Re}\,f(\omega_n)\bigr\}(t_n)\bigr\}+C\text{.}
\end{equation}
\end{subequations}
Here $C\in\mathds{R}$ is an arbitrary constant, up to which the Kramers-Kronig transform is defined, so it has to be
chosen from some physical considerations (e.g. from the condition $\mathrm{Im}\,f(0)=0$). The algorithm
(\ref{Eq:KKFFT})
has the computational complexity of the order of $N\,\mathrm{ln}\,N$ in contrast to the straightforward calculation of
the convolution, which requires $N^2$ elementary operations and is therefore less effective.

One has to remember two important properties of the FFT transform that are essential. (i) The accuracy of the Fourier
(and therefore Kramers-Kronig) transformations does not depend on the number of points $N$, but only on the interval
size $\mathit{\Omega}\kern.3pt_\text{max}$, which means that even for small $N$ the values of the Kramers-Kronig
transform at every point $\omega_n$ will be as accurate as for large $N$. (ii) The discrete Fourier transform
performed
within a finite interval accurately represents the Fourier transform of the original function only far from the
interval's borders, therefore $\mathit{\Omega}\kern.3pt_\text{max}$ should be much larger than the maximal energy, up
to
which the Kramers-Kronig counterpart of the original function needs to be calculated. To illustrate this second
property
and to provide a rule of thumb for choosing an appropriate $\mathit{\Omega}\kern.3pt_\text{max}$ for a particular
problem, the following example will be considered.

\hvFloat[floatPos=t, capWidth=1.0, capPos=b, capVPos=t, objectAngle=0]{figure}
        {\includegraphics[width=0.9\textwidth]{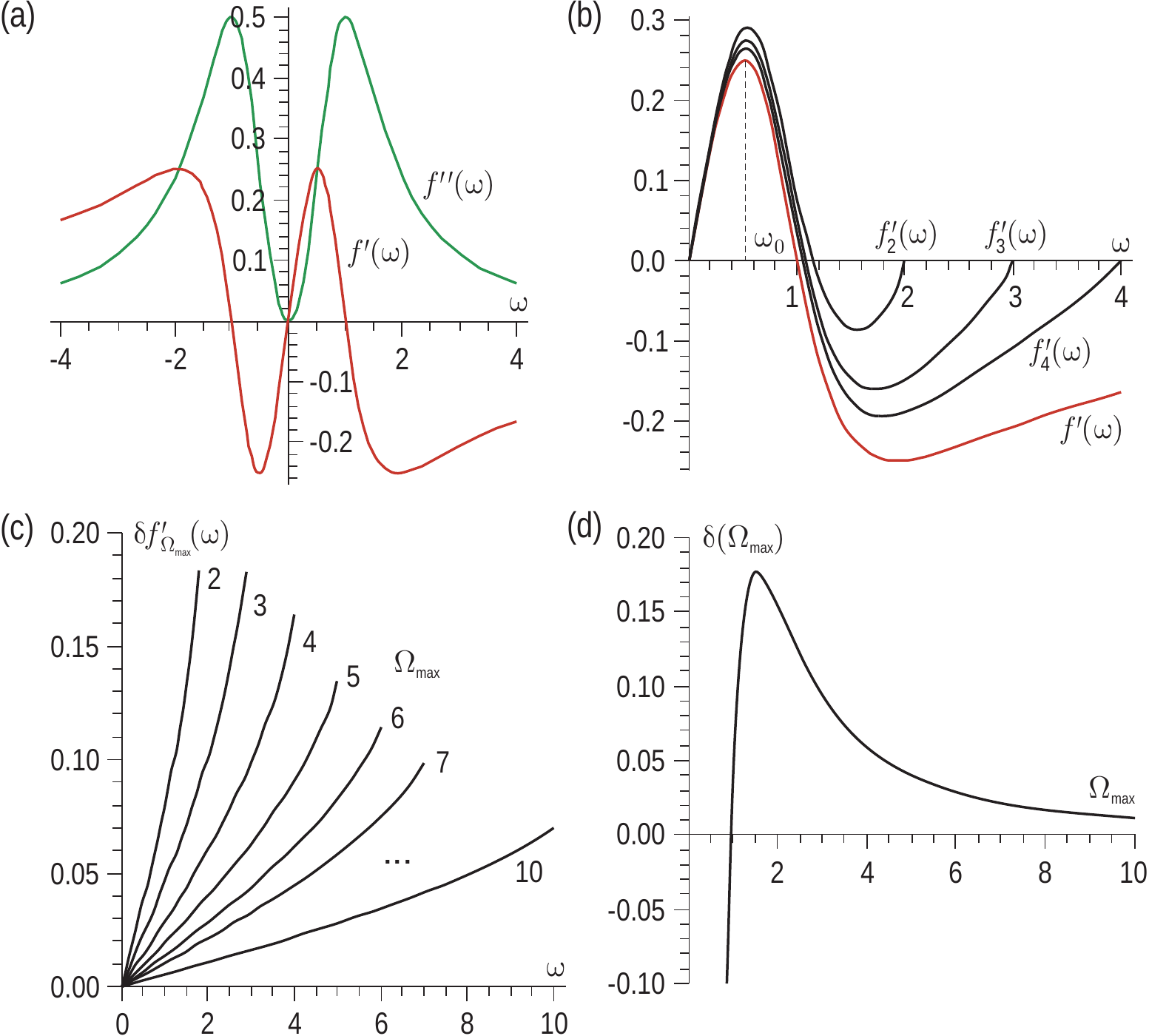}}
        {\textbf{(a)}~Real and imaginary parts of $f\text{(}\omega\text{)}$ (see example given in text).
        \textbf{(b)}~Comparison of the $f'_{\!\mathit{\Omega}\kern.3pt_\text{max}}\text{(}\omega\text{)}$ for different
        $\mathit{\Omega}\kern.3pt_\text{max}$ with $f'\text{(}\omega\text{)}$.
        \textbf{(c)}~$\delta\!f'_{\!\mathit{\Omega}\kern.3pt_\text{max}}\text{(}\omega\text{)}$ for different
        $\mathit{\Omega}\kern.3pt_\text{max}$, as indicated beside each curve. \textbf{(d)}~The relative deviation
        $\delta(\mathit{\Omega}\kern.3pt_\text{max})$ of the FFT-transformed function
        $f'_{\!\mathit{\Omega}\kern.3pt_\text{max}}\text{(}\omega\text{)}$ from $f'\text{(}\omega\text{)}$ as a
        function of $\mathit{\Omega}\kern.3pt_\text{max}$.} {Fig:AccuracyKK}

\textbf{Example.} Let us consider a simple rational function $f''\text{(}\omega\text{)}=\omega^2/(1+\omega^4)$,
similar
to the one used in \S\ref{SubSec:SelfEnergyModel} for modeling the real part of the self-energy. The Kramers-Kronig
transformation of this function can be calculated analytically to be
$f'\text{(}\omega\text{)}=\omega\,(1-\omega)/[\sqrt{2}\,(1+\omega^4)]$. It reaches its maximum at
$\omega_0=\sqrt{2-\sqrt{3}}$. Both functions are plotted in Fig.\,\ref{Fig:AccuracyKK}~(a). It can be easily checked
that
$f\text{(}\omega\text{)}=f'\text{(}\omega\text{)}+\mathrm{i}\,f''\text{(}\omega\text{)}=-x/[2\,\mathrm{i}\,x+\sqrt{2}\,(x^2-1)]$
is a Hardy function. Let us restrict the function $f''\text{(}\omega\text{)}$ to an interval
$[-\mathit{\Omega}\kern.3pt_\text{max};~\mathit{\Omega}\kern.3pt_\text{max}]$ and define
$f'_{\!\mathit{\Omega}\kern.3pt_\text{max}}\text{(}\omega\text{)}$ as its FFT-based Kramers-Kronig transformation
according to (\ref{Eq:KKFFTRe}). Fig.\,\ref{Fig:AccuracyKK}~(b) shows
$f'_{\!\mathit{\Omega}\kern.3pt_\text{max}}\text{(}\omega\text{)}$ in comparison with $f'\text{(}\omega\text{)}$ for
different values of $\mathit{\Omega}\kern.3pt_\text{max}$. As expected, the deviation of the two functions $\delta\!
f'_{\!\mathit{\Omega}\kern.3pt_\text{max}}\text{(}\omega\text{)}=f'_{\!\mathit{\Omega}\kern.3pt_\text{max}}\text{(}\omega\text{)}\,-f'\text{(}\omega\text{)}$
increases with $\omega$ and decreases with $\mathit{\Omega}\kern.3pt_\text{max}$, as shown in
Fig.\,\ref{Fig:AccuracyKK}~(c). As our usual region of interest is $\omega<\omega_0$, we can characterize the accuracy
of the Kramers-Kronig transformation by the value $\delta(\mathit{\Omega}\kern.3pt_\text{max})=\delta\!
f'_{\!\mathit{\Omega}\kern.3pt_\text{max}}(\omega_0)/f'_{\!\mathit{\Omega}\kern.3pt_\text{max}}(\omega_0)$, i.e. the
relative deviation of the two functions at their maximum, which is plotted in Fig.\,\ref{Fig:AccuracyKK}~(d) versus
$\mathit{\Omega}\kern.3pt_\text{max}$. From this plot we can derive the rule of thumb for choosing the size of the
interval needed for accurate numerical calculation of the Kramers-Kronig transformations: 5\% and 1\% accuracies at
frequencies $\omega<\omega_0$ are achieved with $\mathit{\Omega}\kern.3pt_\text{max} = 10\,\omega_0$ and
$\mathit{\Omega}\kern.3pt_\text{max} = 20\,\omega_0$ respectively. This result depends, of course, on the rate with
which the tails of the function decay at $|\omega|\rightarrow\infty$. For the function $\omega^2/(1+\omega^6)$,
decaying
like $1/\omega^4$, the 5\% accuracy is reached already for $\mathit{\Omega}\kern.3pt_\text{max} = 7\,\omega_0$.

\prefacesection{Publication List} \fancyhead[LO]{Publication List} \fancyhead[RE]{\quad}

\begin{list}{$\circ$}{\leftmargin 2.2ex}

\item{\underline{D.~S.~Inosov}, J.~Fink, A.~A.~Kordyuk, S.~V.~Borisenko, V.~B.~Zabolotnyy, R.~Schuster, M.~Knupfer,
B.~Büchner, R.~Follath, H.~A.~Dürr, W.~Eberhardt, V.~Hinkov, B.~Keimer, H.~Berger; \textit{Momentum and Energy
Dependence of the Anomalous High-Energy Dispersion in the Electronic Structure of High Temperature Superconductors},
\href{http://link.aps.org/abstract/PRL/v99/e237002}{Phys.~Rev.~Lett. \textbf{99}, 237002} (2007).}

\item{\underline{D.~S.~Inosov}, S.~V.~Borisenko, I.~Eremin, A.~A.~Kordyuk, V.~B.~Zabolotnyy, J.~Geck, A.~Koitzsch,
J.~Fink, M.~Knupfer, B.~Büchner; \textit{Relation between the one-particle spectral function and dynamic spin
susceptibility in superconducting Bi$_2$Sr$_2$CaCu$_2$O$_{8+\delta}$},
\href{http://link.aps.org/abstract/PRB/v75/e172505}{Phys.~Rev.~B \textbf{75}, 172505} (2007).}

\item{\underline{D.~S.~Inosov}, R.~ Schuster, A.~A.~Kordyuk, J.~Fink, S.~V.~Borisenko, V.~B.~Zabolotnyy,
D.~V.~Evtushinsky, M.~Knupfer, B.~Büchner, R.~Follath, H.~Berger; \/\textit{Excitation energy map of the high-energy
dispersion anomalies in cuprates}, \href{http://link.aps.org/abstract/PRB/v77/e212504}{Phys.~Rev.~B \textbf{77}, 212504}
(2008).}

\item{S.~V.~Borisenko, A.~A.~Kordyuk, A.~N.~Yaresko, V.~B.~Zabolotnyy, \underline{D.~S.~Inosov}, R.~Schuster,
B.~Büchner, R.~Weber, R.~Follath, L.~Patthey, H.~Berger; \/\textit{Pseudogap and charge density waves in two
dimensions}, \href{http://link.aps.org/abstract/PRL/v100/e196402}{Phys.~Rev.~Lett. \textbf{100}, 196402} (2008).}

\item{D.~V.~Evtushinsky, S.~V.~Borisenko, A.~A.~Kordyuk, V.~B.~Zabolotnyy, \underline{D.~S.~Inosov}, B.~Büchner,
H.~Berger, L.~Patthey, R.~Follath; \/\textit{Pseudogap-driven Hall effect sign reversal},
\href{http://link.aps.org/abstract/PRL/v100/e236402}{Phys.~Rev.~Lett. \textbf{100}, 236402} (2008).}

\item{A.~Grüneis, C.~Attaccalite, T.~Pichler, V.~Zabolotnyy, H.~Shiozawa, S.~L.~Molodtsov, \underline{D.~S.~Inosov},
A.~Koitzsch, M.~Knupfer, J.~Schiessling, R.~Follath, R.~Weber, P.~Rudolf, L.~Wirtz, A.~Rubio;
\/\textit{Electron-electron correlation in graphite},
\href{http://link.aps.org/abstract/PRL/v100/e037601}{Phys.~Rev.~Lett. \textbf{100}, 037601} (2008).}

\item{S.~Borisenko, A.~Kordyuk, V.~Zabolotnyy, J.~Geck, \underline{D.~S.~Inosov}, A.~Koitzsch, J.~Fink, M.~Knupfer,
B.~Büchner, V.~Hinkov, C.~T.~Lin, B.~Keimer, T.~Wolf, S.~G.~Chiuzbaian, L.~Patthey, R.~Follath; \textit{Kinks, nodal
bilayer splitting, interband scattering in YBa$_2$Cu$_3$O$_{6+x}$},
\href{http://link.aps.org/abstract/PRL/v96/e117004}{Phys.~Rev.~Lett. \textbf{96}, 117004} (2006).}

\item{A.~Koitzsch, S.~V.~Borisenko, \underline{D.~Inosov}, J.~Geck, V.~B.~Zabolotnyy, H.~Shiozawa, M.~Knupfer, J.~Fink,
B.~Büchner, E.~D.~Bauer, J.~L.~Sarrao, R.~Follath; \/\textit{Hybridization effects in CeCoIn$_5$ observed by
angle-resolved photoemission}, \href{http://link.aps.org/abstract/PRB/v77/e155128}{Phys.~Rev.~B \textbf{77}, 155128}
(2008).}

\item{V.~B.~Zabolotnyy, S.~V.~Borisenko, A.~A.~Kordyuk, \underline{D.~S.~Inosov}, J.~Geck, A.~Koitzsch, J.~Fink, M.~Knupfer,
B.~Büchner, S.-L.~Drechsler, V.~Hinkov, B.~Keimer, L.~Patthey; \textit{Disentangling surface and bulk photoemission
using circularly polarized light}, \href{http://link.aps.org/abstract/PRB/v76/e024502}{Phys.~Rev.~B \textbf{76}, 024502}
(2007).}

\item{V.~B.~Zabolotnyy, S.~V.~Borisenko, A.~A.~Kordyuk, J.~Geck, \underline{D.~S.~Inosov}, A.~Koitzsch, J.~Fink, M.~Knupfer,
B.~Büchner, S.-L.~Drechsler, L.~Patthey, V.~Hinkov, B.~Keimer; \textit{Momentum and temperature dependence of
renormalization effects in the high-temperature superconductor YBa$_2$Cu$_3$O$_{7-\delta}$},
\href{http://link.aps.org/abstract/PRB/v76/e064519}{Phys.~Rev.~B \textbf{76}, 064519} (2007).}

\item{A.~A.~Kordyuk, V.~B.~Zabolotnyy, \underline{D.~S.~Inosov}, S.~V.~Borisenko; \textit{From tunneling to
photoemission: correlating two spaces}, \href{http://dx.doi.org/10.1016/j.elspec.2007.02.017}{J.~Electr.
Spectr.~Rel.~Phen. \textbf{159}, 91} (2007)}.

\item{\underline{D.~S.~Inosov}, S.~V.~Borisenko, I.~Eremin, A.~A.~Kordyuk, V.~B.~Zabolotnyy, J.~Geck, A.~Koitzsch,
J.~Fink, M.~Knupfer, B.~Büchner; \textit{About the relation between the quasiparticle Green's function in cuprates
obtained from ARPES data and the magnetic susceptibility},
\href{http://www.sciencedirect.com/science?_ob=ArticleURL&_udi=B6TVJ-4NDDM21-N&_user=770391&_coverDate=09\%2F01\%2F2007&_alid=659958301&_rdoc=2&_fmt=summary&_orig=search&_cdi=5536&_sort=d&_docanchor=&view=c&_ct=5&_acct=C000042599&_version=1&_urlVersion=0&_userid=770391&md5=f89c5a3acd1294b30333613365a531a2}
{Physica~C \textbf{460\,--\,462}, 939} (2007)}

\item{V.~B.~Zabolotnyy, S.~V.~Borisenko, A.~A.~Kordyuk, J.~Geck, \underline{D.~S.~Inosov}, A.~Koitzsch, J.~Fink,
M.~Knupfer, B.~Büchner, V.~Hinkov, B.~Keimer, R.~Follath; \textit{Anomalous surface overdoping as a clue to the puzzling
electronic structure of YBCO-123},
\href{http://www.sciencedirect.com/science?_ob=ArticleURL&_udi=B6TVJ-4ND60KS-8&_user=770391&_coverDate=09\%2F01\%2F2007&_alid=659960553&_rdoc=1&_fmt=summary&_orig=search&_cdi=5536&_sort=d&_docanchor=&view=c&_ct=5&_acct=C000042599&_version=1&_urlVersion=0&_userid=770391&md5=2ec88f12f74e87053c3e8b201decc9a2}
{Physica~C \textbf{460\,--\,462}, 888} (2007)}

\item{A.~Koitzsch, S.~V.~Borisenko, \underline{D.~S.~Inosov}, J.~Geck, V.~B.~Zabolotnyy, H.~Shiozawa, M.~Knupfer,
J.~Fink, B.~Büchner, E.~D.~Bauer, J.~L.~Sarrao, R.~Follath; \textit{Observing the heavy fermions in CeCoIn$_5$ by
angle-resolved photoemission},
\href{http://www.sciencedirect.com/science?_ob=ArticleURL&_udi=B6TVJ-4NCSGPD-2&_user=770391&_coverDate=09\%2F01\%2F2007&_alid=659960553&_rdoc=3&_fmt=summary&_orig=search&_cdi=5536&_sort=d&_docanchor=&view=c&_ct=5&_acct=C000042599&_version=1&_urlVersion=0&_userid=770391&md5=1846f6d5dad1b34225ef7be948c70035}
{Physica~C \textbf{460\,--\,462}, 666} (2007)}

\item{\underline{D.~S.~Inosov}, S.~V.~Borisenko, V.~B.~Zabolotnyy, D.~V.~Evtushinsky, A.~A.~Kordyuk, B.~Büchner,
R.~Follath, H.~Berger; \/\textit{Fermi surface nesting in several transition metal dichalcogenides}, accepted to
New~J.~Phys., preprint available from \href{http://arxiv.org/pdf/0805.4105}{arXiv:0805.4105} (2008).}

\item{A.~A.~Kordyuk, S.~V.~Borisenko, V.~B.~Zabolotnyy, R.~Schuster, \underline{D.~S.~Inosov}, R.~Follath,
A.~Varykhalov, L.~Patthey, H.~Berger; \/\textit{Non-monotonic pseudo-gap in high-$T_\textup{c}$ cuprates}, submitted to
Phys.~Rev.~Lett., preprint available from \href{http://arxiv.org/pdf/0801.2546}{arXiv:0801.2546} (2007).}

\item{A.~Koitzsch, I.~Opahle, S.~Elgazzar, S.~V.~Borisenko, J.~Geck, V.~B.~Zabolotnyy, \underline{D.~Inosov}, H.~Shiozawa,
M.~Richter, M.~Knupfer, J.~Fink, B.~Büchner, E.~D.~Bauer, J.~L.~Sarrao, R.~Follath; \/\textit{The electronic structure
of CeCoIn$_5$ from angle-resolved photoemission spectroscopy}, submitted to Phys.~Rev.~B.}

\item{A.~Koitzsch, \underline{D.~Inosov}, J.~Fink, M.~Knupfer, H.~Eschrig, S.~V.~Borisenko, G.~Behr, A.~Kohler, J.~Werner,
B.~Buchner, R.~Follath, H.~A.~Dürr; \textit{Electronic structure of LaO$_{1-x}$F$_x$FeAs from photoemission
spectroscopy}, submitted to Phys.~Rev.~Lett., preprint available from
\href{http://arxiv.org/pdf/0806.0833}{arXiv:0806.0833} (2008).}

\end{list}

\bibliographystyle{phaip} \bibliography{PhD}

\prefacesection{Acknowledgements} \fancyhead[LO]{Acknowledgements} \fancyhead[RE]{\quad}

\noindent I would like to thank Prof. Bernd Büchner, my supervisor, and Dr. Sergey Borisenko, the leader of the ARPES
group at the Institute for Solid State Research, for the possibility to work at IFW Dresden and for the three fruitful
years of my life spent in a group of talented scientists from whom I had much to learn. I thank all my colleagues, and
especially Alexander Kordyuk, Martin Knupfer, Jörg Fink, Volodya Zabolotnyy, Roman Schuster, Daniil Evtushinsky,
Andreas Koitzsch, and Jochen Geck, for their many suggestions, helpful criticism, stimulating discussions, and constant
support during my studies.

I acknowledge helpful discussions with N.\,M.\,Plakida from the Max Planck Institute of Complex Systems, M.\,Lindroos
from the Tampere University of Technology, and my reader Ilya Yeremin from the Technical University Carolo-Wilhelmina
in Braunschweig.

Our experiments would not be possible without technical support of Roland Hübel and Ronny Schönfelder, whose
persistence in fighting the second law of thermodynamics in the labs is praiseworthy. I express my gratitude to the
synchrotron teams at BESSY and SLS, in particular to Rolf Follath, Luc Patthey, Gheorghe Chiuzb\u{a}ian, Ramona Weber,
Jorge Lobo, Fritz Dubi, and Christoph Hess. The high-quality single crystals were provided by Vladimir Hinkov, Bernhard
Keimer and Andreas Erb.

Special thanks to my wife Kate for bearing the difficult lot of a physicist's spouse and for her help in designing some
of the nicest figures for this dissertation.

Finally, I wish to thank my parents and friends in Kiev for their patience and friendly encouragement, and my German
friends in Dresden, in particular Julia Garten, for not letting me feel lonely in a foreign country.

My work in IFW Dresden is part of the Forschergruppe FOR538 and is supported by the DFG under Grants No. KN393/4.
The experimental data were acquired at the Berliner Elektronenspeicherring-Gesellschaft für Synchrotron Strahlung
m.b.H. (BESSY) and at the Swiss Light Source (SLS).

This document was produced using MiK\TeX~v.2.7 typesetting system.

\end{document}